\pdfoutput=1
\documentclass[journal]{IEEEtran} 
\IEEEoverridecommandlockouts

\usepackage{cite}
\usepackage{amsmath,amssymb,amsfonts}
\usepackage{graphicx}
\usepackage{textcomp}
\usepackage{xcolor, colortbl}
\usepackage{textpos} 
\usepackage[breaklinks=true]{hyperref} 
\def\UrlBreaks{\do\/\do-} 
\usepackage{etoolbox}
\appto\UrlBreaks{\do\-} 

\usepackage{float} 
\usepackage{pifont} 
\usepackage{amsmath} 
\usepackage{longtable} 
\usepackage{footnote} 
\usepackage{footmisc} 
\usepackage{threeparttable} 
\usepackage{gensymb} 
\usepackage{pgfplots} 
    \pgfplotsset{compat=1.17} 
\usepackage{pbox} 
\usepackage{stackengine} 
\usepackage{environ} 
\usepackage{makecell}
\usepackage{multirow} 
\usepackage{arydshln} 
\usepackage{hyperref} 
\usepackage[acronym]{glossaries}
\usepackage{glossary-longbooktabs}
\usepackage{glossary-mcols}
\usepackage{scalerel} 
\usepackage{tikz} 
\usetikzlibrary{svg.path} 
\usepackage{soul} 
\usepackage{xltabular}

\definecolor{orcidlogocol}{HTML}{A6CE39}
\tikzset{
  orcidlogo/.pic={
    \fill[orcidlogocol] svg{M256,128c0,70.7-57.3,128-128,128C57.3,256,0,198.7,0,128C0,57.3,57.3,0,128,0C198.7,0,256,57.3,256,128z};
    \fill[white] svg{M86.3,186.2H70.9V79.1h15.4v48.4V186.2z}
                 svg{M108.9,79.1h41.6c39.6,0,57,28.3,57,53.6c0,27.5-21.5,53.6-56.8,53.6h-41.8V79.1z M124.3,172.4h24.5c34.9,0,42.9-26.5,42.9-39.7c0-21.5-13.7-39.7-43.7-39.7h-23.7V172.4z}
                 svg{M88.7,56.8c0,5.5-4.5,10.1-10.1,10.1c-5.6,0-10.1-4.6-10.1-10.1c0-5.6,4.5-10.1,10.1-10.1C84.2,46.7,88.7,51.3,88.7,56.8z};
  }
}

\newcommand\orcidicon[1]{\href{https://orcid.org/#1}{\mbox{\scalerel*{
\begin{tikzpicture}[yscale=-1,transform shape]
\pic{orcidlogo};
\end{tikzpicture}
}{|}}}}

\definecolor{ForestGreen}{rgb}{.13,.545,.13}
\definecolor{Red}{rgb}{1,0,0}
\definecolor{Gray}{rgb}{.745,.745,.745}
\definecolor{Orange}{rgb}{1,.647,0}
\definecolor{Yellow}{rgb}{1,1,0}
\newcommand{\comment}[1]{} 

\hypersetup{
    colorlinks,
    linkcolor={black!90!black},
    citecolor={black!50!black},
    urlcolor={black!80!black}
}

\makeatletter
\def\thickhline{%
  \noalign{\ifnum0=`}\fi\hrule \@height \thickarrayrulewidth \futurelet
   \reserved@a\@xthickhline}
\def\@xthickhline{\ifx\reserved@a\thickhline
               \vskip\doublerulesep
               \vskip-\thickarrayrulewidth
             \fi
      \ifnum0=`{\fi}}
\makeatother

\newlength{\thickarrayrulewidth}
\iftrue
	\newcommand{\aygun}[1]{\textcolor{ForestGreen}{#1}} 
	\newcommand{\aygundel}[1]{\textcolor{red}{\st{#1}}} 
	\newcommand{\ergin}[1]{\textcolor{red}{Ergin: #1}} 
	\newcommand{\mustafa}[1]{\textcolor{blue}{Mustafa: #1}} 
	\newcommand{\abbasi}[1]{\textcolor{orange}{Abbasi: #1}} 
\else
	\newcommand{\aygun}[1]{}
	\newcommand{\aygundel}[1]{}
	\newcommand{\ergin}[1]{}
	\newcommand{\mustafa}[1]{}
	\newcommand{\abbasi}[1]{}
\fi

\makeatletter

\def\Autoref#1{%
  \begingroup
  \edef\reserved@a{\cpttrimspaces{#1}}%
  \ifcsndefTF{r@#1}{%
    \xaftercsname{\expandafter\testreftype\@fourthoffive}
      {r@\reserved@a}.\\{#1}%
  }{%
    \ref{#1}%
  }%
  \endgroup
}
\def\testreftype#1.#2\\#3{%
  \ifcsndefTF{#1autorefname}{%
    \def\reserved@a##1##2\@nil{%
      \uppercase{\def\ref@name{##1}}%
      \csn@edef{#1autorefname}{\ref@name##2}%
      \autoref{#3}%
    }%
    \reserved@a#1\@nil
  }{%
    \autoref{#3}%
  }%
}
\makeatother

\newcolumntype{e}{>{\raggedright}p{7.5cm}} 
\newcolumntype{f}{>{\raggedright}p{2.5cm}} 
\newcolumntype{g}{>{\raggedright}p{2cm}} 
\newcolumntype{h}{>{\raggedright}p{1.25cm}} 
\newcolumntype{j}{>{\raggedright}p{1.5cm}} 
\newcolumntype{Q}{>{\raggedright\arraybackslash}p{4.5cm}} 
\newcolumntype{S}{>{\raggedright\arraybackslash}p{2.85cm}} 
\newcolumntype{T}{>{\raggedright\arraybackslash}p{3.5cm}} 
\newcolumntype{t}{>{\raggedright\arraybackslash}p{3cm}} 
\newcolumntype{U}{>{\raggedright\arraybackslash}p{2.5cm}} 
\newcolumntype{V}{>{\raggedright\arraybackslash}p{2cm}} 
\newcolumntype{Y}{>{\raggedright\arraybackslash}p{1cm}} 
\newcolumntype{Z}{>{\raggedright\arraybackslash}p{0.5cm}} 

\newcommand\clearrow{\global\let\rowmac\relax} \clearrow 
\newcolumntype{C}[1]{>{\hsize=#1\hsize\rowmac\centering\arraybackslash}X} 
\newcolumntype{L}[1]{>{\hsize=#1\hsize\rowmac\raggedright\arraybackslash}X} 
\newcolumntype{R}[1]{>{\hsize=#1\hsize\rowmac\raggedleft\arraybackslash}X} 

\newglossarystyle{tabular}{%
     {\begin{tabular}{lp{\glsdescwidth}}}%
     {\bottomrule\end{tabular}}%
  \ifglsnogroupskip
  \else
  \fi
}

\makenoidxglossaries
\newacronym{3G}{3G}{3rd Generation Partnership Project} 
\newacronym{3GPP}{3GPP}{3rd Generation for Mobile Communication} 
\newacronym{5G}{5G}{5th Generation} 
\newacronym{6G}{6G}{6th Generation}
\newacronym{A2A}{A2A}{Air-to-air} 
\newacronym{A2G}{A2G}{Air-to-ground} 
\newacronym{AANET}{AANET}{Aeronautical Ad-hoc Networks}  
\newacronym{aBS}{aBS}{Aerial Base Station} 
\newacronym{ACAS}{ACAS}{Airborne Collision Avoidance System} 
\newacronym{ACARS}{ACARS}{Aircraft Communication Addressing and Reporting System} 
\newacronym{ADS-B}{ADS-B}{Automatic Dependent Surveillance Broadcasting}
\newacronym{AFDX}{AFDX}{Avionics Full-Duplex Ethernet}
\newacronym{AI}{AI}{Artificial Inteligence}
\newacronym{AeroMACS}{AeroMACS}{Airport Communication System} 
\newacronym{AMHS}{AMHS}{Air Traffic Services Message Handling System}
\newacronym{AMSS}{AMSS}{Aeronautical Mobile-Satellite Service}
\newacronym{AOC}{AOC}{Aeronautical Operation Control} 
\newacronym{AoI}{AoI}{Age of Information}
\newacronym{AP}{AP}{Access Point}
\newacronym{API}{API}{Application Programming Interface}
\newacronym{ARQ}{ARQ}{Automatic Repeat Request} 
\newacronym{ATSC}{ATSC}{Air Traffic Services Communication}  
\newacronym{ATM}{ATM}{Air Traffic Management} 
\newacronym{ATN}{ATN}{Aerial Telecommunication Network} 
\newacronym{AV}{AV}{Aerial Vehicle} 
\newacronym{BS}{BS}{Base Station} 
\newacronym{BGAN}{BGAN}{Broadband Global Area Network Service}  
\newacronym{BVLoS}{BVLoS}{Beyond Visual Line-of-sight}
\newacronym{C2}{C2}{Command and Control}
\newacronym{C-V2V}{C-V2V}{Cellular Vehicle-to-vehicle Communication}
\newacronym{CA}{CA}{Carrier Aggregation} 
\newacronym{CDMA}{CDMA}{Code Division Multiple Access}
\newacronym{CGC}{CGC}{Complementary Ground Component}
\newacronym{CNPC}{CNPC}{Control and Non-payload Communication} 
\newacronym{CoMP}{CoMP}{Coordinated Multipoint} 
\newacronym{CTOL}{CTOL}{Conventional Take-off and Landing} 
\newacronym{DA2GC}{DA2GC}{Direct Air-to-ground Communication}
\newacronym{DAA}{DAA}{Detect and Avoid} 
\newacronym{DAL}{DAL}{Development Assurance Level} 
\newacronym{E}{E}{Emulator}
\newacronym{EAN}{EAN}{European Aviation Network} 
\newacronym{EASA}{EASA}{European Union Aviation Safety Agency} 
\newacronym{ECC}{ECC}{Electronic Communications Committee} 
\newacronym{eMBB}{eMBB}{Enhanced Mobile Broadband}
\newacronym{EL}{EL}{Egress Link}
\newacronym{EPC}{EPC}{Evolved Packet Core}
\newacronym{EUROCAE}{EUROCAE}{European Organisation for Civil Aviation Equipment}
\newacronym{eVTOL}{eVTOL}{Electrical Vertical Take-off and Landing} 
\newacronym{FAA}{FAA}{US Federal Aviation Administration} 
\newacronym{FAO}{FAO}{Fully Autonomous Operation} 
\newacronym{FACOM}{FACOM}{Future Aerial Communications} 
\newacronym{FANET}{FANET}{Flying Ad-hoc Network}  
\newacronym{FMTP}{FMTP}{Flight Message Transfer Protocol} 
\newacronym{FlyBS}{FlyBS}{Flying Base Station} 
\newacronym{FPV}{FPV}{First-person View}
\newacronym{FSS}{FSS}{Fixed-Satellite Services}
\newacronym{G2G}{G2G}{Gate-to-gate} 
\newacronym{GEO}{GEO}{Geostationary Earth Orbit} 
\newacronym{GLSR}{GLSR}{Geographic Load Share Routine}
\newacronym{GMM}{GMM}{Gaussian Mixture Modeling}
\newacronym{GNN}{GNN}{Generative neural network}
\newacronym{GPS}{GPS}{Global Position System} 
\newacronym{GSM}{GSM}{Global System for Mobile} 
\newacronym{HAP}{HAP}{High Altitude Platform} 
\newacronym{HAPCom}{HAPCom}{High Altitude Platform Communications} 
\newacronym{HAPS}{HAPS}{High Altitude Platform System} 
\newacronym{HIL}{HIL}{Hardware-in-the-loop} 
\newacronym{HTS}{HTS}{High Throughput Satellite}  
\newacronym{ICAO}{ICAO}{International Civil Aviation Organization}
\newacronym{ICIC}{ICIC}{Inter-Cell Interference Coordination}
\newacronym{IEEE}{IEEE}{Institute of Electrical and Electronics Engineers} 
\newacronym{IETF}{IETF}{Internet Engineering Task Force} 
\newacronym{IFEC}{IFEC}{In-flight Entertainment and Communications} 
\newacronym{IMU}{IMU}{Inertial Measurement Unit}
\newacronym{IL}{IL}{Ingress Link} 
\newacronym{IoT}{IoT}{Internet of Things} 
\newacronym{IoFT}{IoFT}{Internet of Flying Things} 
\newacronym{IRS}{IRS}{Intelligent Reflecting Surface} 
\newacronym{ITU}{ITU}{International Telecommunication Union}
\newacronym{IP}{IP}{Internet Protocol} 
\newacronym{IPS}{IPS}{Internet Protocol Suit} 
\newacronym{IPv6}{IPv6}{Internet Protocol Version 6} 
\newacronym{ISL}{ISL}{Inter-satellite Link}
\newacronym{KNN}{KNN}{K-Nearest Neighbour}
\newacronym{KPI}{KPI}{Key Performance Indicator} 
\newacronym{LAA}{LAA}{License Assisted Access} 
\newacronym{LAP}{LAP}{Low Altitude Platform} 
\newacronym{LDACS}{LDACS}{L-band Digital Aeronautical Communications System} 
\newacronym{LEO}{LEO}{Low Earth Orbit} 
\newacronym{LIDAR}{LIDAR}{Light Detection And Ranging}
\newacronym{LiFi}{LiFi}{Light-Fidelity}
\newacronym{LMDS}{LMDS}{Local Multipoint Distribution System} 
\newacronym{LoRa}{LoRa}{Long Range} 
\newacronym{LoS}{LoS}{Line-of-sight} 
\newacronym{LTE}{LTE}{Long Term Evolution} 
\newacronym{MAC}{MAC}{Medium Access Control} 
\newacronym{MANET}{MANET}{Mobile Ad-hoc Network} 
\newacronym{MBB}{MBB}{Mobile Broadband} 
\newacronym{MEC}{MEC}{Mobile Edge Computing} 
\newacronym{MEO}{MEO}{Medium Earth Orbit} 
\newacronym{MIMO}{MIMO}{Multiple-input Multiple-output} 
\newacronym{ML}{ML}{Machine Learning} 
\newacronym{mMTC}{mMTC}{Massive Machine Type Communication} 
\newacronym{mmWave}{mmWave}{Milli-meter Wave} 
\newacronym{MNO}{MNO}{Mobile Network Operator} 
\newacronym{MSS}{MSS}{Mobile Satellite Service}
\newacronym{MPTCP}{MPTCP}{Multipath Transmission Control Protocol}
\newacronym{MTC}{MTC}{Machine Type Communications}
\newacronym{MTOM}{MTOM}{Maximum Take-Off Mass}
\newacronym{MU-MIMO}{MU-MIMO}{Multi-user Multiple-input Multiple-output}
\newacronym{NASA}{NASA}{National Aeronautics and Space Administration}
\newacronym{NB-IoT}{NB-IoT}{Narrow-band Internet of Things} 
\newacronym{NEMO}{NEMO}{Network Mobility} 
\newacronym{NF}{NF}{Network Function} 
\newacronym{NGMN}{NGMN}{Next Generation Mobile Networks}  
\newacronym{NOMA}{NOMA}{Non-orthogonal Medium Access}
\newacronym{NTN}{NTN}{Non-Terrestrial Networks}
\newacronym{OBP}{OBP}{On-Board Processing System} 
\newacronym{OG2}{OG2}{Orbcomm Generation 2} 
\newacronym{OFDM}{OFDM}{Orthogonal Frequency Division Multiplexing} 
\newacronym{OLDI}{OLDI}{On-line Data Interchange} 
\newacronym{QAM}{QAM}{Quadrature Amplitude Modulation} 
\newacronym{QoS}{QoS}{Quality of Service} 
\newacronym{PER}{PER}{Packet Error Ratio} 
\newacronym{RA}{RA}{Radio Altimeter}
\newacronym{RAS}{RAS}{Radio Astronomy Services}
\newacronym{RCO}{RCO}{Reduced Crew Operations} 
\newacronym{RL}{RL}{Reinforcement Learning} 
\newacronym{RF}{RF}{Radio Frequency}
\newacronym{RNN}{RNN}{Recurrent Neural Network}
\newacronym{RPO}{RPO}{Remote Piloting Operation} 
\newacronym{RPAS}{RPAS}{Remotely Piloted Aircraft System} 
\newacronym{RSRP}{RSRP}{Reference Signal Received Power} 
\newacronym{RSSI}{RSSI}{Received Signal Strength Indicator}
\newacronym{RTCA}{RTCA}{Radio Technical Commission for Aeronautics}
\newacronym{RTP}{RTP}{Real Time Protocol} 
\newacronym{RTT}{RTT}{Round-trip Time} 
\newacronym{S}{S}{Simulator} 
\newacronym{SA2GC}{SA2GC}{Satellite Air-to-ground Communication}  
\newacronym{SAGIN}{SAGIN}{Space-air-ground Integrated Networks} 
\newacronym{SatCom}{SatCom}{Satellite Communication} 
\newacronym{SAR}{SAR}{Search and Rescue} 
\newacronym{SDN}{SDN}{Software-defined Networking} 
\newacronym{SDR}{SDR}{Software-defined Radio}
\newacronym{SIL}{SIL}{Software-in-the-loop} 
\newacronym{SINR}{SINR}{Signal-to-interference-and-noise Ratio} 
\newacronym{SNR}{SNR}{Signal-to-noise Ratio} 
\newacronym{SIR}{SIR}{Signal-to-interference Ratio} 
\newacronym{SPP}{SPP}{Shortest Path Problem} 
\newacronym{SPO}{SPO}{Single Pilot Operation}
\newacronym{STDMA}{STDMA}{Spatial Time Division Multiple Access}
\newacronym{STOL}{STOL}{Short Takeoff and Landing} 
\newacronym{SWaP}{SWaP}{Size, Weight and Power}
\newacronym{SWIM}{SWIM}{System Wide Information Management} 
\newacronym{TBR}{TBR}{Trajectory-based Routing} 
\newacronym{TCP}{TCP}{Transmission Control Protocol} 
\newacronym{TDD}{TDD}{Time Division Duplexing} 
\newacronym{TELCO}{TELCO}{Telecommunication Operator} 
\newacronym{TP}{TP}{Trajectory Planning} 
\newacronym{TWT}{TWT}{Target Wake Time} 
\newacronym{U2I}{U2I}{Unmanned Aerial Vehicle-to-infrastructure} 
\newacronym{U2X}{U2X}{UAV-to-everything}
\newacronym{UAM}{UAM}{Urban Air Mobility}
\newacronym{UAS}{UAS}{Unmanned Aircraft System} 
\newacronym{UAV}{UAV}{Unmanned Aerial Vehicle} 
\newacronym{UDP}{UDP}{User Datagram Protocol} 
\newacronym{UE}{UE}{User Equipment} 
\newacronym{UHF}{UHF}{Ultra High Frequency} 
\newacronym{uRLLC}{uRLLC}{Ultra-reliable Low-latency Communication}
\newacronym{USS}{USS}{UAS Service Supplier}
\newacronym{UTM}{UTM}{Unmanned Traffic Management} 
\newacronym{UWB}{UWB}{Ultra-wideband}
\newacronym{V2V}{V2V}{Vehicle-to-vehicle} 
\newacronym{V2X}{V2X}{Vehicle-to-everything} 
\newacronym{VANET}{VANET}{Vehicular Ad-hoc Networks} 
\newacronym{VDL-2}{VDL-2}{VHF Datalink-2}
\newacronym{VHF}{VHF}{Very High Frequency} 
\newacronym{VLoS}{VLoS}{Visual Line-of-sight} 
\newacronym{VoIP}{VoIP}{Voice over IP} 
\newacronym{VR}{VR}{Virtual Reality} 
\newacronym{VTOL}{VTOL}{Vertical Take-off and Landing} 
\newacronym{WAIC}{WAIC}{Wireless Avionics Intra-communications} 
\newacronym{WiFi}{WiFi}{Wireless Fidelity} 
\newacronym{WiMAX}{WiMAX}{Worldwide Interoperability for Microwave Access} 
\newacronym{WLAN}{WLAN}{Wireless Local Access Network} 
\newacronym{WSN}{WSN}{Wireless Sensor Network}

\begin{document}
\title{A Survey of Wireless Networks for Future Aerial COMmunications (FACOM) \\
\thanks{A. Baltaci and D. Schupke are with Airbus, Central Research and Technology, Taufkirchen, 82024 Germany. A. Baltaci is also with Technical University of Munich, Department of Informatics, Garching, 85748, Germany. (email: ayguen.baltaci@airbus.com; dominic.schupke@airbus.com). \\
E. Dinc is currently with the Cavendish Laboratory, University of Cambridge, CB3 0HE, UK. During the time of this study, he was a postdoc at KTH Royal Institute of Technology, Stockholm 164 40, Sweden. (e-mail: ed502@cam.ac.uk).\\
M. Ozger and C. Cavdar are with the School of Electrical Engineering and Computer Science, KTH Royal Institute of Technology, Stockholm 164 40, Sweden. (email: ozger@kth.se; cavdar@kth.se).\\
A. Alabbasi is with Ericsson. (e-mail: abdulrahman.alabbasi@ericsson.com).
}
}

\author{Aygün Baltaci\textsuperscript{\orcidicon{0000-0003-0316-4038}\,}, Ergin Dinc\textsuperscript{\orcidicon{0000-0001-6982-206X}\,}, \IEEEmembership{Member, IEEE}, Mustafa Ozger\textsuperscript{\orcidicon{0000-0001-8517-7996}\,}, \IEEEmembership{Member, IEEE}, Abdulrahman Alabbasi\textsuperscript{\orcidicon{0000-0002-6614-5208}\,}, Cicek Cavdar\textsuperscript{\orcidicon{0000-0003-0525-4491}\,}, Dominic Schupke\textsuperscript{\orcidicon{0000-0001-6562-7363}\,}, \IEEEmembership{Senior Member, IEEE}
}

\maketitle 

\vspace{-0.25 cm}
\begin{abstract}
Electrification turned over a new leaf in aviation by introducing new types of aerial vehicles along with new means of transportation. Addressing a plethora of use cases, drones are gaining attention in the industry and increasingly appear in the sky. Emerging concepts of flying taxi enable passengers to be transported over several tens of kilometers. Therefore, unmanned traffic management systems are under development to cope with the complexity of future airspace, thereby resulting in unprecedented communication needs. Moreover, the long-term increase in the number of commercial airplanes pushes the limits of voice-oriented communications, and future options such as single-pilot operations demand robust connectivity. In this survey, we provide a comprehensive review and vision for enabling the connectivity applications of aerial vehicles utilizing current and future communication technologies. We begin by categorizing the connectivity use cases per aerial vehicle and analyzing their connectivity requirements. By reviewing more than 500 related studies, we aim for a comprehensive approach to cover wireless communication technologies, and provide an overview of recent findings from the literature toward the possibilities and challenges of employing the wireless communication standards. After analyzing the proposed network architectures, we list the open-source testbed platforms to facilitate future investigations by researchers. This study helped us observe that while numerous works focused on cellular technologies to enable connectivity for aerial platforms, a single wireless technology is not sufficient to meet the stringent connectivity demands of the aerial use cases, especially for the piloting operations. We identified the need of further investigations on multi-technology heterogeneous network architectures to enable robust and real-time connectivity in the sky. Future works should consider suitable technology combinations to develop unified aerial networks that can meet the diverse quality of service demands of the aerial use cases.
\end{abstract}

\section{Introduction}
\label{section: introduction}

Recent advances in the aviation industry toward electrified \glspl{AV} have led to carbon dioxide-friendly and cost-efficient concepts for aerial transportation. While \glspl{UAV} introduced a variety of aerial applications such as surveillance and disaster relief \cite{7463007}, emerging \gls{eVTOL} vehicles are making urban aerial passenger transportation into a reality. The concept of \textit{\gls{UAM}} comprises these looming operations of \glspl{eVTOL} and \glspl{UAV}. Furthermore, the increasing demands of international passenger travel push the limits of the current communication capacity of commercial airplanes \cite{plass12}. These concepts captivate business players, and hence, the airspace operations are undergoing an evolution to handle the coexistence of all the emerging aerial use cases. 

As the number and variety of \glspl{AV} increase, several challenges need to be addressed to ensure safe operations in the sky. The \glspl{AV} should coordinate with each other to share the airspace efficiently. In this regard, we observe the introduction of \gls{UTM} to bring digitalization and regulation of air traffic in the low-altitude airspace \cite{DLR_Uspace}. Besides, the safety-oriented nature of aviation imposes challenging certification requirements to ensure the airworthiness of \glspl{AV} \cite{icao_annex8}. As connectivity has a vital role in the operations of emerging \glspl{AV}, it emphasizes the demand for resilient connectivity hardware, robust connectivity links as well as the use of aviation safety spectrum in particular applications. 

\glspl{AV} have different vehicle and flight characteristics as well as diverse connectivity use cases inducing heterogeneity in \gls{QoS} requirements such as end-to-end data rate, latency and communication reliability. Although we can consider \gls{DA2GC} systems using existing terrestrial network infrastructures to provide connectivity for \glspl{UAV} at low altitudes, \gls{DA2GC} for airplanes at an altitude of approximately 10 km requires sophisticated architectures to provide coverage. Moreover, \glspl{RPO} of the \glspl{AV} requires real-time and very reliable connectivity to ensure safe operations, which is challenging for the state-of-the-art connectivity technologies. Most of the current \glspl{RPO} of \glspl{UAV} take place with the \gls{WiFi} technology due to hardware availability and cost-efficiency \cite{8746290}. However, the limited range of \gls{WiFi} prevents its use in \gls{BVLoS} applications. 

Emerging digital communication between airplanes and \gls{ATM} entities has introduced new \gls{IP}-based solutions instead of the conventional analog voice connectivity \cite{sita_costeffectiveip}.

\onecolumn
	\begin{table}[t]
		\begin{center}
		    \caption {List of Acronyms\vspace{-0.225cm}}
				\begin{tabular}{l l l l}
					\hline
					\rule{0pt}{1.1em}\textbf{Notation} & \textbf{Description} & \textbf{Notation} & \textbf{Description} \\ \hline
					\rule{0pt}{1.25em}\glsentryshort{3GPP} & \glsentrylong{3GPP} & \glsentryshort{MANET} & \glsentrylong{IoT} \\
					\glsentryshort{A2A} & \glsentrylong{A2A} & \glsentryshort{MEC} & \glsentrylong{IoT} \\
					\glsentryshort{A2G} & \glsentrylong{A2G} & \glsentryshort{MEO} & \glsentrylong{MEO} \\
					\glsentryshort{ACARS} & \glsentrylong{ACARS} & \glsentryshort{MIMO} & \glsentrylong{MIMO} \\
					\glsentryshort{ACAS} & \glsentrylong{ACAS} & \glsentryshort{mMTC} & \glsentrylong{mMTC} \\
					\glsentryshort{ADS-B} & \glsentrylong{ADS-B} & \glsentryshort{mmWave} & \glsentrylong{mmWave} \\
					\glsentryshort{AeroMACS} & \glsentrylong{AeroMACS} & \glsentryshort{MNO} & \glsentrylong{MNO} \\
					\glsentryshort{AI} & \glsentrylong{AI} & \glsentryshort{MPTCP} & \glsentrylong{MPTCP} \\
					\glsentryshort{AMHS} & \glsentrylong{AMHS} & \glsentryshort{MTC} & \glsentrylong{MTC} \\
					\glsentryshort{AMSS} & \glsentrylong{AMSS} & \glsentryshort{MTOM} & \glsentrylong{MTOM} \\
					\glsentryshort{ATM} & \glsentrylong{ATM} & \glsentryshort{MU-MIMO} & \glsentrylong{MU-MIMO} \\
					\glsentryshort{AV} & \glsentrylong{AV} & \glsentryshort{NASA} & \glsentrylong{NASA} \\
					\glsentryshort{BS} & \glsentrylong{BS} & \glsentryshort{NB-IoT} & \glsentrylong{NB-IoT} \\
					\glsentryshort{BVLoS} & \glsentrylong{BVLoS} & \glsentryshort{NGMN} & \glsentrylong{NGMN} \\
					\glsentryshort{C-V2V} & \glsentrylong{C-V2V} & \glsentryshort{OFDM} & \glsentrylong{OFDM} \\
					\glsentryshort{C2} & \glsentrylong{C2} & \glsentryshort{OLDI} & \glsentrylong{OLDI} \\
					\glsentryshort{CGC} & \glsentrylong{CGC} & \glsentryshort{PER} & \glsentrylong{PER} \\
					\glsentryshort{CNPC} & \glsentrylong{CNPC} & \glsentryshort{QAM} & \glsentrylong{QAM} \\
					\glsentryshort{CoMP} & \glsentrylong{CoMP} & \glsentryshort{QoS} & \glsentrylong{QoS} \\
					\glsentryshort{CTOL} & \glsentrylong{CTOL} & \glsentryshort{RAS} & \glsentrylong{RAS} \\
					\glsentryshort{DA2GC} & \glsentrylong{DA2GC} & \glsentryshort{RF} & \glsentrylong{RF} \\
					\glsentryshort{DAA} & \glsentrylong{DAA} & \glsentryshort{RL} & \glsentrylong{RL} \\
					\glsentryshort{DAL} & \glsentrylong{DAL} & \glsentryshort{RPAS} & \glsentrylong{RPAS} \\
					\glsentryshort{EAN} & \glsentrylong{EAN} & \glsentryshort{RPO} & \glsentrylong{RPO} \\
					\glsentryshort{EASA} & \glsentrylong{EASA} & \glsentryshort{RSRP} & \glsentrylong{RSRP} \\
					\glsentryshort{ECC} & \glsentrylong{ECC} & \glsentryshort{RTCA} & \glsentrylong{RTCA} \\
					\glsentryshort{EL} & \glsentrylong{EL} & \glsentryshort{RTP} & \glsentrylong{RTP} \\
					\glsentryshort{eMBB} & \glsentrylong{eMBB} & \glsentryshort{RTT} & \glsentrylong{RTT} \\
					\glsentryshort{EUROCAE} & \glsentrylong{EUROCAE} & \glsentryshort{SatCom} & \glsentrylong{SatCom} \\
					\glsentryshort{eVTOL} & \glsentrylong{eVTOL} & \glsentryshort{SDN} & \glsentrylong{SDN} \\
					\glsentryshort{FAA} & \glsentrylong{FAA} & \glsentryshort{SDR} & \glsentrylong{SDR} \\
					\glsentryshort{FACOM} & \glsentrylong{FACOM} & \glsentryshort{SINR} & \glsentrylong{SINR} \\
					\glsentryshort{FANET} & \glsentrylong{FANET} & \glsentryshort{SIR} & \glsentrylong{SIR} \\
					\glsentryshort{FAO} & \glsentrylong{FAO} & \glsentryshort{SNR} & \glsentrylong{SNR} \\
					\glsentryshort{FMTP} & \glsentrylong{FMTP} & \glsentryshort{SPO} & \glsentrylong{SPO} \\
					\glsentryshort{FSS} & \glsentrylong{FSS} & \glsentryshort{STDMA} & \glsentrylong{STDMA} \\
					\glsentryshort{GEO} & \glsentrylong{GEO} & \glsentryshort{SWaP} & \glsentrylong{SWaP} \\
					\glsentryshort{GPS} & \glsentrylong{GPS} & \glsentryshort{SWIM} & \glsentrylong{SWIM} \\
					\glsentryshort{HAP} & \glsentrylong{HAP} & \glsentryshort{TCP} & \glsentrylong{TCP} \\
					\glsentryshort{HAPCom} & \glsentrylong{HAPCom} & \glsentryshort{TDD} & \glsentrylong{TDD} \\
					\glsentryshort{HTS} & \glsentrylong{HTS} & \glsentryshort{TP} & \glsentrylong{TP} \\
					\glsentryshort{ICAO} & \glsentrylong{ICAO} & \glsentryshort{U2X} & \glsentrylong{U2X} \\
					\glsentryshort{ICIC} & \glsentrylong{ICIC} & \glsentryshort{UAM} & \glsentrylong{UAM} \\
					\glsentryshort{IETF} & \glsentrylong{IETF} & \glsentryshort{UAS} & \glsentrylong{UAS} \\
					\glsentryshort{IFEC} & \glsentrylong{IFEC} & \glsentryshort{UAV} & \glsentrylong{UAV} \\
					\glsentryshort{IL} & \glsentrylong{IL} & \glsentryshort{UDP} & \glsentrylong{UDP} \\
					\glsentryshort{IMU} & \glsentrylong{IMU} & \glsentryshort{UE} & \glsentrylong{UE} \\
					\glsentryshort{IoT} & \glsentrylong{IoT} & \glsentryshort{uRLLC} & \glsentrylong{uRLLC} \\
					\glsentryshort{IP} & \glsentrylong{IP} & \glsentryshort{UTM} & \glsentrylong{UTM} \\
					\glsentryshort{IRS} & \glsentrylong{IRS} & \glsentryshort{UWB} & \glsentrylong{UWB} \\
					\glsentryshort{ISL} & \glsentrylong{ISL} & \glsentryshort{V2X} & \glsentrylong{V2X} \\
					\glsentryshort{ITU} & \glsentrylong{ITU} & \glsentryshort{VDL-2} & \glsentrylong{VDL-2} \\
					\glsentryshort{LAA} & \glsentrylong{LAA} & \glsentryshort{VHF} & \glsentrylong{VHF} \\
					\glsentryshort{LDACS} & \glsentrylong{LDACS} & \glsentryshort{VLoS} & \glsentrylong{VLoS} \\
					\glsentryshort{LEO} & \glsentrylong{LEO} & \glsentryshort{VoIP} & \glsentrylong{VoIP} \\
					\glsentryshort{LIDAR} & \glsentrylong{LIDAR} & \glsentryshort{WAIC} & \glsentrylong{WAIC} \\
					\glsentryshort{LiFi} & \glsentrylong{LiFi} & \glsentryshort{WiFi} & \glsentrylong{WiFi} \\
					\glsentryshort{LoRa} & \glsentrylong{LoRa} & \glsentryshort{WiMAX} & \glsentrylong{WiMAX} \\
					\glsentryshort{LoS} & \glsentrylong{LoS} & \glsentryshort{WSN} & \glsentrylong{WSN} \\
					\glsentryshort{MAC} & \glsentrylong{MAC} &  &  \\
					\hline
				\end{tabular}
				\label{table: list_of_acronyms}
		\end{center}
	\end{table}
\twocolumn

\begin{figure*}[t]
\begin{center}
	\centerline{\includegraphics[width=0.75\textwidth,keepaspectratio]{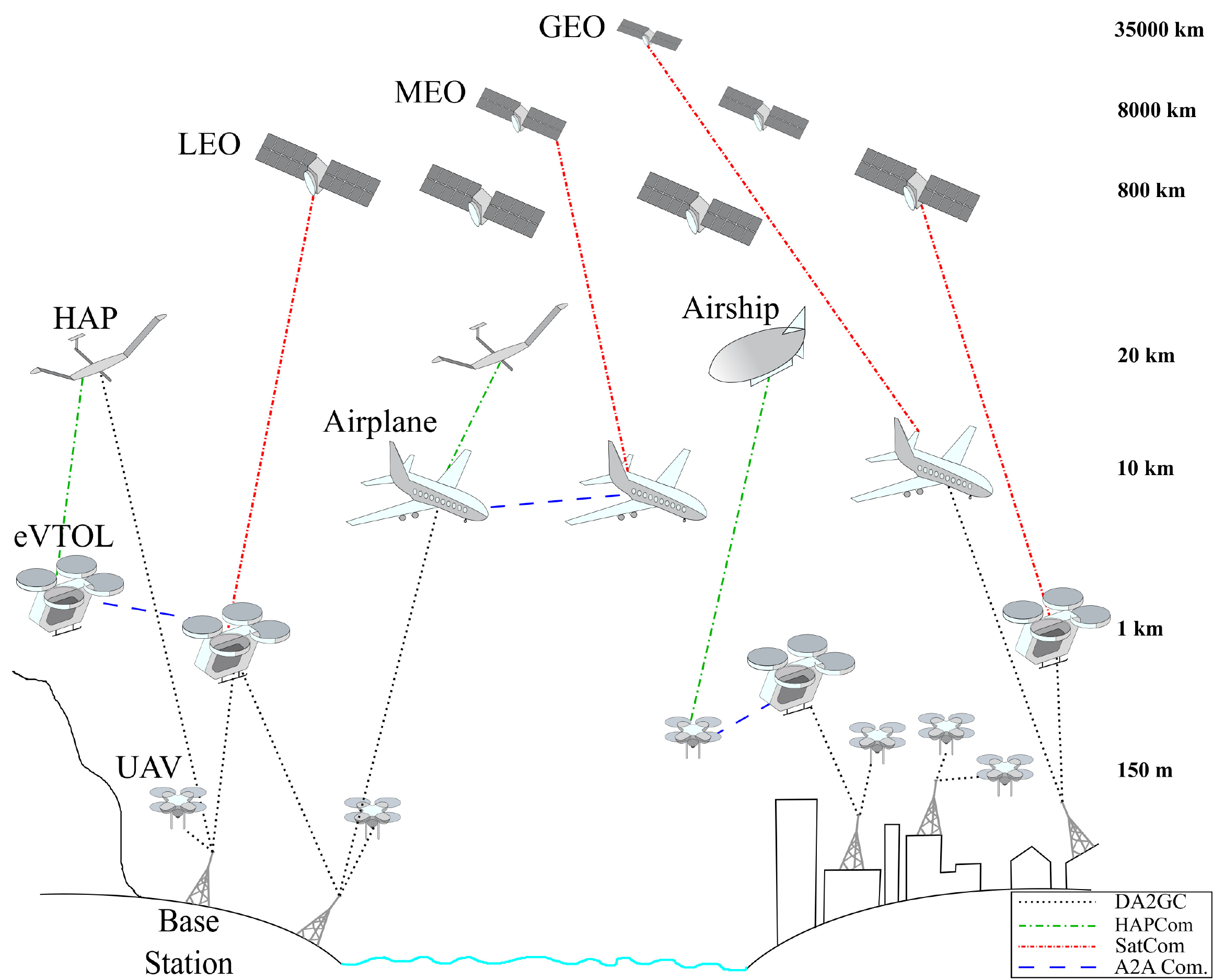}}
	\caption {Our vision of \gls{FACOM}, in which we foresee a mixture of heterogeneous connectivity. While \glsentryshort{DA2GC} technologies provide connectivity in urban areas, \glsentryshortpl{HAP} enable extended coverage along with \glsentryshort{A2A} in rural areas. \glsentryshort{A2A} also supports the collision avoidance systems. Utilizing \glsentryshort{LEO} for high-rate services, \glsentryshort{MEO} and \glsentryshort{GEO} constellations can provide global coverage. The airship, \glsentryshort{HAP}, and satellite objects can be interlinked and have feeder links with ground. These links, which may share resources with any of the shown links, are excluded for clarity.}  
	\label{fig: facom_connectivity}
\end{center}
\end{figure*}

Future \gls{SPO} of airplanes requires robust connectivity to enable cooperation between onboard and ground operators \cite{nasa_spo_conops}. All these applications of \glspl{AV} pose a diverse range of connectivity requirements. This challenge requires a holistic approach, considering all the connectivity technologies, to enable the diverse aerial use cases in the sky. We should evaluate the capabilities of different connectivity platforms such as \gls{DA2GC}, \gls{SatCom}, \gls{HAPCom} and \gls{A2A}, as shown in \autoref{fig: facom_connectivity}. Therefore, in our study, we introduce the term, \textit{\glsentryfull{FACOM}}, as the connectivity ecosystem that incorporates all these looming aerial connectivity use cases and their potential connectivity solutions. 

\gls{FACOM} services have significant demands from the industry due to emerging business markets. Hence, \gls{FACOM} can become a remarkable venture for communication operators, particularly with the increasing number of \glspl{UAV} in the near future. In this regard, connectivity technologies are under development to provide services for the \gls{FACOM} use cases. 

The \gls{3GPP} standardization body includes non-terrestrial networks as part of the future 5G architectures \cite{3GPP_36777, 3GPP_38811, 3GPP_22125, 3GPP_23754, 3GPP_23755, 3GPP_22829}, and embraces \gls{MTC} along with its conventional human-centric networks \cite{3GPP_36785, 3GPP_22261}. Meanwhile, emerging satellite networks pave the way for high capacity and robust communication on a global scale. \gls{LEO} constellations promote cost-efficient and real-time connectivity over satellites \cite{oneweb, McDowell_2020}. Moreover, future IEEE standards adopt more flexible and reliable architectures to widen its deployment in different application scenarios \cite{wifi6, 9090146, wifi_alliance_halow}. 

\begin{figure*}[t]
\begin{center}
	\centerline{\includegraphics[width=\textwidth,keepaspectratio]{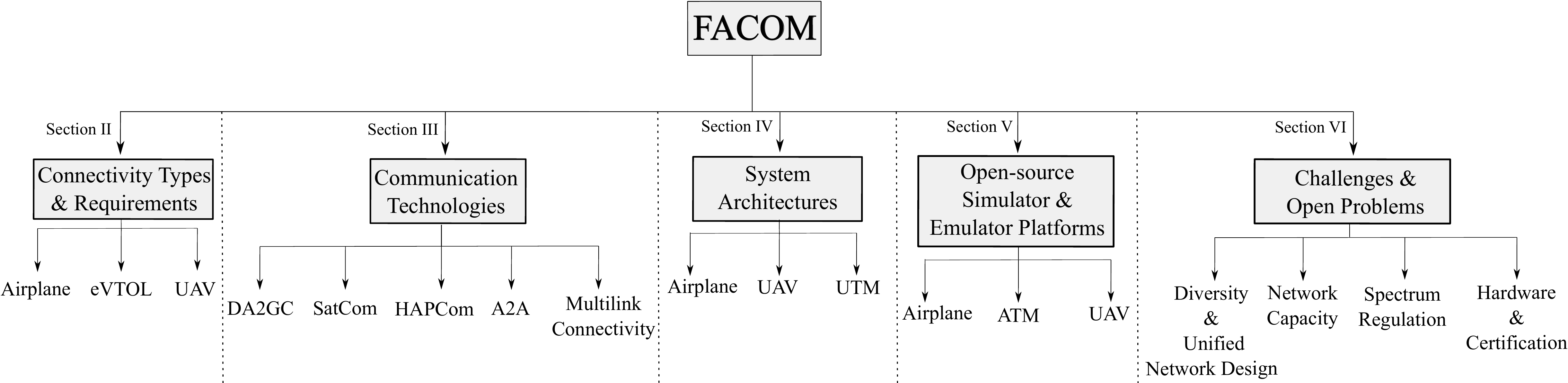}}
	\caption {Outline of the survey.}
	\label{fig: surveyoutline_seminar}
\end{center}
\end{figure*}

\begin{savenotes}
	\begin{table}[t]
		\caption {Ranges of Communication Reliability in this Study, based on \cite{5gpp_reliabilityranges} \vspace{-0.225cm}}
		\begin{center}
			{\renewcommand{\arraystretch}{1.5} 
				\begin{tabular}{l c}
					\hline
					\textbf{\makecell[l]{Communication\\Reliability}} & \textbf{Range} \\ \hline
					Low & $<99.9$\% \\
					Medium & $99.9$\% - $99.999$\% \\
					High & $>99.999$\%  \\ 
					\hline
				\end{tabular}}
				\label{table: reliability_range}
		\end{center}
	\end{table}
\end{savenotes}

We need to evaluate the capabilities of each technology to determine its feasibility for future applications in the sky. For instance, although \gls{GEO} constellations can provide wide coverage and reliable connectivity, they are unsuitable for latency-sensitive applications. Instead, upcoming \gls{LEO} constellations and \gls{HAPCom} promote real-time connectivity solutions, where the service interruption due to rapid handovers poses another challenge. In addition, coverage of \gls{DA2GC} is primarily limited to flights over the ground, and \gls{A2A} communication requires a certain extent of vehicle proximity to ensure end-to-end connectivity. Therefore, we analyze the advantages and drawbacks of each communication technology individually with respect to the \gls{QoS} requirements of the aerial use cases, and determine the suitable matches.  

In this paper, we aim to provide a comprehensive overview of the connectivity use cases of \glspl{AV}, their connectivity demands, and potential connectivity solutions to realize \gls{FACOM}. We focus on the \gls{RF} wireless technologies for non-military \glspl{AV}, excluding wired as well as optical connectivity solutions. We only cover the external communication of \glspl{AV}, leaving the onboard communication such as \gls{IFEC} and \gls{WAIC} outside the scope of this survey. In addition, we study the external communication that take place from take-off until landing. The types of communication while an \gls{AV} is on the ground, e.g., at the airport or under maintenance, are also outside the scope. 

We begin with categorizing the use cases per type of \glspl{AV}, and specify their connectivity demands and challenges from the aviation perspective. As concepts of flying taxis are of the recent venture, the literature has, to our best knowledge, not yet covered the connectivity needs and potential solutions for these platforms. Thus, we also provide our vision toward the future connectivity demands of \glspl{eVTOL}. Afterward, we investigate the literature to evaluate the capabilities of available and near-future communication technologies with respect to the connectivity demands of \gls{FACOM} use cases. We present a broad survey by analyzing more than 500 papers and discussing the solutions provided in the literature as well as open research questions. Based on the networking capabilities of each connectivity technology, we provide our match study presenting suitable technologies with respect to the \gls{QoS} demands of the use cases. After analyzing potential heterogeneous multi-connectivity solutions for robust remote piloting operations, we review the proposed \gls{FACOM} network and 5G system architectures in the literature. We also provide a comprehensive list of the open-source flight and networking testbed platforms that can facilitate future investigations toward the realization of \gls{FACOM}. Finally, we conclude by summarizing the key findings and open challenges for future studies. We illustrate the outline of this survey in \autoref{fig: surveyoutline_seminar}. 

The rest of this paper is organized as follows. In \autoref{section: usecases}, we provide a background on the \gls{FACOM} use cases and their connectivity requirements. Thereafter, we present a literature survey regarding the potentials and challenges of the current wireless technologies for \gls{FACOM} in \autoref{section: wirelesstechnologies}. We demonstrate the proposed network and 5G system architectures in \autoref{section: networkarchitectures}. We also provide a list of open-source flight and network testbeds in \autoref{section: simulators}. Finally, we discuss the open challenges in \autoref{sec: openresearchchallenges} and present the conclusions in \autoref{section: conclusion}. 

\subsection{Nomenclature and Definitions}
\label{subsection: nomenclature}

\begin{figure*}[t]
\begin{center}
	\centerline{\includegraphics[width=\textwidth,keepaspectratio]{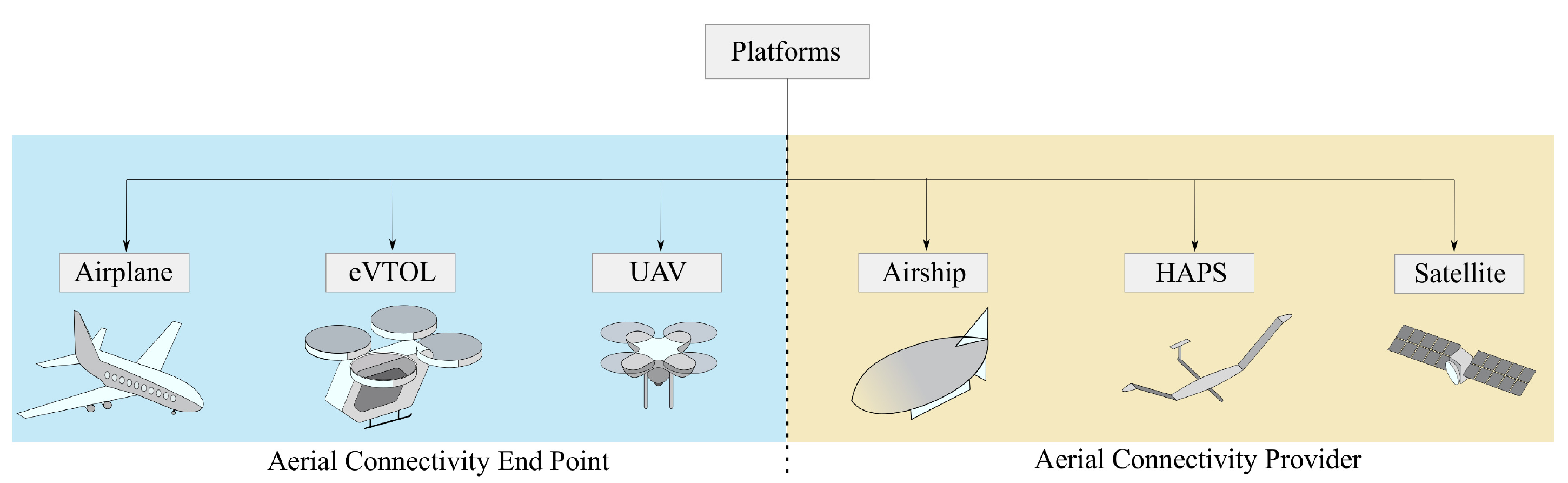}}
	\caption {Categorization of the aerial platforms with respect to their relationship to wireless connectivity in this study}
	\label{fig: aerialvehicles}
\end{center}
\end{figure*}

We categorize the aerial platforms based on their relation to the connectivity, as shown in \autoref{fig: aerialvehicles}. Airplanes and \glspl{UAV} can also play a role in providing connectivity \cite{9149163, dt_airplaneprovideconnectivity}. However, such concepts are beyond the scope of this study since they deserve an own survey. Consequently, we do not cover the studies related to aerial \gls{BS} in our literature review. Furthermore, \glspl{HAP} are also \glspl{UAV}; however, we treat them as a separate platform due to their distinct flight characteristics at very high altitudes. We mainly focus on the \glspl{AV} that demand connectivity, and we study the platforms of connectivity providers in \autoref{section: wirelesstechnologies}. We define the \glspl{AV} as follows: 

\begin{itemize}
\item \textbf{Airplane}: They are high altitude, fixed-wing, civil passenger, or cargo transportation platforms. 
\item \textbf{\gls{eVTOL}}: These \glspl{AV} are  electrical urban aircraft with vertical take-off and landing capabilities and designed for passenger transportation. We also include \gls{CTOL} vehicles such as helicopters in this category. 
\item \textbf{\gls{UAV}}: \glspl{UAV} are fixed or rotary-wing, unmanned-type of flying platforms, also known as \textit{drones}. 
\item \textbf{\gls{AV}}: It includes airplane, \gls{eVTOL} and \gls{UAV} under one common term. 
\end{itemize}


For \gls{FACOM}, we define connectivity directives as follows:

\begin{itemize}
\item \textbf{\gls{EL}}: It is the data transmission link in which the packets travel from an \gls{AV} to the ground. It also refers to the \textit{uplink} channel of conventional mobile networks. 
\item \textbf{\gls{IL}}: It is the data transmission link in the opposite direction of \gls{EL}, which also corresponds to the \textit{downlink} channel. 
\end{itemize}

We define \textit{communication reliability} as the overall probability of accomplishing an end-to-end message transmission within a predefined time constraint. From the communication perspective, it contains the following elements \cite{3GPP_22261}: 
\begin{itemize}
\item \textbf{Service Availability}: It is defined as the ratio of the amount of the time a network can provide an end-to-end connectivity service with the required \gls{QoS} to the total amount of time the network is expected to deliver service. Another definition could be the ability to perform a connectivity function at a required time instant within a required time interval \cite{ITU_G911}. 
\item \textbf{Packet Delivery}: It refers to the probability of successful end-to-end packet delivery in the network layer within a specified latency constraint. 
\end{itemize}

From the aviation perspective, the communication reliability includes the following elements \cite{easa_airtoground_evolution}: 
\begin{itemize}
\item \textbf{Continuity}: It is the probability of completing an operational communication transaction within the transaction time. 
\item \textbf{Integrity}: It is the probability of having undetected error(s) in a completed communication transaction. 
\end{itemize}

We evaluate the demands of the communication reliability for \gls{FACOM} use cases in \autoref{section: usecases}, and the references thereof consider only the service availability or packet delivery. However, all four elements have specific roles in providing the reliable and robust connectivity in \gls{FACOM} and thus, our definition of communication reliability comprises all these elements. Furthermore, similar to the approach in \cite{5gpp_reliabilityranges}, we categorize the ranges of communication reliability throughout this study as \textit{low, medium} and \textit{high}. As concrete reliability requirements for particular aerial applications are not yet defined, we state their requirements within the ranges specified in \autoref{table: reliability_range}. 

We also provide the list of the abbreviations in \autoref{table: list_of_acronyms} at the end of the paper.

\subsection{Related Works}
\label{subsection: relatedworks}

Recently, numerous studies have surveyed network-related topics for low-altitude \glspl{AV}, such as the networking challenges of \glspl{UAV} \cite{wirelesscomm_book, Vinogradov_2018, 7463007, 7317490, 8682048, 8660516}, feasibility of using cellular networks for \glspl{UAV} \cite{8579209, 8470897, 8675384}, and routing challenges in \glspl{FANET} \cite{8359865, 8741010, 9044378}. We contribute to the literature in several different and novel aspects as follows: 

\begin{itemize}
    \item We follow a top-down approach, specifying the applications stemming from airplanes, \glspl{eVTOL} and \glspl{UAV} and deriving the connectivity requirements for each application.
    \item We provide a vision of the connectivity use cases and demands of \glspl{eVTOL} for the future passenger transportation. Although \gls{UAV}-related studies are recently on the rise, we highlight the unique communication requirements of \glspl{eVTOL}, that are not yet prominent in the literature.
    \item We take a holistic approach to evaluate of the communication technologies for \glspl{AV}. We consider not only the \gls{DA2GC} technologies such as those based on cellular and IEEE, but also non-terrestrial systems such as satellites and \glspl{HAP}. We further provide a match-study by determining suitable technologies for each aerial application.
    \item In the literature review, we mainly focus on the recent studies from 2018 to 2021 to provide an up-to-date overview of the literature since research on \gls{FACOM} has significantly increased during this period.
    \item We study the heterogeneous and multi-link connectivity options for \gls{BVLoS} \gls{RPO} of \glspl{AV} since the pilots require stringent connectivity requirements, which are difficult to achieve using a single-link connectivity.
    \item We study the proposed network and 5G system architectures in the literature for \glspl{AV} and \gls{UTM}. We also provide a comprehensive list of open-source flight and networking platforms for future studies on \gls{FACOM}.
\end{itemize}

Among the existing surveys, we selected the ones that are most similar to our survey in context, and we present our novel aspects compared to them in \autoref{table: related_work}.

\begin{savenotes}
	\begin{table*}[t]
		\caption {Comparison of the Relevant Surveys with Our Study \vspace{-0.25cm}}
		\begin{center}
			{\renewcommand{\arraystretch}{1.5} 
				\begin{tabular}{l l c c c c c}
					\hline
					\multicolumn{2}{l}{\textbf{Topics/Surveys}} & \textbf{\gls{FACOM}} & \textbf{Liu \emph{et al.} \cite{8368236}} & \textbf{Cao \emph{et al.} \cite{8438489}} & \textbf{Zolanvari \emph{et al.} \cite{8935306}} & \textbf{Hayat \emph{et al.} \cite{7463007}} \\ \hline
					 \multirow{5}{*}{Platforms} & Satellite & \checkmark & \checkmark & \checkmark & \ding{55} & \ding{55} \\
					 & \glsentryshort{HAP} & \checkmark & \checkmark & \checkmark & \ding{55} & \ding{55} \\
					 & Airplane & \checkmark & \checkmark & \ding{55} & \ding{55} & \ding{55} \\
					 & \glsentryshort{eVTOL} & \checkmark & \ding{55} & \ding{55} & \ding{55} & \ding{55} \\
					 & \glsentryshort{UAV} & \checkmark & \checkmark & \checkmark & \checkmark & \checkmark \\ \hline
					 \multicolumn{2}{l}{Connectivity Requirements} & \checkmark & \ding{55} & \ding{55} & \ding{55} & \checkmark \\
					 \multicolumn{2}{l}{Connectivity use cases and Demands of \glspl{eVTOL}} & \checkmark & \ding{55} & \ding{55} & \ding{55} & \ding{55} \\
					 \multicolumn{2}{l}{\gls{UTM}} & \checkmark & \ding{55} & \ding{55} & \ding{55} & \ding{55} \\ \hline
					 \multirow{8}{*}{Wireless Networks} & Cellular & \checkmark & \checkmark & \checkmark & \checkmark & \checkmark \\
					 & IEEE & \checkmark & \ding{55} & \ding{55} & \checkmark & \checkmark \\
					 & \glsentryshort{SatCom} & \checkmark & \checkmark & \checkmark & \checkmark & \ding{55} \\
					 & \glsentryshort{HAPCom} & \checkmark & \ding{55} & \checkmark & \ding{55} & \ding{55} \\
					 & \glsentryshort{A2A} & \checkmark & \checkmark & \checkmark & \ding{55} & \checkmark \\
					 & \glsentryshort{LDACS} & \checkmark & \ding{55} & \ding{55} & \checkmark & \ding{55} \\
					 & \glsentryshort{AeroMACS} & \checkmark & \ding{55} & \ding{55} & \checkmark & \ding{55} \\
					 & VDL-2 & \checkmark & \ding{55} & \ding{55} & \checkmark & \ding{55} \\\hline
					 \multicolumn{2}{l}{Heterogeneous Connectivity} & \checkmark & \checkmark & \checkmark & \ding{55} & \ding{55} \\
					 \multicolumn{2}{l}{Network Design} & \checkmark & \ding{55} & \ding{55} & \ding{55} & \ding{55} \\
					 \multicolumn{2}{l}{Simulator/Emulator Platforms} & \checkmark & \ding{55} & \ding{55} & \ding{55} & \ding{55} \\
					 \multicolumn{2}{l}{Year} & 2021 & 2018 & 2018 & 2020 & 2016 \\
					\hline
				\end{tabular}}
				\label{table: related_work}
		\end{center}
	\end{table*}
\end{savenotes}

\section{Use Cases and Connectivity Requirements}
\label{section: usecases}

We discuss different connectivity use cases, stemming from each type of \glspl{AV}, and analyze their connectivity requirements in detail. Our findings include the demands of both \gls{EL} and \gls{IL}.

\subsection{Airplanes}
\label{subsec: usecases_airplanes}

\begin{figure}[t]
\begin{center}
	\centerline{\includegraphics[width=0.4\textwidth,keepaspectratio]{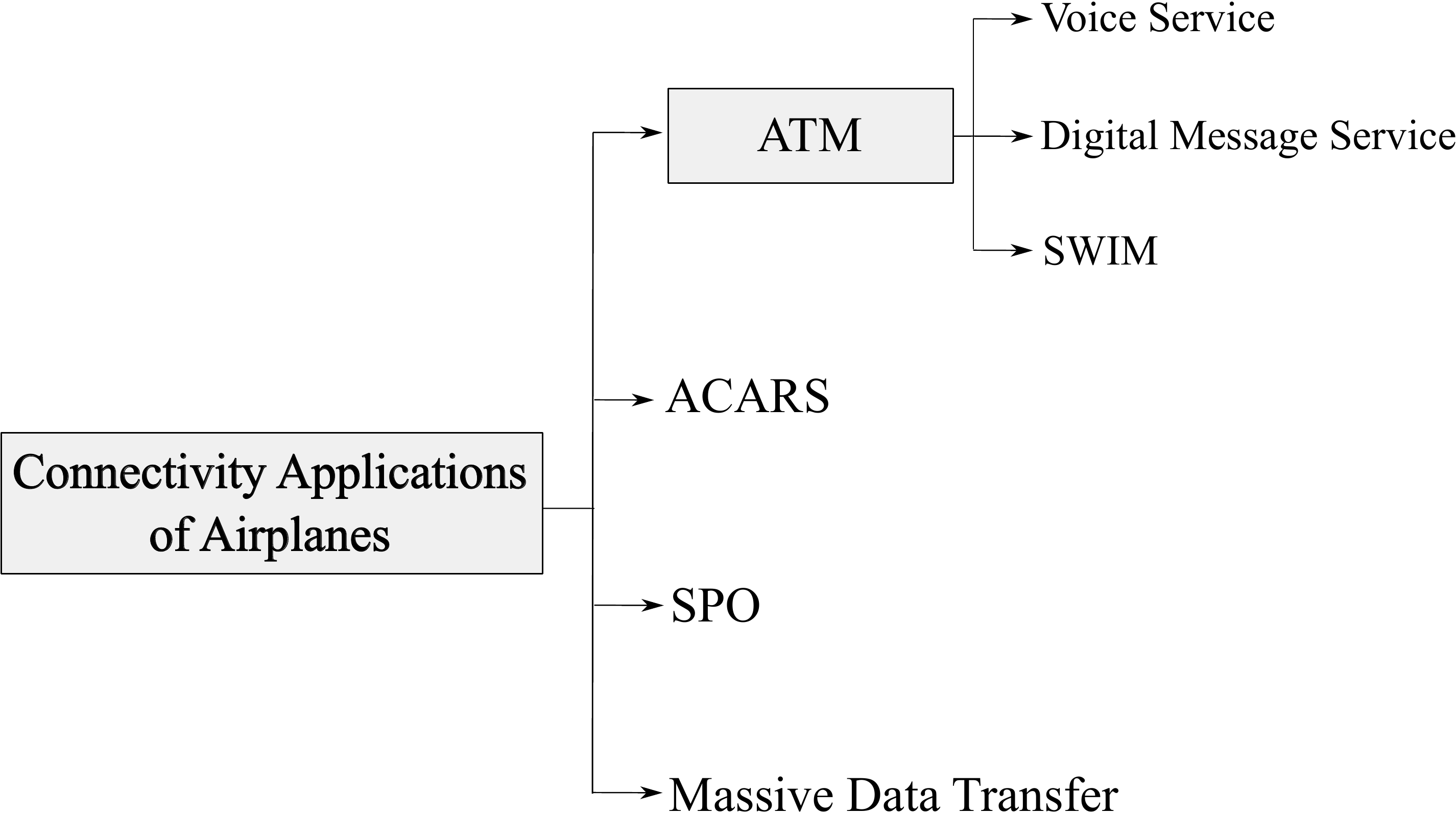}}
	\caption {Connectivity applications of airplanes, excluding onboard commmunications.}
	\label{fig: airplane_connectivitytypes}
\end{center}
\end{figure}

\begin{figure}[t]
\begin{center}
	\centerline{\includegraphics[width=0.5\textwidth,keepaspectratio]{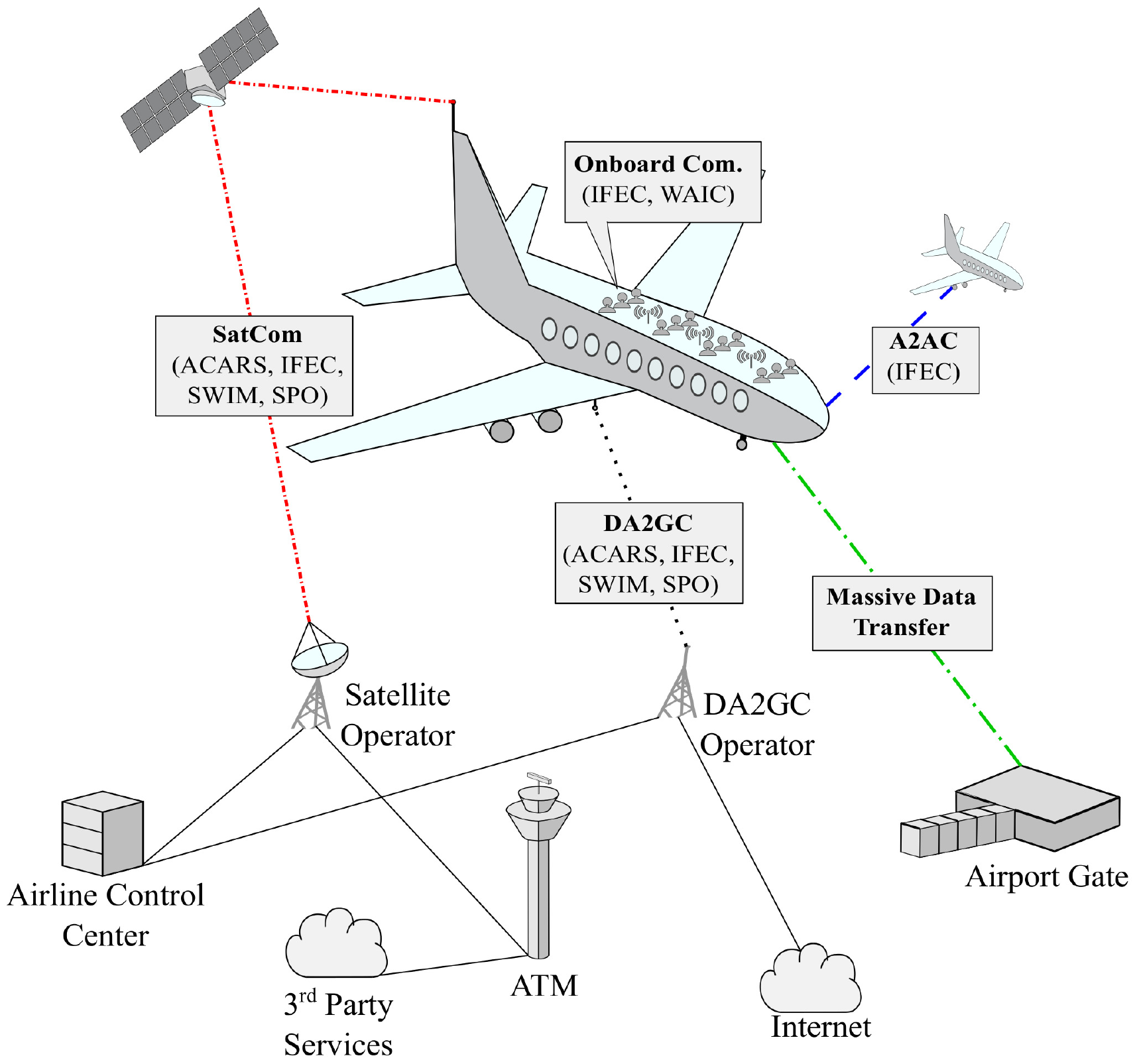}}
\caption {Overall connectivity architecture of an airplane. While \gls{SatCom} and \gls{DA2GC} can support various applications, onboard connectivity in an airplane mainly hosts \gls{IFEC} and \gls{WAIC} applications. \gls{A2A} communication can further support \gls{IFEC}, and massive data transfer takes place just before landing, after take off or when an airplane is at the airport. Airline control centers utilize \gls{ACARS} to fetch data from an airplane, and the \gls{ATM} participants intercommunicate using, e.g., \gls{SWIM}.}
	\label{fig: airplane_communication_illustration}
\end{center}
\end{figure}

\gls{FACOM} has an essential role in future airplane operations and passenger connectivity offerings. We categorize the connectivity use cases associated with airplanes in \autoref{fig: airplane_connectivitytypes}, and illustrate the overall connectivity architecture of an airplane in \autoref{fig: airplane_communication_illustration}. 

\subsubsection{ATM}
\label{subsec: usecases_airplanes_atm}

\gls{ATM} systems employed \gls{VHF} communication for more than 50 years, and voice communication are the primary use cases of \gls{VHF} systems \cite{5935265}. The state-of-the-art \textit{\gls{VDL-2}} enables \gls{A2G} data transfer up to 31.5 kbps \cite{helfrick07}. Aviation industry expected \gls{VHF} systems to saturate by 2020-2025 before the occurrence of coronavirus disease \cite{plass12}. The industry needs to consider alternative solutions to \gls{VHF} by the time the airplane operations come back to the routine.

Towards future \gls{ATM} solutions, national initiatives launched various projects such as European SESAR \cite{sesar_euatmmasterplan}, SANDRA \cite{plass12}, and American NextGen \cite{schnell04}. They design future \gls{ATM} systems to provide increased safety, 4D (i.e., latitude, longitude, altitude, and time \cite{sesar_4dtrajectory}) trajectory-based navigation, situational awareness, and the connectivity between the cockpit and \gls{ATM} \cite{mayr14,bauer11}. The new techniques can be backward compatible and utilize \gls{VDL-2} as a back-up connectivity \cite{sesar_euatmmasterplan}.

\gls{ICAO} published standards regarding aerial network protocols, and they define three different types of services for future \glspl{ATM} \cite{ICAO_9896}: 
\begin{enumerate}
    \item \textbf{\gls{OLDI}/\gls{FMTP}}: \gls{OLDI} is the messaging protocol for the communication between adjacent \glspl{ATM}, sharing the coordination and transfer of flight data. \gls{FMTP} is the \gls{TCP}/\gls{IP}v6 communication stack to support the \gls{OLDI}.
    \item \textbf{\gls{VoIP}}: It supports the voice services of \gls{ATM} over \gls{IP}.
    \item \textbf{\gls{AMHS}}: It provides generic messaging services via \gls{TCP}/\gls{IP}v4 (or \gls{IP}v6) over the aerial networks.
\end{enumerate}

Besides, \gls{SWIM} is a developing standard for information exchange such as flight, weather, or surveillance between \gls{ATM} service providers \cite{eurocontrol_swim}. Hence, it hosts \gls{A2G} as well as ground-to-ground infrastructure. \gls{SWIM} aims at a cost-efficient, flexible, and service-oriented architecture \cite{icao_swim}.

\gls{ICAO} published a global air navigation plan in 2016 to anticipate future developments in aviation for the next 15-year period \cite{icao_ganp}. It foresees the utilization of Internet-based protocols in \gls{SWIM} to maximize the interoperability, integration of airplanes, and artificial intelligence use in the \gls{SWIM} environment \cite{icao_ganp, icao_ganp_2019}. In this regard, \gls{AeroMACS} is a candidate technology to enable wireless connectivity between airplanes and the \gls{ATM} service providers at airports \cite{9155746}.

Future \gls{ATM} systems envision an all-\gls{IP} system instead of a specialized closed network \cite{rula16}, since an all-\gls{IP} solution is cost-effective for the aviation industry \cite{sita_costeffectiveip}. However, one of the main concerns is the specific mobility of airplanes. To this end, the authors of \cite{liu11} investigated the network mobility problems in terms of end-to-end latency, overhead, and network load. Although significant efforts attempt to provide improved safety and security to \gls{ATM} systems, high capacity and reliable connectivity during all phases of a flight are still in demand. Therefore, \gls{FACOM} can be a promising alternative to meet the requirements of \gls{ATM} with the emerging communication technologies and advanced network virtualization/slicing techniques. 

\subsubsection{\glsentryfull{ACARS}}
\label{subsec: usecases_airplanes_acars}
\gls{ACARS} is a data protocol developed to provide connectivity between airplanes and airline operation control centers along with different ground entities such as flight explorer, oceanic clearance, and government agencies \cite{acars_commsgroup, easa_airtoground_evolution}. Current operations support connectivity via both \gls{VHF} and satellites in the \gls{AMSS} bands 1.1610-1.6265 GHz and 5-5.15 GHz \cite{easa_airtoground_evolution, icao_handbook_freqspectrum}, and the message length is limited to 1960 bytes \cite{acars_commsgroup}. While \gls{EL} comprises request messages such as weather updates and clearance requests, \gls{IL} exchanges the requested information \cite{acars_commsgroup}. In addition, both links provide text messaging service \cite{acars_commsgroup}. 

\gls{ICAO} foresees the extension of \gls{ACARS} into \gls{VDL-2} and broadband satellite systems to meet the increasing aviation demand \cite{icao_ganp}. Furthermore, the \gls{IP}v6 protocol can be part of the future \gls{ACARS}; however, datalink security remains a major concern \cite{8735356}. Therefore, the study, \cite{9256659}, proposed a detection and authentication-based safety architecture.

\subsubsection{Single-Pilot Operations (SPOs)}
\label{subsec: usecases_airplanes_spo}

\begin{figure}[t]
\begin{center}
	\centerline{\includegraphics[width=0.5\textwidth,keepaspectratio]{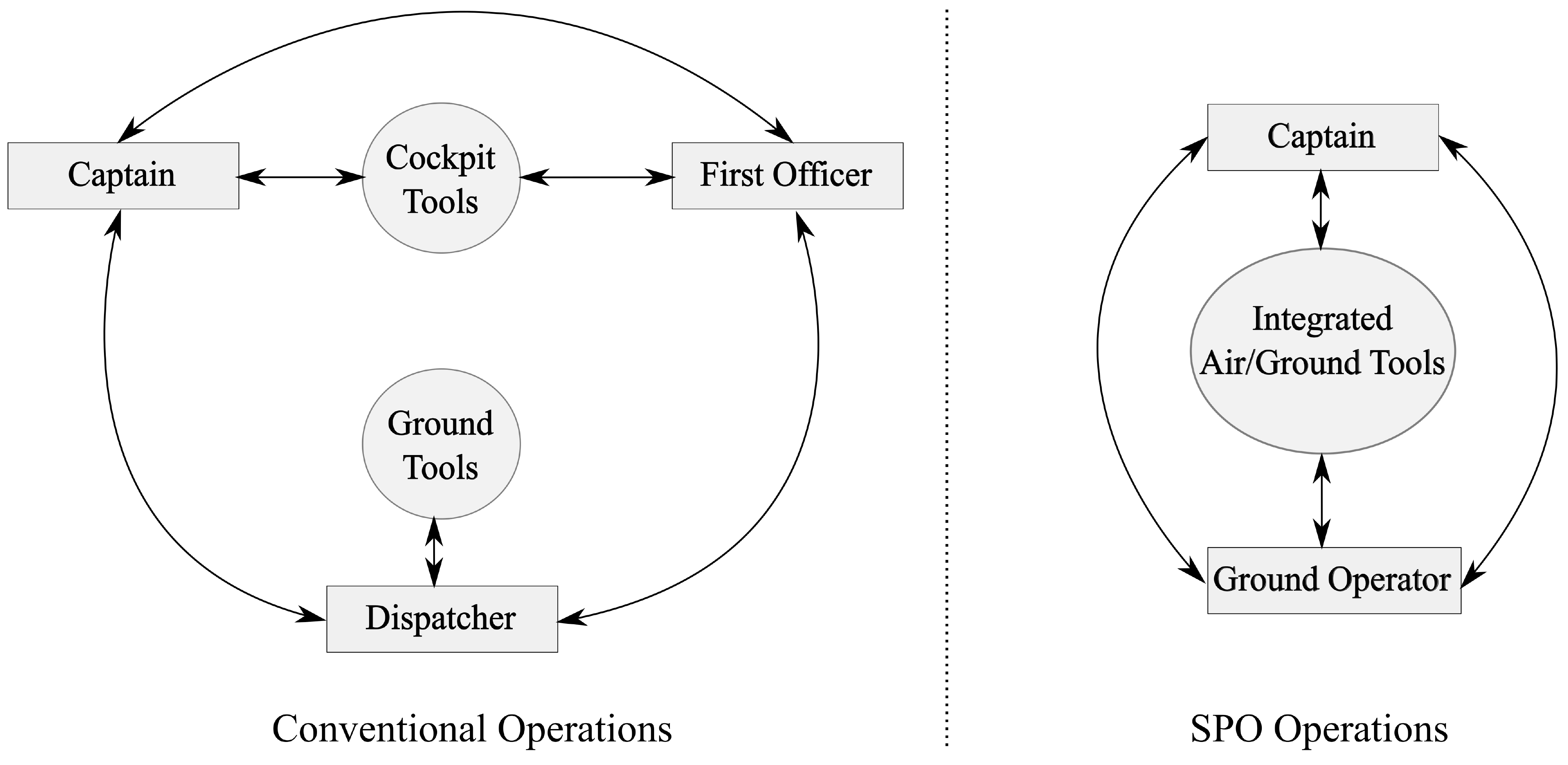}}
	\caption {\gls{NASA}'s vision in transitioning from conventional two-piloted airplane operations to \gls{SPO}. They merge the ground and air tools and replace the onboard first officer with a ground operator \cite{nasa_spo_conops}.}
	\label{fig: spo_operations}
\end{center}
\end{figure}

Future airplane operations may embrace \glspl{SPO}, where only a single pilot is onboard and operates the airplane with the assistance of an onboard automation and/or a ground operator \cite{nasa_spo_conops}. In such scenarios, the connectivity can become a complementary technology to support the safe operation of airplanes in the air. Reliable and robust airplane-ground connectivity is the key for monitoring and supporting the operation of an airplane from the ground. The \gls{NASA} studies the concept of future \glspl{SPO} and visualizes the evolution of flight operations for \glspl{SPO}, as shown in \autoref{fig: spo_operations} \cite{nasa_spo_conops}. Instead of conventional separated air and ground tasks, they integrate air and ground tools, and share the flight-related tasks between the onboard captain and a ground operator. In such scenarios, aviation industry may also employ virtual reality technologies to enable smooth interactions \cite{8735185}. Furthermore, appropriately defining the roles of the pilot and the ground operator is vital to ensure a coherent coordination \cite{8569803}. As a result, \glspl{SPO} have diverse research aspects, and the readers may refer to \cite{8039185} for a comprehensive overview on the system design of \gls{SPO}.

\subsubsection{Massive Data Transfer}
\label{subsec: usecases_airplanes_sensordataoffloading}

Airplanes utilize massive number of sensors to monitor and predict maintenance actions \cite{ITUM2197}. Sensors help decrease the cost of maintenance and maximize the lifetime of airplanes. As an example, Airbus launched an open data platform \textit{Skywise} to provide improved fleet operation through predictive and preventive maintenance \cite{skywise}. 

Massive data transfer can take place just before landing, after taking off or on ground, e.g., at the airport or during airplane maintenance. It includes the transmission of collected sensor data as well as the data from the onboard computers. For instance, an Airbus A320 generates 10 GB data per flight hour \cite{airbus_flightdata}, and thus, we must provide high-rate \gls{EL} connectivity to transfer the onboard data within a short time interval until the next flight.

\subsubsection{Connectivity Requirements}
\label{subsec: usecases_airplanes_connectivityrequirements}

We present the summary of the connectivity requirements of each airplane use cases in \autoref{table: connectivityrequirements_airplane}. The requirements of \gls{ATM} applications are based on ICAO's specifications \cite{ICAO_9896}. As for \gls{SPO}, we observe an increasing \gls{QoS} demands toward full \gls{SPO}. Although \gls{NASA} requires up to 10$^{-8}$ communication reliability, wireless technologies can face extreme challenges to meet such high requirements due to the nature of varying \gls{RF} channel conditions. In practice, we can meet this demand by utilizing 8 links in parallel assuming them uncorrelated with one another and each having 99.9\% communication reliability. Thus, such high requirements can pave the way for novel optimization methodologies to reduce the number of required links in parallel and to efficiently use the available channel resources.  

As for the full \gls{SPO} use case, providing 20 Mbps with 300 ms latency threshold \cite{nasa_spo_conops} is also a challenging requirement at altitudes above 10 km under high mobility rates such as 1000 km/h. Finally, we found out 45 Mbps for massive data transfer based on the study of \cite{sandraspaper_mdt}. For instance, we can offload a 4 hour flight data 320 Gb \cite{skywise}) of an airplane in less than two hours using a 45 Mbps link. 100 ms latency is our assumption since a massive amount of data transfer must be completed within a short time interval while the airplane is on the ground \cite{sandraspaper_mdt}.


\begin{savenotes}
    \begin{table*}[t]
		\caption {Connectivity Requirements of Airplanes. \vspace{-0.225cm}}
		    \begin{center}
		    \begin{threeparttable}
			{\renewcommand{\arraystretch}{1.75} 
				\begin{tabular}{l c c c c c}
					\hline
					\multicolumn{2}{l}{\textbf{Application}} & \textbf{Data Rate (Mbps)} & \textbf{\makecell{End-to-end\\Latency (ms)}} & \makecell[c]{\textbf{Communication}\\\textbf{Reliability}} & \textbf{References} \\ \hline
					 \multirow{3}{*}{\makecell[l]{ATM}} & \gls{VoIP} & 0.012 & $<$100 & High & \multirow{3}{*}{\cite{ICAO_9896}}\\
					  & \gls{OLDI}/\gls{FMTP} & 0.01 & $<$1000 & 99.95\% &\\
					  & \gls{AMHS} & 0.02 & $<$5000 & 99.9999\% &\\ \hline
					  \multirow{3}{*}{\makecell[l]{SPO}}& Current \gls{SPO}$^\ast$ & 0.03 & 40000 & 99.999\% & \multirow{3}{*}{\cite{nasa_spo_conops}} \\
					  & Cargo \gls{SPO}$^\dagger$ & 15 & 10000 & 99.999\% & \\
					  & Full \gls{SPO}$^\ddagger$ & 20 & 300 & 99.999999\% &\\ \hline
					 \multicolumn{2}{l}{Massive Data Transfer} & 45 & 100 & Medium & \cite{sandraspaper_mdt} \\
					\hline
				\end{tabular}}
				\label{table: connectivityrequirements_airplane}
				\begin{tablenotes}[flushleft]
    				\scriptsize
    				\item \textit{$^\ast$Crew operations in military airplanes.}
    				\item \textit{$^\dagger$In 10 years.}
    				\item \textit{$^\ddagger$In 20 years.}
				\end{tablenotes}
		\end{threeparttable}
		\end{center}
	\end{table*}
\end{savenotes}

\subsection{\glsentryshortpl{eVTOL}}
\label{subsec: usecases_evtols}

\begin{figure}[t]
\begin{center}
	\centerline{\includegraphics[width=0.4\textwidth,keepaspectratio]{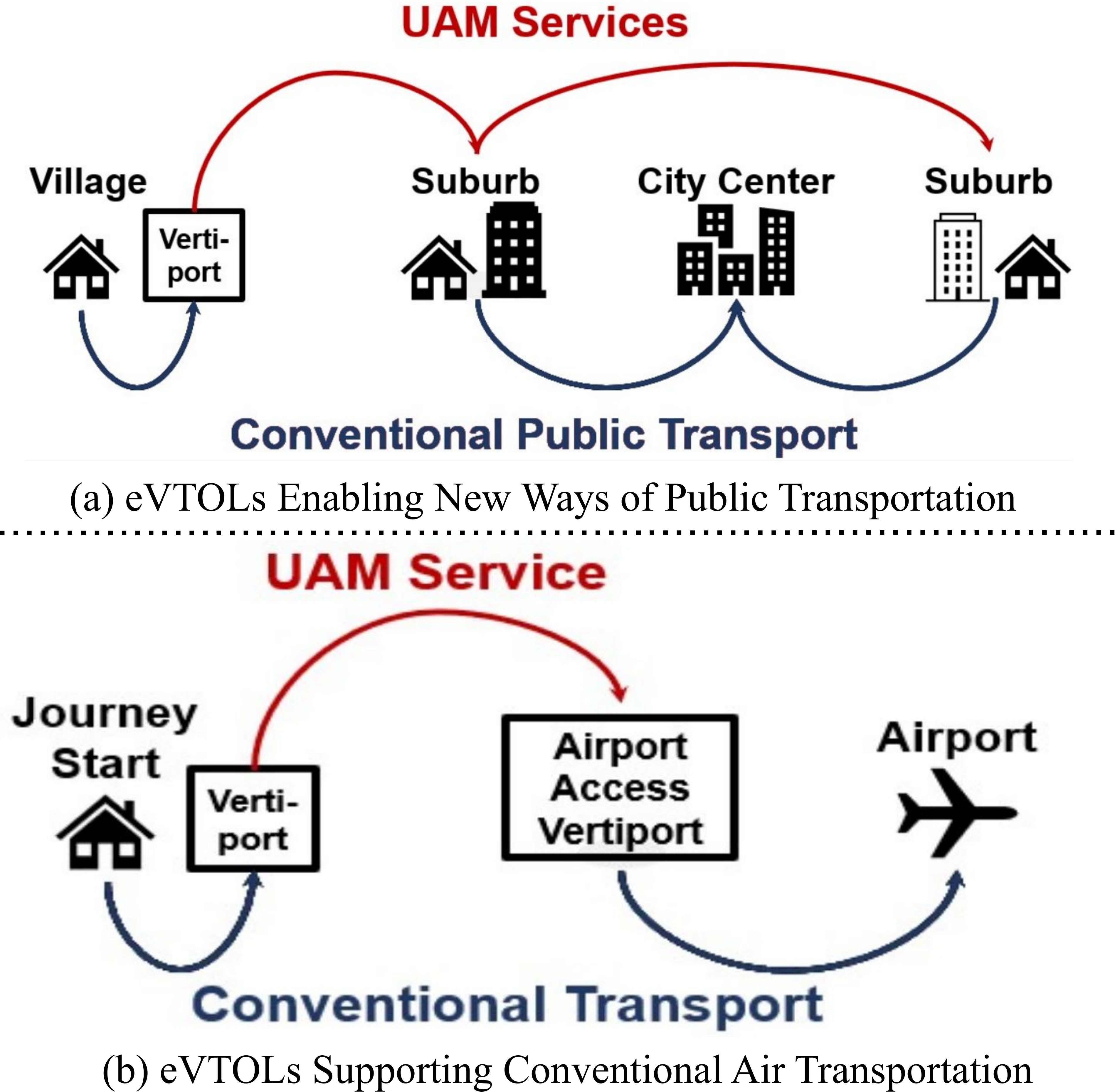}}
	\caption {New ways of transportation \glspl{eVTOL} enable with \gls{UAM}, according to \cite{luftfahrt_uam_2030}. In (a), \glspl{eVTOL} provides direct transportation between vertiports in suburban and rural areas, and in (b), it serves as an airport shuttle for the airline passengers.}
	\label{fig: evtol_transportation}
\end{center}
\end{figure}

\glspl{eVTOL} are the flying vehicles with vertical take-off and landing capabilities aimed mainly at passenger transportation, usually in short to medium range flight distances ($<$80 km) \cite{evtol_range}. We can consider \glspl{eVTOL} as the spin-off of conventional helicopters equipped with electrical engines, therefore compromising in flight distances. Additionally, there is a pronounced tendency to operate \glspl{eVTOL} without a pilot on board, which enables, e.g., new flying-taxi concepts. Thus, they complement the existing ground passenger transportation means in urban and suburban areas. \autoref{fig: evtol_transportation} shows the new ways of transportation \glspl{eVTOL} introduce \cite{luftfahrt_uam_2030}. 

The \gls{EASA} categorizes \glspl{eVTOL} with respect to the number of passenger seating: a) 0-1; b) 2-6; c) 7-9 passengers and their maximum weight can be up to 3175 kg \cite{easa_SC-VTOL-01} with a maximum flight altitude of 1 km \cite{nasa_uam}. \gls{NASA} foresees three main use cases for \glspl{eVTOL}: 1) airport shuttle; 2) air taxi; 3) air ambulance \cite{nasa_uam}. Although \gls{eVTOL} concepts are of the recent venture, several works already cover various research aspects such as the future requirements and the enabler technologies \cite{intel_evtol_dasc, uberair_vehiclerequirements}, collision prevention, and collision mitigation techniques \cite{8941429}, safety risks arising from the interactions between the \gls{eVTOL} pilot and onboard automation systems \cite{9081685}, and the optimization of the required time of arrival with respect to battery constraints and vertiport capacity \cite{8569645, 8569225}. 

\begin{figure}[t]
\begin{center}
	\centerline{\includegraphics[width=0.45\textwidth,keepaspectratio]{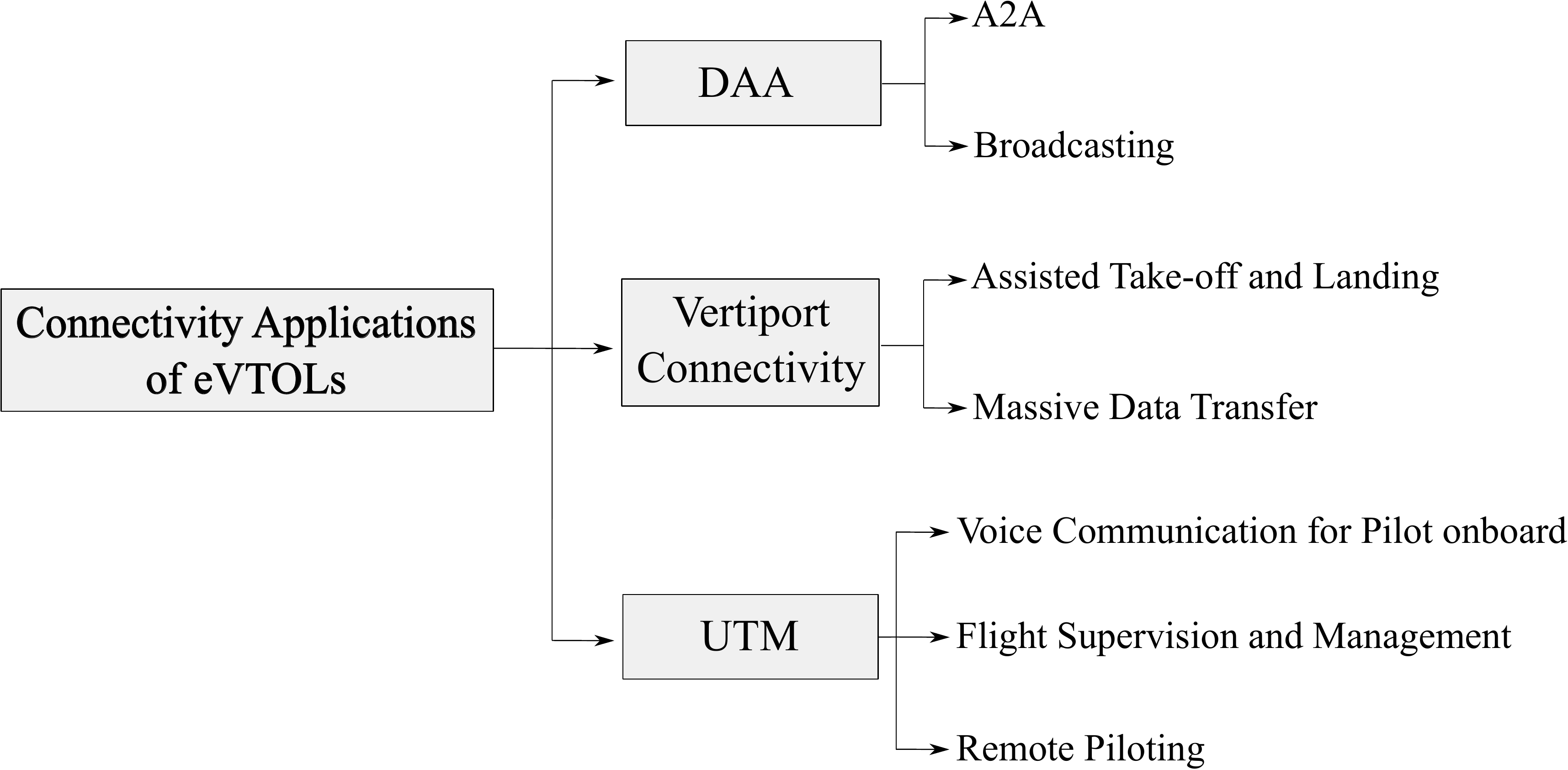}}
	\caption {Connectivity use cases of \glspl{eVTOL}, excluding the onboard communication.}
	\label{fig: evtol_connectivitytypes}
\end{center}
\end{figure}

We present the connectivity use cases of \glspl{eVTOL} in \autoref{fig: evtol_connectivitytypes}. We will first describe the \gls{DAA} and vertiport connectivity, afterwards we will comprehensively elaborate on the \gls{UTM} and its concept of operations.

\subsubsection{Detect and Avoid}
\label{subsec: usecases_evtols_daa}

The \glspl{eVTOL} employ radar-based systems to detect obstacles and dangers during flights. \gls{DAA} has two primary functions \cite{icao_rpas_rules}: 

\begin{enumerate}
    \item Detect conflicting \glspl{AV} or other safety-threatening obstacles, events, and weather conditions.
    \item Determine and perform danger-preventive maneuvers.
\end{enumerate}

Depending on the level of autonomy, the \gls{DAA} system can notify the operator or it can take preventive actions by itself, which are computed by onboard systems. Thus, certain \gls{DAA} systems have a control station on the ground \cite{rtca_daa}. In order to enable timely reactions against dangers, the \gls{DAA} demands robust connectivity to send notifications to the remote pilot or to the conflicting \gls{AV}. Broadcast communication can also be a part of the \gls{DAA}, such that the \glspl{eVTOL} periodically report their \gls{AV} identification and location to their surroundings as well as to the \gls{UTM} in real-time.

We can realize such a system with a conventional radar, i.e., an \gls{ACAS} \cite{9081631}, \gls{ADS-B} \cite{8904324} and vision-based or sound-based detection systems e.g., \gls{LIDAR} \cite{8539587} or ultrasonic \cite{9166607}. \gls{DAA} establishes \gls{A2A} links to communicate with the conflicting \glspl{AV}, and the \gls{A2A} reachability is an essential metric to ensure that the onboard system can detect an unknown object well in advance to avoid collisions. Therefore, its reachability requirement depends on the type of \glspl{AV}, due to the varying speed and flight characteristics of \glspl{AV}.

\subsubsection{Vertiport Connectivity}
\label{subsec: usecases_evtols_vertiport}

Besides rural and airport areas, we expect vertiports to be a part of the metropolitan infrastructures such as main bus/train stations and shopping centers \cite{deloitte_uam} to provide convenient pick-up/drop-off locations to the \gls{eVTOL} passengers. In such environments, \glspl{eVTOL} require assistance from the ground to enable precise take-off and landing \cite{MITRE_conops}. Hence, this demand takes place when the \gls{eVTOL} is nearby a vertiport zone, requiring short-range and robust connectivity. 

Similar to airplanes, \gls{eVTOL} operators offload accumulated sensor as well as onboard computer data between the flights for predictive and preventive maintenance. This use case also demands short-range and high-rate connectivity.

\subsubsection{Unmanned Traffic Management (UTM)}
\label{subsec: usecases_evtols_UTM}

As \glspl{eVTOL} and \glspl{UAV} occupy the sky, they must coordinate with one another, helicopters, airplanes as well as \glspl{HAP} to efficiently share the low-altitude sky, which is within the class G airspace (uncontrolled airspace that begins from the surface and extends until the base of the controlled airspace \cite{faa_airspaceregulations}) \cite{DLR_Uspace}. Conventional \gls{ATM} systems cannot meet the emerging \gls{eVTOL} and \gls{UAV} operations due to: 1) the increased number of total operations; 2) the increased vehicle density; 3) cruising operations at lower altitudes than that of today; 4) varying levels of pilot, automation and vehicle capabilities \cite{uam_icrat}. Thus, \glspl{UTM} introduce the regulation of these vehicles in a more-autonomous manner compared with the \gls{ATM}. \gls{MTC} can become the dominant connectivity type in \gls{UTM} rather than the human-centric \gls{ATM} communication in the future \cite{intel_evtol_dasc}. 

Number of different \gls{UTM} initiatives take place all over the world, such as U-Space in Europe \cite{sesar_corus}, \gls{UTM} in the USA \cite{nasa_utm_uasOperations} and JUTM in Japan \cite{JUTM}. Early research tackles various topics, e.g., airspace conflict resolution \cite{airbusutm2, airbusutm3} and flight planning \cite{airbusutm4}. 

We can summarize the list of main \gls{UTM} services as follows: 

\begin{itemize}
    \item \textbf{Vehicle Registration}: This service allows the operators to register their \glspl{AV} in the \gls{UTM} ecosystem to obtain flight permissions \cite{faa_nextgen_conops}.
    \item \textbf{Vehicle Identification and Tracking}: During a flight, an \gls{AV} must broadcast its identification along with the current position information to its surrounding and to the \gls{UTM}, which enables the real-time tracking of the \gls{AV} \cite{sesar_corus}. 
    \item \textbf{Communication Services}: \gls{UTM} can also provide the connectivity means to \gls{AV} operators to establish \gls{CNPC}, which is also known as \textit{\gls{C2} Communication}, with the \glspl{AV} \cite{faa_nextgen_conops}. 
    \item \textbf{Flight Planning}: \gls{UTM} can design the trajectory of the flight considering airspace flight rules, restrictions, etc. and send it to the \gls{AV} \cite{faa_nextgen_conops}. 
    \item \textbf{Airspace Authorization and Restrictions} It provides permissions to use the airspace for the corresponding \gls{AV} \cite{faa_nextgen_conops}, and regulates geofencing and geocaging.
    \item \textbf{Assisted Take-off and Landing Services}: \gls{UTM} can support the management of the departures and arrivals at vertiports \cite{MITRE_conops}, which is an area of land, water or structure for \glspl{eVTOL} to take-off and land \cite{easa_SC-VTOL-01}.
    \item \textbf{Conflict, Separation and Emergency Management}: By utilizing the real-time tracking information from the \gls{AV}, \gls{UTM} can ensure separation between the \glspl{AV} and avoid conflicts in advance with dynamic rerouting updates. \cite{faa_nextgen_conops}. In case of emergencies, \gls{UTM} can guide an \gls{AV} to an emergency landing zone to avoid danger in the air. The \glspl{eVTOL} can also have \gls{DAA} capabilities to detect obstacles nearby and take timely actions to avoid conflicts \cite{sesar_corus}.
    \item \textbf{Weather and Mapping Services}: \gls{UTM} can provide weather and terrain related information to the \gls{AV} operator to increase the safety of the flight \cite{faa_nextgen_conops}. 
    \item \textbf{Cooperation with \gls{ATM}}: \gls{UTM} must cooperate with \gls{ATM} near the airport zones, where the \glspl{AV} share the low-altitude airspace\cite{sesar_corus}.
\end{itemize}

\gls{NASA} developed a \gls{UTM} architecture \cite[Figure 3]{nasa_utm_conopsv2}, and located \gls{UAS} Service Supplier, which is similar to the air navigation service supplier in manned aviation, at the center of the \gls{UTM} ecosystem coordinating the interactions between different services. They included the 3$^\text{rd}$ party data services, remote operators, public safety as well as a gateway to the national airspace system to support the operations of \gls{UTM} \cite{nasa_uasServiceSupplierDevelopment}. Thus, the involvement of these entities and the interactions between one another influence the overall \gls{UTM} connectivity architecture. 

As \glspl{eVTOL} and \glspl{UAV} operate within short distances, national entities can establish \gls{UTM} systems at regional levels \cite{icao_rpas_rules}. In the event of inter-region or cross-border flights, different \glspl{UTM} can cooperate with each other \cite{3GPP_22825}. This implies the demand for handoff mechanisms between \glspl{UTM}. Therefore, the functions of \gls{UTM} strongly rely on the information exchange between different entities. Consequently, the connectivity is an essential function for the operations of \glspl{UTM}. Based on all these interactions, we demonstrate the connectivity use cases of \glspl{UTM} in \autoref{fig: utm_connectivitytypes}. 

\gls{UTM} communicates with \glspl{eVTOL} for the airspace regulation as well as the conflict resolution, and \gls{AV} operators contact the \gls{UTM} for flight requests, trajectory updates and other $3^\text{rd}$ party services. The connectivity between \gls{UTM} and \glspl{eVTOL} must be robust to enable real-time and reliable information exchange. The connectivity between \gls{UTM} and the ground pilot usually takes place before the take-off phase to acquire relevant flight permissions. Thus, available networks on ground can easily meet the connectivity requirements. Finally, wired connectivity enables the communication to \gls{ATM}, neighbor \gls{UTM} and other $3^\text{rd}$ party services. 

\comment{
\begin{figure}[t]
\begin{center}
	\centerline{\includegraphics[width=0.45\textwidth,keepaspectratio]{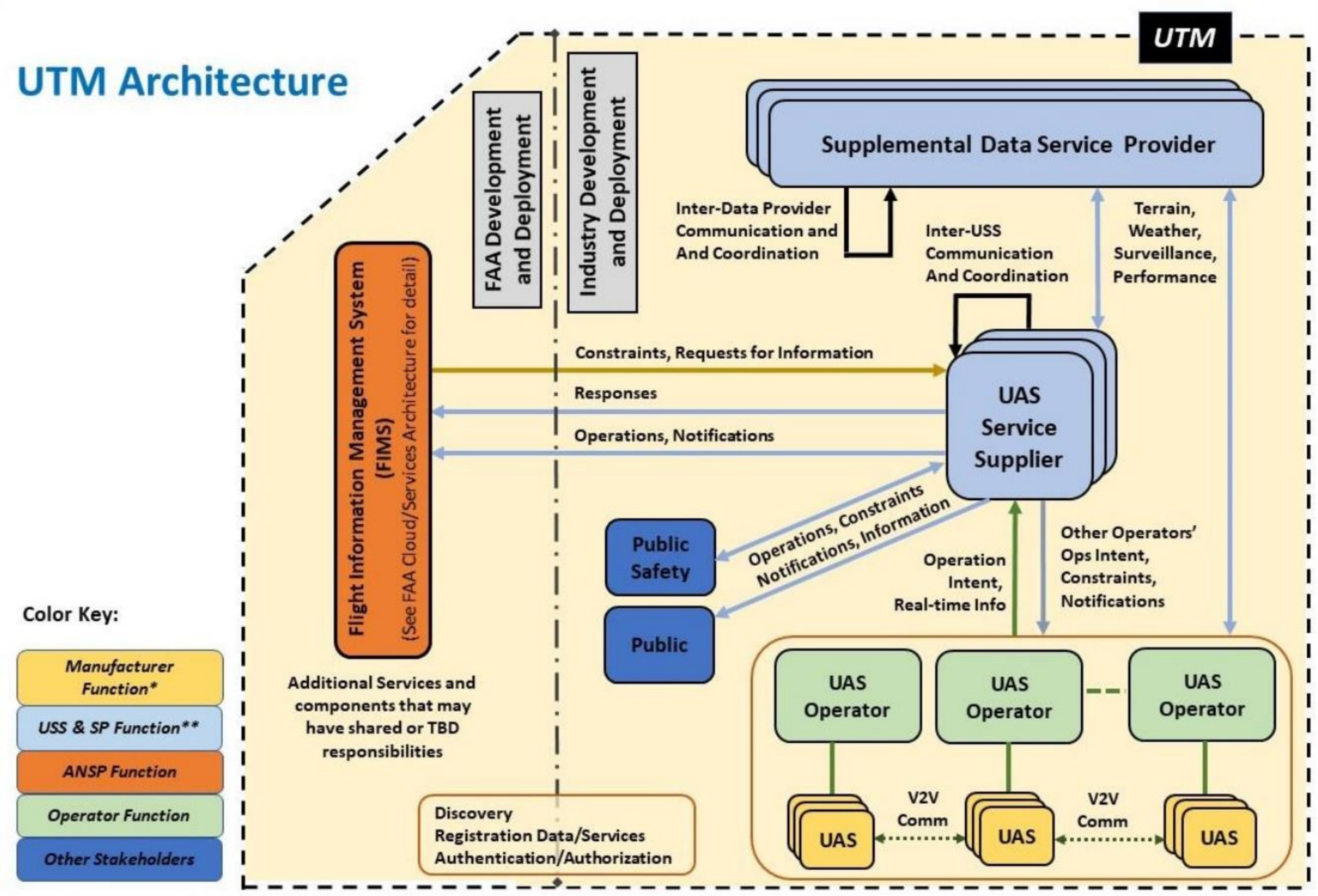}}
	\caption{\gls{UTM} architecture defined by \gls{NASA} and FAA \cite{nasa_utm_conopsv2}. \gls{UAS} Service Supplier coordinates the interactions between different \gls{UTM} participants and service providers.}
	\label{fig: nasa_uasServiceSupplierDevelopment}
\end{center}
\end{figure}
}

\subsubsection{Piloting}
\label{subsec: usecases_evtols_piloting}

\begin{figure}[t]
\begin{center}
	\centerline{\includegraphics[width=0.45\textwidth,keepaspectratio]{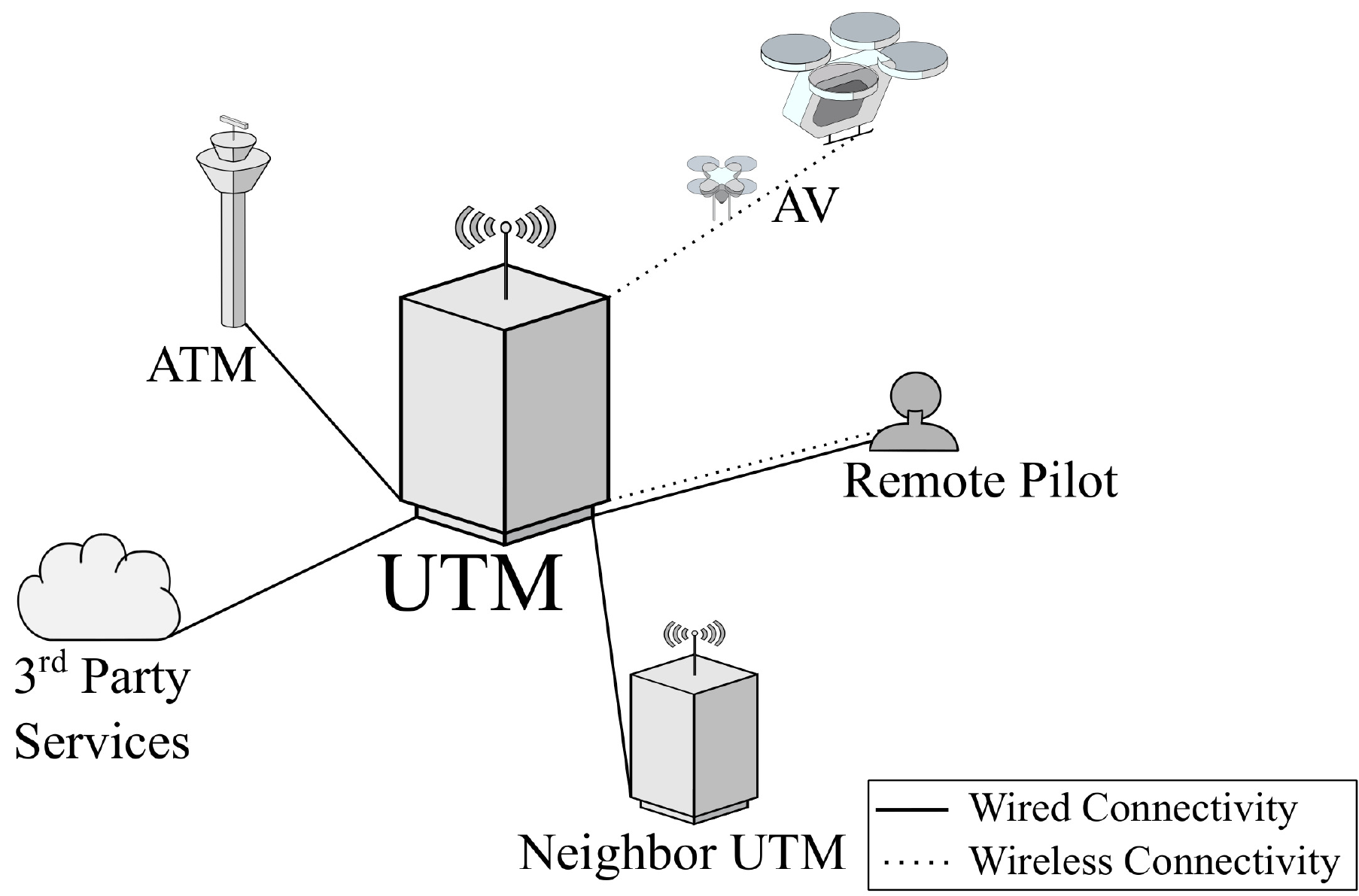}}
	\caption{Connectivity use cases of \gls{UTM}.}
	\label{fig: utm_connectivitytypes}
\end{center}
\end{figure}

\begin{figure*}[t]
\begin{center}
	\centerline{\includegraphics[width=0.9\textwidth,keepaspectratio]{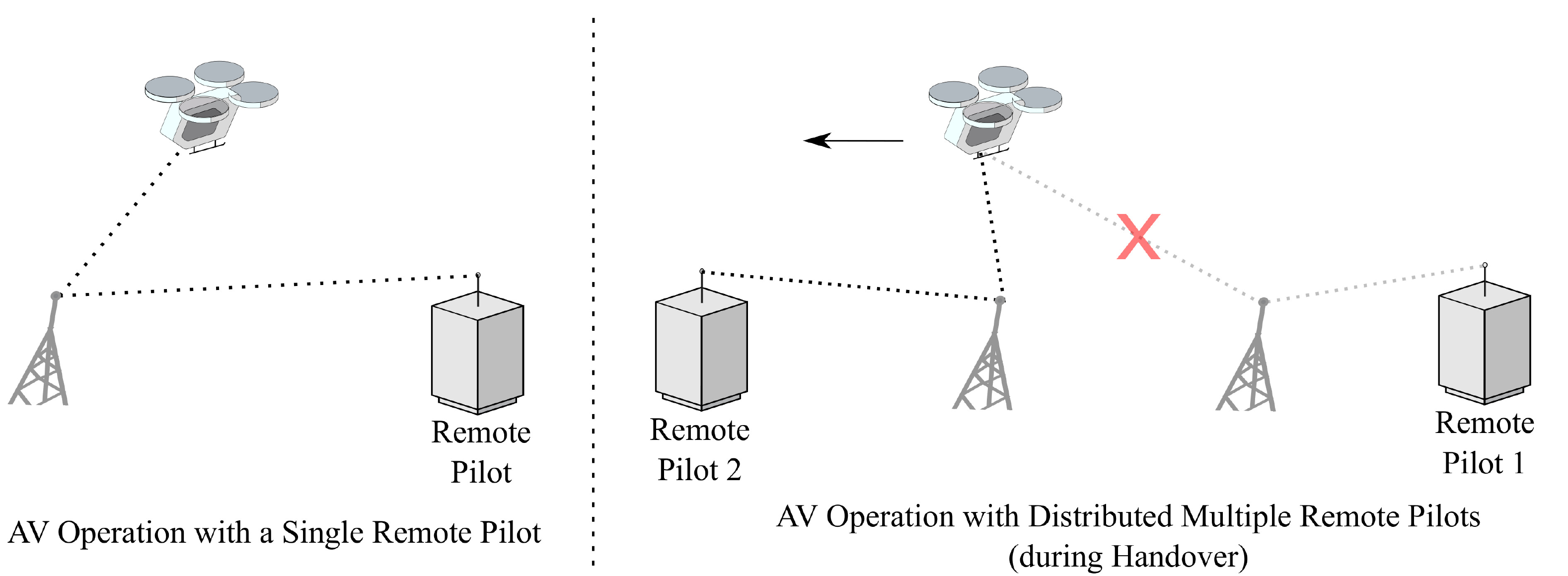}}
    \caption {Single vs. Multiple \gls{RPO} Model, based on the description from \gls{ICAO} \cite{icao_rpas_rules}. Although we demonstrate a single-link connectivity from a remote pilot, the handover can become more complex in multi-link operations.}
    \label{fig: evtol_piloting}
\end{center}
\end{figure*}
        
Before detailing the type of piloting scenarios, it is worth describing the well-known terms \textit{\glsentryfull{UAS}} and \textit{\gls{RPAS}} in this context: 
\begin{itemize}
    \item \textbf{\glsentryfull{UAS}}: It is an airborne system without a pilot onboard \cite{easa_drone_operator_pilot}, and an equipment to remotely control the aircraft \cite{EASA_UAS_definition}: It can be a ground control station, a remote pilot, a \gls{UTM} as well as a communication system \cite{FAA_UAS_definition, ICAO_UAS_RPAS_definition, 3GPP_UAS_definition, DroneRegulations}. We can divide the \gls{UAS} into two subcategories: 1) \gls{RPAS}; 2) fully-autonomous \gls{UAS}, which can operate by itself without requiring any pilot intervention \cite{easa_drone_operator_pilot}.  
    \item \textbf{\glsentryfull{RPAS}}: It is an airborne system that consists of a remotely piloted \gls{AV}, its remote operator and a communication link \cite{icao_rpas_rules}. Although \gls{RPAS} is a subcategory within the \gls{UAS} ecosystem \cite{easa_drone_operator_pilot, EASA_UAS_definition}, \gls{ICAO} distinguishes the \gls{UAS} and \gls{RPAS} in a way that \gls{RPAS} has the same certification standards as manned aircraft. Yet, \gls{ICAO} poses different certification requirements for \gls{UAS} and separates it from the manned aircraft in the airspace \cite{ICAO_UAS_RPAS_definition}.
\end{itemize}

\begin{savenotes}
	\begin{table*}[t]
		\caption {Connectivity Requirements of \glspl{eVTOL}. \vspace{-0.225cm}}
		\begin{center}
		\begin{threeparttable}
			{\renewcommand{\arraystretch}{1.5} 
				\begin{tabular}{l c c c c c}
					\hline
					\multicolumn{2}{l}{\textbf{Application}} & \makecell[c]{\textbf{Data Rate}\\\textbf{(Mbps) (\gls{IL}/\gls{EL})}} & \makecell[c]{\textbf{Latency}\\\textbf{(ms)}} & \makecell[c]{\textbf{Communication}\\\textbf{Reliability}} & \textbf{References} \\ \hline
					 \gls{UTM} & \makecell{Flight Supervision\\and Management} & 0.01-0.1 & $<$500 & 99.999\% & \cite{3GPP_22125, sesar_corus, ngmn_urllc} \\ \hline
					 \multirow{3}{*}{Piloting} & \rule{0pt}{1.5em}Pilot onboard & 0.012 & $<$100 & High & \cite{ICAO_9896} \\ 
					 & \rule{0pt}{1.5em}\gls{RPO} & \makecell{10-100$\dagger$/0.25-1$^\ast$} & 10-150 & High & \cite{3GPP_22125} \\
					 & \glsentryshort{FAO} & 0.1-1 & 100-500 & Medium & \cite{3GPP_22125, baiocchi15} \\ \hline
					 \multirow{2}{*}{Vertiport} & \rule{0pt}{1.5em}\makecell[c]{Assisted Take-off\\and Landing} & 0.01-0.1 & 10$^\ddagger$, 140$^\lozenge$ & High & \cite{3GPP_22125} \\
					 & Massive Data Transfer & 25 & 100 & Medium & \cite{sandraspaper_mdt} \\ \hline
					 \multicolumn{2}{l}{\gls{DAA}} & 0.01-0.1 & 1000 & Medium-High & \cite{rtca_daa, 8636976} \\ 
					\hline
				\end{tabular}}
				\label{table: connectivityrequirements_evtols}
				\begin{tablenotes}[flushleft]
				        \scriptsize
				        \item \textit{$^\ast$For control/telemetry traffic.}
				        \item \textit{$^\dagger$Video streaming for first-person view.}
				        \item \textit{$^\ddagger$For \gls{IL} link.}
				        \item \textit{$^\lozenge$For \gls{EL} link.}
				\end{tablenotes}
		\end{threeparttable}
		\end{center}
	\end{table*}
\end{savenotes}

We grouped the piloting scenarios of \glspl{eVTOL} into three categories based on the level of vehicle autonomy:

\begin{enumerate}
    \item \textbf{Pilot onboard \gls{eVTOL}}: We can expect that the first integration of \glspl{eVTOL} into airspace requires a pilot onboard to operate the vehicle. We assume that the pilot can communicate with \gls{ATM} to obtain flight permissions, updates and other flight-related information in this first phase. Later, \gls{UTM} can take over the management of the low-altitude airspace and directly communicate the \gls{AV}. Lilium Jet \cite{lilium} is an instance of a pilot onboard \glspl{eVTOL}. Even though the onboard pilot handles the safety-critical functions, he/she nevertheless demands robust connectivity to coordinate the flight with the \gls{ATM} as well as the $3^\text{rd}$ party services.
    
    \item \textbf{\glsentryfull{RPO}}: In this concept, ground pilots remotely operate an \gls{AV}, which can supply pilots with a first-person view by onboard cameras and other useful sensor data. Thus, these operations fall into the \gls{RPAS} category and a technical report from \gls{ICAO} defines a number of rules for \gls{RPO} \cite{icao_rpas_rules}:
       \begin{itemize}
            \item Only one remote pilot can control an \gls{AV} at a time,
            \item Multiple remote pilots located at different regions can handle the operation of an \gls{AV} during international or long flights, as illustrated in \autoref{fig: evtol_piloting}.
            \item The transfer of \gls{CNPC} between different network service providers is possible.
        \end{itemize}
       Therefore, \gls{RPO} demands robust \gls{CNPC} with seamless handover mechanism to enable control data exchange and multiple video streams between one or multiple end users. CityAirbus demonstration is an example type of this category \cite{airbus_uam_vehicles}. 
    \item \textbf{\glsentryfull{FAO}}: As the level of autonomy increases, the \glspl{AV} can operate without a remote pilot. A ground supervisor can still be present to monitor the flight depending on the level of autonomy \cite{Airbus_BluePrint}. They can obtain the flight routes and updates from \gls{UTM} and perform self-operations. Nevertheless, the remote supervision ensures safe operations to handle unexpected events in the air. Thus, \gls{FAO} falls into the fully-autonomous \gls{UAS} category. Airbus Vahana demonstration \cite{airbus_uam_vehicles} showed concepts of such operations. 
\end{enumerate}

\gls{CNPC} is vital in \glspl{RPO} and \glspl{FAO}. On the \gls{EL}, an \gls{AV} transmits video stream along with telemetry-related information such as \gls{GPS}, \gls{IMU} and other flight-related status information, and the remote pilot sends control commands on the  \gls{IL} \cite{ITU_M2171}. Due to the critical role of the communication in the safe operation as well as the human involvement, the connectivity demands for \glspl{RPO} are very stringent. The connectivity must support robust and real-time data exchange of asymmetric traffic on the \gls{EL} and \gls{IL}. While \gls{EL} delivers the video streaming traffic, the \gls{IL} hosts the periodic control data exchange to remotely operate the vehicle \cite{baltaci_trafficgeneratorpaper}. 

In general, we can expect an inverse correlation between the level of vehicle autonomy and the connectivity demand of vehicles during flight \cite{commVsAutonomy}. As \glspl{AV} gain the capability of onboard computing, the controllers can process the video information onboard and perform piloting decisions in real-time. As a result, \glspl{AV} utilize the ground communication only for status and flight-related information updates, and the communication reliability requirements are alleviated along with lower data rate and latency demands. 
\comment{
\subsubsection{Onboard Connectivity in eVTOLs}
\label{subsec: usecases_evtols_mtc}

\glspl{eVTOL} host different types of sensors to monitor the status of the engines, onboard systems, vehicle structures, etc. Flight control systems can utilize sensor data in real-time to operate critical functions of the vehicle. Although real-time applications may demand low rates, the safety-critical functions imply high communication reliability, just as the \gls{WAIC} use cases of airplanes \cite{ITUM2197}. 

Similar to conventional airplanes, \gls{eVTOL} passengers demand seamless \gls{IFEC} during their journey for web surfing, video conferencing and in-flight entertainment. Due to the low altitude operations of \glspl{eVTOL}, we consider ground connectivity options, if available, more feasible than the satellite services for \gls{IFEC}. Employing ground networks can also help passengers seamlessly continue using their already-existing cellular services onboard the \gls{eVTOL}.
}
\subsubsection{Connectivity Requirements}
\label{subsec: usecases_evtols_connectivityrequirements}

\begin{savenotes}
	\begin{table*}[t]
		\caption {Safety Objectives for \glspl{eVTOL} as defined by \gls{EASA} \cite{easa_SC-VTOL-01, easa_moc_sc-vtol}. \vspace{-0.225cm}}
		\begin{center}
		\begin{threeparttable}
			{\renewcommand{\arraystretch}{1.5} 
				\begin{tabular}{l c c c c}
					\hline
					\textbf{\makecell{\glsentryshort{eVTOL} Type/\\Failure Condition Classification}} & \textbf{Minor} & \textbf{Major} & \textbf{Hazardous} & \textbf{Catastrophic} \\ \hline
					0-1 passenger & $\leq 10^{-3}$ (\glsentryshort{DAL} D) & $\leq 10^{-5}$ (\glsentryshort{DAL} C) & $\leq 10^{-6}$ (\glsentryshort{DAL} C) & $\leq 10^{-7}$ (\glsentryshort{DAL} C) \\
					2-6 passengers & $\leq 10^{-3}$ (\glsentryshort{DAL} D) & $\leq 10^{-5}$ (\glsentryshort{DAL} C) & $\leq 10^{-7}$ (\glsentryshort{DAL} C) & $\leq 10^{-8}$ (\glsentryshort{DAL} B) \\
					7-9 passengers & $\leq 10^{-3}$ (\glsentryshort{DAL} D) & $\leq 10^{-5}$ (\glsentryshort{DAL} C) & $\leq 10^{-7}$ (\glsentryshort{DAL} B) & $\leq 10^{-9}$ (\glsentryshort{DAL} A) \\
					\hline
				\end{tabular}}
				\label{table: easa_evtol_safetyobjectives}
		\end{threeparttable}
		\begin{tablenotes}
		        \item \textit{\hspace{1.5cm}The failure rates are the average probabilities per flight hour.}
		\end{tablenotes}
		\end{center}
	\end{table*}
\end{savenotes}

\begin{savenotes}
	\begin{table}[t]
		\caption {\glsentryshortpl{DAL} for Commercial Airplanes \cite{cadence_do254}. \vspace{-0.225cm}}
		\begin{center}
			{\renewcommand{\arraystretch}{1.5} 
				\begin{tabular}{l l}
					\hline
					\textbf{\glsentryshort{DAL}} & \textbf{Consequence of Failure} \\ \hline
					Level A & Crash, deaths \\
					Level B & May cause crash, deaths \\
					Level C & May cause stress, injuries \\
					Level D & May cause inconvenience \\
					Level E & No safety effect \\
					\hline
				\end{tabular}}
				\label{table: dals}
		\end{center}
	\end{table}
\end{savenotes}

We analyze the communication requirements of \gls{eVTOL} use cases and present the results in \autoref{table: connectivityrequirements_evtols}. As \gls{eVTOL} concepts are of the recent venture, we mainly make realistic assumptions by relating its use cases to the similar ones from airplanes and \glspl{UAV} \autoref{subsec: usecases_uavs}.

Flight supervision and management requires a low-rate communication between \gls{UTM} and \glspl{AV} with a lenient latency bound, as defined by 3GPP \cite{3GPP_22125}. For the pilot onboard scenario, we only assumed the voice connectivity to \gls{ATM} to obtain flight permissions, updates and other flight-related information. Therefore, we assumed the same requirements as the \gls{VoIP} use case of the airplanes in \autoref{table: connectivityrequirements_airplane}.  

The \glspl{RPO} are the most connectivity-critical scenarios as the \gls{AV} control merely relies on the communication from the ground. The data rate demands of \glspl{eVTOL} are larger compared with that of \glspl{UAV}. One of the reasons is that we can expect the double or triple redundancy architectures of conventional airplane flight systems to be employed for \glspl{eVTOL} as well. Then, the number of sensors and avionic systems also increase in a similar magnitude. We also expect \glspl{eVTOL} to be equipped with multiple or 360$\degree$ cameras to provide a full-vision to the remote pilot. The latency demands are also tight as the remote pilot needs to operate the vehicle in real-time. Furthermore, the data path must be robust and secure to prevent unauthorized use of the link. 

As the autonomy of the \gls{eVTOL} increases, we assume that the onboard processors begin to take the major tasks of the remote pilot. Depending on the level of autonomy, connectivity can enable flight supervision/delegation or remote assistance in case of emergency \cite{Airbus_BluePrint}. 

Regarding the vertiport applications, we provided our assumptions based on the following reasoning. Assisted take-off and landing is a low-rate connectivity since it only contains bidirectional localization-related information exchanges. However, its communication reliability must be high with a tight latency threshold since it may be directly involved in the safe take-off and landing procedure of \glspl{eVTOL}. For massive data transfer, we derived our assumption based on \cite{sandraspaper_mdt} for airplanes, and scaled it down since the data rate demands of \glspl{eVTOL} can be lower due to the less number of onboard sensors and flight-related logs. As for \gls{DAA}, it must also support robust communication when the remote pilot must perform maneuvers by herself/himself to avoid conflicts.


As for the communication reliability, we provide a perspective different from the available literature on \glspl{eVTOL}, based on the \glspl{DAL} in aviation. In conventional airplane systems, \glspl{DAL} define the criticality and the influence of each airplane function to the safe operation \cite{cadence_do254}. The \glspl{DAL} are mapped onto the system failure rates 10$^{-3}$-10$^{-9}$, based on the impact of that function to the operation of the vehicle \cite{cadence_do254}. \gls{EASA} took a similar approach for \glspl{eVTOL} and studied the required minimum failure rates of the safety functions of \glspl{eVTOL} under different safety objectives, based on the number of passengers \cite{easa_SC-VTOL-01, easa_moc_sc-vtol}. We present the outcome of EASA's study in \autoref{table: easa_evtol_safetyobjectives}. For instance, if the failure condition of connectivity function causes a  \textit{catastrophic} event on a single passenger \gls{eVTOL}, then the failure rate must be a minimum of 10$^{-7}$ per flight hour. 

In this table, some of the same failure rates are mapped onto different \glspl{DAL}. This can be due to the fact that EASA derived the failure rates based on the report \cite{faa_failuredescriptions} and the \glspl{DAL} according to the report \cite{sae_ARP4754A}. Additionally, \autoref{table: dals} shows the consequence of failure of functions based on their \gls{DAL} levels in commercial airplanes \cite{cadence_do254}. 

We can consider the wireless communication link as one of the safety functions of \glspl{eVTOL}, especially in \glspl{RPO}. Thus, we can interpret the communication requirements based on \autoref{table: easa_evtol_safetyobjectives}. At the same time, we also need to determine whether wireless links contribute to the communication reliability calculation in parallel with other functions or in series with a particular function. Considering the parallel scenario, the communication reliability must meet the rate of 10$^{-7}$ per flight hour to avoid catastrophic events (e.g., fatal injury \cite{faa_failuredescriptions}) in \glspl{eVTOL} with a single passenger. However, such high rates are difficult to achieve in wireless connectivity environment due to the unpredictable \gls{RF} channel conditions (e.g. fading and shadowing) that can degrade the wireless link quality. Therefore, a comprehensive elaboration on the role of connectivity to the overall safety of the \gls{eVTOL} is vital to determine the reliability demands accordingly. Nonetheless, we can still expect the communication reliability requirements to be up to 10$^{-5}$ in \glspl{RPO} due to the direct involvement of humans in the loop. 

\begin{figure}[t]
\begin{center}
	\centerline{\includegraphics[width=0.5\textwidth,keepaspectratio]{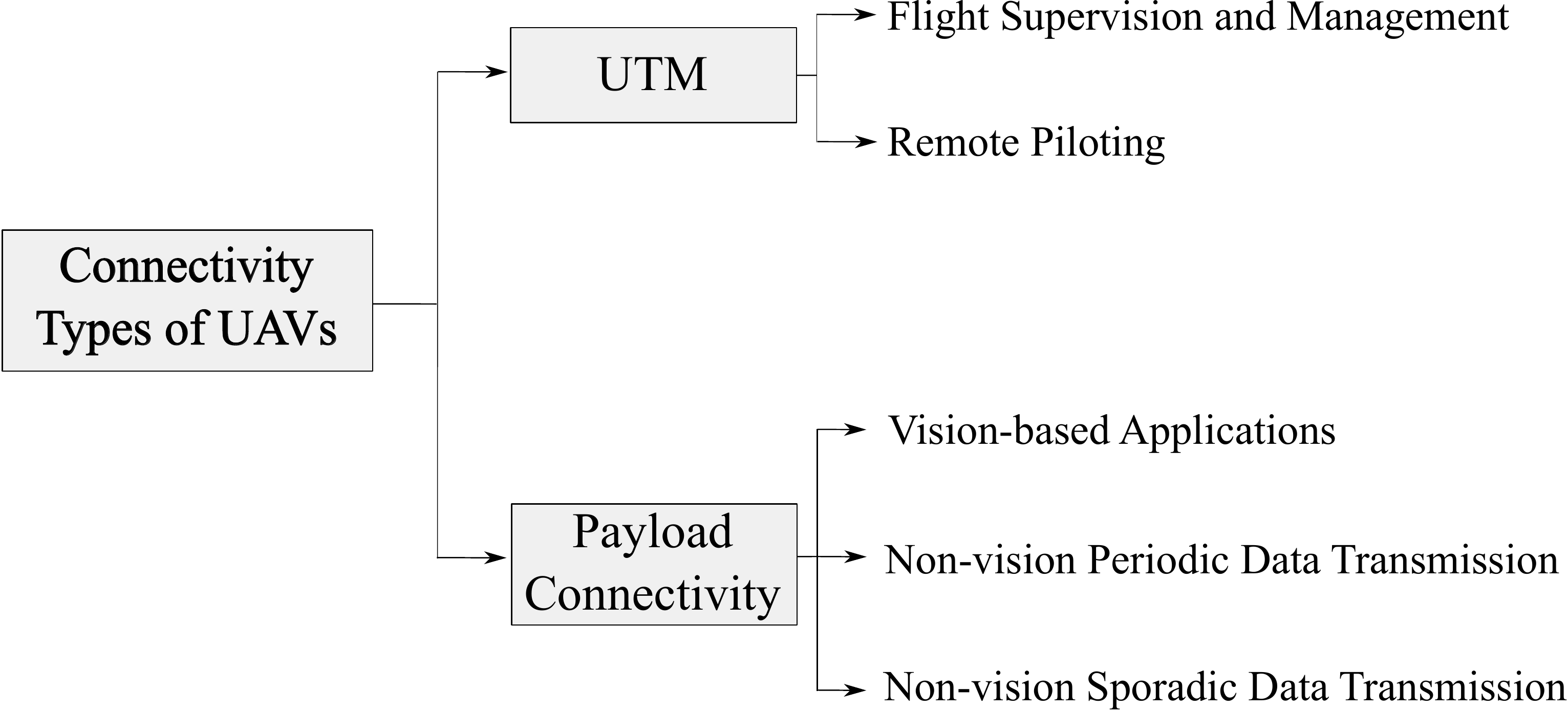}}
	\caption {Connectivity use cases of \glspl{UAV}.}
	\label{fig: uav_connectivitytypes}
\end{center}
\end{figure}

\begin{savenotes}
	\begin{table*}[t]
		\caption {Specifications and Communication Requirements of \gls{UAV} Applications \vspace{-0.225cm}}
		\begin{center}
		\begin{threeparttable}
			{\renewcommand{\arraystretch}{1.5} 
				\begin{tabular}{l c c c c c c c c}
					\hline
					\multirow{2}{*}{\textbf{Application}} & \multicolumn{4}{l}{\textbf{Flight Characteristics}} & \multicolumn{3}{l}{\textbf{Connectivity Requirements}} & \textbf{References} \\
					& \makecell{Altitude\\(m)} & \makecell{Speed\\(m/s)$^\ast$} & \makecell{Deployment\\Density$^\ast$} & \makecell{Area of\\Operations$^\ast$} & \makecell{Data Rate\\(Mbps)} & \makecell{Latency\\(ms)} & \makecell{Communication\\Reliability$^\ast$} \\ \hline
					Vision-based & $<$300 & 15 & Medium & Urban/Rural & 0.3-120 & 20-200 & Medium & \cite{3GPP_22125, 8918497, 8470897} \\
					Delivery & 100 & 15 & High & Urban/Rural & 0.2-0.3 & 500 & Low & \cite{8918497} \\
					\gls{IoT} & 300 & 15 & Low & Urban/Rural & 0.05-0.25 & $<$10 & Low & \cite{iot_datarate, iot_datarate2} \\
					Agriculture & 300 & 15 & Low & Rural & 0.2-0.3 & 500 & Medium & \cite{8918497} \\
					\hline
				\end{tabular}}
				\label{table: connectivityrequirements_uav_applications}
				\begin{tablenotes}[flushleft]
				        \item \textit{$^\ast$Our assumptions.}
				\end{tablenotes}
		\end{threeparttable}
		\end{center}
	\end{table*}
\end{savenotes}

Regarding the \gls{FAO}, we consider the connectivity requirements to be more relaxed compared with the \glspl{RPO}, since flight decisions rely more on the onboard computation. We can expect up to 1 Mbps for periodic flight updates, vehicle status messages, and other operation-related information exchange \cite{baiocchi15}. The upper latency bound is 500 ms since the involvement of a ground supervisor is rare during flight and real-time operations should be handled onboard the \gls{AV}. The communication reliability demands are also lower compared with the \gls{RPO} as the connectivity is not a safety-critical function anymore. Regarding the broadband connectivity, we specify the capacity as 15 Mbps based on \cite{ngmn}. The total data rate per \gls{eVTOL} depends on the number of onboard passengers.

Overall, various use cases require connectivity for \glspl{eVTOL} and their requirements are diverse depending on the amount of information exchange and their role on the operation of the vehicle. The \gls{RPO} is the most challenging scenario to provide robust connectivity. The communication reliability rate of 10$^{-5}$ is beyond what the current wireless technologies can provide in aerial environments. Due to the nature of \gls{RF} propagation, the \gls{RF} channel conditions fluctuate depending on the environment and eventually, ensuring high communication reliability and seamless connectivity become difficult during an entire flight. Therefore, we foresee the demand for heterogeneous, multi-link network architectures for \gls{RPO}. We discuss it further in \autoref{subsection: wirelesstechnologies_heterogeneous}. 

\subsection{UAV}
\label{subsec: usecases_uavs}

\glspl{UAV}, also known as \textit{drones}, are fixed- or rotary-wing unmanned \glspl{AV}, which fall into the \gls{UAS} category for non-human transportation. \gls{EASA} categorizes these vehicles into three types: 1) 1 m maximum dimension and 5 kg \gls{MTOM}; 2) 3 m maximum dimension and 200 kg \gls{MTOM}; 3) 8 m maximum dimension and 600 kg MTOM \cite{easa_SCLight-UAS}. The maximum operation altitude of \glspl{UAV} is dependent on the national regulations, usually 90 - 150 m \cite{DroneRegulations}. We categorize the connectivity use cases of \glspl{UAV} as shown in \autoref{fig: uav_connectivitytypes}.

\gls{UTM} handles the \gls{CNPC} and coordinates the operations of \glspl{UAV}, similar to the \glspl{eVTOL} as described in \autoref{subsec: usecases_evtols_connectivityrequirements}. The payload connectivity serves three main scenarios. Vision-based applications include all the use cases that involve data transmission with cameras. Afterwards, we divide non-vision applications into two categories: 1) periodic communication occur in a deterministic pattern such as \gls{IoT} data collection; 2) sporadic communication are rather event-based, such as agriculture applications. We further discuss about the \gls{UAV} applications in the following section. 

\begin{figure}[t]
\begin{center}
	\centerline{\includegraphics[width=0.4\textwidth,keepaspectratio]{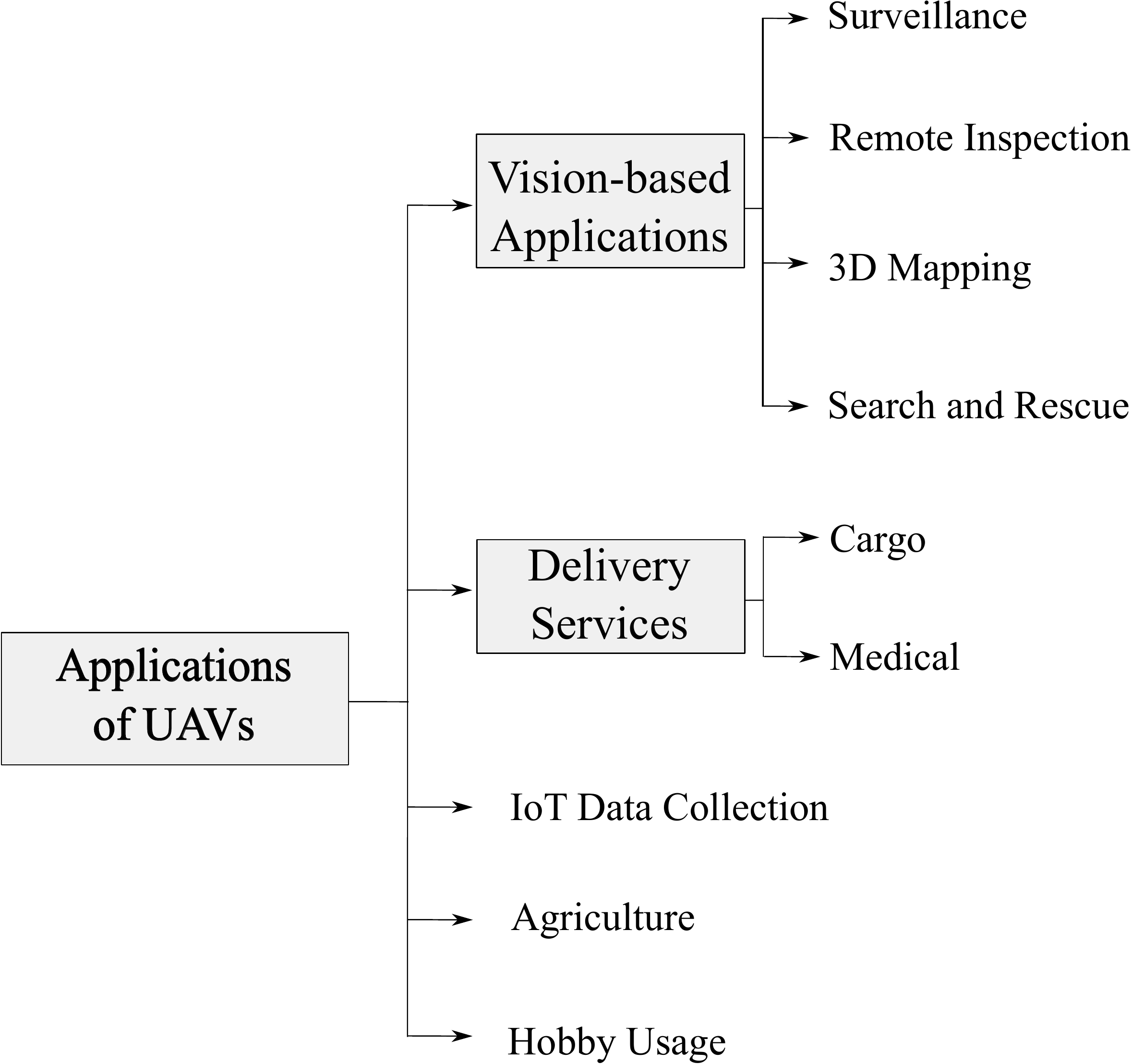}}
	\caption {Application scenarios of \glspl{UAV}.}
	\label{fig: uav_applications}
\end{center}
\end{figure}

\subsubsection{Application Scenarios of UAVs}
\label{subsec: usecases_uav}

Although originally developed for the military use, civil applications acquired \glspl{UAV} for different scenarios such as delivery, surveillance, agriculture, remote data collection and many more \cite{Liu_2020}. \glspl{UAV} can operate in different environments, deployment densities, altitudes, and speeds depending on the requirements of the particular application. We categorize the application areas of \glspl{UAV} as shown in \autoref{fig: uav_applications}.

\begin{savenotes}
	\begin{table*}[t]
		\caption {Safety Objectives for \gls{UAS} as defined by \gls{EASA} \cite{easa_SCLight-UAS}. \vspace{-0.225cm}}
		\begin{center}
			{\renewcommand{\arraystretch}{1.5} 
				\begin{tabular}{l c c c c}
					\hline
					\textbf{\makecell{UAV Type/\\Failure Condition Classification}} & \textbf{Minor} & \textbf{Major} & \textbf{Hazardous} & \textbf{Catastrophic} \\ \hline
					Size: $<$1 m, \gls{MTOM}: $<$5 kg & $<$10$^{-2}$ (\gls{DAL} D) & $<$10$^{-4}$ (\gls{DAL} C) & $<$10$^{-6}$ (\gls{DAL} C) & $<$10$^{-8}$ (\gls{DAL} B) \\
					Size: $<$3 m, \gls{MTOM}: $<$200 kg & $<$10$^{-3}$ (\gls{DAL} D) & $<$10$^{-5}$ (\gls{DAL} C) & $<$10$^{-7}$ (\gls{DAL} C) & $<$10$^{-9}$ (\gls{DAL} B) \\
					Size: $<$8 m, \gls{MTOM}: $<$600 kg & (\gls{DAL} D) & (\gls{DAL} C) & (\gls{DAL} B) & (\gls{DAL} A) \\
					\hline
				\end{tabular}}
				\label{table: easa_uav_safetyobjectives}
				\begin{tablenotes}
		        \item \textit{\hspace{1.5cm}The failure rates are the average probabilities per flight hour.}
		\end{tablenotes}
		\end{center}
	\end{table*}
\end{savenotes}

\begin{savenotes}
	\begin{table}[t]
		\caption {Connectivity Requirements of \gls{UAV} \gls{CNPC}. \vspace{-0.225cm}}
		\begin{center}
		\begin{threeparttable}
			{\renewcommand{\arraystretch}{1.5} 
				\begin{tabular}{l c c c c}
					\hline
					\textbf{\gls{RPO} Type} & \makecell{\textbf{Data Rate} \\ \textbf{(Mbps)}} & \makecell{\textbf{Latency}\\\textbf{(ms)}} & \makecell{\textbf{Communication}\\\textbf{Reliability}} & \textbf{References} \\ \hline
					\rule{0pt}{2em}\glsentryshort{VLoS} & \makecell{0.05-0.15$^\ast$ \\ 2$^\dagger$} & \makecell{10-250$^\ast$\\1$^\dagger$} & 99-99.99\% & \makecell{\cite{3GPP_36777, 8470897}\\\cite{baltaci_trafficgeneratorpaper, 3GPP_22125}} \\
					\rule{0pt}{2.5em}\gls{BVLoS} & \makecell{5} & $<$300 & 99-99.99\% & \makecell{\cite{eurocae_rpaslatency, rtca_c2link }\\\cite{3GPP_22125, ngmn_urllc}\\\cite{baltaci_trafficgeneratorpaper}} \\
					\rule{0pt}{1.5em}FAO & $<$0.08 & 100-1000 & High & \cite{3GPP_22125} \\
					\hline
				\end{tabular}}
				\label{table: connectivityrequirements_uav_cnpc}
				\begin{tablenotes}[flushleft]
				        \scriptsize
				        \item \textit{$^\ast$Control and telemetry stream.}
				        \item \textit{$^\dagger$Video stream.}
				\end{tablenotes}
		\end{threeparttable}
		\end{center}
	\end{table}
\end{savenotes}

\paragraph{Vision-based Applications}
\glspl{UAV} can collect vision-based data such as the inspection of hardly accessible places or construction of critical infrastructures with high risk for workers \cite{9107776, 9142230, 8853298, 8756125, 10.5555/3237383.3237462}. \glspl{UAV} can also serve 3D mapping \cite{9059403, 9072891, 8949363} road traffic surveillance \cite{motlagh16A}, cinematography \cite{10.1145/3347713}, monitoring areas, or people \cite{9096075}, sports events \cite{wang17} and also search and rescue in case of natural disasters or calamities \cite{7463007}. 

\glspl{UAV} for vision-based application usually fly below 300 m \cite{3GPP_22125} with a low-speed or in a hovering condition to provide high-quality pictures, probably except for the events of news and sports \cite{wang17}. Operating in both urban and rural areas, their connectivity requirements solely depend on the quality of video stream.

\paragraph{Delivery Services}
Since the first successful cargo delivery in July 2015 \cite{lim15}, different industries such as grocery \cite{layne15}, postal services \cite{DHL_UAV}. Research regarding the delivery services focus on the monitoring systems \cite{7557769}, delivery scheduling \cite{8972296} and the optimized routing \cite{8972170, 9126258, 9014596}. 

In delivery services, \glspl{UAV} fly at low altitudes with a medium speed for an optimal delivery time. The authors of \cite{app10124362} presented a delivery service model using the rooftops of the buildings. Delivery services can demand connectivity for uploading destination waypoints and the remote tracking of the package \cite{DHL_UAV}.

\paragraph{IoT Data Collection}

\glspl{UAV} can collect also non-visual information such as sensor data about temperature, humidity, lighting or noise level \cite{motlagh17}. A \gls{UAV} can either sense the data directly with the onboard sensors or remotely collect it from distributed sensors \cite{9114970}. 

In order to achieve energy-efficient \gls{UAV} operations, a number of studies try to optimize the trajectory \cite{9054054, 9114970, 8842600}, minimize the data collection time \cite{9098850, 9098850, 8432487}, minimize the transmit power of \gls{IoT} devices \cite{9121255}, device positioning \cite{8894454}, wireless charging solutions to remote \gls{IoT} devices \cite{7036813, WPT_IoT}, providing edge computing to \gls{IoT} devices \cite{10.1145/3382756} or even underwater data collection \cite{9031732}. 

For the collection of non-visual data, \glspl{UAV} can operate at higher altitudes than \glspl{UAV} for vision-based applications, while the maximum operational speed can be similar. Their communication requirements are more lenient than that of vision-based applications since they transfer lower amount of latency-tolerant data. 

\paragraph{Agriculture}
Agriculture use cases owned 80\% of the global \gls{UAV} revenues in 2017 \cite{nixon17}. To enable \textit{smart farming}, \glspl{UAV} can collect field-level data such as plant count, soil H$_{2}$O level, temperature or imagery data for plant and animal monitoring \cite{8077607, 8959613, 9128695, 9140588}. \glspl{UAV} can also detect the growth of the plants \cite{8995212}, identify diseases in advance, reduce crop damage \cite{8805152}, support emergency situations that can harm the farms, such as wildfire \cite{9086204}, help with planting, crop spraying and irrigation \cite{8921148, 9107681}, fruit counting \cite{8934989, 8988024} and planned harvesting \cite{8943842}. All these use cases can advance agriculture in the years to come.

\subsubsection{Connectivity Requirements and Specifications of UAV Applications}
\label{subsec: usecases_uav_connectivityrequirements1}

We summarize the flight characteristics and the communication requirements of \gls{UAV} applications in \autoref{table: connectivityrequirements_uav_applications}. While the flying altitudes may go up to 300 m, the flying speed is mainly 10-15 m/s. We can realize the majority of the use cases in both urban and rural areas. 

As for communication requirements, vision-based use cases often require real-time data transmission and thus, the latency can be as low as 10 ms \cite{3GPP_22125}. In general, \gls{IoT} applications do not demand real-time transmission, and thus, we can relax the latency up to a range of a couple of seconds \cite{iot_datarate}. 

\subsubsection{Connectivity Requirements for CNPC}
\label{subsec: usecases_uav_connectivityrequirements2}

We categorize the \glspl{RPO} of \glspl{UAV} into three types: 

\begin{enumerate}
    \item \textbf{\gls{VLoS}}: A remote pilot operates the \gls{UAV} within his/her range of vision with a direct link to the \gls{UAV}.
    \item \textbf{\glsentryfull{BVLoS}}: A remote pilot operates the \gls{UAV} beyond his/her range of vision, mostly via multi-hop connectivity.
    \item \textbf{\glsentryfull{FAO}}: A \gls{UAV} can autonomously operate from the take-off until it reaches the destination, and a remote pilot/supervisor may take over the control when necessary \cite{Airbus_BluePrint}.
\end{enumerate}

We present the connectivity requirements for \gls{CNPC} of \glspl{UAV} in \autoref{table: connectivityrequirements_uav_cnpc}. The data rate demands depend on the application type. \gls{VLoS} operations may require data rates in the range of 5 - 150 kbps \cite{3GPP_36777, 8470897, baltaci_trafficgeneratorpaper} for control and telemetry traffic, depending on the type of \glspl{UAV} and the number of exchange parameters. We expect slightly less rates for \glspl{FAO} \cite{baiocchi15} since the rate and amount of control data exchange is lower. As for \gls{BVLoS} operations, the data rate demands on \gls{IL} can be high due to the video stream for the first-person view \cite{3GPP_22125}.  

The latency requirements of \gls{VLoS} and \gls{BVLoS} operations depend on the application scenario. While joystick-like piloting demand real-time communication, dynamic waypointing can relax this requirement. According to \gls{3GPP} reports, the latency requirements can vary 50-100 ms \cite{3GPP_22125, 3GPP_36777}, and \gls{RTCA} specifies 155 ms as the upper bound \cite{rtca_c2link}. However, the study of \gls{EUROCAE} assumes a rather relaxed latency threshold, which bounds to 300 ms \cite{eurocae_rpaslatency}. In \cite{baltaci_trafficgeneratorpaper}, where we experimentally modeled and measured the \gls{UAV} data traffic, we find 250 ms to be the upper bound. As for \gls{FAO}, the latency demand again depends on the type of operations. We must enable low-latency communication between an airborne controller and the remote pilot if ground components take role in flight decision processing. We can relax the latency threshold up to 5 s if the onboard controllers perform the flight-related decisions \cite{3GPP_22125}.

Remote piloting requires high communication reliability to ensure safe operations. Literature has not comprehensively evaluated the communication reliability requirements. While \gls{3GPP} states 99.9\% reliability requirement for \gls{CNPC} \cite{3GPP_36777}, \gls{RTCA} requires $>$99.976\% availability and $>$99.9\% communication continuity \cite{rtca_c2link}. 

Similar to our communication reliability analysis for \glspl{eVTOL}, we evaluate the  communication reliability requirements of \glspl{UAV} based on its \gls{DAL} specifications from \gls{EASA}, which is shown in \autoref{table: easa_uav_safetyobjectives} \cite{easa_SCLight-UAS}. From this table, we can expect the communication reliability requirements for \gls{VLoS} and \gls{BVLoS} operations to be 10$^{-3}$ - 10$^{-4}$, depending on the size and the \gls{MTOM} of the \gls{UAV}. In \glspl{FAO} scenarios, we can relax this requirement to the range 10$^{-2}$ - 10$^{-3}$.

All in all, we categorized the connectivity use cases of \glspl{AV} and identified their \gls{QoS} demands emerging from these use cases. The \gls{QoS} demands are diverse and stringent depending on the mission criticality of the particular application and thus, single technology may not be sufficient to meet the connectivity demands. In the next section, we will elaborate on the recent works regarding the capabilities of the wireless technologies to provide required \gls{QoS} meets of the aerial applications.
\section{Communication Technologies}
\label{section: wirelesstechnologies}

\begin{figure*}[t]
\begin{center}
	\centerline{\includegraphics[width=\textwidth,keepaspectratio]{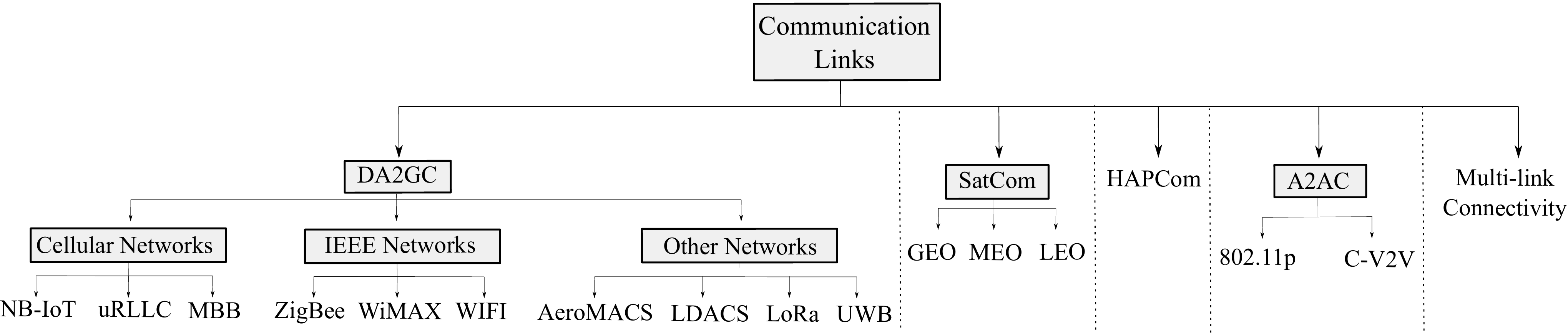}}
	\caption {Categorization of the communication technologies in this survey.}
	\label{fig: communicationtechnologies}
\end{center}
\end{figure*}

\gls{FACOM} networks utilize various communication technologies to satisfy diverse connectivity requirements as outlined in the previous section. In this section, we extensively review the literature regarding the studies that evaluated the capabilities of wireless communication technologies in the context of \gls{FACOM}. We categorize the wireless technologies as: 1) \gls{DA2GC}; 2) \gls{SatCom}; 3) \gls{HAPCom}; 4) \gls{A2A}. We group the studies according to the types of \glspl{AV} they studied: We first show the works related to airplanes, then \glspl{eVTOL}/helicopters and finally the \glspl{UAV} in each subsection. We illustrate the categorization of this section in \autoref{fig: communicationtechnologies}. 

\subsection{Direct Air-to-ground Communication}
\label{subsection: wirelesstechnologies_da2gc}

\gls{DA2GC} includes all the terrestrial technologies, which provide direct \gls{A2G} connectivity to the \glspl{AV}. We investigate \gls{DA2GC} under three categories: 1) cellular networks; 2) IEEE 802.11 networks; 3) other networks. We present the state-of-the-art studies that tackle various research topics regarding the wireless connectivity demands of \glspl{AV} in the following subsections. 

\subsubsection{Cellular Networks}
\label{subsection: wirelesstechnologies_da2gc_cellular}

The recent evolution in cellular networks brings along the possibility of deploying them for \glspl{AV}, as they inherited cutting-edge technologies to address the particular demands of different application scenarios. Especially, the emerging 5G technologies advertise prominent solutions toward facilitating the communication demands of diverse use cases. Network virtualization, network slicing and multi-access edge computing methods in 5G support flexible network architectures and traffic prioritization for mission-critical applications \cite{RysavyResearch}. Furthermore, \gls{3GPP} discusses topics related to the connectivity in the sky, such as enhanced 4G support for aerial vehicles \cite{3GPP_36777, 3GPP_38811}, remote identification and tracking of aerial vehicles \cite{3GPP_22125, 3GPP_23754}, 5G enhancement for \glspl{UAV} \cite{3GPP_22261, 3GPP_22829} and the application layer support \cite{3GPP_23755}. All these studies show the initial attempts to fit cellular technologies as a candidate solution for the connectivity demands in the sky. 

We show the summary of the recent works in the scope of cellular networks for aerial communication in \autoref{table: cellular_literaturereview}. We categorized these works following the 5G terminology and the potential networking challenges. The majority of the studies focus only on the use cases of \glspl{UAV} as the wireless connectivity applications of \glspl{eVTOL} are of the recent venture. 

\begin{savenotes}
	\begin{table*}[t]
		\caption {Categorization of the Literature regarding Cellular Networks for \gls{FACOM}. \vspace{-0.225cm}}
		\begin{center}
			{\renewcommand{\arraystretch}{1.5} 
				\begin{tabular}{l c l c}
					\hline
					\multicolumn{2}{l}{\textbf{Cellular Domain}} & \textbf{Subject} & \textbf{References}\\ \hline		
					\multirow{9}{*}{\makecell[l]{Network\\Services}} & & \makecell[l]{eMBB-baased \gls{A2G} network design and performance evaluation} & \cite{chen20205g, 8911712, 8030548, 7107747} \\
					& & \makecell[l]{Feasibility of MBB for \gls{A2G} connectivity of helicopters/\glspl{eVTOL}} & \cite{8384882} \\ 
					& & \makecell[l]{Evaluation of (e)MBB for CNPC via \gls{UAV} field measurements} & \cite{10.1145/3325421.3329765, 8746290, 8911712, 8858453, 8539117, 8568838} \\
					& & \makecell[l]{\gls{A2G} channel measurements and modeling for \gls{UAV} \gls{CNPC}} & \cite{8377373, 8891526, 8891422} \\
					& & \makecell[l]{Evaluation of MBB for \gls{UAV}-based video streaming} & \cite{9001975, 7396417} \\
					& & \makecell[l]{Suitability of MBB for \gls{UTM}-\gls{UAV} connectivity} & \cite{8102112, 8756738} \\ \cline{2-4}
					& \multirow{2}{*}{uRLLC} & \makecell[l]{Multi-operator connectivity to improve communication reliability} & \cite{8746579, 9145477} \\
					& & \makecell[l]{\gls{CoMP} to improve aerial coverage} & \cite{8998329, 9151343} \\ \hline
					\multirow{4}{*}{Technology} & \multirow{2}{*}{\makecell{Network\\Softwarization}} & \makecell[l]{SDN-based networking framework to support \glspl{UAV}} & \cite{8406959, 7983162} \\
					& & \makecell[l]{Performance evaluation of network slicing for \glspl{UAV}} &  \cite{8756738} \\ \cline{2-4}
					& \glsentryshort{mmWave} & \makecell[l]{Beamforming and beamtracking to improve \gls{UAV} \gls{A2G} channel quality} & \cite{8385471, 9149027} \\ \cline{2-4}
					& \multirow{2}{*}{\glsentryshort{MEC}} & \glsentryshort{MEC} for end-to-end latency optimization & \cite{9162928, 9093794} \\ 
					& & \makecell[l]{\glsentryshort{MEC} for the optimization of energy consumption of \glspl{UAV}} & \cite{9162853, 9093794} \\ \hline
					\multirow{13}{*}{\makecell[l]{Issues\\and\\Challenges}} & Coverage & \makecell[l]{Coverage analysis for \gls{UAV} \gls{A2G} connectivity} & \cite{9249665, 9149403, 9148841, 8998329, 8969181, 8692749} \\ \cline{2-4}
					& \multirow{5}{*}{\makecell{Trajectory\\Planning\\(TP)}} & \makecell[l]{\gls{UAV} \gls{TP} and optimization with coverage constraints} & \cite{6391922, 9149190, 9005434, 9013177, 9145194, 9014041} \\
					& & \gls{TP} to maximize \gls{A2G} channel quality & \cite{9253515} \\
					& & \makecell[l]{Joint \gls{TP} and resource allocation for \glspl{UAV}} & \cite{8685130, 8647671} \\
					& & \makecell[l]{\gls{UAV} \gls{TP} with the consideration of antenna radiation pattern} & \cite{10.1145/3376897.3377860, 9037325} \\
					& & \makecell[l]{\gls{TP} to optimize the energy consumption of \glspl{UAV}} & \cite{9162853, 9149835} \\ \cline{2-4}
					& \multirow{3}{*}{\makecell{Resource\\Allocation}} & \makecell[l]{Interference analysis and mitigation techniques for \gls{CNPC}} & \cite{8675384, 8647820, 8913615, 8964329, 7470934, Qualcomm_ltetrials, 9128682, 9099899, 9069873, 8976147, 8641388, 8906143, 8301389, 8369158} \\
					& & \makecell[l]{Aerial \glsentryshort{UE}-\gls{BS} association and handover schemes to mitigate interference} & \makecell{\cite{9238986, 9221328}} \\
					\rule{0pt}{2em}& & \makecell[l]{Evaluation of the coexistence of ground and aerial \glspl{UE}} & \makecell{\cite{9295084,  9261835, 9249665, 9151343}\\\cite{ 9129453, 8761670, 8692749, 8301389}} \\ \cline{2-4}
					& \rule{0pt}{2em}\multirow{3}{*}{\makecell{Antenna\\Design}} & \makecell[l]{Antenna design to maximize \gls{UAV} \gls{A2G} channel quality} & \makecell{\cite{9148841, 8969181, 10.1145/3376897.3377860, 8636798}\\\cite{8692749, 9145089}} \\
					& & \makecell[l]{3D beamforming and beamtracking for \gls{UAV} \gls{A2G} communication} & \cite{9296324, 9214955, 9093884, 8964589} \\
					& & \makecell[l]{\gls{IRS} to improve \gls{UAV} \gls{A2G} performance} & \cite{9356531, 9120632, 9145273} \\ \cline{2-4}
					& \multirow{2}{*}{\makecell{Frequency\\Spectrum}} & Spectrum demand for \gls{UAV} \gls{CNPC} & \cite{8902713} \\ 
					& & \makecell[l]{Spectrum reuse techniques to improve spectrum efficiency} & \cite{8493128, 9045290} \\ 
					\hline
				\end{tabular}}
				\label{table: cellular_literaturereview}
		\end{center}
	\end{table*}
\end{savenotes}

\paragraph{eMBB}
\label{subsection: wirelesstechnologies_da2gc_cellular_mbb}

\gls{eMBB} is the main network services of cellular networks with human-centric high-rate communication characteristics. While articles \cite{8911712, 7107747} experimentally study the 4G \gls{A2G} network performance for airplanes, the authors of \cite{8911712, 8030548, dinc17} theoretically evaluate the achievable \gls{DA2GC} capacity. In \cite{8030548}, the authors compare the performance of 4G and 5G-based \gls{DA2GC} as well as \gls{LEO} satellite using a 20 MHz channel bandwidth. 5G \gls{DA2GC} and \gls{LEO} satellites significantly outperform 4G with the help of multi-user beamforming and spot beam technologies. Furthermore, we show that large planar antenna arrays (500-1000 antenna elements) can provide an average 1.2 Gbps \gls{EL} to airplanes in the European airspace using 50-100 MHz bandwidth \cite{dinc17,dinc2020total}. 

The authors of \cite{7107747} experimentally evaluate the 4G \gls{A2G} performance on a research airplane and observed signal losses due to the directional antenna pattern. Although they measure a maximum of 40 Mbps on the \gls{EL}, which is sufficient for \gls{ATM} and \gls{SPO} use cases, a signal loss at 3 km implies even worse performance for over 10 km cruising altitudes of airplanes. In addition, another study proposes a deep-learning based 5G \gls{A2G} network design \cite{chen20205g}. It consists of two deep neural networks, where the first one approximates the network behavior with respect to the user data rate, and the second one optimizes antenna up-tilt angles as well as the inter-site distances to achieve the optimal data rate. \comment{We also performed 4G \gls{A2G} measurements using \glspl{UAV} and then scaled the results to the airplane scenarios \cite{8911712}. By using carrier aggregation and beamforming techniques, we extrapolated achievable data rates up to 257 Mbps at 13 km altitude.}


Cellular \gls{DA2GC} technologies gained maturity to be deployed commercially, and Gogo \cite{atg4} as well as Smartsky \cite{smartskywhite} provide \gls{DA2GC} in the North America. Gogo ATG-4 can achieve up to 9.8 Mbps per cell with more than 200 ground base stations. Deutsche Telekom, Nokia and Airbus tested a 4G-based ground station having a 100 km inter-site distance \cite{ecc,nokia}. According to their results, \gls{A2G} link at 2.6 GHz with 20 MHz provides 26-30 Mbps on the \gls{EL} and 17 Mbps on the \gls{IL} with less than 60 ms latency for an airplane at 10 km altitude with 800 km/h speed. However, \gls{DA2GC} requires increased spectrum resources to provide high achievable data rates. 


As for helicopters, the authors of \cite{8384882} propose a heterogeneous datalink architecture with 4G, \gls{VHF} and \gls{SatCom} to support \gls{ATM}, airline operations, and control-related tasks. \comment{They foresee 4G as feasible technology due to the low-altitude operation of helicopters.} A number of studies performed experimental measurements with \glspl{UAV} to evaluate the performance of \gls{eMBB} in urban and suburban environments. While the authors of \cite{8746290, 8911712, 8858453} analyze the data rate performance along with signal quality, the article, \cite{10.1145/3325421.3329765}, also consider the handover conditions at different altitudes and evaluate how the handovers influence the data rate performance. Measurements report the maximum average data rate to be 40 Mbps on the \gls{IL} and 20 Mbps on the \gls{EL}  \cite{10.1145/3325421.3329765}. Furthermore, \cite{10.1145/3325421.3329765, 8539117, 8568838} measure \gls{RSRP} and \gls{SIR} with \glspl{UAV} flying at different altitudes and find out that multipath in urban areas causes severe signal fluctuations \cite{8568838}. 

The authors of \cite{telecomperspective_lte5G} measure the latency performance of 4G networks at 2.6 GHz with a \gls{UAV}. While the average latency is 200-300 ms at 50-100 m altitude, the maximum latency is measured to be 2.5-3 s at 300 m. Such large latency values exceed the requirements for the remote piloting of \glspl{UAV} (\autoref{table: connectivityrequirements_uav_cnpc}) and show the performance degradation of current cellular networks at high altitudes. Differently, the authors of \cite{8858453} measure the average \gls{RTT} in the range 76-92 ms up to a 150 m altitude. 

Regarding \gls{A2G} channel measurements, the authors of \cite{8377373, 8891422} derive a model with pathloss exponent and shadow fading in urban scenarios for the altitudes up to 40 m. The results show that the paths between the \gls{UAV} and the \glspl{BS} become more clear as the altitude of the \gls{UAV} increases and the shadow fading is uncorrelated to the altitude. They further extend these studies by evaluating the feasibility of ray tracing models to predict the variations in the shadow fading \cite{8891526}. By comparing the ray tracing model with the field measurements, they conclude a well match between the shadowing predictions of the model and the actual measurements.  

\gls{eMBB} technology can also support the connectivity demands of \gls{UTM}. The authors of \cite{8102112} evaluate the suitability of 4G for \gls{UTM}-\gls{UAV} communication and simulations show that 4G can support up to 200 \glspl{UAV}/km$^{2}$ using a 5 MHz channel with a message delivery ratio of 95\% of 300 byte messages.

\paragraph{uRLLC}
\label{subsection: wirelesstechnologies_da2gc_cellular_urllc}

With the emergence of safety-critical robotic and machinery operations, the cellular ecosystem introduces \gls{uRLLC} to provide robust connectivity solutions. \gls{uRLLC} targets up to 10$^{-5}$ communication reliability of 32 bytes with 1-10 ms latency \cite{GSMA_5GGuide, qualcomm_urllc}. This technology is especially captivating for the \gls{BVLoS} \gls{RPO} in \gls{FACOM} \cite{8647652}, and the authors of \cite{9295186} discuss how the 5G ecosystem can provide reliable \gls{CNPC}. 

One of the methodologies to achieve \gls{uRLLC} is to utilize multi-connectivity. Introducing link or network diversity can provide improved latency and reliability performance. In this regard, articles \cite{8746579, 9145477}, evaluate the performance improvement from network diversity using a single \gls{UAV} connected to two different public 4G networks. We discuss these works in detail in \autoref{subsection: wirelesstechnologies_heterogeneous}.

\gls{CoMP} improves the network performance, especially at cell edges, by jointly coordinating multiple \glspl{BS} from the same network to serve a single \gls{UE}. In the context of \gls{FACOM}, we can consider \gls{CoMP} to increase the cellular coverage in the air with the aid of multi-\gls{BS} connectivity. In a recent study, \cite{8998329}, the authors achieve 32\% coverage improvement with a \gls{UAV} as an aerial \gls{UE} using \gls{CoMP} in their numerical analysis. Frequency allocation schemes between neighbor \glspl{BS}, and the orchestration of data transmission can also influence the \gls{ICIC}, which the authors of \cite{9151343} studied for \glspl{UAV} having equal distances to its serving \glspl{BS}. 

\comment{
\paragraph{mMTC}
\label{subsection: wirelesstechnologies_da2gc_cellular_nbiot}

Cellular networks recently added \gls{MTC} services to support the communication demands of low-rate, low-power machinery services. Thus, we consider this technology for the \gls{MTC} use cases of \glspl{AV}. Our recent studies, \cite{8911660, 8569540}, consider 4G \gls{MTC} technology for onboard airplane cabin applications. We evaluate the performance of 4G \gls{MTC} systems for low-rate \gls{WAIC} applications and find out that it can support the connectivity demands of non-safety critical applications \cite{8911660}. Additionally, we propose 4G \gls{MTC} for cabin monitoring and cargo tracking use cases onboard airplanes \cite{8569540}. 

We observe a gap in literature about cellular \gls{MTC} technologies for the use cases of \glspl{AV}. As airplanes have over thousands of sensors, future works should evaluate potential network architectures and supportable node density onboard cabin networks. Furthermore, analyzing the feasibility of supporting autonomous \gls{eVTOL}/\gls{UAV} operations as well as \gls{UTM} communication with \gls{MTC} are also relevant research topics in \gls{FACOM}. 
}

\paragraph{Network Softwarization}
\label{subsection: wirelesstechnologies_da2gc_cellular_networksoftwarization}

Softwarization recently gained attention in the cellular ecosystem as it can provide flexibility and adaptability to the networks for particular \gls{QoS} objectives. The \glspl{AV} have numerous use cases with diverse \gls{QoS} demands in \gls{FACOM}, and thus, network softwarization can become a key element to support the aerial applications. In literature, several works study the softwarization methods with \gls{SDN} architectures for monitoring and anomaly detection services of \glspl{UAV} \cite{8406959, 7983162}. Furthermore, the separation of control and user planes in the \gls{SDN} can help cellular networks provide robust communication. 

\begin{figure}[t]
\begin{center}
	\centerline{\includegraphics[width=0.5\textwidth,keepaspectratio]{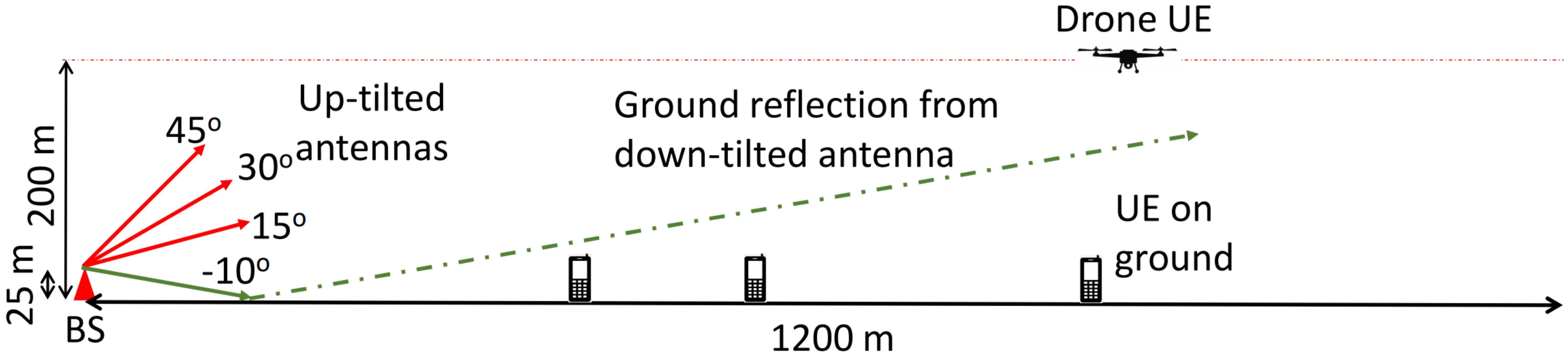}}
	\caption {Two-antenna \gls{BS} Model for the simulation study of \cite{9045290}. While the up-tilted antenna (one of the red lines) serves for aerial \glspl{UE}, down-tilted antenna (green line) serves both for ground and also for aerial \glspl{UE} with the ground reflection.}
	\label{fig: 9045290}
\end{center}
\end{figure}


Network virtualization techniques can be beneficial to provide guaranteed \gls{QoS} for the \gls{RPO} in \gls{FACOM}. We tested the resource isolation performance of network slicing with a single \gls{UAV} on a 5G testbed \cite{8756738}. Allocating dedicated resources for the \gls{CNPC} link of the \gls{UAV}, we show that the \gls{CNPC} slice maintain the data rate and the \gls{RTT} while we congest the payload slice in the same network. Furthermore, the authors of \cite{9183797} combine network softwarization with blockchain technology in multi-\gls{UAV} swarming scheme with the goal of increasing the network flexibility as well as data security. 

Overall, these studies show that we can employ softwarization in a diverse set of applications, and we expect prospective studies also to exploit the benefits of softwarization on the use cases of airplanes and \glspl{eVTOL}.

\paragraph{mmWave}
\label{subsection: wirelesstechnologies_da2gc_cellular_mmwave}

This technology exploits the spectrum over 6 GHz for capacity-demanding applications with stationary end devices. In aerial communication, \gls{mmWave} can be suitable for the massive data transfer scenarios of airplanes and \glspl{eVTOL} at air-/vertiports. Employing \gls{mmWave} can be challenging in other scenarios, where \glspl{AV} are in motion. Continuous beam alignment under 3D mobility requires sophisticated beamforming and beamtracking techniques. Nevertheless, a number of studies propose various beamforming techniques to provide stable \gls{A2G} connectivity \cite{8741314, 8385471, 9149027}. While the aim of \cite{8741314} is to jointly optimize the beamwidth and beampower of the \gls{EL} signal from \glspl{UAV}, the authors of \cite{9149027} develop a location-assisted beamforming technique to improve beam tracking and inter-beam interference cancellation. In addition, the article, \cite{8385471}, evaluates the beamtracking performance at 28 GHz for \gls{A2G} channels and finds out that the channel quality is stable using aligned antennas within a particular error margin.

In \cite{9045290}, the authors consider a ground \gls{mmWave} \gls{BS} model with two antennas to simultaneously serve aerial and ground \glspl{UE} with efficient spectrum sharing, as shown in \autoref{fig: 9045290}. Via simulation studies, they find out that the 30$\degree$ up and 10$\degree$ down as optimal antenna tilting angles to simultaneously serve ground \glspl{UE} and flying \glspl{UAV} at 200 m altitude. From our perspective, we consider the usage of \gls{mmWave} technologies challenging for \glspl{eVTOL} and \glspl{UAV} due to the difficulties of constant beam alignment, unless they follow deterministic flight routes (e.g. using corridors or following ground highways) like airplanes. Future works should also consider achievable data rates to offload high volume of flight data from \glspl{eVTOL} at vertiports. 

\comment{
\paragraph{LAA}
\label{subsection: wirelesstechnologies_da2gc_cellular_laa}

\gls{LAA} utilizes the unlicensed spectrum to boost the network capacity. Unlicensed spectrum is in use mostly in indoor environments; thus, \gls{LAA} can accommodate this spectrum for \glspl{AV} in outside environments without interfering with already-existing applications in the unlicensed spectrum. 

Cabin applications of airplanes/\glspl{eVTOL} can also employ \gls{LAA}. However, we must consider the coexistence scenarios with the \gls{WiFi} networks. We presented the performance of \gls{LAA} when it simultaneously operates with a varying number of \gls{WiFi} access points on a mockup cabin \cite{8569540}. When only \gls{LAA} is present, we measured 1.68\% \gls{PER}; however, the rate increases up to 45\% when 12 \gls{WiFi} access points are present in the cabin. Furthermore, we also proposed a seamless gate-to-gate connectivity architecture utilizing \gls{LAA} when an airplane is on the ground or below 357 m altitude \cite{8288087}. With numerical studies, we reported that an onboard 4G network utilizing \gls{LAA} can provide 5 Mbps data rate for 240 passengers. Future studies can also elaborate on employing \gls{LAA} for \gls{IFEC} services of \glspl{eVTOL}.
}

\paragraph{Mobile Edge Computing (MEC)}
\label{subsection: wirelesstechnologies_da2gc_cellular_mec}

\gls{MEC} hosts certain applications on the edge of networks to provide enhanced \gls{QoS} and to reduce network congestion. In \cite{9162853}, the authors aim at minimizing the energy consumption of \glspl{UAV} by partly offloading their tasks to a \gls{MEC}-enabled ground \glspl{BS}. Other studies rather utilize \glspl{UAV} to support the ground networks, such as \gls{MEC}-enabled \glspl{UAV} to meet the latency demands of ground \glspl{UE} \cite{9162928, 9093794}, or \glspl{UAV} as cache servers \cite{9133208}.

\paragraph{Coverage}
\label{subsection: wirelesstechnologies_da2gc_cellular_coverage}

The ground-centric nature of cellular networks and their major deployments being in urban regions bring challenges to provide a seamless coverage in \gls{FACOM}. Furthermore, the diverse altitudes of the flying platforms require a special consideration in the network design. In literature, a number of works study the coverage probability in the 3$^\text{rd}$ dimension to enable \gls{CoMP} transmission \cite{8998329}, to measure the network performance with different antenna configurations \cite{9148841, 8969181}, to evaluate the \gls{SINR} performance \cite{8368062} and to study the influence from the distribution of charging stations \cite{9153823}.

We can define the \textit{coverage probability} as the probability of receiving a \gls{SIR} or \gls{SINR} higher than a target threshold \cite{9148841, 8969181}. Comparing omni-directional, doughnut-shaped sine, doughnut-shaped cosine and directional antenna radiation patterns, the authors of \cite{8969181} conclude that the directional beam pattern with tilting angle toward ground \glspl{BS} provides the best \gls{SINR} coverage. They also find out that the optimal flight altitude to maximize the coverage depends on both the tilting angle of the antenna at the ground \gls{BS} as well as the multipath effects caused by the flight environment. 

The \gls{UAV} altitude and pathloss exponent play  significant roles in coverage probability \cite{8368062}. When \glspl{UAV} fly at 100 m altitude with an omnidirectional antenna, the authors of \cite{8692749} report a coverage drop from 76\% to 30\% compared with the ground \glspl{UE} according to their numerical analysis. However, using a directional antenna with the optimal tilting angle on \glspl{UAV} can increase the ratio from 23\% to 89\%. In \cite{9149403}, the authors provide algorithms to generate large-scale blockage and pathloss maps to locate poor coverage areas in the air. 

These studies show the severity of coverage problems in the air. Although they study the coverage scenario for \glspl{UAV}, the operations of \glspl{eVTOL} take place even at higher altitudes (up to 1 km \cite{nasa_uam}). In this case, we should also consider the trade-off between employing the existing ground infrastructure versus building dedicated \gls{A2G} networks. Although dedicated network architectures can significantly eliminate the coverage issues in the air, it requires a novel network design to provide an interference-free coverage at varying altitudes along with a costly network deployment. Furthermore, the \gls{ECC} mentions that the network operators do not intend to develop specific network planning for the use cases of \gls{FACOM} \cite{ecc_report309}. Therefore, future studies should also elaborate on the trade-off between the operating public versus dedicated aerial networks for \gls{FACOM}.

\paragraph{Trajectory Planning}
\label{subsection: wirelesstechnologies_da2gc_cellular_tp}

Trajectory optimization problems are applicable especially for \glspl{UAV}, as airplanes have predefined end-to-end flight paths. We expect also \glspl{eVTOL} to follow predefined trajectories. Path planning is vital to avoid coverage holes \cite{6391922} and to maximize the data rates by optimizing the resource allocation \cite{9149835, 8647671} as well as to maximize the \gls{A2G} channel quality for the routes of \glspl{eVTOL} \cite{9253515}.

Several studies utilize the graph theory with different purposes such as the trajectory design using radio maps \cite{9149190, 9013177}, the tradeoff between the trajectory length and connectivity outage ratio \cite{9037325} as well as the minimization of trajectory duration with a maximum tolerable outage duration \cite{9005434}. Moreover, a number of works propose machine learning-based algorithms to capture the dynamic 3D environment for \gls{UAV} Trajectory Planning. For instance, the authors propose deep \gls{RL} methods to plan the path of \glspl{UAV} based on its connectivity constraints \cite{9145194, 8871183}, to minimize \gls{UAV} mission completion time \cite{9014041} and to minimize interference to ground \glspl{UE} \cite{8654727}. 

\paragraph{Resource Allocation}
\label{subsection: wirelesstechnologies_da2gc_cellular_interference_resourceallocation}

Interference is the most investigated metric for \gls{UAV} communication due to the altitude dependency on the \gls{LoS} \gls{A2G} channels. Resource allocation is a fundamental methodology to control the interference and helps \glspl{UAV} and ground \glspl{UE} coexist in cellular networks. According to a recent study from \gls{3GPP} \cite{3GPP_36777}, current terrestrial networks can support the aerial \glspl{UE} up to 300 m altitude, if the percentile of the aerial \glspl{UE} remain below 33\% per cell. 

While the authors of \cite{8647820, 9128682} study the \gls{EL} transmission of \glspl{UAV} to estimate the probability of outage rate and to minimize the interference for increasing the sum-rate, another study, \cite{8692749}, analyzes the \gls{IL} coverage and spectral efficiency with the coexistence of \glspl{UAV} and ground \glspl{UE}.

Interference mitigation techniques play a key role to provide robust connectivity in \gls{FACOM}, and we can categorize these techniques as \cite{9069873, 8301389}: 
\begin{enumerate}
    \item Interference cancellation; 
    \item Inter-cell interference coordination; 
    \item Beam switching;
    \item Power control.
\end{enumerate}

In \cite{9099899}, the authors propose an interference cancellation scheme that adjacent \glspl{BS} cooperate with the co-channel \gls{BS} by not serving any \gls{UE} in the channel of \gls{UAV}  to cancel the interference. Similarly, they also propose a \gls{UAV}-assisted inter cell interference coordination scheme, where the \gls{UAV} senses the transmission of ground users to determine the allocation of resource blocks \cite{8976147}. Hence, they can avoid co-channel interference with the ground \glspl{UE}. 

In \cite{8913615}, the authors propose a cooperative beamforming technique at ground \glspl{BS} to mitigate the interference to co-channel transmissions from the \gls{BS} to the \glspl{UAV}. Moreover, the authors of \cite{9129453} jointly optimize the handover decisions and interference to ground \glspl{UE} by a Q-learning algorithm. Differently, the authors of \cite{8636798} employ directional antennas to improve the handover occurrences. 


Overall, we can observe the particular attention of literature to interference subject for cellular networks. This challenge is one of the main drawback of the cellular ecosystem to host the \gls{FACOM} applications with the already-existing infrastructure. Furthermore, the interference conditions may become even more problematic at higher altitudes (up to 1 km), where the \glspl{eVTOL} operate. Similar to coverage challenges, acquiring a dedicated network infrastructure for \gls{FACOM} can help avoid the interference problems in the air. The readers may refer to \cite{8675384, 8964329} for detailed surveys regarding the interference solutions. 

\paragraph{Antenna Design}
\label{subsection: wirelesstechnologies_da2gc_cellular_antennadesign}

Connectivity in \gls{FACOM} requires novel antenna design to enable reliable communication. The design of the antenna beam depends on the use case. For instance, while directional beam patterns can enable coverage at higher altitudes, we can consider rather omni-directional patterns for \gls{A2A} links for collision avoidance. In literature, the authors of \cite{9148841, 8969181, 10.1145/3376897.3377860} evaluate the performance of different antenna configurations, and the results reveal that the directional pattern provides the best \gls{SINR} performance to \glspl{UAV}. Furthermore, the authors of \cite{8636798} consider the directional pattern at both \gls{BS} and \gls{UAV} antennas and reported a handover reduction rate of 50-75\%. All these studies bring us to a conclusion that directional patterns can help improve the \gls{A2G} channel performance. 

Beamforming and beam tracking are relevant subjects in the antenna design. They can help alleviate the inter-cell interference issues introduced by direct \gls{LoS} links from multiple \glspl{BS}. However, unpredictable mobility of \glspl{UAV} pose challenges to track the beams. Therefore, the authors of \cite{9093884, 8964589} utilize the mobility information of \glspl{UAV} to form and track the beams. Furthermore, optimal antenna tilting is also an essential parameter in the antenna design \cite{8692749, 9145089}. 

Beamforming techniques can be practical also for the applications of \glspl{eVTOL}. Different than free route mobility patterns of \glspl{UAV}, \gls{eVTOL} may operate along flight corridors (e.g., along the ground highways) and their flight routes can be more deterministic. For instance, we can equip up-tilting antennas on the already-existing ground \gls{BS} towers along the highways to provide connectivity in the air.

\begin{figure}[t]
\begin{center}
	\centerline{\includegraphics[width=0.5\textwidth,keepaspectratio]{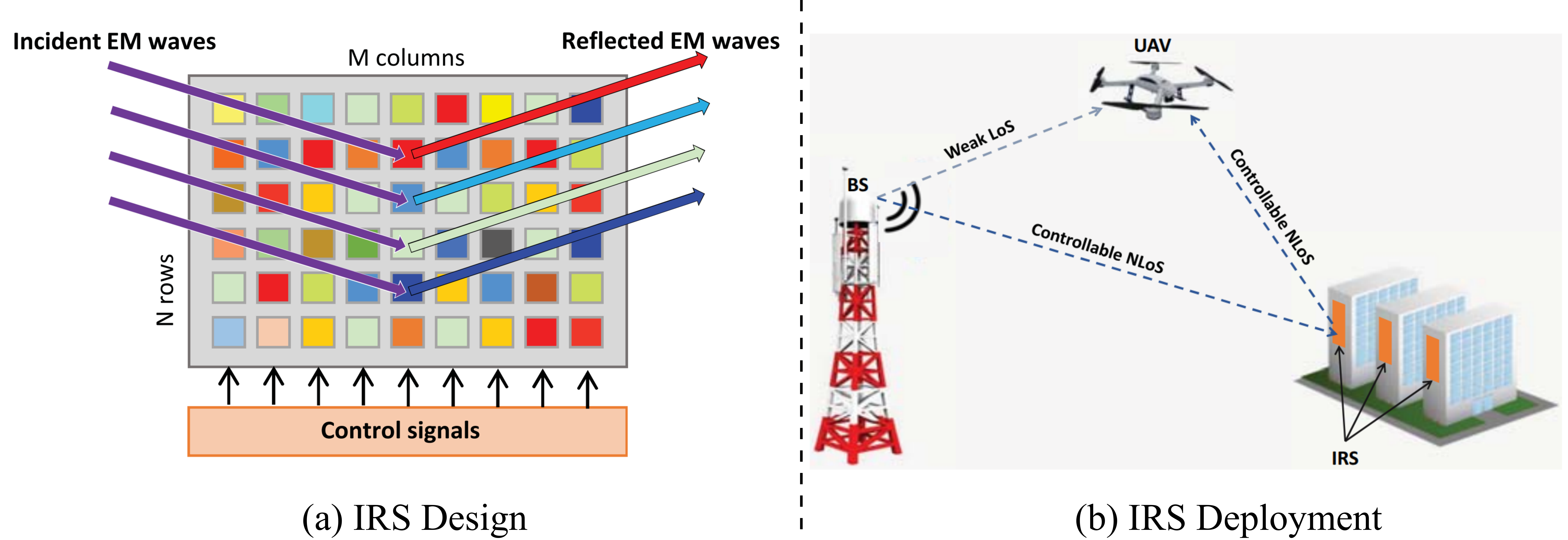}}
	\caption {Demonstration of \gls{IRS} by the authors of \cite{9120632}. In (a), they show a design of \gls{IRS}, where they control the absorption of an M x N size \gls{IRS} by feeding a control signal. In (b), they present a way to send configurable non \gls{LoS} signal by reflecting a signal from the \gls{IRS}-integrated building walls.} 
	\label{fig: 9120632}
\end{center}
\end{figure}

The \gls{IRS} technology recently gained attention in \gls{RF} systems. It is a reconfigurable surface in real-time in terms of electromagnetic absorption and reflections \cite{9120632, 9145273}. While the authors of \cite{9120632} evaluates the signal gain by deploying \gls{IRS} on building walls, the article, \cite{9145273} utilizes \gls{IRS} on \glspl{UAV} to increase the energy efficiency of the communication system. They achieve efficiency gain up to 50\% by optimizing the beamforming vector at the \gls{BS} and the phase shift matrix of the reflecting elements at the \gls{UAV}. 

\paragraph{Frequency Spectrum}
\label{subsection: wirelesstechnologies_da2gc_cellular_frequencyspectrum}

The cellular spectrum are licensed and regulated for the ground \glspl{UE}. Therefore, the aviation regulators must allow the usage of these bands for \gls{FACOM} services. \gls{LAA} technology operates under unlicensed spectrum, and it can be favorable due to the global harmonization of the unlicensed bands. However, we can employ unlicensed spectrum only for certain non-safety applications due to the lack of regulation. 

\gls{ITU} reported that 34 MHz is sufficient to enable the \gls{RPO} over terrestrial networks \cite{ITU_M2171}. As for airplanes and \glspl{eVTOL}, we can also consider spectrum sharing with mobile and fixed networks since they can follow predefined flight routes. In this regard, \gls{ECC} recently study the compatibility of aerial \glspl{UE} within the licensed bands of mobile and fixed networks up to an altitude of 10 km using a single aerial \gls{UE} \cite{ecc_report309}. While they conclude that regulations must limit the density of aerial \glspl{UE} along with no-fly zones to avoid interference to ground \glspl{UE}, they list several bands, in which aerial \glspl{UE} can coexist in the existing mobile and fixed networks. We present the list of these bands with the restrictions of their usage in \autoref{table: ecc_frequencybandsforaerialues}. 

\begin{savenotes}
	\begin{table*}[t]
		\caption {List of the Frequency Bands ECC Allowed Aerial \glspl{UE} to Operate in Mobile and Fixed Communication Networks \cite{ecc_report309}.\vspace{-0.225cm}}
		\begin{center}
			{\renewcommand{\arraystretch}{1.5} 
				\begin{tabular}{l l l}
					\hline
					\textbf{\makecell[l]{Frequency\\Range\\(MHz)}} & \textbf{Coexisting Service} &  \textbf{Restrictions} \\ \hline
					\rule{0pt}{2em}703-733 & \makecell[l]{- Broadcasting Receivers\\- \glspl{RAS}} &  \makecell[l]{- Aerial \glspl{UE} must fly $>$30 m above the ground.\\- No-fly zones or alternative measures around \gls{RAS} sites for aerial operations in\\700-713.5 MHz.} \\ \cline{2-3}
					\rule{0pt}{1.75em}832-862 & \makecell[l]{- Aeronautical Radio Navigation Services\\- \glspl{RAS}} &  \makecell[l]{- No-fly zones or alternative measures around \gls{RAS} sites for aerial operations in\\832-835 MHz.} \\\cline{2-3}
					880-915 & - Railway Mobile Radios & - N/A. \\\cline{2-3}
					\rule{0pt}{1.75em}1710-1785 & - Meteorological Satellites & \makecell[l]{- Emission limit of -40 dBm/MHz in 1675-1710 MHz for aerial \glspl{UE} operating in\\1710-1785 MHz.} \\\cline{2-3}	\rule{0pt}{4em}1920-1980 & \makecell[l]{- Mobile-Satellite Service\\\gls{CGC}\\Aeronautical Systems\\- Future Railway Mobile Communication\\System Cab-radio Receiver} & \makecell[l]{- Minimum separation distance of 15 km between \gls{CGC} \glspl{BS} and aerial \glspl{UE} operating\\ below 1980 MHz with out-of-band emission limit of -7 dBm/4.5 MHz.\\ - Minimum separation distance of 2.5 km between \gls{CGC} \glspl{BS} and aerial \glspl{UE} operating\\below 1980 MHz with out-of-band emission limit of -30 dBm/MHz (spurious).\\- Zero minimum separation between \gls{CGC} \glspl{BS} and aerial \glspl{UE} operating below\\1980 MHz with out-of-band emission limit of -30 dBm/MHz.} \\\cline{2-3}
					\rule{0pt}{1.75em}2500-2570 & \makecell[l]{- Radio Astronomy\\- \gls{ATM} Radars}&  \makecell[l]{- No-fly zones or alternative measures to protect the radars operating above 2700 MHz.}\\\cline{2-3}
					\rule{0pt}{1.75em}2570-2620 & \makecell[l]{- \glspl{RAS}\\- \gls{ATM} Radars} & \makecell[l]{- No-fly zones or alternative measures to protect the radars operating above 2700 MHz.} \\\cline{2-3}
					\rule{0pt}{2.25em}3400-3800 & \makecell[l]{- \gls{FSS}\\- Radiolocation Services\\- \glspl{RAS}} & \makecell[l]{- Separation distance of 26.7-290 km between \gls{FSS} earth stations and aerial \glspl{UE}.\\- Unwanted emissions of an aerial \gls{UE} limited to $<$-60 dBm/MHz.\\- Unwanted emissions of an aerial \gls{UE} limited to -50 dBm/MHz below 3400 MHz.} \\
					\hline
				\end{tabular}}
				\label{table: ecc_frequencybandsforaerialues}
		\end{center}
	\end{table*}
\end{savenotes}

To satisfy 99.9\% communication reliability requirements of the \gls{CNPC} of \glspl{UAV}, the authors of \cite{8902713} find out that a reservation of 1.4 MHz is sufficient for the current demand and can extend to 5 MHz spectrum in the next 20 years. Furthermore, the authors propose an aerial control system to increase the spectrum efficiency in the air in \cite{8493128}. Their scheme separates the control plane and the data plane of \glspl{UAV} and performs the control plane on \gls{A2A} links to allocate the empty spectrum to \gls{A2G} channels. Additionally, a recent study, \cite{9045290}, presents a network design with two antennas, dedicating one antenna per ground and aerial \glspl{UE}, to enable spectrum reuse in cellular networks. We further discuss the spectrum challenges in \autoref{sec: openresearchchallenges_spectrumregulation} in detail. 

\comment{
\paragraph{Vehicle Identification, Detection and Localization}
\label{subsection: wirelesstechnologies_da2gc_cellular_vehicleid}

Cellular networks can also provide localization and identification services for \glspl{AV}. These services are especially useful for \gls{UTM} to regulate the airspace. In this regard, \gls{3GPP} has already initiated a number of studies to provide such services using cellular network infrastructure \cite{3GPP_22825, 3GPP_23754}. In \cite{3GPP_22825}, they present the potential use cases and service requirements for remote identification of \glspl{UAV}. They expand this study later in Release 17 to present solutions to \gls{UAV}-related issues such as \gls{UAV}\&\gls{UAV} controller identification, tracking and authorization by the \gls{UTM}. 
The authors of \cite{8255736} utilize \glspl{mmWave} to detect small-size amateur \glspl{UAV}. In our study, \cite{9120588}, we investigate the localizability probability of \glspl{UAV} over cellular infrastructure and concluded that the localization performance is better at higher altitudes. 
}

\begin{savenotes}
	\begin{table*}[t]
		\caption {Categorization of the Literature regarding IEEE and Other Networks for \gls{FACOM}. \vspace{-0.225cm}}
		\begin{center}
			{\renewcommand{\arraystretch}{1.5} 
				\begin{tabular}{l l c}
					\hline
					\textbf{Wireless Technology} & \textbf{Subject} & \textbf{References}\\ \hline
					\multirow{4}{*}{\makecell[l]{WiFi}} & \makecell[l]{Evaluation of data rate performance onboard airplane cabin} & \cite{8417707} \\
					& \makecell[l]{Coexistence analysis with 802.11.4e onboard airplane cabin} & \cite{7503863} \\ 
					& \makecell[l]{Performance evaluation for \gls{UAV} \gls{A2A} communications} & \cite{7986413} \\
					& \makecell[l]{Performance evaluation to support live video streaming for \glspl{UAV}} & \cite{multicastvideostreaming_uav} \\ \hline
					WiMAX & \makecell[l]{Suitability for rescue and mission operations of \glspl{UAV}} & \cite{rahman14} \\ \hline
					\multirow{2}{*}{ZigBee} & \makecell[l]{Suitability for MTC communications onboard airplanes} & \cite{6928291, 7830305} \\
					& \makecell[l]{Connectivity for flight formation control of \glspl{UAV}} & \cite{7625728} \\ \hline
					\multirow{4}{*}{UWB} & \makecell[l]{Suitability for MTC applications onboard airplanes} & \cite{Ankit_WAIC} \\ 
					& \makecell[l]{\gls{A2G} channel measurements} & \cite{9119755, 8741964, 9153116, 9148768} \\
					& \makecell[l]{Analysis of electromagnetic interference to \gls{UAV} communications} & \cite{8986049} \\
					& \makecell[l]{Antenna design to lower air drag on \glspl{UAV}} & \cite{8824115} \\ \hline
					\multirow{5}{*}{LoRa} & \makecell[l]{Suitability for MTC communications onboard airplanes} & \cite{8661597} \\ 
					& \makecell[l]{Performance evaluation for V2X communications of \glspl{UAV}} & \cite{8471484} \\
					& \makecell[l]{Performance evaluation for \glspl{RPO} of \glspl{UAV}} & \cite{7593502} \\
					& \makecell[l]{Suitability for \gls{UAV} swarm communications} & \cite{8570043} \\
					& \makecell[l]{Coexistence study with \gls{ATM} radars} & \cite{8094705} \\ \hline
					AeroMACS & \makecell[l]{Performance evaluation for \gls{UAV} applications} & \cite{8453394} \\
					\hline
				\end{tabular}}
				\label{table: ieee_others_literaturereview}
		\end{center}
	\end{table*}
\end{savenotes}

\subsubsection{IEEE Networks}
\label{subsection: wirelesstechnologies_da2gc_ieee}

As IEEE technologies are vastly available on the market and exploit the unlicensed frequency spectrum, they can provide flexible and cost-efficient solutions in \gls{FACOM}. However, the unlicensed spectrum raises several issues regarding the communication reliability and certifiability due to aviation safety requirements. Additionally, IEEE protocols are mainly suitable for the applications with short communication distance and therefore, their applicability for \gls{BVLoS} operations can be challenging. We present the studies from literature with respect to the IEEE standards and we summarize the related works of IEEE as well as other \gls{A2G} technologies in \autoref{table: ieee_others_literaturereview}.

\paragraph{WiFi}
\label{subsection: wirelesstechnologies_da2gc_ieee_wifi}

\gls{WiFi} is currently the most-widely used technology for \gls{VLoS} \gls{UAV} operations, thanks to the unlicensed spectrum and high data rates in \gls{LoS} conditions. Conventional \gls{IFEC} systems onboard airplanes utilize \gls{WiFi} via Satellite backhaul \cite{8288087}. Number of works evaluate this technology for the future use cases in \gls{FACOM}. 

In regards to airplanes, the authors of \cite{8417707} compare the data rate performance of \gls{WiFi} with 4G and \gls{LiFi} onboard an aircraft cabin and found out that \gls{WiFi} provides the worst data rate performance. Another work evaluates the coexistence of \gls{WiFi} with a 802.11.4e \gls{WSN} in an aircraft cabin \cite{7503863} and proposes a whitelisted hopping method to increase the packet delivery ratio. As for \glspl{UAV}, another study, \cite{7986413}, evaluates the performance of \gls{WiFi} at 2.4 GHz for \gls{A2A} links. Their experimental results indicate that the use of \gls{WiFi} is not possible for \gls{A2A} applications due to the large number of remote controls working in the same band. The authors of \cite{multicastvideostreaming_uav} aim to enable live multicast video streaming over multiple \glspl{UAV} and thus, they develop \gls{RTP}-based rate-adaptive point-to-multipoint streaming framework using the IEEE 802.11a protocol. Experimental measurements show up to 30\% gain in the video quality compared with the legacy multicast. 
 

Emerging technologies in \gls{WiFi} can also alleviate the connectivity demands in \gls{FACOM}. IEEE 802.11 ah standard paves the way for exploiting sub 1 GHz spectrum with coverage rate up to 1 km for \gls{IoT} devices \cite{wifihalow}, which we can consider for low-rate non-safety \gls{MTC} in \gls{FACOM}. Recent IEEE 802.11 ax standard promotes over 6 Gbps data rates using a 160 MHz channel bandwidth. The 1024-\gls{QAM} scheme helps \gls{WiFi} achieve a high spectral efficiency in the unlicensed band. Additionally, the target wake time scheme enables a planned scheduling for power-constrained devices \cite{wifi6}. All these features in \gls{WiFi} standards keep the rigorous competition between \gls{3GPP} and IEEE technologies for the use cases of \gls{FACOM}. Furthermore, the upcoming IEEE 802.11 be standard even utilizes 4K-\gls{QAM} schemes to maximize achievable data rates using 320 MHz bandwidth. It also enables multi-link operations along with multi-band aggregation to support low-latency applications \cite{garciarodriguez2020ieee}. Therefore, IEEE standards can embrace the emerging \gls{uRLLC} use cases in the near term. 

Overall, current \gls{WiFi} solutions can be feasible mainly for the cabin applications of airplanes and \glspl{eVTOL}. In the case of safety-critical cabin applications, IEEE protocols require certain modifications to achieve higher reliability rates. Although the utilization of unlicensed spectrum leads to cost-efficient solutions, the available spectrum is scarce and therefore, the supportable number of applications are limited. 

\paragraph{WiMAX}
\label{subsection: wirelesstechnologies_da2gc_ieee_wimax}

\gls{WiMAX} is another option to provide a local network for \glspl{AV}. It can provide large coverage rates up to 50 km with a maximum of 70 Mbps and 28 Mbps data rates on the \gls{IL} and the \gls{EL}, respectively \cite{Song_2014, WiMAX_uldatarate}. It can be practical especially for the use cases of \glspl{UAV}, such as agriculture and disaster relief/public safety applications. In \cite{rahman14}, the authors considered the utilization of \gls{WiMAX} technology for connecting \glspl{UAV} in rescue and monitoring missions in Alpine environments. 

\paragraph{ZigBee}
\label{subsection: wirelesstechnologies_da2gc_ieee_zigbee}

ZigBee is designed for the \gls{MTC} applications and thus, a number of studies evaluated it for various \gls{MTC} use cases of \gls{FACOM}, such as the anomaly detection on the runways at airports in \cite{6928291}, hydraulic leakage detection on airplanes \cite{7830305} and flight formation control of \glspl{UAV} \cite{7625728}. 

In general, ZigBee can be a cost-efficient solution for the \gls{MTC} use cases in \gls{FACOM}; however, their short coverage range ($<$100 m \cite{element14_zigbee}) along with the use of unlicensed spectrum limits their deployment scenarios in \gls{FACOM}.

\subsubsection{Other Networks}
\label{subsection: wirelesstechnologies_da2gc_others}

\paragraph{Ultra-wide Band}
\label{subsection: wirelesstechnologies_da2gc_others_uwb}

\gls{UWB} technology diversifies from the aforementioned narrow-band technologies thanks to its anti-multipath capabilities and the low-power consumption \cite{8242220}. It can be suitable to support the communication demands of low-power applications onboard airplanes \cite{Ankit_WAIC} as well as \glspl{UAV}. The recent IEEE standard, \textit{802.15.4z}, propose enhanced positioning capabilities along with lower on-air transmission times \cite{8733537}. 



Several studies perform measurements to characterize the \gls{A2G} channels for \gls{UWB} at between 3.1-4.8 GHz \cite{9119755, 8741964} and at 6.5 GHz \cite{9153116, 9148768}. Furthermore, another study analyzes the effects of \gls{UWB} electromagnetic pulse on \gls{UAV} communication links \cite{8986049}. They state that the \gls{UAV} link can be exposed to the strong electromagnetic pulse interference and therefore, it is necessary to install a protection module on the \gls{RF} front-end. In this regard, \gls{UWB} systems also require special antenna design to accommodate them in \gls{UAV} environments. In \cite{8824115}, the authors present an antenna design with a 29 x 39 mm low-profile structure to lower the air drag during \gls{UAV} flights.


\paragraph{LoRa}
\label{subsection: wirelesstechnologies_da2gc_others_lora}

\gls{LoRa} is low-power low-rate technology designed with the consideration of long-range \gls{MTC} applications. An instance use case can be the vibration monitoring of aircraft structure, where the authors of \cite{8661597} deploy a multi-hop \gls{LoRa} network for this purpose. Another use case can be the \gls{V2X} communication of \glspl{UAV}. Utilizing \gls{LoRa}, the authors of \cite{8471484} manage to send data at 10 km range with 0 dBm transmission power, which is an energy-efficient transmission rate to cover such distance. 

Although \gls{RPO} use cases demand stringent communication reliability, authors of \cite{7593502} propose the combination of \gls{LoRa} and 3G modems to provide secure \gls{BVLoS} \gls{RPO} links. The authors of \cite{8570043} also consider \gls{LoRa} along with a low-latency \gls{MAC} layer for robust connectivity, but rather for the \gls{UAV} swarms. Their tests with 10 \glspl{UAV} and 20 robots lead to improved reliability than \gls{WiFi} with the allowance of high swarm density as well as larger coverage.

\gls{LoRa} operates in the 800 MHz frequency spectrum in Europe, in which the \gls{ATM} radars also coexist and therefore, the authors of \cite{8094705} evaluate this coexistence scenario. With an experimental evaluation, they suggest to increase the permissible interference level of dispatcher radars by 3 dB relative to the 20 dB$\micro$V/m for 859 MHz and above in order to ensure the coexistence.

\paragraph{AeroMACS}
\label{subsection: wirelesstechnologies_da2gc_others_aeromacs}

\gls{AeroMACS} provides safe and secure wireless communication in the C-band at the airports, and it can provide up to 18.2 Mbps and 6.8 Mbps on the \gls{IL} and the \gls{EL}, respectively \cite{AeroMACS_specifications}. \gls{WiMAX} forum investigates the possible global employment of this system as a flexible and scalable solution for the connectivity demands of \glspl{ATM} \cite{aeromacswimax}. In addition, \gls{AeroMACS} can work with the mobile satellite systems without creating significant interference \cite{wilson10}. In literature, the authors of \cite{8453394} evaluated \gls{AeroMACS} for the use cases of \glspl{UAV} and analyzed its \gls{OFDM} structure to evaluate its suitability. Their coherence time calculations showed that \gls{AeroMACS} can support mobility rates up to 130 km/h. 

\paragraph{LDACS}
\label{subsection: wirelesstechnologies_da2gc_others_ldacs}

\gls{LDACS} is an L-band terrestrial communication platform to provide a reliable connectivity service to future \gls{FACOM} use cases in the aviation safety spectrum. \gls{LDACS} has an \gls{OFDM}-based architecture and can achieve up to 1.43 Mbps using 500 kHz channel bandwidth \cite{ietf_raw_ldacs}. 

\gls{IETF} currently proposes an informational document about the \gls{LDACS} system \cite{ietf_raw_ldacs}. They define a 58.32 ms long frame and 250 ms super frame structures to provide robust connectivity. They aim to support the voice and data communication demands of the future civil aviation and they include not only \gls{A2G}, but also \gls{A2A} connectivity \cite{ietf_raw_ldacs}. Recent flight tests with \gls{LDACS} technology present the maturity of the technology for aviation \cite{ldacs_flighttest}. Nevertheless, \gls{LDACS} can also complement the satellite-based communication in the future \cite{ietf_raw_ldacs}. Finally, \gls{ICAO} also anticipates \gls{LDACS} as one of the future wireless technologies for \gls{A2G} communication \cite{icao_ganp}.   

\begin{savenotes}
	\begin{table*}[t]
		\caption {Categorization of the Literature regarding \gls{SatCom} and \gls{HAPCom} for \gls{FACOM}. \vspace{-0.225cm}}
		\begin{center}
			{\renewcommand{\arraystretch}{1.5} 
				\begin{tabular}{l l c}
					\hline
					\textbf{Wireless Technology} & \textbf{Subject} & \textbf{References}\\ \hline
					\multirow{9}{*}{\makecell[l]{SatCom}} & \makecell[l]{Antenna development studies for radar and aircraft-to-satellite communications} & \cite{8496839} \\
					& \makecell[l]{Beam tracking and channel design for \gls{AV}-to-satellite communications} & \cite{8286975, 7482031, 7026641, 9130836} \\
					& \makecell[l]{Phased-array antenna development for \gls{AV}-to-satellite communications} & \cite{8713061, Catalani_2020} \\
				    & \makecell[l]{Achievable throughput analysis in \gls{A2G} communications} & \cite{8761108} \\
				    & \makecell[l]{Impact of the antenna view angle on airplane-to-satellite communications} & \cite{5286353} \\
					& \makecell[l]{Development of random linear coding for airplane-\gls{GEO} satellite connectivity} & \cite{8690876} \\
					& \makecell[l]{L-band channel model development for \gls{AV}-to-satellite communications} & \cite{8886850} \\ 
					& \makecell[l]{Development of novel methods to avoid satellite link blockage from helicopter blades} & \cite{8587496, 8587471, 8924539} \\ 
					& \makecell[l]{Link adaptation techniques for \gls{AV}-to-satellite communications} & \cite{7601547} \\ \hline
					\multirow{3}{*}{\makecell[l]{HAPCom}} & \makecell[l]{\glsentryshort{HAPS} to ground RF propagation modeling studies} & \cite{9391325, 9391216, 9391285} \\
					& \makecell[l]{Antenna and beamforming design to support airplane-\glsentryshort{HAPS} communications} & \cite{9378524} \\
					& \makecell[l]{Design of a novel \glsentryshort{MAC} layer to enable joint \gls{FANET} and \gls{HAPCom}} & \cite{6581252} \\
					\hline
				\end{tabular}}
				\label{table: satcom_hapcom_literaturereview}
		\end{center}
	\end{table*}
\end{savenotes}

\paragraph{ADS-B}
\label{subsection: wirelesstechnologies_da2gc_others_adsb}

\gls{ADS-B} is an \gls{RF}-based surveillance system utilized in various aerial applications such as situational awareness, visual separation and inter-aircraft spacing \cite{faa_adsb}. \glspl{AV} broadcast their position information (latitude, longitude), altitude, velocity, \gls{AV} identification and other information to inform the other nearby \glspl{AV} \cite{faa_adsb}. \gls{ADS-B} is part of the NextGen program, which aims to modernize the U.S. air transportation system \cite{aopa_adsb}, and which recently began to mandate the installation of \gls{ADS-B} systems on airplanes \cite{aopa_adsb, faa_adsb}. \gls{ADS-B} operates at 1090 MHz carrier frequency, reaching 1 Mbps data rate with a frame size of 56 or 112 bits \cite{kim16, icao_adsb}. \gls{ADS-B} can have transmission ranges up to 474 km \cite{sesar_adsb}. \gls{ADS-B} has two components: 1) ground terminal; 2) air terminal. Ground terminals are passive receivers and collect all \gls{ADS-B} signals coming from different airplanes in the transmission range. By receiving its location from the onboard \gls{GPS} system, airplanes broadcast their position, identity, altitude and velocity to the ground station \cite{sesar_adsb}. Furthermore, air terminals communicate with bidirectional \gls{ADS-B} signals. Airplanes periodically transmit the broadcast \gls{ADS-B} signals, usually at a rate of 2 Hz \cite{icao_adsb}. 

\gls{ADS-B} systems lately gain popularity and they can also provision collision avoidance systems for \glspl{eVTOL}. As current \gls{ADS-B} hardware challenges the \gls{SWaP} requirements of \glspl{eVTOL}, uAvionix recently developed a modular \gls{ADS-B} terminal to integrate this technology into \glspl{eVTOL} \cite{uavionix_adsb}. Although \glspl{UAV} can also host \gls{ADS-B} systems \cite{mitre_adsb}, the \gls{ADS-B} spectrum becomes saturated due to the excessive number of \glspl{UAV} in the sky, limiting its scalability. Our approach to overcome this challenge is to utilize a ground based sense and avoid mechanism such that \glspl{UAV} utilize cellular connectivity to report their position to a ground \gls{BS}, which informs the other \glspl{AV} and the \gls{ATM} \cite{8904324}. 

The \gls{FAA} regulates the mandatory use of \gls{ADS-B} systems on airplanes since the beginning of 2020 \cite{ADSB2020}. Other variants of \gls{ADS-B} systems are also under investigation, such as a multi-channel \gls{ADS-B} receiver to enhance satellite-based airplane surveillance \cite{8714514} as well as an \gls{ADS-B} receiver payload design for nanosatellites to increase the current coverage of \gls{ADS-B} systems \cite{8712093}. Thus, \gls{ADS-B} is one of the key enabler technologies for collision avoidance use case of future airplanes.  

\subsection{Satellite Networks}
\label{subsection: wirelesstechnologies_satellite}

In this section, we begin by providing a background on the \gls{SatCom}. Followed by the state of the art in the satellite services, we then discuss some of the challenges utilizing \gls{SatCom} to support the connectivity use cases of \gls{FACOM}. We present the summary of the literature review of \gls{SatCom} in \autoref{table: satcom_hapcom_literaturereview}.

\subsubsection{Background}
\label{subsection: wirelesstechnologies_satellite_background}

We can classify the satellites operations with respect to their frequency spectrum, as shown in \autoref{table: satellite_frequencybands} \cite{inmarsat_satfrequencies}. While \gls{GPS} carriers and mobile phone satellites operate on the L-Band, the S-Band is used for radars, weather systems as well as \gls{SatCom} \cite{8496839, inmarsat_sband_satellite}. The satellite TV network occupies the C-Band since it is less susceptible to rain-fade than Ku-Band due to longer signal wavelength. The spectrum is less congested at Ka-band, whereas, it is a major constraint for satellite operators at C- and Ku-band \cite{9124728}. Ka-band satellites are highly efficient with a lower bandwidth cost \cite{9124728}. As the available spectrum is limited in lower frequency bands, satellite operators began to investigate the Q/V band to provide high-rate broadband services \cite{10.1117/12.2536091}. For instance, Hughes network aims to launch its JUPITER 3 system, which can provide up to 500 Gbps using Ka-, Q- and V- band \cite{hughes_qvband}. However, the main challenge of Q/V band is the high sensitivity to the atmospheric conditions, such as rain \cite{hughes_qvband}.  


The size and the weight of user terminals are essential parameters in \gls{FACOM} due to stringent \gls{SWaP} requirements of \glspl{AV}. Large antennas, as specified in \autoref{table: satellite_frequencybands}, can be challenging to deploy in \glspl{eVTOL} as well as \glspl{UAV}. Additionally, satellites terminals are usually power-starving due to the large communication distances. For instance, current Inmarsat satellite terminals for machine-to-machine communication can demand up to 50 W during data transmission \cite{inmarsat_powerconsumption}. This is another drawback to realize \gls{SatCom} in \gls{FACOM} due to the limited onboard power resources, especially on \glspl{eVTOL} and \glspl{UAV}. Finally, although we find more available spectrum at \glspl{mmWave}, it further increases the power requirements to compensate for the link budget. The 3D mobility of the \glspl{AV} requires advanced beamtracking and beamsteering techniques in \gls{mmWave} scenarios. 

We can classify the types of satellite handovers as follows \cite{4062836}: 

\begin{itemize}
    \item \textbf{Link Layer Handovers}: It occurs when the \gls{UE} switches one or more links between the communication endpoints due to the dynamic connectivity patterns of \gls{LEO} satellites.
    \begin{enumerate}
        \item \textbf{Spotbeam Handover}: It takes place when the \gls{UE} crosses the boundary between the neighboring spotbeams of a satellite, an intra-satellite or spotbeam handover occurs.
        \item \textbf{Satellite Handover}: It happens when the satellite transfers one of its existing connection of a \gls{UE}.
        \item \textbf{\gls{ISL} Handover}: This type of handover happens when interplane ISLs are temporarily off due to the change in distance and viewing angle between satellites in neighboring orbits. It causes rerouting of the current connection, which results in a handover.
    \end{enumerate}
    \item \textbf{Network Layer Handovers}: It occurs when one of the \glspl{UE} changes the \gls{IP} address due to change in coverage or mobility. 
    \begin{enumerate}
        \item \textbf{Hard Handover}: It is the \textit{break-before-make} scheme, where the previous link is released before establishing the connection to the new link. 
        \item \textbf{Soft Handover}: This is the \textit{make-before-break} scheme, where the previous link is released after connecting to the new link. 
    \end{enumerate}
    \item

\begin{savenotes}
	\begin{table}[t]
		\caption {Satellite Frequency Bands and Their Characteristics for Civilian Use \cite{inmarsat_satfrequencies, satellite_bands, hps_qbandantennasize}.\vspace{-0.225cm}}
		\begin{center}
			{\renewcommand{\arraystretch}{2.25} 
				\begin{tabular}{l c c c}
					\hline
					\textbf{\makecell[l]{Frequency\\Band}} & \textbf{\makecell[c]{Frequency\\Range\\(GHz)}} & \textbf{\makecell[c]{Available\\Spectrum\\(MHz)}} & \textbf{\makecell[c]{Antenna Type\\and Diameter\\(m)}} \\ \hline
					L-band & 1-2 & 15 & \makecell{Omnidirectional\\$<$0.2-0.6}\\
					S-band & 2-4 & 70 & \makecell{Omnidirectional\\0.2-0.6} \\
					C-band & 4-8 & 500 & \makecell{Directional\\$>$1.8}\\
					Ku-band & 12-18 & 500 & \makecell{Directional\\0.9-1.2}\\
					Ka-band & 26-40 & 3500 & \makecell{Directional\\0.25-1.2}\\
					Q/V-band & 40-75 & $>$5000 & \makecell{Directional\\0.8-1.2}\\
					\hline
				\end{tabular}}
				\label{table: satellite_frequencybands}
		\end{center}
	\end{table}
\end{savenotes}

\end{itemize}

The effects of these handovers on the connectivity are important to evaluate potential service interruption scenarios in \gls{FACOM}. As the communication reliability is an essential metric for \gls{CNPC}, we should observe different ways to mitigate the effects of handovers, such as \textit{make-before-break} method to avoid interruption before a handover takes place.

\subsubsection{Satellite Constellations}
\label{subsection: wirelesstechnologies_satellite_constellations}

In this part, we provide the characteristics of the satellite constellations. We start with the capabilities of \gls{GEO} and \gls{MEO} constellations presenting their advantages and drawbacks for the use cases of \gls{FACOM}. Afterwards, we discuss the recent advancement in \gls{LEO} constellations along with the upcoming \gls{LEO} services. We present the communication capabilities of these constellations in \autoref{table: satellite_performance}.

\paragraph{GEO}
\label{subsection: wirelesstechnologies_satellite_constellations_geo}

The \gls{GEO} constellations are located 35786 km away from the Earth’s surface at the geosynchronous orbit to match Earth's rotation. Its frequency reuse factors are low due to the large beam coverage and thus, it provides low-data rate per user within the beam \cite{8761108}. However, \gls{GEO} satellites compensate this by utilizing Ka-band to achieve lower beam width, and thus less beam coverage. ViaSat and Hughes are the instances of \gls{GEO} broadband service providers \cite{ASatTELEBBS}. With large latency characteristics, \gls{GEO} broadband services can normally provide below $<$1 Mbps, with certain exceptional high-rate plans \cite{inmarsat_servicescomparison}. 

\gls{GEO} links with usually in 700-800 ms \cite{inmarsat_servicescomparison} latency fail to serve mission-critical applications in \gls{FACOM} that need low end-to-end latency. Additionally, the typically large and heavy \gls{GEO} airborne terminals (up to 30 kg \cite{inmarsat_satfrequencies}) challenge \gls{SWaP} requirements of \glspl{AV}. Large-scale antennas are problematic for the low air drag requirements, and the connectivity cost is also high. A satellite terminal can cost 50-200 thousand \$ for the aeronautical use \cite{inmarsat_satfrequencies}. This further limits its deployment in large-scale. Nevertheless, the global coverage is its main advantage, and we can consider \gls{GEO} as a primary link in certain applications.

\paragraph{MEO}
\label{subsection: wirelesstechnologies_satellite_constellations_meo}

\gls{MEO} constellations serve at the altitudes between \gls{LEO} and \gls{GEO} constellations. Although they are the least common constellation types, a number of upcoming satellite services occupy the space such as the new O3b, Audacy, Methera Global, Laser Light, ESA and CNSA constellations \cite{Curzi_2020}. Current O3b networks utilize the Ka-band, and their beam coverage is around 700 km in diameter \cite{o3bnetworks}. The end-to-end latency can be around 130 ms \cite{o3bnetworks2}, which is a reasonable range for particular scenarios in \gls{FACOM}. In \gls{FACOM}, we can consider \gls{MEO} to find an optimal balance between latency and coverage, although their large \gls{UE} terminals can be still problematic to be integrated into \glspl{eVTOL} and \glspl{UAV}. 

\comment{
\paragraph{Little LEO}
\label{subsection: wirelesstechnologies_satellite_constellations_littleleo}

The term, \textit{little}, specifies the \gls{LEO} systems, which operate below 1 GHz \cite{itu_littleleo}. They provide data services \cite{itu_littleleo} and their capacity varies 4.8-24 kbps on the \gls{IL} and 2.4-9.6 kbps on the \gls{EL} for Orbcomm and \gls{LEO} One \cite{LEOSATOV}. They operate at \gls{VHF} and UHF bands \cite{itu_littleleo} without \gls{ISL}. In \gls{FACOM}, they can be favorable as they operate at VHF/UHF bands and they can support the low-rate digital messaging services of \gls{ATM}/\gls{UTM}.

\paragraph{Big LEO}
\label{subsection: wirelesstechnologies_satellite_constellations_bigleo}

Big \gls{LEO} satellites provide voice and limited data services above 1 GHz, such as Iridium and Globalstar \cite{MITISSCSATCons, LEOSATOV}. They can support up to 9.6 kbps voice and 7.2 kbps data services and the constellations are at the altitude 780-1410 km \cite{LEOSATOV}. Big \glspl{LEO} utilize multi-beam technology with 16-48 beams per satellite \cite{LEOSATOV} and may use \glspl{ISL}. We can consider this technology for the voice communication demands of the pilot onboard \gls{eVTOL} operations.  
}
\paragraph{LEO}
\label{subsection: wirelesstechnologies_satellite_constellations_broadbandleo}

\begin{figure}[t]
\begin{center}
	\centerline{\includegraphics[width=0.5\textwidth,keepaspectratio]{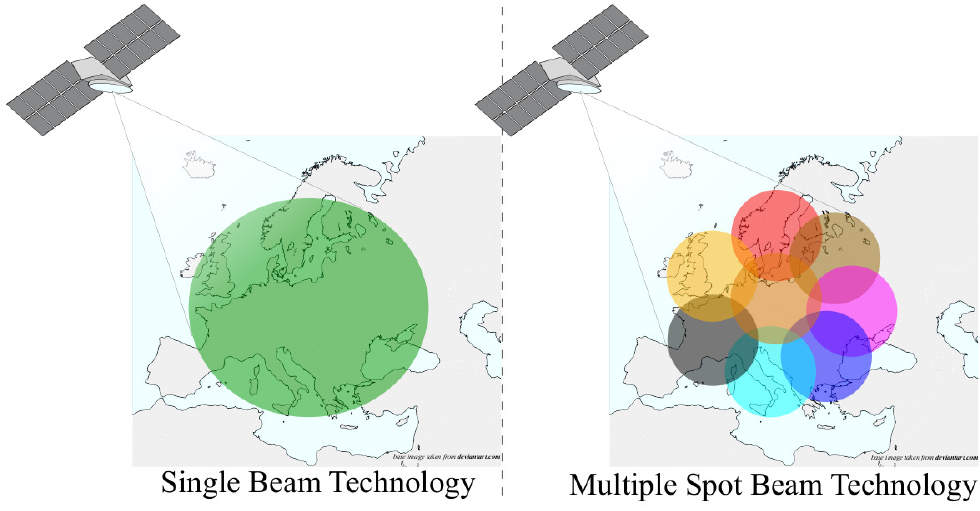}}
	\caption {Comparison of single beam satellites with high throughput multiple spot beam satellites. The conventional single beam operations provide seamless large coverage; however, multiple beams promote user-centric, \gls{QoS}-aware services.}
	\label{fig: satellite_hts}
\end{center}
\end{figure}

\gls{LEO} satellites occupy the low-earth orbit with massive constellations to globally provide high-speed data services such as Starlink \cite{McDowell_2020}, OneWeb \cite{oneweb} and Telesat \cite{DELPORTILLO2019123}. The number of \gls{LEO} satellites can be as high as 12000 for Starlink constellations \cite{McDowell_2020} and the constellation altitude for OneWeb is at 1200 km \cite{oneweb}. OneWeb promises to provide up to 400 Mbps per user \cite{oneweb}. Starlink and OneWeb majorly operate at the Ka/Ku-band utilizing a channel bandwidth of 250 MHz \cite{DELPORTILLO2019123}. 

\begin{savenotes}
	\begin{table*}[t]
		\caption {Networking Capabilities of the Satellite Constellations.\vspace{-0.225cm}}
		\begin{center}
			{\renewcommand{\arraystretch}{1.75} 
				\begin{tabular}{l c c c c c c}
					\hline
					\textbf{Constellation} & \textbf{\makecell{Altitude\\(km)}} &  \textbf{\makecell{Coverage\\(km)}} & \textbf{\makecell{User/System\\Data Rate (Gbps)}} & \textbf{\makecell{Latency\\(ms)}} & \textbf{\makecell{Communication\\Reliability}} & \textbf{References}\\ \hline
					\gls{GEO} & 36000 & 4000 & $<$0.00085/300 & 700-800 & 99.5\% & \cite{viasat_geocapacity, 8935306, ses_meo, inmarsat_servicescomparison, geo_availability, inmarsat_satfrequencies} \\
					\gls{MEO} & 2000-20000 & 700 & $<$0.8/84 & 130 & 99.9\% & \cite{o3bnetworks, o3bnetworks2, ses_meo, inmarsat_satfrequencies, meo_datarate} \\
					\makecell[l]{\gls{LEO}} & 100-2000 & 450 & 0.75/7000 & 10-35 & 99.9\% & \cite{8933127, 8700141, iridiumdataplan, ISNOV, inmarsat_satfrequencies, LEOSATOV} \\
					\hline
				\end{tabular}}
				\label{table: satellite_performance}
		\end{center}
	\end{table*}
\end{savenotes}

Most of the broadband \gls{LEO} constellations adopt multiple spot beam technology. This enables the \gls{HTS} concept based on a frequency reuse and high directive beam spot, as shown in \autoref{fig: satellite_hts}. The number of spot beams per satellite is an essential factor in deciding the satellite delivered capacity. The number of spot beams per satellite can be as high as 50 (SkyBridge) with an average diameter of 2000 km \cite{spacexnongeo}. \gls{HTS} is ideal for point-to-point services \cite{inmarsat_satfrequencies}. One disadvantage of \gls{HTS} is small beam elevation angles at high altitudes, which cause performance degradation \cite{8700141}. OneWeb and SpaceX satellites may also be capable of steerable beams to enable user-centric services. Finally, OneWeb and SpaceX aim to enable user terminals with phased-array antennas with the minimum size of 30 cm \cite{onewebnongeo, spacexnongeo}.

In \gls{FACOM}, we can list several advantages of employing \gls{LEO} constellations. The end-to-end latency rates are similar to that of conventional ground networks and thus, it can play a critical role to support the piloting applications of all  \glspl{AV}. Furthermore, small and light \gls{UE} terminals with the possibility of flat array antennas can be suitable to the \gls{SWaP} requirements of the \glspl{AV}. Upcoming massive constellations promise high-rates with cost-efficient services and hence, they can be a part of the \gls{FACOM} ecosystem to support high-rate and mission-critical services.

The main disadvantage of \gls{LEO} is the rapid handover rate (as low as a few minutes \cite{8700141}), which can frequently interrupt the connectivity and reduce the overall network availability. Thus, minimizing the service interruption time is one of the main challenges to tackle to integrate \gls{LEO} in \gls{FACOM}. For instance, phased-array antennas can mitigate this condition by enabling simultaneous connectivity to different spot beams before a handover takes place on the main beam.

The constant high-speed movement (i.e., 7 km/s) of the \gls{LEO} satellites also pose challenges to provide reliable \glspl{ISL}, which also introduce Doppler effects \cite{8700141}. Additionally, the limited lifespan of \gls{LEO} satellites, which can be 5-15 years \cite{leo_lifeexpectancy}, brings doubts on the service continuity on a long term. 

\subsubsection{Challenges and Future Solutions}
\label{subsection: wirelesstechnologies_satellite_challenges_futuresolutions}

The constellation altitude and the number of satellites covering the Earth affect supportable node density in the covered area. We can achieve a higher satellite capacity while maintaining a wide coverage by increasing the number of spot beams as well as the frequency reuse factor. However, increasing the number of satellites and the spot beams result in a higher number of satellite handovers \cite{CHOLEOSATMNET}. In this case, we can employ beam tracking technologies to reduce the handover rates, such as the blind method \cite{8286975}, step-tracking \cite{7482031} as well as multi-mode-monopulse methods \cite{7482031}. Similarly, the authors of \cite{7026641} develop an antenna tracking system for \glspl{UAV} on the Ka-band, and the authors of \cite{9130836} investigate an algorithm for recursive 3D channel tracking for the \gls{UAV}-\gls{LEO} connectivity. 

The emerging phased-array antenna solutions in \gls{LEO} constellations can be promising for \gls{FACOM}. In this regard, several works developed phased-array airborne antennas to enable \gls{SatCom} at the Ka-band \cite{8713061} and the Ku-band \cite{Catalani_2020}. Additionally, Gogo claims that it can provide up to 100 Mbps with steerable phased array antennas, which are suitable for airplanes \cite{2kuGOGO, ElbertABCA2KU}. Furthermore, antenna view angles from airplane to \gls{GEO} satellites also impact the overall system performance, as investigated in \cite{5286353}.

The large end-to-end latency rates of \gls{GEO} constellations make it impractical for the time-critical use cases of \gls{FACOM}, and retransmissions further worsen the condition. Thus, the authors of \cite{8690876} propose using random linear coding instead of retransmissions to avoid overhead latency on airplane communication. We can utilize \gls{LEO} satellites for time-critical applications as its latency rates are in the same magnitude to that of \gls{DA2GC} networks \cite{8700141}. Furthermore, certain applications may demand high-rates such as the \gls{RPO} as well as the vision-based applications of \glspl{UAV}. 

As for helicopter-to-\gls{SatCom}, the blades of helicopters cause a periodic blockage, which worsens the \gls{RF} channel. Several works introduce different methods to tackle this problem, such as Walsh-Hadamard code division multiplexing \cite{8587496}, time diversity with a novel channel estimation scheme \cite{8587471}, and also time diversity but with maximal ratio combining as well as zero-forcing \cite{8924539}. These studies are beneficial for the communication between \glspl{eVTOL} and satellites since their rotor blades can also influence the \gls{SatCom}.

\begin{figure*}[t]
\begin{center}
	\centerline{\includegraphics[width=0.75\textwidth,keepaspectratio]{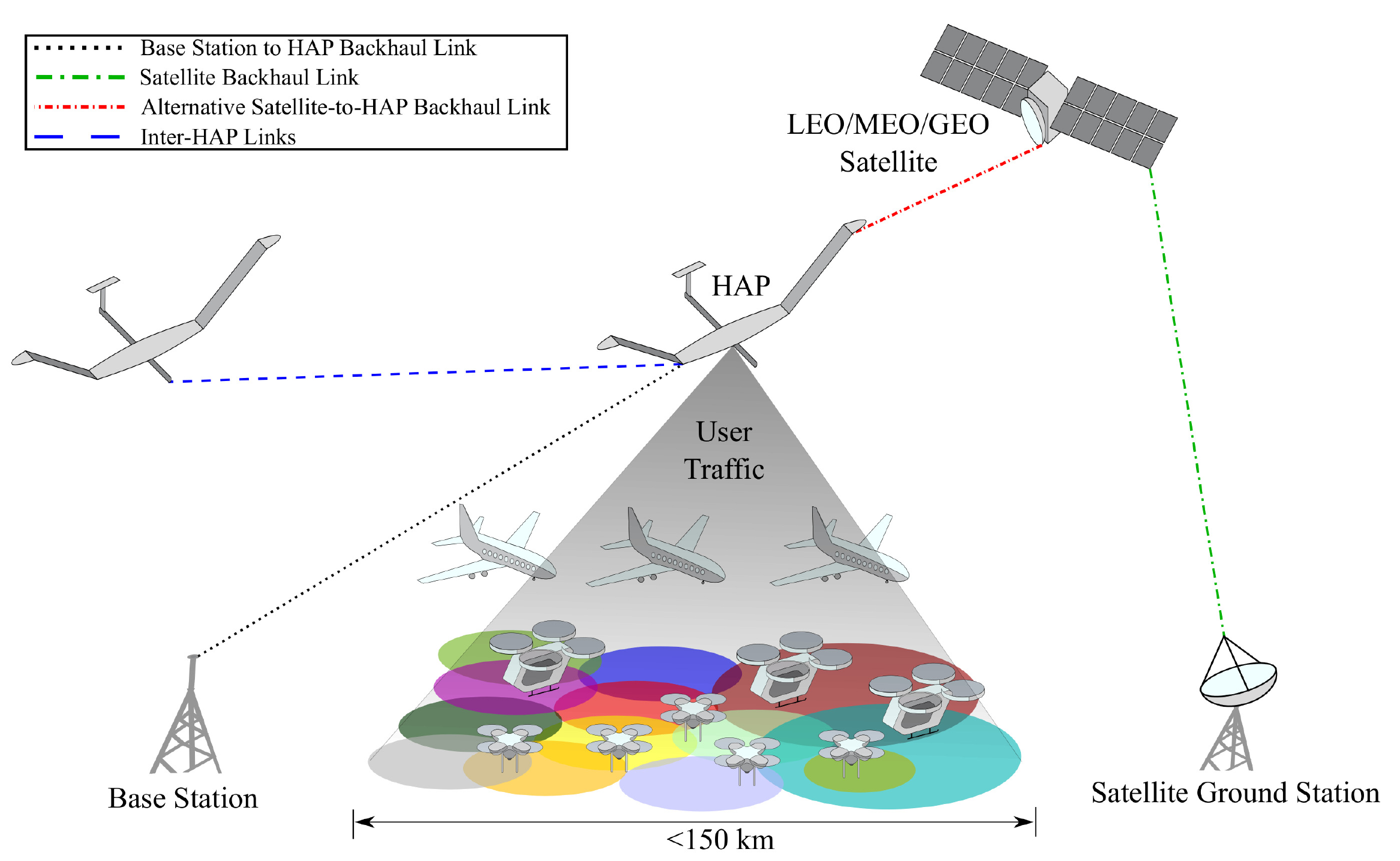}}
	\caption {Proposed \gls{HAP} architecture, based on \cite{HAPslides}. It can provide on-demand services for \glspl{AV}, ground \glspl{BS} and satellites can provide the backhaul connectivity. \glspl{HAP} can also establish \gls{A2A} links between each other for data relaying.}
	\label{fig: hap_architecture}
\end{center}
\end{figure*}

In regards to \gls{RF} channel, the authors of \cite{8886850} mention that available channel models do not realistically represent all the propagation impairments and thus, they derived a channel model on L-band with utilizing \gls{GPS} signals. The attenuation due to tropospheric effects for air-to-satellite links is significant in \gls{mmWave} frequencies \cite{8439347}. Hence, the authors propose link adaptation algorithms and implemented them in \gls{MEO} constellations for \glspl{UAV}. They also further demonstrate the possibility of employing \glspl{SDR} for adaptive \gls{SatCom} systems \cite{7601547}.

Overall, \gls{SatCom} has a key role to enable \gls{FACOM} in the future. We should determine the suitable \gls{SatCom} technologies per each aerial use case with respect to their \gls{QoS} demands. Nonetheless, the aforementioned challenges require research efforts to fit \gls{SatCom} into the \gls{FACOM} ecosystem. 

\subsection{HAP Networks}
\label{subsection: wirelesstechnologies_hapcom}

In this section, we investigate the utilization of \glspl{HAP} for the use cases of \gls{FACOM}. We discuss the background and state-of-the-art for \glspl{HAP} and then we outline the challenges and open issues for their deployment in \gls{FACOM}. We summarize the literature review of \gls{HAPCom} in \autoref{table: satcom_hapcom_literaturereview}. 

\subsubsection{Background}
\label{subsection: wirelesstechnologies_hapcom_background}

\glspl{HAP} are quasi-stationary aerial platforms operating in the stratosphere, and located 17-25 km above the Earth’s surface \cite{1423332, ITU_F1500}. Their ease of deployment, incremental expanding, and high elevation angles, flexibility and reconfigurability, low-cost operation, as well as low propagation delay are their prominent features to provide robust connectivity to the sky. Furthermore, they have the capabilities of providing wide coverage, broadcast or multicast transmissions and the ability to move rapidly in emergency situations. Nevertheless, they pose certain disadvantages such as, the necessity of monitoring the station, the immature airship technology, and the stabilization of the onboard antenna \cite{1423332}. 

The \gls{RTT} propagation of \glspl{HAP} is 0.13-0.33 ms \cite{kurt2020vision}, which is lower than the 4G subframe length of 0.5 ms. This enables the provisioning of cellular service while maintaining the original frame size. \gls{ITU} regulates the transmit power limitations of \glspl{HAP} to to avoid harmful interference \cite{ITU_WRC19} and hence, they do not pose threat to the human health. Similar to high throughput satellite systems, they can provide connectivity with spot beams antennas, and this enables many opportunities for scanning beams to follow the traffic. It also allows a smooth system growth, where the spot beam resizing can increase the network capacity, depending on the flexibility of equipment upgrade. Furthermore, \gls{HAPCom} systems are relatively less complex than the \gls{LEO} satellites systems, because of their proven stability against the motion \cite{1423332} and they can even provide indoor coverage \cite{1423332}. Total cost of a \gls{HAP} system reaches about 50 - 300 million USD, depending on features of the system \cite{HAPPSTT}.

We can classify the \glspl{HAP} based on the type of vehicle: 
\begin{enumerate}
    \item \textbf{Unmanned Airship (Air balloon)}: It has a semi-rigid propulsion system and can carry a mission payload of 1000-2000 kg \cite{1423332}. They are mainly solar-powered balloons \cite{GRENA2013215}. Their design allows them to stay about more than 5 years aloft up maintaining their position within a 1 km cube \cite{1423332}. Google Loon is an instance of this type \cite{googleloon}.  
    \item \textbf{Solar-powered Unmanned Aircraft}: It adopts electric engines and propellers as the propulsion system \cite{kurt2020vision}. Solar cells mounted on the wings and stabilizers power the aircraft \cite{ARUM2020232}. The flying time is several months and the aircraft can usually carry 50-300 kg of payload. They are able to maintain their position within 1-3 km radius \cite{1423332}. Airbus Zephyr is an example of solar-powered aircraft \cite{airbus_zephyr}. 
\end{enumerate}

Besides the flight time, airships and solar-powered aircraft are also favorable as they can carry radio units to act as a relaying or base station equipment. 

\subsubsection{Spectrum Allocation}
\label{subsection: wirelesstechnologies_hapcom_spectrumallocation}

\gls{ITU} allocated a number of frequency spectrum in L-, S-, C-, K-, Ka- as well as V-band for \gls{HAPCom} \cite{kurt2020vision}. \gls{HAPCom} may also receive the allocation of 18-32 GHz for fixed services since this spectrum is less sensitive to rain attenuation in comparison to the 47 GHz \cite{1423332}. The bands around 47 GHz require twice the amount of power to guarantee the service availability to that of 28 GHz band \cite{EHELINETBCAPP03, ITURIULTxFSHAPs}.

\begin{savenotes}
	\begin{table*}[t]
		\caption {Networking Capabilities of the \gls{HAPCom}.\vspace{-0.225cm}}
		\begin{center}
			{\renewcommand{\arraystretch}{1.5} 
				\begin{tabular}{l c c c c c c c c c}
					\hline
					\textbf{\makecell[l]{Altitude\\(km)}} & \textbf{Spectrum} & \textbf{Mobility} & \textbf{\makecell{Coverage\\(km)}} & \textbf{\makecell{User Data Rate\\(Mbps) (\gls{IL}/\gls{EL})}} & \textbf{\makecell{System Data\\Rate (Mbps)}} & \textbf{\makecell{Latency\\(ms)}} & \textbf{\makecell{Communication\\Reliability}} & \textbf{References}\\ \hline
					17-25 &  \makecell[c]{L-, S-, C-,\\K-, Ka-, V-band} & Yes & 150 & 11/2 & 11000-33000 & 0.13-0.33 & 99-99.9\% & \cite{1423332, ITU_F1500, kurt2020vision, HAP_book} \\
					\hline
				\end{tabular}}
				\label{table: hap_performance}
		\end{center}
	\end{table*}
\end{savenotes}

\subsubsection{Architecture and Design}
\label{subsection: wirelesstechnologies_hapcom_architectureanddesign}

In general, \gls{HAPCom} systems can employ two different architectures: 1) relay station, which pushes the computation and complexity away from the airship; 2) base station, where \gls{HAPCom} has an onboard processing system to process the relayed signal and also to apply beamforming techniques. Besides, \glspl{HAP} can also serve for \gls{IoT} services, as an aerial data center or as an alternative link to the \gls{UE} to minimize the interruption of \gls{LEO} handovers \cite{kurt2020vision}.

In literature, the authors of \cite{WCSSPRDPTELEBSHAP01} mentioned that 16 \glspl{HAP} can cover Japan and 18 \glspl{HAP} are sufficient for the entire Greece. A \gls{HAPCom} positioned at 20 km altitude can cover a radius of 50-200 km  \cite{WCSSPRDPTELEBSHAP01}. Moreover, \gls{HAPCom} can host cellular architectures to enable non-terrestrial communication. We present an instance of a \gls{HAP} architecture with cellular and satellite backhauling in \autoref{fig: hap_architecture} \cite{HAPslides}. With this architecture, we can provide backhaul links either directly from the ground station or indirectly via \gls{A2A} links. We can also utilize satellites as redundant backhaul links \cite{HAPSWCOMM01, CFHAPEUROP00}. Moreover, we can provide a flexible cell design to meet event-based temporal hotspot coverage with micro and macro cells \cite{CCAP01}. As \glspl{HAP} can provide large coverage in rural areas, multi-cell configuration with narrow beamwidth can help improve the link budget \cite{9178753}. The authors of \cite{EWCSHAPCWTHC97} propose ring-shaped cell clustering architecture and \cite{IHAPTUMTSNIACD03, HALONet00} report that sectoring the inner and outer cells can help improve the network performance. 


The displacement of a \gls{HAP} poses two challenges: 
\begin{enumerate}
    \item The displacement interrupts the backhaul link to ground station, which can affect the network performance.
    \item Cells on the far edge of the coverage area may no longer have an acceptable link budget for reliable connectivity \cite{CPHAPNOmmWB00, PMCSHAP01}.
\end{enumerate}

We can tackle the first challenge by acquiring multiple links to different ground stations so that \gls{HAP} can connect to the one with the shortest \gls{LoS} path. For the latter case, we must limit the maximum displacement distance so that the link budget remains sufficient or we can realize sophisticated user resource allocation techniques between different \gls{HAP} units. Nonetheless, vertical and horizontal displacement on \glspl{HAP} can cause coverage instability, interrupt the handover procedure and degrade the network performance \cite{8891546}. 

\glspl{HAP} should have the capability to control the beam to cancel the influence of altitude/position variation. Both multi-beam horn and digital beamforming antennas achieve the aforementioned requirements at different frequencies \cite{PFTPTBUHAPS01, WCSSPRDPTELEBSHAP01}. As \gls{MIMO} systems are deployable on \glspl{HAP}, they can provide directional 3D beams with multi-antenna arrays to improve the link quality \cite{kurt2020vision}.

\subsubsection{The Role of HAPCom in FACOM}
\label{subsection: wirelesstechnologies_hapcom_hapcominfacom}

In the context of \gls{FACOM}, we can employ \glspl{HAP} for various purposes, such as urban area support for extended coverage or application-specific deployment to increase network capacity or communication reliability. \glspl{HAP} can also collect sensing data from the sensors of other \glspl{AV} or the sensors on its own platform. \glspl{HAP} can also help in navigation and positioning services \cite{WNBHAPPINCS02}. 

We summarize the general communication characteristics of \gls{HAPCom} systems in \autoref{table: hap_performance}.

In literature, the authors of \cite{6581252} present a network architecture composed of \glspl{HAP} and \glspl{FANET}. They utilize \glspl{HAP} to aid the neighbor discovery of \gls{FANET} nodes due to narrow beams of directional antennas. Furthermore, in \cite{8746381}, the authors study a \gls{HAP}-aided relaying satellite free space optics quantum key-distribution scheme to support secure communication of ground nodes and \glspl{AV}. Finally, number of studies, \cite{9391325, 9391216, 9391285}, studied the \gls{RF} propagation to correctly model the \gls{HAP}-to-ground channels.

\subsubsection{Future Challenges}
\label{subsection: wirelesstechnologies_hapcom_futurechallenges}

The movement of the \glspl{AV} imposes a constraint on the maximum data rate that can be transferred. For fixed stations, the displacement of the \gls{HAP} may cause a deviation of the main lobe. The required data rate determines the choice of steerable or fixed antennas. High elevation angles can be also another limiting factor due to the rapid changes in the angle. If fixed antennas are desirable to minimize the cost, a wider beamwidth antenna can serve to areas directly under the platform, while narrow beamwidth antennas can serve for long-distance coverage. 

The authors of \cite{TDWCSSPK02} provided a decent review of the research on the technology for a stratospheric communication system in Korea. From a networking perspective, \gls{IP} routing can pose challenges over hybrid \gls{HAP}-satellite systems. Topology information can rapidly become obsolete, especially when we employ a \gls{LEO} backhauling to provide global connectivity. Multicasting can be an alternative option; however, developing efficient multicast protocols for time-variant \gls{HAP} links can be challenging. Handover scenarios in \glspl{HAP} demand a particular attention. Considering the varying cruising speeds of the \glspl{AV} as well as the displacement of \glspl{HAP}, handovers can take place very often, thus further studies should also cover the handover aspects. 

Overall, while \glspl{HAP} can alleviate the provision of high-rate and low-latency services for \glspl{AV}, they also pose particular challenges to provide reliable links. Nevertheless, \glspl{HAP} can provide cost-efficient and dynamic solutions for \gls{FACOM}. For a comprehensive overview about \glspl{HAP}, the readers may refer to the survey \cite{kurt2020vision, 9356529}. 

\subsection{A2A}
\label{subsection: wirelesstechnologies_a2a}

\begin{figure}[t]
\begin{center}
	\centerline{\includegraphics[width=0.45\textwidth,keepaspectratio]{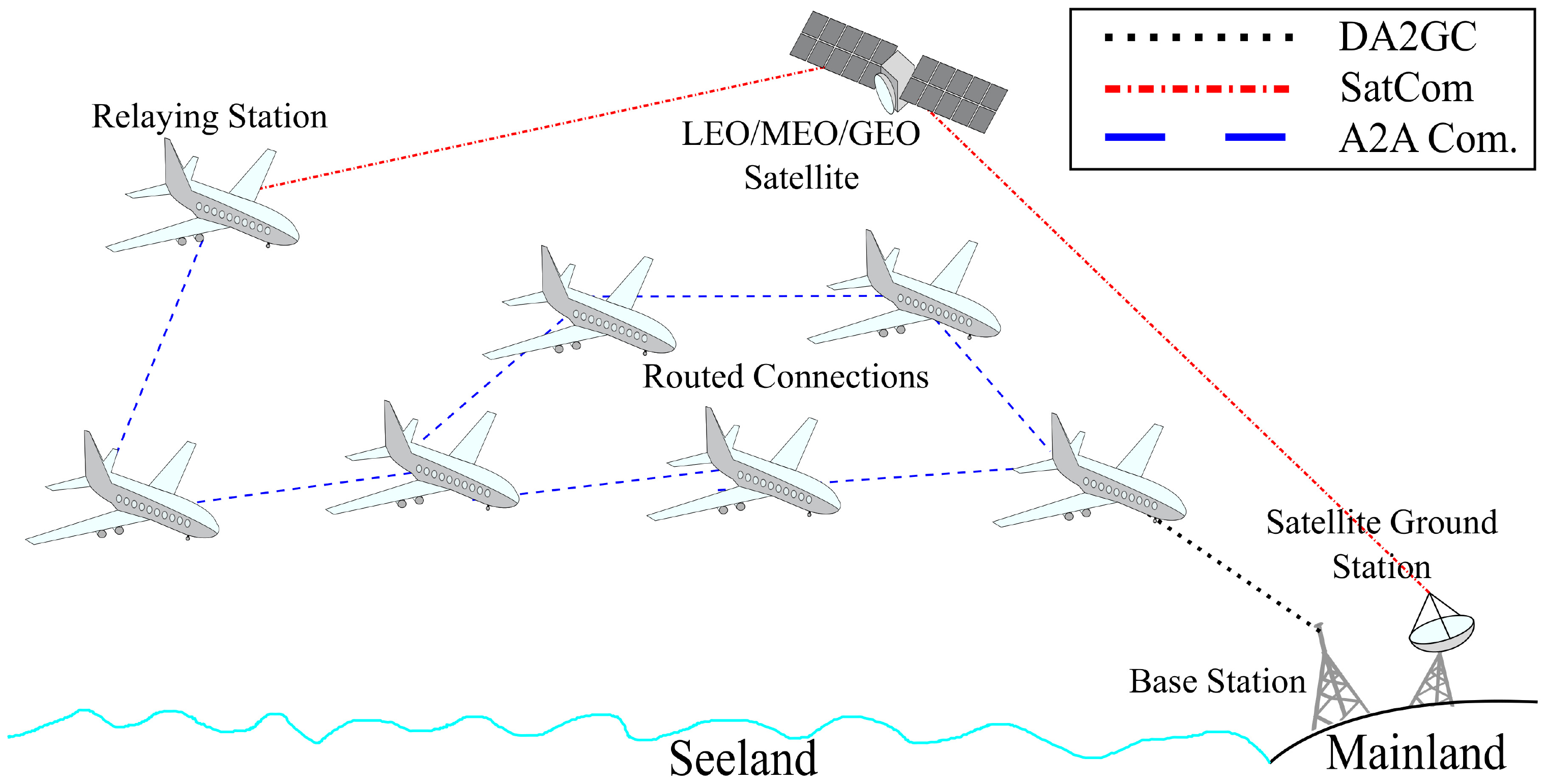}}
	\caption {\gls{FANET} topology \cite{O2}, where airplanes establish \gls{A2A} connectivity with each other and one of the airplanes set up \gls{A2G} or \gls{SatCom} connectivity to deliver the data to the ground.}
	\label{fig: a2a_topology}
\end{center}
\end{figure}

\glspl{FANET} are a subclass of \gls{MANET}, in which the mobile nodes are the \glspl{AV} \cite{O1, FANETS}. \glspl{MANET} are formed by inter-connecting the \glspl{AV} and routers so that \glspl{AV} can directly communicate with each other or relay the information without the involvement of a ground system, as shown in \autoref{fig: a2a_topology}. 

\gls{FANET} nodes move with higher speed and have more degree of movements than that of \glspl{MANET} but rather have more predictable mobility models with less node density \cite{O4-14, O4-13}.

\gls{FANET} poses particular advantages to become a vital part of \gls{FACOM} and we can list a few of them as follows \cite{O2}:

\begin{itemize}
    \item \gls{FANET} addresses the resource management issues that may arise from using satellite resources to support the rapid increase of \gls{IFEC}.
    \item It also addresses the problem of provisioning delay sensitive applications as an advantage over satellite links.
    \item In comparison to satellite or ground stations systems, \gls{FANET} systems require less time and cost for deployment. Hence, it gives mobile network operator a feasible solution for broadband data provisioning in comparison to \gls{IFEC} via satellite. 
\end{itemize}

While airplanes can realize \gls{A2A} links to increase network capacity for \gls{IFEC}, we foresee two potential use cases for the \gls{A2A} links of \glspl{eVTOL}: 
\begin{enumerate}
    \item Collision avoidance; 
    \item Data relaying.
\end{enumerate}

The former one requires low data rates, and it also inherits broadcast-based signalling to detect and avoid safety-threatening conditions in the air. The data relaying can be useful to provide connectivity for the \glspl{AV} outside the coverage area or to deliver time-critical data to the ground network in a timely manner. We can realize this use case in hybrid with another technology, such as \gls{DA2GC} or \gls{SatCom}. This way, \gls{A2A} can support the primary link to increase the overall data rate or to extend the coverage in rural areas. However, the data relaying use case can be less likely for deployment with the consideration of the \gls{SWaP} and cost requirements of the \glspl{AV}.

\subsubsection{A2A Standards}
\label{subsection: wirelesstechnologies_a2a_standards}

Standardization regulatory recently exposed \glspl{FANET} to address the connectivity issues of ground and aerial vehicles. In this regard, \gls{3GPP} and IEEE proposed new protocols, which paved the way for commercializing the \gls{A2A} systems.

\paragraph{3GPP V2V}
\label{subsection: wirelesstechnologies_a2a_3gpp_v2v}

Although cellular systems are naturally human-centric networks, \gls{3GPP} initiated novel releases to enable direct communication between vehicles \cite{3GPP_36785}, Recent articles in literature also consider the \gls{3GPP} \gls{C-V2V} standards for aerial applications. The authors of \cite{9115898, 8904448} evaluate a scenario, where \gls{A2A} and \gls{EL} channel of ground \glspl{UE} share the same spectrum. They compare the interference mitigation performance of underlay and overlay spectrum sharing mechanisms \cite{9115898}. 

In \cite{9055054}, the authors present a concept of \gls{U2X} communication that include \gls{A2A} links. They considered \gls{A2A} links for relaying the sensor data collected by \glspl{UAV} on the ground. The authors of \cite{9151343} propose a 3D \gls{CoMP} model for \gls{A2A} communication, where \glspl{UAV} were employed both as aerial \glspl{BS} and as aerial \glspl{UE}. 

\paragraph{IEEE 802.11p}
\label{subsection: wirelesstechnologies_a2a_ieee_80211p}

Although a number of works study this technology for the inter-connectivity of ground vehicles, its extension to aerial use cases require a particular attention. The authors of \cite{8735159} evaluated the performance of this technology with flying and non-flying \glspl{UAV}, and they achieve up to a 2.9 km communication distance at a 20 m altitude. Such communication range can enable multiple \gls{A2A} connectivity options in dense airspace scenarios. Moreover, authors of \cite{9198731} analyze video traffic transmission over 802.11p links, and they conclude that it cannot support video transmissions due to the bursty nature of the video traffic. 

Number of studies experimentally modeled the 802.11 \gls{A2A} channel at 2.4 GHz \cite{8766809, 7414180}. They find out that power attenuation is similar to the free space pathloss model, and the multipath effect degrades at higher altitudes. However, the authors of \cite{10.1145/3412060.3418431} observe the pathloss exponent to be slightly higher than that of the free space at 60 GHz.  

Several works compare the performance of IEEE 802.11p and \gls{C-V2V} for vehicular use cases. While the authors of \cite{9133075} report that 802.11p outperforms cellular \gls{V2X} using different message lengths and inter-packet intervals, the numerical analysis study from \cite{8891313} concludes that NR-\gls{V2X} has superior communication reliability, latency and data rate performance compared with that of 802.11p. 

\begin{savenotes}
	\begin{table*}[t]
		\caption {Networking Capabilities of the Wireless Communication Technologies.\vspace{-0.225cm}}
		\begin{center}
		\begin{threeparttable}
			{\renewcommand{\arraystretch}{1.5} 
				\begin{tabular}{l c c c c c c c}
					\hline
					\textbf{Technology} & \textbf{Standard} & \textbf{\makecell{User Data\\Rate (Mbps)\\(\gls{IL}/\gls{EL})}} & \textbf{\makecell{Latency\\(ms)}} & \textbf{\makecell{Communication\\Reliability}} & \textbf{\makecell{Cell Coverage\\(km)}} & \textbf{Spectrum} & \textbf{References}\\ \hline
					\multirow{3}{*}{Cellular} & \gls{eMBB} & 100/50 & 20 & 99.9\% & 5-32 & \multirow{3}{*}{Licensed$^\ast$} & \multirow{2}{*}{\cite{GSMA_5GGuide, qualcomm_urllc, mbb_coverage, ngmn_urllc}}\\
					& \gls{uRLLC} & 0.1-1000$^\dagger$/8$^\ddagger$ & $<$10 & $<$99.9999\% & $<$1 & & \\
					& mMTC & 0.1/0.15$^\lozenge$ & 890-2340$^\lozenge$ & 99.99\% & $<$100 & & \cite{nbiot_ratelatency, qualcomm_urllc, gsma_ltem} \\ \hline
					\multirow{3}{*}{\gls{SatCom}} & \gls{GEO} & 0.85/0.49$^\circledast$ & 700-800 & 99.5\% & 4000 & \multirow{3}{*}{Licensed} & \cite{inmarsat_servicescomparison, geo_availability, ses_meo} \\
					& \gls{MEO} & $<$800/$<$800 & 130 & 99.9\% & 700 & & \cite{o3bnetworks, o3bnetworks2, meo_datarate, ses_meo} \\
					& \gls{LEO} & 750/375 & 10-35 & 99.9\% & 450 & & \cite{8933127, 8700141, ISNOV} \\ \hline
					\multirow{5}{*}{IEEE} & \makecell{\gls{WiFi} 6} & 1200 & 10 & \multirow{2}{*}{N/A$^\bullet$} & \multirow{2}{*}{$<$0.3} & \multirow{5}{*}{Unlicensed} & \cite{huawei_wifi6, wifi6, oughton2020revisiting} \\
					& \makecell{\gls{WiFi} 7} & 30000 & 5 & & & & \cite{9090146, oughton2020revisiting} \\
					& \makecell{HaLoW} & 0.15-3.9 & 100-3000 & 99\% & $<$1 & & \cite{wifi_halow_datarate, wifi_alliance_halow, wifihalow} \\
					& \makecell{\gls{WiMAX}} & 70/28 & 37-77 & - & 50 & & \cite{Song_2014, 5654506, WiMAX_uldatarate} \\
					& \makecell{ZigBee} & 0.25 & 15 & - & $<$0.1 & & \cite{element14_zigbee} \\ \hline
					\gls{LoRa} & & 0.003-0.027 & 56-1400 & & $<$15 & Unlicensed & \cite{itu_lora, orangelabs_lora} \\ 
					\gls{AeroMACS} & & 18.2/6.8 & $<$60 & 99.9-99.95\% & $<$8.3 & \multirow{2}{*}{Aviation} & \cite{AeroMACS_specifications} \\ 
					\gls{LDACS} & & \makecell{0.315-1.428/\\0.294-1.390} & 100 & - & $<$370 &  & \cite{ietf_raw_ldacs, ldacs_latency, LDACS_specifications} \\
					\multirow{2}{*}{\gls{A2A}} & \gls{C-V2V} & 3-27 & 20-100 & 67-99\% & $<$1 & Licensed & \cite{3GPP_36785, ETSI_122185, v2x_80211p_reliability, v2x_coverage} \\
					& 802.11p & 3-54 & $<$100 & 59-98\% & 0.4-2.5 & Unlicensed & \cite{3GPP_36785, 80211p_reliability, v2x_80211p_reliability, 7222906} \\
					\hline
				\end{tabular}}
				\label{table: technologies_performance}
				\begin{tablenotes}[flushleft]
				        \scriptsize
				        \item \textit{$^\ast$\gls{LAA} technology operates in the unlicensed spectrum on \gls{IL}.} 
				        \item \textit{$^\dagger$According to the study of \gls{NGMN} \cite{ngmn_urllc}, the data rate capabilities of uRLLC can be diverse, due to different application requirements.} 
				        \item \textit{$^\ddagger$Only given value in \cite{ngmn_urllc}, which is for automated guided vehicle applications of uRLLC. Similar to \gls{IL}, the data rate capabilities of \gls{EL} can depend}\\\textit{on the application scenario.} 
				        \item \textit{$^\lozenge$Although these values are for the \gls{NB-IoT} of 4G, we assume similar capabilities in mMTC with the support of higher node density.} 
				        \item \textit{$^\circledast$The Global Xpress service of Inmarsat claims up to 50 Mbps on \gls{IL} and 5 Mbps on \gls{EL}, respectively \cite{inmarsat_servicescomparison}.} 
				        \item \textit{$^\bullet$\gls{WiFi} 6/7 provides improved communication reliability thanks to the introduction of \gls{MU-MIMO} and longer \gls{OFDM}}\\\textit{symbols \cite{huawei_wifi6}.}
				\end{tablenotes}
		\end{threeparttable}
		\end{center}
	\end{table*}
\end{savenotes}

\subsubsection{Challenges and Potential Solutions of A2A for FACOM}
\label{subsection: wirelesstechnologies_a2a_challenges_potentialsolutions}

As the literature comprehensively covered the \gls{FANET} protocols for \glspl{UAV}, we briefly provide the main challenges and potential solutions, especially toward airplane and \glspl{eVTOL}. The readers may refer to the \gls{FANET}-focused surveys such as \cite{9045408, 9044378, 8741010} to read more about the studies related to \glspl{UAV} in \gls{FANET}. 

\paragraph{Coverage and Connectivity}
\label{subsection: wirelesstechnologies_a2a_challenges_potentialsolutions_coverage_connectivity}

Certain number of airplanes are in the sky at any given time that can form a \gls{FANET} among each other. According to the air traffic data records of FlightAware \cite{O2-6}, the number of airplanes in the sky at anytime were on average 9728, at maximum 12586 and at minimum 3354 in 2016. The range of communication between two airplanes should be within the minimum horizontal separation range, which is 9.3 km when a radar or an \gls{ADS-B} system is in use \cite{ICAO_airplaneseparation}. In \cite{8473483}, the authors evaluate the capabilities of \gls{A2A} communication to extend the \gls{DA2GC} coverage. As an alternative to \glspl{SatCom}, \gls{A2A} links can provide a 37 Mbps throughput up to 432 km \gls{DA2GC} and 340 km \gls{A2A} coverage. 

Airplanes can establish \gls{LoS} \gls{A2A} connectivity up to 740 km distance due to the Earth's curvature \cite{O7}. In order to provide connectivity at such a large distance, the authors of \cite{O7} design a geographic load share routing and access protocol, which utilizes the information about both the airplane's position and the buffer size. It exploits the total \gls{A2G} available capacity by balancing the traffic aggregated by all air nodes. Furthermore, we considered combining \gls{A2A} and \gls{DA2GC} links to maximize the number of airplanes that can be connected with a specific data rate threshold \cite{8761882}. Our simulation studies show that over 90\% of aircraft can have at least 50 Mbps in the region of North Atlantic. The authors of \cite{O6-18} also address the connectivity in between two airplanes based on their distance, transmission range, and airplane density in a one-flight route, whereas the study, \cite{O6}, analyzes two-way flight route connectivity.

The authors of \cite{9081678} evaluate the \gls{LDACS} system for \gls{A2A} communication between airplanes and compare the performance of ALOHA-based and the \gls{STDMA} \gls{MAC} mechanisms for the \gls{LDACS} \gls{A2A} system. They report that \gls{STDMA} is a better candidate because \gls{LDACS} requires the available bandwidth to split into 0.5 MHz sub-channels. In \cite{8932658}, the authors consider \gls{A2A} links as secondary applications in the \gls{A2G} frequency spectrum and let \gls{A2A} applications transmit data on \gls{A2G} spectrum whenever the channel is free. 

\paragraph{Routing}
\label{subsection: wirelesstechnologies_a2a_challenges_potentialsolutions_routing}

Routing is a vital element of \gls{FANET} for efficient and robust networking. Due to the different mobility characteristics of each type of \glspl{AV}, they demand different routing mechanism to optimize the end-to-end connectivity. Several works proposed routing schemes for various purposes. Flat hierarchical structure of routing might be inefficient for \gls{FANET} due to the large node density. Hence, the authors of \cite[Page 122]{O2-9} propose a model for clustering of airplanes to produce a scalable system for a large global network of airplanes.

The high speed of the airplanes also influences the routing. Link stability and path interruption are essential parameters when designing the routing protocol. Moreover, designing a stable routing protocol via a clustering scheme must ensure that member airplanes of clusters do not frequently leave their associated clusters. The authors of \cite{O3} present an efficient method for constructing stable routes and clusters for \gls{ATM} and operational-related communication. They also propose two routing schemes, based on random packet code division multiple access and trajectory based routing scheme, where the sender computes trajectory of each packet \cite{O3-14, O3}. This scheme enables simple path diversity without the need of geo-location exchange messages efficient broadcasting, etc. Furthermore, the authors of \cite{O5} design their solution based on finding the shortest path routing for \gls{FANET} in an oceanic versus a continental study. In the oceanic one, they measured 68.2 kbps with a latency of 184 ms, whereas in the continental scenario, the data rate dropped to 38.3 kbps with a latency of 401 ms. 

As for \glspl{UAV}, we observe numerous novel routing mechanisms recently proposed in literature. Thus, the readers may refer to the recent surveys, \cite{9044378, 8741010}, for a comprehensive overview on the routing of \glspl{UAV}. 

\paragraph{Air Interface}
\label{subsection: wirelesstechnologies_a2a_challenges_potentialsolutions_airinterface}

Airborne antennas must be omnidirectional or mechanically steerable to steer the beam pattern both in azimuth and in elevation. Beamsteering in azimuth and elevation is necessary in order to permit the relatively narrow antenna beam to be pointed at the satellite, so that the antenna can successfully receive a relatively weak signal \cite{MPMBSACAAS}. Additionally, the antenna should be lightweight and have a low air drag due to the \gls{SWaP} constraints of \glspl{AV}. 

Beamforming and beamtracking can be suitable for airplanes and \glspl{eVTOL} since their mobility models are more predictable. The authors of \cite{8739859} present an antenna array design that consists of 9 dipole antennas for \glspl{FANET}. They aim to optimize the beamforming efficiency along with low side lobes. Besides the antenna design, dual-radio design can also improve the networking performance, as suggested in \cite{8737798, 8676034}. The authors of \cite{8676034} combine omnidirectional and multi-beam directional antennas, and use location predicting algorithms to enable beamtracking. They report a 35.8\% increase in link robustness, which is measured by the ratio of the number of links that are not disrupted until the end of data transmission to the total number of established links between \glspl{UAV} in the simulation.  

\subsection{Multi-link Connectivity}
\label{subsection: wirelesstechnologies_heterogeneous}

\begin{savenotes}
	\begin{table*}[!htbp]
		\caption {Match study between the use cases of \gls{FACOM} and the communication technologies. Green highlights indicate our preferred technologies per use case.\vspace{-0.225cm}}
		\begin{center}
		\begin{threeparttable}
			{\renewcommand{\arraystretch}{1.8} 
			\scalebox{1}{
				\begin{tabular}{l c c | c c c | c c c | c c c c | c c c | c c}
					\hline
					\textbf{\gls{AV}} & \multicolumn{2}{l}{\textbf{Use Case}} & \multicolumn{3}{c}{\textbf{Cellular}} & \multicolumn{3}{c}{\textbf{\gls{SatCom}}} & \multicolumn{4}{c}{\textbf{IEEE}} & \multirow{2}{*}{\textbf{\makecell{L\\o\\R\\a}}} & \multirow{2}{*}{\textbf{\makecell{A\\e\\r\\o\\M\\A\\C\\S}}} & \multirow{2}{*}{\textbf{\makecell{L\\D\\A\\C\\S}}} & \multicolumn{2}{c}{\textbf{\gls{A2A}}}\\
					& & & \makecell{e\\M\\B\\B} & \makecell{u\\R\\L\\L\\C} & \makecell{m\\M\\T\\C} & \makecell{L\\E\\O} & \makecell{M\\E\\O} & \makecell{G\\E\\O} & \makecell{W\\I\\F\\I} & \makecell{W\\i\\M\\A\\X} & \makecell{H\\a\\L\\o\\W} & \makecell{Z\\i\\g\\B\\e\\e} & & & & \makecell{C-\\V\\2\\V} & \makecell{8\\0\\2.\\1\\1\\p} \\ 
					& & & & & & & & & & & & & & & & & \\\hline
					\multirow{7}{*}{\makecell{A\\i\\r\\p\\l\\a\\n\\e}} & \multirow{3}{*}{\makecell{\gls{ATM}}} & \gls{VoIP} & \checkmark & \checkmark & & \cellcolor{green!30}\checkmark$^\star$ & & & & \checkmark & & \checkmark & \checkmark & \checkmark & \cellcolor{green!30} \checkmark & \\
					& & \makecell{OLDI} & \checkmark & \checkmark & & \cellcolor{green!30}\checkmark$^\star$ & \checkmark & \checkmark & & \checkmark & & & \checkmark & \checkmark & \cellcolor{green!30}\checkmark & & \\
					& & \makecell{\gls{AMHS}} & \checkmark & \checkmark & & \cellcolor{green!30}\checkmark$^\star$ & \checkmark & \checkmark & & \checkmark & & & \checkmark & \checkmark & \cellcolor{green!30}\checkmark & & \\ \cline{2-18}
					& \multirow{3}{*}{\makecell{\gls{SPO}}} & \makecell{Current\\\gls{SPO}} & \cellcolor{green!30}\checkmark & \checkmark & & \cellcolor{green!30}\checkmark$^\star$ & \checkmark & \checkmark & & & & & \checkmark & \checkmark & \checkmark & & \\
					& & \makecell{Cargo\\\gls{SPO}} & \cellcolor{green!30}\checkmark & & & \cellcolor{green!30}\checkmark$^\star$ & \checkmark & & & & & & & & & & \\
					& & \makecell{Full\\\gls{SPO}} & \cellcolor{green!30}\checkmark & & & \cellcolor{green!30}\checkmark$^\star$ & \checkmark & & & & & & & & & & \\ \cline{2-18}
					& & \makecell{Mass. Dat.\\Trans.} & \cellcolor{green!30}\checkmark$^\ddagger$ & & & & & & \cellcolor{green!30}\checkmark$^\ddagger$ & & & & & & & & \\ \hline
					\multirow{5}{*}{\makecell{e\\V\\T\\O\\L}} & \gls{UTM} & \makecell{Flight\\Mgnt.} & \cellcolor{green!30}\checkmark & \checkmark & & \checkmark & \checkmark & & & \checkmark & & & & \checkmark & \checkmark & & \\ \cline{2-18}
					& \multirow{3}{*}{\makecell{Pilot\\ing}} & \makecell{On-\\board} & \cellcolor{green!30}\checkmark & \checkmark & & \checkmark & \checkmark & \checkmark & & \checkmark & & & & & & & \\
					& & \gls{RPO} & \cellcolor{green!30}\checkmark & \checkmark & & \cellcolor{green!30}\checkmark & \checkmark & & & \checkmark & & & & & & & \\
					& & \gls{FAO} & \cellcolor{green!30}\checkmark & \checkmark & & \checkmark & \checkmark & & & \checkmark & & &  \checkmark & & \checkmark & & \\ \cline{2-18}
					& \multirow{2}{*}{\makecell{Vertiport}} & \makecell{Assis.\\T.O./Land.} & \checkmark & \checkmark & & \checkmark & & & \checkmark & \checkmark & & & & \cellcolor{green!30}\checkmark & \checkmark & & \\
					& & \makecell{Data Off\\loading} & \cellcolor{green!30}\checkmark$^\ddagger$ & & & & & & \checkmark & & & & & & & & \\ \cline{2-18}
					& \gls{DAA} & & & & & & & & & & & & & & \checkmark & \cellcolor{green!30}\checkmark & \checkmark \\ \hline
					\multirow{8}{*}{\makecell{U\\A\\V}} & \multirow{3}{*}{\makecell{Pilot\\ing}} & \gls{VLoS} & \checkmark & \checkmark & & \checkmark & \checkmark & & \cellcolor{green!30}\checkmark & \checkmark & & & & & \checkmark & & \\
					& & \makecell{BV\\\gls{LoS}} & \cellcolor{green!30}\checkmark & \checkmark & & \checkmark & \checkmark & & & \checkmark & & & & & & & \\
					& & \gls{FAO} & \checkmark & \checkmark & \cellcolor{green!30}\checkmark & \checkmark & \checkmark & \checkmark & & \checkmark & & & \checkmark & \checkmark & \checkmark & & \\ \cline{2-18}
					& \multirow{4}{*}{\makecell{\makecell{Pay\\load}}} & \makecell{Vision\\-based} & \cellcolor{green!30}\checkmark & & & \checkmark & \checkmark & & & & & & & & & & \\
					& & \makecell{Delivery} & \cellcolor{green!30}\checkmark & & & \checkmark & \checkmark & & & & & & & & & & \\
					& & \gls{IoT} & & \checkmark & \cellcolor{green!30}\checkmark & \checkmark & \checkmark & \checkmark & & & \checkmark & & \checkmark & & & & \\ 
					& & \makecell{Agri\\culture} & \cellcolor{green!30}\checkmark & \checkmark & & \checkmark & \checkmark & & & & & & \checkmark & & & & \\
					\hline
				\end{tabular}}}
				\label{table: technologies_matchstudy}
				\begin{tablenotes}
				        \scriptsize
				        \item \hspace{1.5cm} \textit{$^\star$When flying over the ocean.}
    				    \item \hspace{1.5cm} \textit{$^\dagger$For safety-critical applications.}
    				    \item \hspace{1.5cm} \textit{$^\ddagger$Potentially using \gls{mmWave} technologies to achieve high-rates in an abundant spectrum.}
				\end{tablenotes}
		\end{threeparttable}
		\end{center}
	\end{table*}
\end{savenotes}

Multi-link heterogeneous network architectures can be the enabler for achieving higher network capacity, reliable and robust networking as well as seamless coverage. In the context of \gls{FACOM}, heterogeneous networking is essential in \gls{RPO} applications. Although a single wireless link can provide sufficient data rate even for video streaming scenarios, it cannot achieve ultra reliable communication as well as ubiquitous coverage in different environments. For these reasons, we can introduce multi-link architectures with different means of diversity, listed as follows: 

\begin{enumerate}
    \item \textbf{Technology Diversity}: As each communication technology has particular networking characteristics, we can employ them together to combine their advantages in a single network. An instance can be the combination of cellular and \gls{LEO} networks. While cellular networks can provide robust connectivity in urban areas, \gls{LEO} can support the operations in suburban areas with seamless coverage. 
    \item \textbf{Network Diversity}: We can achieve this by utilizing multiple network infrastructures of the same or different wireless technologies. This architecture can ensure isolated end-to-end paths to improve the overall network availability. 
    \item \textbf{\gls{BS} Diversity}: Multiple \glspl{BS} can simultaneously serve an \gls{AV} to improve the overall communication reliability by introducing path diversity. If the \glspl{BS} belong to the same network, then we can achieve diversity only until the first hop at the \glspl{BS}. Thus, the overall capacity stays the same. \gls{CoMP} architecture of cellular standard falls in this category. 
    \item \textbf{Frequency Diversity}: A single \gls{BS} can serve an \gls{AV} with multiple links operating at different center frequency. Such operation can introduce different \gls{RF} channel conditions in the same environment; thus, it can help increase the reliability until the first hop. 
\end{enumerate}

An early consideration of multilink architectures for \glspl{UAV} is \cite{6006004}, in which they propose a novel link layer called \textit{FlowCode} to enable the usage of conventional transport protocols in high mobility scenarios. FlowCode utilizes beam diversity from multiple transceivers to increase data and reception rate. 

The authors of \cite{9145477, 8746579} conduct field measurements with a \gls{UAV} to evaluate the improvements in reliability and latency over multiple \glspl{MNO}. The common notion of observing higher \gls{RSRP} does not hold in urban environments according to the results in \cite{9145477}. Furthermore, they report a performance gain with multiple links compared with single link due to the performance variations of the \glspl{MNO} at different altitudes and environments. Furthermore, the authors of \cite{8746579} show that while the minimum reliability rate with a single \gls{MNO} is 97.6\%, connectivity over dual operators can maintain 99.9\% \gls{PER} at all altitudes on the \gls{IL}. Although these studies show overall reliability improvement, guaranteeing reliability at any time instance can be challenging as the individual \gls{MNO} designed their networks independently with the consideration of the ground use cases.

Besides utilizing dual \glspl{MNO}, a combination of a public and dedicated cellular network can be another option, as the authors of \cite{8923123} propose it with \gls{MPTCP} for maritime search and rescue missions of \glspl{UAV}. They conclude that the multi-link protocol increases the range and improves the data rate performance.

We can also consider multi-link connectivity using different wireless technologies. As each communication technology targets different use cases and owns different link characteristics, they can complement each other to ensure robust connectivity at all times during the mission of \glspl{AV}. For instance, Deutsche Telekom and Inmarsat commercialized the \gls{EAN}, which provides in-flight connectivity for airplanes using \gls{DA2GC} as well as \gls{SatCom} \cite{ean_dt_inmarsat_nokia}. \gls{EAN} can provide up to 75 Mbps on \gls{IL} per cell with 300 base stations. In literature, the authors of \cite{8700598} perform a modeling study with a satellite link and \gls{WiFi} access points using \gls{MPTCP} to control a convoy of \glspl{UAV}. They aim to provide robust bandwidth allocation as well as high communication reliability.

In the cases where dual-connectivity is insufficient, we can increase the number of parallel links to increase the overall robustness. The authors of \cite{8735265} present an example triple-redundant multi-link architecture employing cellular, \gls{WiFi} and \gls{LoRa} for the \gls{UAV}-ground station connectivity. Their redundancy design employs a cellular network as the primary link and the other two as fallback links when there is no cellular coverage. Furthermore, the authors of \cite{9162943} develop a novel application layer protocol called \textit{NECTOR}, which is a \gls{UDP}-based architecture with two 4G and one satellite links. The receiver controls the packet reception rate with a torrent-based methodology and they improve reliability by employing network coding. In the end, increasing the number of parallel links can usually help achieve higher reliability. However, it naturally comes with extra networking cost. Thus, we should optimize the required number of links per the intended use case. 

From these studies, we observe that the majority of the studies focus on increasing the overall communication reliability and the robustness of wireless links. However, we should also consider heterogeneity in the backhaul to increase the reliability for end-to-end packet delivery. We should also evaluate the different combinations of the communication technologies to determine how to achieve 10$^{-5}$ communication reliability for \glspl{RPO} of \glspl{eVTOL} and \glspl{SPO} of airplanes. 

\subsection{Connectivity Options vs. FACOM use cases: Match Study}
\label{subsection: wirelesstechnologies_matchstudy}

\begin{savenotes}
	\begin{table*}[t]
		\caption {Network and 5G System Architecture Studies from Literature for \gls{FACOM}.\vspace{-0.225cm}}
		\begin{center}
			{\renewcommand{\arraystretch}{2} 
				\begin{tabular}{l l l c}
					\hline
					\textbf{Vehicle/Entity} & \textbf{Design Type} & \textbf{Study Goal} & \textbf{References} \\ \hline
					\multirow{5}{*}{Airplane} & \multirow{3}{*}{\gls{A2G}} & Proposal of various \gls{A2G} architectures to support IFBC & \cite{8010762, 7073483, 8552136} \\
					& & An IP-based \gls{A2G} architecture for future ATM-cockpit communications & \cite{5172851} \\
					& & An \gls{SDN}-based \gls{A2G} architecture to enable multilink heterogeneous networking & \cite{8569862} \\ \cline{2-4}
					& \multirow{2}{*}{Cabin} & A gate-to-gate connectivity architecture for seamless passenger connectivity & \cite{8288087} \\ 
					& & Performance evaluation of an integrated cellular network architecture & \cite{8569540} \\ \hline
					\multirow{9}{*}{\gls{UAV}} & \multirow{3}{*}{Cellular} & Integration of \gls{UTM} services into 5G system architecture & \cite{9142706, 9076122} \\
					& & A 4G-based network architecture to support \glspl{CNPC} of \glspl{UAV} & \cite{7763289} \\
					& & An LTE-based network architecture to enable aerial monitoring of remote locations & \cite{7396417} \\ \cline{2-4}
					& \multirow{3}{*}{\gls{A2A}} & Comparison of the advantages and disadvantages of various \gls{FANET} architectures & \cite{8550873, 6825193, 8277614} \\
					& & Proposal of a new \gls{FANET} architecture to avoid single point of failure & \cite{9199627} \\
					& & Proposal of a layered architecture with a low latency routing algorithm to minimize latency & \cite{8599015} \\ \cline{2-4}
					& \multirow{3}{*}{\makecell{Space-air-ground\\Architecture}} & Blockchain-based 6G space-air-ground architecture for secure \gls{UAV} communications & \cite{9184022} \\
					& & An \gls{SDN}-based space-air-ground architecture for high-throughput and robust communications & \cite{9213109} \\
					& & Survey on different space-air-ground network architectures & \cite{8368236} \\
					\hline
				\end{tabular}}
				\label{table: networkarchitectures}
		\end{center}
	\end{table*}
\end{savenotes}

In this section, we numerically compare the capabilities of the wireless technologies against the connectivity requirements of \gls{FACOM} use cases, as we presented it \autoref{section: usecases}. For this reason, we provide the performance metrics of most of the evaluated wireless technologies in \autoref{table: technologies_performance}. We consider the user data rate, latency, communication reliability, cell coverage as well as the spectrum as the most relevant metrics for the match study. Overall, each communication technology has unique performance characteristics and poses particular challenges to adapt themselves for the skies. 

We present the conclusion of our match study in \autoref{table: technologies_matchstudy}. Comparing the connectivity requirements of each aerial use case with the performance metrics of the communication technologies, we outline the candidate technologies that can support each aerial application. Furthermore, we highlight the most suitable technology for each use case in green. We do not include \glspl{HAP} in wireless technologies as it does not have a specific connectivity standard. We also do not select \gls{WiFi} technologies for the majority of the use cases due to their limited coverage and the single-hop nature. Similarly, we can consider \gls{C-V2V} and IEEE 802.11p only for the \gls{DAA} links of \glspl{eVTOL} since their communication range is less than 1 km. 

In regards to airplanes, \gls{LDACS} is the candidate technology for the \gls{VoIP} as well as the digital messaging services since it employs the aviation spectrum and has long frame durations to increase the robustness of the link. It also promises high communication reliability. However, \gls{LDACS} cannot meet the high data rate requirements of future \glspl{SPO}. Thus, we foresee that \glspl{SPO} require a heterogeneous connectivity comprising \gls{eMBB}/\gls{eMBB} and \gls{LEO} connectivity. While cellular networks can enable the connectivity during continental flights, \gls{LEO} can complement it during intercontinental flights. 

As for future \gls{eVTOL} use cases, we expect that \gls{UTM} communication can adapt \gls{LDACS} to communicate with \glspl{AV}. Due to human involvement in passenger transportation, connectivity is one of the critical functions. Utilizing aviation safety spectrum in \gls{LDACS} can further enhance the overall communication integrity. Moreover, \gls{LDACS} can ensure global spectrum harmony when \glspl{UTM} in different regions need to coordinate with each other. However, \gls{LDACS} is not sufficient for \glspl{RPO} of \glspl{eVTOL} due to the high rate video transmission. For this reason, we rather recommend cellular connectivity, if available, to ensure large network capacity. Although we do not expect transoceanic applications for \glspl{eVTOL}, \gls{LEO} can complement \gls{eMBB} to improve the overall coverage.  

\begin{figure*}[t]
\begin{center}
	\centerline{\includegraphics[width=0.75\textwidth,keepaspectratio]{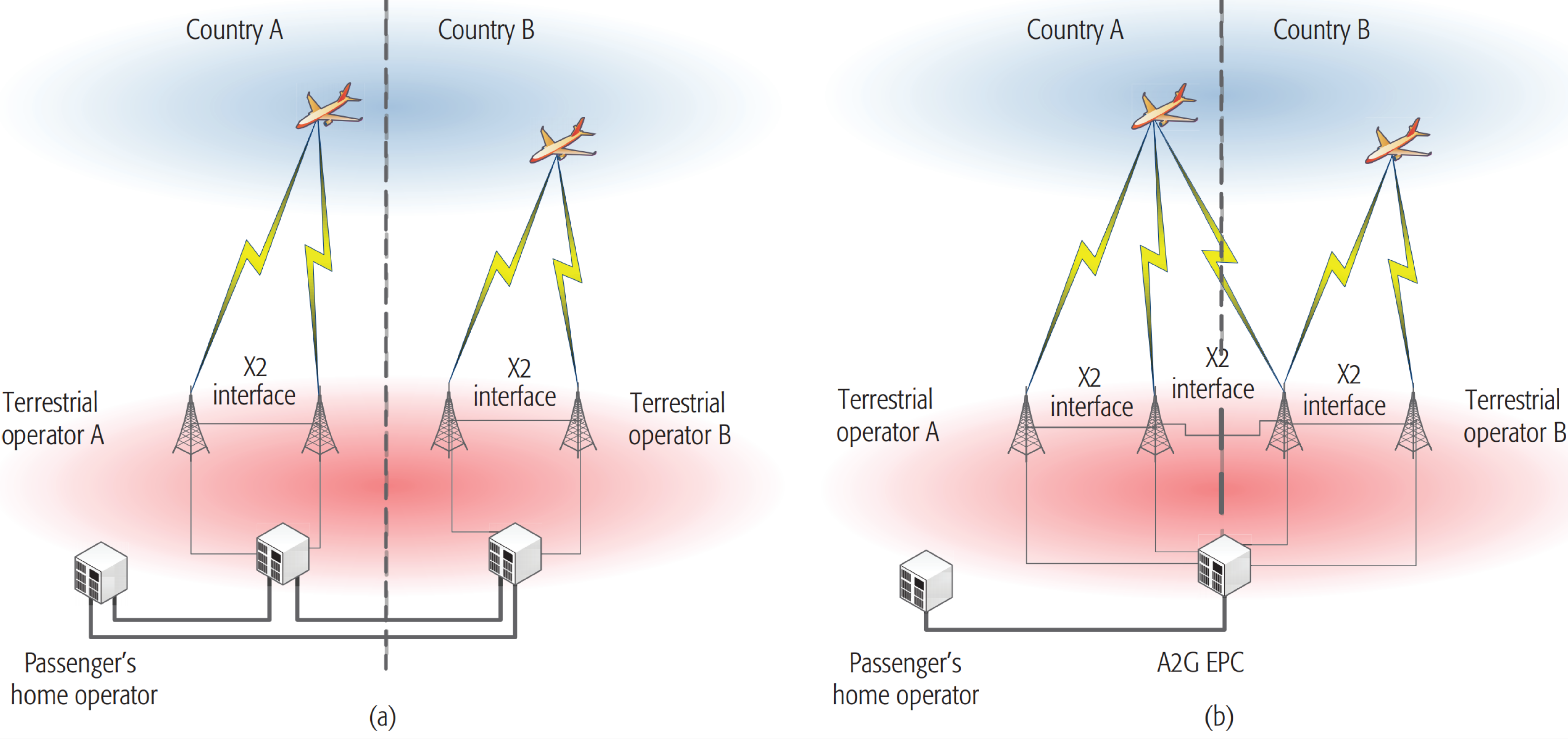}}
	\caption {Proposed network architectures in \cite{8010762}: (a) Decentralized core per \gls{MNO}, (b) Centralized core operator with an \gls{A2G} \gls{MNO}.}
	\label{fig: 8010762}
\end{center}
\end{figure*}

\gls{AeroMACS} can be suitable for the connectivity demands at vertiports as its development particularly targets the connectivity applications at airports. While the \gls{mMTC} is sufficient for the MTC onboard \gls{eVTOL} cabin, \gls{eMBB} can support the data offloading at vertiport gates. As for \gls{DAA}, we select \gls{C-V2V} rather than IEEE 802.11p due to the advantages of licensed spectrum compared with the unlicensed spectrum, which we discuss in \autoref{sec: openresearchchallenges_spectrumregulation}. Although \gls{LDACS} is another option for \gls{DAA}, we prioritize it for \glspl{RPO} due the limited bandwidth on the aviation spectrum. \gls{eMBB} can also support the connectivity for piloting operations as well as the vision-based payload communication of \glspl{UAV}. \gls{LEO} can be the alternative solution when the operations take place in suburban regions. 


Selecting the most suitable technology for each use case introduces a diverse set of wireless ecosystems in \gls{FACOM}. These networks need to coexist together and operate without interrupting each other. Additionally, the deployment and operation cost of different networks further make it difficult to realize a large mixture of communication technologies in \gls{FACOM}. Thus, it is also significant to utilize the technologies that can meet the demands of multiple applications so that we can minimize the heterogeneity of the wireless technologies. 
\section{System Architectures for FACOM}
\label{section: networkarchitectures}

In this section, we present the recent studies that proposed various network and system architectures to enable the applications of \gls{FACOM} in the sky. Dividing the studies with respect to their target \gls{AV} platforms, we present the summary of the works in \autoref{table: networkarchitectures}. 

\subsection{Network Architectures for Airplanes}
\label{section: networkarchitectures_airplanes}

In \cite{8010762}, we compare decentralized and centralized \gls{DA2GC} architectures on international routes, as shown in \autoref{fig: 8010762}. We present advantages of the latter with a dedicated \gls{A2G} network with a centralized core since this can enable better harmonization en-route and flexibility in providing \gls{QoS} for different applications and resource allocation. However, the authors of \cite{8552136} focused on the required modifications on the radio access part of the already-existing ground cellular infrastructures to provide \gls{A2G} connectivity. They suggested placing directional antenna arrays facing upward on the \glspl{BS} for airplanes, so that they could maximize the \gls{SNR} with beamforming.

\begin{figure*}[t]
\begin{center}
	\centerline{\includegraphics[width=0.75\textwidth,keepaspectratio]{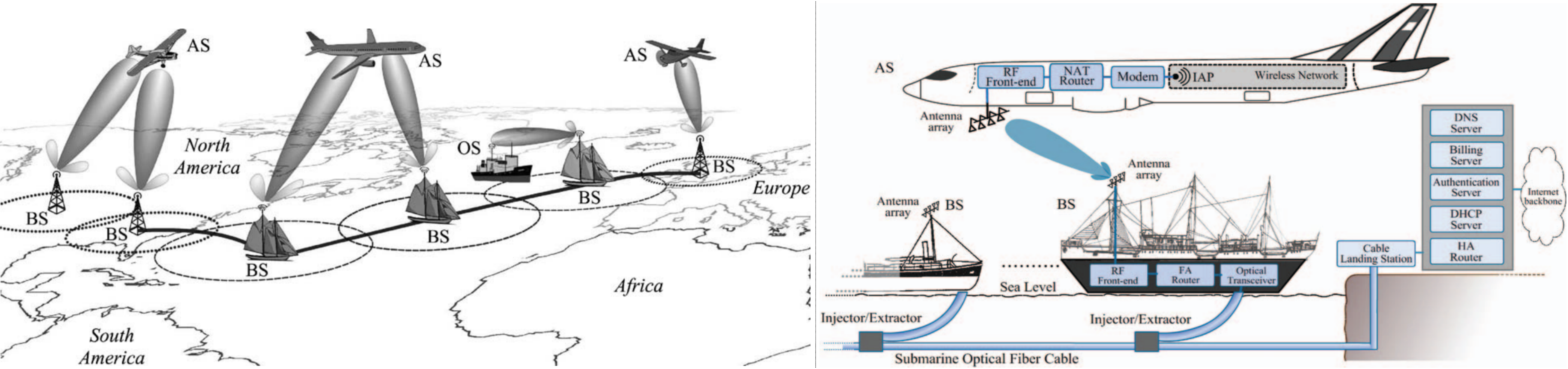}}
	\caption {Proposed network architecture in \cite{7073483} to provide connectivity over the North Atlantic ocean by placing static ships as \glspl{BS} along the already-existing underwater fiber cables.}
	\label{fig: 7073483}
\end{center}
\end{figure*}

Providing connectivity over the ocean is the most problematic scenario for airplane connectivity in international routes. Although conventional networks utilize \gls{SatCom} over the ocean, they cannot provide enough capacity to deliver high-bandwidth rates along with low end-to-end latency. To avoid this problem, the authors of \cite{7073483} came up with a novel design of a ground network architecture with stationary ships located along the fiber lines in the north Atlantic ocean. We show the proposed architecture of \cite{7073483} in \autoref{fig: 7073483}. They suggested integrating ground \glspl{BS} in these ships as well as in other remote islands. The fiber optic cables could serve as the backbone network to the \glspl{BS}. This way, stationary ships could serve as communication as well as navigation service to airplanes, each \gls{BS} covering a radius of approximately 370 km. This is such an interesting solution for transoceanic connectivity; however, the cost of operation of such a network with stationary ships in the ocean can be unfeasible from our perspective. Additionally, the waves in the ocean can continuously swing the ships causing the beam directions to alter.

The authors of \cite{8569862} present a novel architecture for passenger airplanes, which consists of airplane, access and ground network segments. They provide four different radio links for \gls{A2G} connectivity using \glspl{SDR}: 1) satellite; 2) \gls{LDACS}; 3) \gls{AeroMACS}; 4) \gls{VDL-2}. They also integrate \gls{SDN} on both the airplane and ground network segments to dictate the link selection, packet scheduling and other network configurations. 

\comment{
\begin{figure}[t]
\begin{center}
	\centerline{\includegraphics[width=0.5\textwidth,keepaspectratio]{Images/9142706.pdf}}
	\caption {Proposed control and user plane architectures to integrate \gls{UTM} into 5G Ecosystem \cite{9142706}. In control plane, \gls{UTM} communicates with \gls{UAV} via a coordinator gate called \textit{\gls{UTM} Gate Function} and in the user plane, \gls{UTM} communicates the \gls{UAV} via \textit{\gls{UTM} Data Network (UTM DN)}, which handles the admission procedure.}
	\label{fig: 9142706}
\end{center}
\end{figure}

\begin{figure}[t]
\begin{center}
	\centerline{\includegraphics[width=0.5\textwidth,keepaspectratio]{Images/9076122.pdf}}
	\caption {Proposed service-based architecture to include \gls{UAV} services in 5G \cite{9076122}. \textit{\gls{UAV}-based Network Service Control} function is responsible for \gls{UAV}-related services and they set up Northbound \gls{API} to establish connectivity between \gls{USS} and the 3$^\text{rd}$ party application functions.}
	\label{fig: 9076122}
\end{center}
\end{figure}
}

\subsection{Network Architectures for UAVs}
\label{section: networkarchitectures_uavs}

As \gls{RTCA} specified only the physical layer for \gls{CNPC}, the authors of \cite{7763289} proposed a 4G-based network architecture for the upper layers. They inherited the conventional 4G functions into \gls{UAV} environments to support mobility, security and access to the public. They also modified the PHY layer of 4G according to the \gls{RTCA}’s \gls{TDD}-based physical layer standards with the proposal of a new type of channel for \gls{CNPC}-specific message exchange.

\begin{figure}[t]
\begin{center}
	\centerline{\includegraphics[width=0.5\textwidth,keepaspectratio]{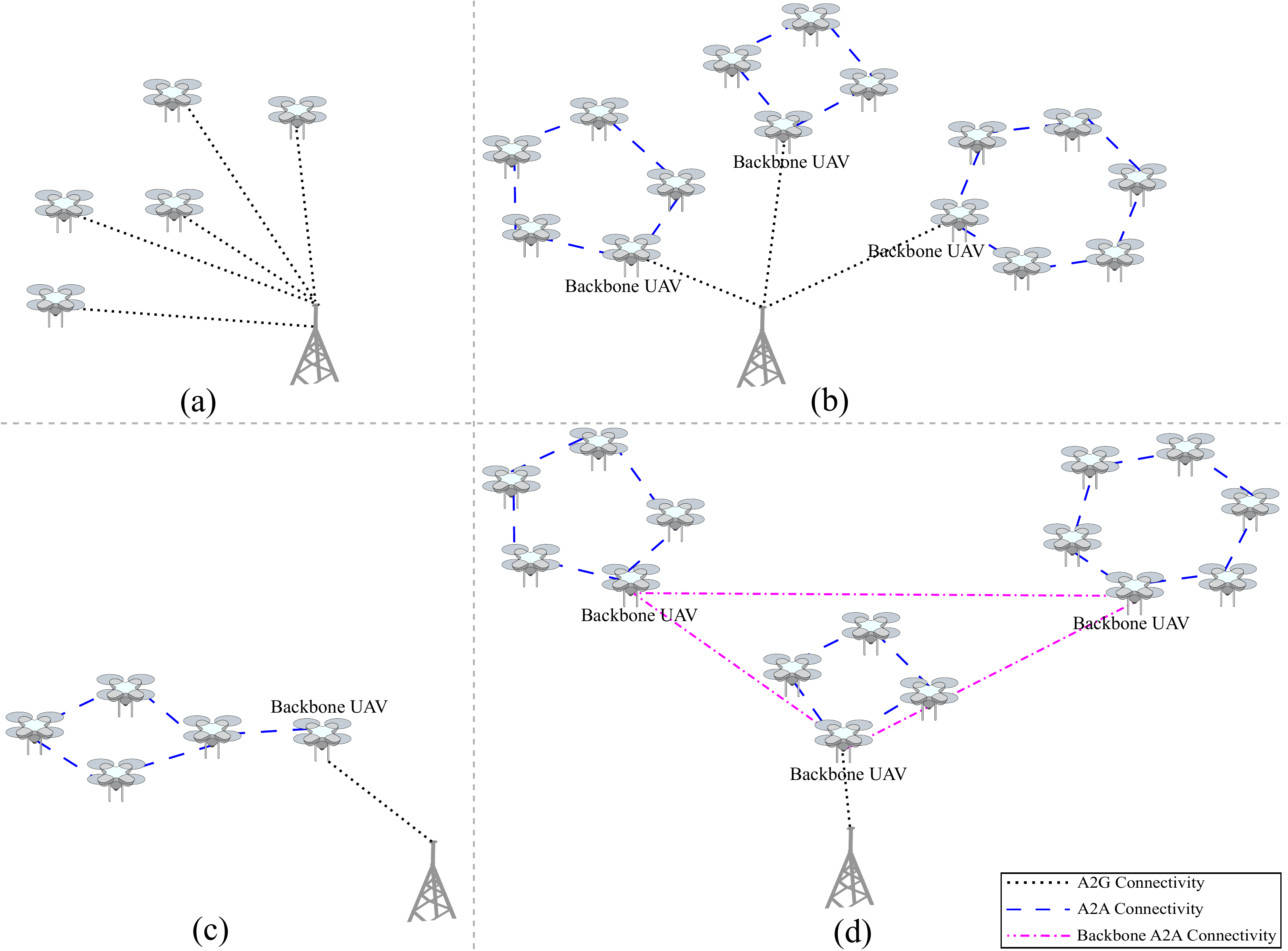}}
	\caption {Four types of conventional ad-hoc architectures: (a) Centralized, (b) Decentralized single cluster, (c) Decentralized multi-cluster, (d) Decentralized multi-cluster with primary cluster heads.}
	\label{fig: a2a_architectures}
\end{center}
\end{figure}

In regards to \gls{A2A}, \cite{9199627, 8550873, 8277614, 6825193} presented the conventional four types of \gls{A2A} architectures for \glspl{UAV}, as demonstrated in \autoref{fig: a2a_architectures}. While the decentralized multi-cluster architecture is practical for the missions that involve \glspl{UAV} with different flight characteristics and mission tasks, single-cluster is preferable for \glspl{UAV} of the same types. In \autoref{fig: a2a_architectures} (d), one \gls{UAV} serves as the backbone network for each cluster to provide connectivity with the ground station. While the centralized architecture minimizes the latency for \gls{CNPC}, decentralized architectures are more robust for \gls{A2A} links.

In order to avoid a single point of failure problem in the above-mentioned ad-hoc architectures, authors of \cite{9199627} proposed having a secondary cluster head besides the primary head. Selecting the second cluster head based on the battery size, \glspl{RSRP} and available battery life, the secondary head replaces the primary head once the primary head falls below a certain battery level threshold.

\begin{figure*}[t]
\begin{center}
	\centerline{\includegraphics[width=1\textwidth,keepaspectratio]{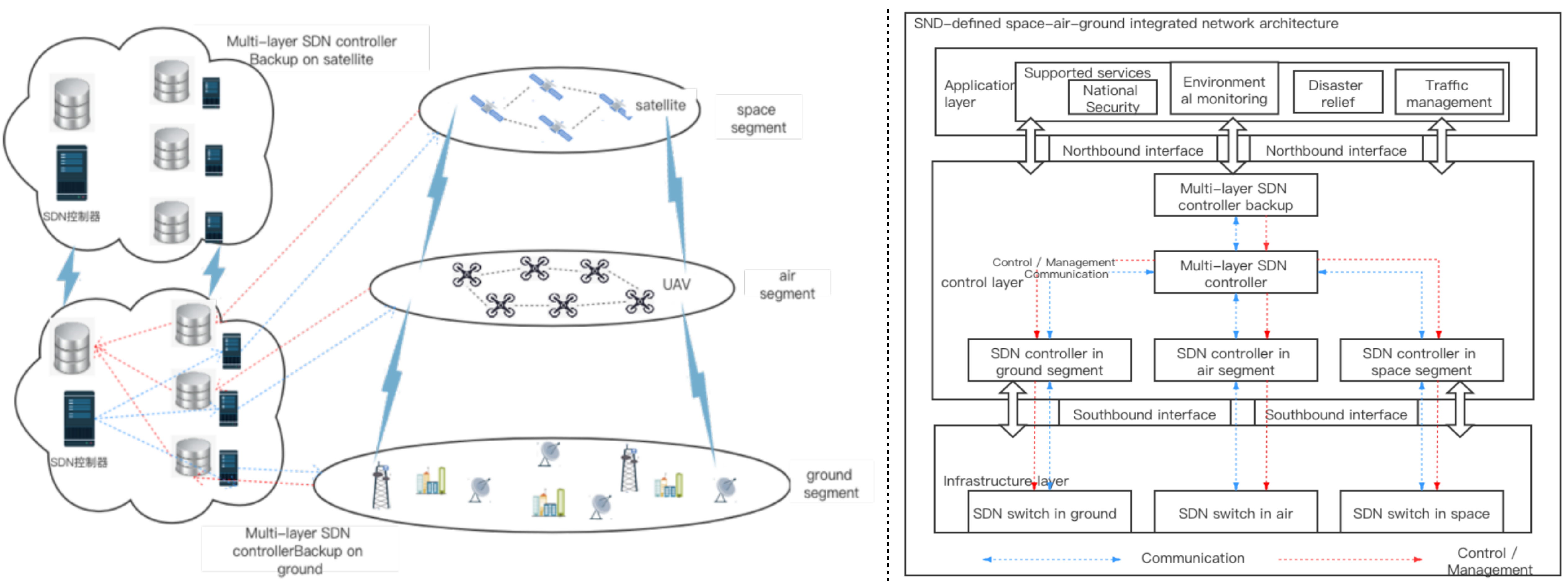}}
	\caption {Space-air-ground architecture proposed by \cite{9213109}. While space and ground elements provide connectivity to the air segment, \gls{SDN} manages the control functions of each layer individually.}
	\label{fig: 9213109}
\end{center}
\end{figure*}

Integration of space-air-ground network segments also gained attention in literature, and the authors of \cite{9213109} demonstrate an \gls{SDN}-based space-air-ground architecture. Each layer has its own \gls{SDN} controller, which are part of a total controller, as shown in \autoref{fig: 9213109}. Upper layer controller manages the network resources and adjusts the network behavior. They store the backup of the upper controller in satellites to improve the security. Furthermore, the authors of \cite{9184022} present a block-chain based 6G space-air-ground network architecture. The architecture is very similar to \autoref{fig: 9213109}, except that they connect all the layers to a common block-chain network instead of \gls{SDN} controllers. They inherit the block-chain technology to regulate the network behavior and manage the network security. They also charge the battery of \glspl{UAV} with the signals from the 6G \glspl{BS}.

\subsection{5G System Architectures for UTMs}
\label{section: networkarchitectures_utms}

The authors of \cite{9142706, 9076122} studied the required modifications on the cellular network architectures to integrate \gls{UTM} services into it. Based on the 5G system architecture of \gls{3GPP} \cite[Fig. 4.2.3-1]{3GPP_23501}, they implemented a special gateway, \textit{North-Bound Interface}, with a \textit{\gls{UTM} Gate Function} for \gls{UTM} \cite{9142706}. The gateway serves as the mediator and coordinator of all the \gls{UTM} functions. For cross-border flights, they proposed using dual sim cards on \glspl{UAV}, since the current de-registration and re-registration processes take longer than 500 ms between \gls{UAS} and \gls{UTM}, as required by \gls{3GPP} \cite{3GPP_22125}. 

Similarly, the authors of \cite{9076122} also modified the 5G architecture to include aerial navigation, airborne user/control plane network functions, connectivity management and control. They introduced a novel control plane network function, \textit{\gls{UAV}-based Network Service Control}, to handle connectivity, resource management and location services related to \glspl{UAV}. 

Overall, these studies focus on the network architectures for airplanes and \glspl{UAV}. Future studies should also elaborate on the dedicated ground network architectures for the \gls{UTM}-\gls{eVTOL} ecosystem, with the consideration of \gls{LDACS} technology. As \glspl{eVTOL} operate at higher altitudes compared with \glspl{UAV}, they require particular attention. The sidelobe radiation from already-existing ground infrastructures may not be sufficient to provide reliable connectivity at high altitudes, up to 1 km. Moreover, space-air-ground architectures are essential to integrate heterogeneous connectivity to the \glspl{AV} and thus, architectures targeting the combination of cellular and \gls{LEO} networks gain importance.

\section{Open Source Simulation and Emulation Platforms}
\label{section: simulators}

In this section, we provide an up-to-date list of open-source simulator and emulator platforms that can be practical in academic research. We divide the simulators into three main categories: 1) \gls{UAV} simulators; 2) airplane simulators; 3) \gls{ATM} simulators. In \autoref{table: uavsimulators} and \autoref{table: airplanesimulators}, we categorize the available sources and compare them with respect to the platform type (simulator or emulator), flight and networking capabilities as well as their particular features, respectively. Some of the platforms have both flight and networking simulation capabilities. These can facilitate the researchers’ network-related experiments with minimal implementation efforts.

Each simulator and emulator platform has particular features that make them stand out toward different use cases. While AVENS \cite{avenssim} can be suitable for \gls{FANET} simulations, UAVSim is rather useful for the simulations with satellite networks due to its integrated OMNeT++ OS3 satellite extension \cite{6825196}. It also provides means to perform cyber security analysis in the simulation. Furthermore, FlyNetSim \cite{baidya2018flynetsim} offers multiple networking interfaces to perform multilink networking emulation in the same \gls{UAV} platform.

Beside the ones with networking capabilities, AirSim is a powerful \gls{UAV} simulator and emulator platform with vehicle, sensor and environment models \cite{airsim2017fsr}. However, it is not scalable for multi-vehicle simulations. In this regard, FlightGoogles can be an alternative option to enable multi-\gls{UAV} simulations yet, it is especially designed for visual applications \cite{Guerra_2019}. Lastly, number of platforms provide both simulation and emulation capabilities such as VENUE \cite{8880644}, FlyNetSim \cite{baidya2018flynetsim} and UB-ANC Emulator \cite{10.1177/1756829319837668}. They can facilitate the deployment of custom algorithms from software to actual hardware environment.

Besides the \gls{AV} simulators, we also have a short list of open-source \gls{UTM} and \gls{ATM} simulators, which can also be practical toward future studies of air traffic management: 1) InterUSS \cite{interusssim}; 2) BlueSky \cite{blueskysim}; 3) ELSA \cite{elsasim}; 4) EuroScope \cite{euroscopesim}; 5) OpenScope \cite{openscopesim}. InterUSS provides an interoperability platform between different \gls{UAS} service suppliers, which is part of the \gls{UTM} architecture \cite[Figure 3]{nasa_utm_conopsv2}. Other simulators are mainly developed to foster the development of future \gls{ATM} solutions, as we described in Section II-A1.

\begin{savenotes}
	\begin{table*}[!htbp]
		\caption {Open-source \gls{UAV} Flight and Network Simulators.\vspace{-0.5cm}}
		\begin{center}
		\begin{threeparttable}
			{\renewcommand{\arraystretch}{1.75} 
			\scalebox{0.8}{
				\begin{tabular}{l c c l l}
					\hline
					\textbf{Name} & \textbf{\makecell{Simulator (S)/\\Emulator (E)}} & \textbf{\makecell{Flight (F)/\\ Network (N)}} & \multicolumn{2}{l}{\textbf{Other Features}} \\ \hline
					\rule{0pt}{2.25em}\makecell[l]{AirSim\\\cite{airsim2017fsr, airsim_link}} & S\&E & F & \makecell[l]{- Collision avoidance capability.\\- Camera, environment, sensor, vehicle models.\\- Deployable on a cloud.} & \makecell[l]{- Realistic environment.\\- Weather and wind effects.} \\ \cline{4-5}
					\rule{0pt}{1.75em}\makecell[l]{AVENS\\\cite{avenssim}} & S & F\&N$^\ast$ & \makecell[l]{- Flight dynamics and mobility.\\- Multi-vehicle support.} & \makecell[l]{- Weather and wind effects.} \\ \cline{4-5}
					\rule{0pt}{2em}\makecell[l]{AVIATOR\\\cite{baltaci_trafficgeneratorpaper}} & S & \makecell[l]{Data Traffic\\Generator} & - Traffic models based on actual \gls{UAV} data traffic. \\ \cline{4-5}
					\rule{0pt}{2.25em}\makecell[l]{CDSSim\\\cite{cdssim}} & S & F\&N$^\dagger$ & \makecell[l]{- Designed for path planning algorithms.\\- Environment and mobility models.\\- Multi-vehicle support.} & \makecell[l]{- Path planning capability.\\- Synchronization between the flight\\and network simulators.}\\ \cline{4-5}
					\rule{0pt}{1.75em}\makecell[l]{CUSCUS\\\cite{7983185, cuscus_link}} & S & F\&N$^\dagger$ & \makecell[l]{- Flight, mobility and sensor models.\\- Multi-vehicle support.} & \makecell[l]{- Path planning capability.} \\ \cline{4-5}
					\rule{0pt}{1.75em}\makecell[l]{DroneVR\\\cite{8942250}} & S & F & \makecell[l]{- Environment and sensor models.\\- Object detection, tracking and avoidance capabilities.} & \makecell[l]{- Path planning and optimization capabilities.\\- Web-based simulator.} \\ \cline{4-5}
					\rule{0pt}{1.75em}\makecell[l]{FL-Air\\\cite{flairsim, flairsim_userguide}} & S & F & \makecell[l]{- Multi-vehicle support.\\- Path planning capability.} & \makecell[l]{- Sensor models.} \\ \cline{4-5}
					\rule{0pt}{2.25em}\makecell[l]{FlightGoogles\\\cite{Guerra_2019}} & S & F & \makecell[l]{- Camera, sensor, motor dynamics and vehicle models.\\- Collision detection capability.\\- Deployable on a cloud.} & \makecell[l]{- Multi-vehicle support.\\- Path planning capability.\\- Realistic environment.} \\ \cline{4-5}
					\rule{0pt}{2.25em}\makecell[l]{FlyNetSim\\\cite{baidya2018flynetsim}} & S\&E & F\&N$^\dagger$ & \makecell[l]{- \gls{IoT} connectivity environment.\\- Multi-networking support\\(e.g. \glspl{UAV} can have both 4G and \gls{WiFi} interfaces.).} & \makecell[l]{- Multi-vehicle support.\\- Synchronization between the flight\\and network simulators.} \\ \cline{4-5}
					\rule{0pt}{1.75em}\makecell[l]{Hector\_quadrator\\\cite{hectorquadratorsim}} & S & F & \makecell[l]{- Camera, flight, motor dynamics,\\sensor and vehicle models.} \\ \cline{4-5}
					\rule{0pt}{1.75em}\makecell[l]{IoD-Sim\\\cite{8880832}} & S & F\&N$^\dagger$ & \makecell[l]{- Mobility models.\\- Multi-vehicle support.} & \makecell[l]{- Path planning capability.} \\ \cline{4-5}
					\rule{0pt}{2em}\makecell[l]{jMAVSim\\\cite{jmavsim_link, jmavsim_multivehiclesim, jmavsim_hilsim}} & S\&E & F & \makecell[l]{- Camera, sensor and vehicle models.} & \makecell[l]{- Multi-vehicle support.} \\ \cline{4-5}
					\makecell[l]{LimoSim\\\cite{Sliwa_2019, limosim_link}} & S & F\&N$^\dagger,^\ddagger$ & \makecell[l]{- Collision avoidance, mobility prediction,\\path planning and situational\\awareness capabilities.} & \makecell[l]{- Environment, \gls{UAV} energy consumption\\and mobility models.\\- Multi-vehicle support.\\- Synchronization between the flight\\and network simulators.} \\ \cline{4-5}
					\makecell[l]{MAVBench\\\cite{8574594}} & S\&E & F & \makecell[l]{- Battery, camera, energy,\\environment and sensor models.} & \makecell[l]{- Designed for performance and\\power optimization-related studies.\\- Object detection and tracking capability.} \\ \cline{4-5}
					\rule{0pt}{1.75em}\makecell[l]{multiUAV\\Simulation \cite{multiuavsimulation_link}} & S & F\&N$^\ddagger$ & \makecell[l]{- Charging stations.\\- Multi-vehicle support.} & \makecell[l]{- \gls{UAV} energy consumption model.} \\ \cline{4-5}
					\rule{0pt}{1.75em}\makecell[l]{NUAV\\\cite{7918926, nuav_link}} & S & F & \makecell[l]{- Environment, flight control and vehicle models.\\- Multi-vehicle support.} & \makecell[l]{- Realistic environment.\\- Weather effects.} \\ \cline{4-5}
					\rule{0pt}{2.5em}\makecell[l]{Obstacle Avoidance\\Simulator for \glspl{UAV}\\\cite{bhagat2020uav, obstacle_avoidance_for_uav_link}} & S & F & \makecell[l]{- Obstacle avoidance, object tracking and\\path planning capabilities.} \\ \cline{4-5}
					\rule{0pt}{2.25em}\makecell[l]{OpenAMASE\\\cite{openamasesim}} & S & F & \makecell[l]{- Camera, flight dynamics, sensor,\\\gls{UAV} autopilot and vehicle models.\\- Multi-vehicle support.} & \makecell[l]{- Object detection and\\path planning capabilities.\\- Wind effects.} \\ \cline{4-5}
					\rule{0pt}{1.75em}\makecell[l]{OpenUAV\\\cite{8443728, openuav_link}} & S & F & \makecell[l]{- Available from a cloud service.\\- Camera model.} & \makecell[l]{- Multi-vehicle support.\\- Path planning capability.} \\ \cline{4-5}
					\rule{0pt}{1.75em}\makecell[l]{PLANE\\\cite{boccadoro2019plane}} & S & F & \makecell[l]{- Environment and vehicle models.\\- Multi-vehicle support.} & \makecell[l]{- Path optimization capability.\\- Weather effects.}\\ \cline{4-5}
					\rule{0pt}{1.75em}\makecell[l]{ROS Quadrotor\\Simulator \cite{rosquadratorsim}} & S & F & \makecell[l]{- Camera, engine, sensor\\and vehicle models.} & \makecell[l]{- Path planning capability.} \\ \cline{4-5}
					\makecell[l]{RotorS\\\cite{rotorssim}} & S & F & \makecell[l]{- Camera, environment, flight dynamics,\\sensor and vehicle models.} & \makecell[l]{- Collision avoidance and\\path planning capabilities.\\- Wind effects.} \\ \cline{4-5}
					\makecell[l]{Simbeeotic\\\cite{6920935}} & S\&E & F & \makecell[l]{- Multi-vehicle support.} \\ \cline{4-5}
					\rule{0pt}{2.25em}\makecell[l]{SwarmLab\\\cite{soria2020swarmlab}} & S & F & \makecell[l]{- Collision avoidance and\\object detection and\\path planning capabilities.} & \makecell[l]{- Environment and vehicle models.\\- Multi-vehicle support.} \\ \cline{4-5}
					\makecell[l]{UB-ANC-\\Emulator \cite{10.1177/1756829319837668}} & S\&E & F\&N$^\dagger$ & \makecell[l]{Energy comsumption, flight dynamics,\\mobility, sensor and vehicle models.\\- Multi-vehicle support.} & \makecell[l]{- Path planning capability.\\- Synchronization between the flight\\emulator and network simulator.} \\ \cline{4-5}
					\rule{0pt}{1.75em}\makecell[l]{UE4\\\cite{10.1007/978-3-319-46448-0_27}} & S & F & \makecell[l]{- Camera and flight dynamics models.\\- Multi-vehicle support.} & \makecell[l]{- Object tracking capability.\\- Realistic environment.} \\ \cline{4-5}
					\rule{0pt}{2em}\makecell[l]{uavEE\\\cite{8607252}} & E & F\&N$^\lozenge$ & \makecell[l]{- Multi-vehicle support.} & \makecell[l]{- Propulsion power model.} \\ \cline{4-5}
					\rule{0pt}{1.75em}\makecell[l]{UAVSim\\\cite{6825196}} & S & F\&N$^\circledast$ & \makecell[l]{- Designed for cyber security analysis.\\- Multi-vehicle support.} & \makecell[l]{- Mobility and vehicle models.} \\ \cline{4-5}
					\rule{0pt}{1.75em}\makecell[l]{VENUE\\\cite{8880644}} & S\&E & F\&N$^\dagger$ & \makecell[l]{- Mobility model.\\- \gls{FANET} routing algorithms available.} & \makecell[l]{- Multi-vehicle support.} \\ 
                \hline
				\end{tabular}}}
				\label{table: uavsimulators}
				\begin{tablenotes}
				        \fontsize{6}{10}
				        \item \hspace{1cm} \textit{$^\ast$Integrated with the OMNeT++ \gls{FANET} simulator.}
				        \item \hspace{1cm} \textit{$^\dagger$Integrated with NS-3 simulator.}
				        \item \hspace{1cm} \textit{$^\ddagger$Integrated with OMNeT++ simulator.}
				        \item \hspace{1cm} \textit{$^\lozenge$Users can select the wireless radio interface.}
				        \item \hspace{1cm} \textit{$^\circledast$Integrated with the OMNeT++ OS3 Satellite Extension.}
				\end{tablenotes}
		\end{threeparttable}
		\end{center}
	\end{table*}
\end{savenotes}

\begin{savenotes}
	\begin{table}[t]
		\caption {Open-source Airplane Simulators without Networking Support.\vspace{-0.225cm}}
		\begin{center}
			{\renewcommand{\arraystretch}{1.75} 
				\begin{tabular}{l c}
					\hline
					\textbf{Name} & \textbf{Notes} \\ \hline
				\rule{0pt}{2.25em}CRRCSim \cite{crrcsim} & \makecell[l]{- Realistic flight model.\\- Support for aircraft control\\using remote control.} \\ \cline{2-2}
				\rule{0pt}{4.5em}FlightGear \cite{flightgearsim} & \makecell[l]{- Aircraft modeling.\\- Flight dynamic models.\\- Multi-vehicle support.\\- Sky model.\\- Vehicle types: A320, 747,\\1903 Wright Flyer.\\- World scenery database.} \\ \cline{2-2} 
				\rule{0pt}{4.5em}Geo-FS \cite{geofssim} & \makecell[l]{- Flight dynamics.\\- Multi-vehicle support.\\- Real-life \gls{ADS-B} in real-time.\\- Real-time atmospheric conditions.\\- Realistic physics engine.\\- Vehicle types: Aircraft, glider,\\helicopter, air balloon.} \\
                \hline
				\end{tabular}}
				\label{table: airplanesimulators}
		\end{center}
	\end{table}
\end{savenotes}

All in all, open-source community has plenty of simulator and emulator platforms to support the future studies of \gls{FACOM} and the researchers can utilize these resources to facilitate their works and to further extend the capabilities of these platforms. As we leave the detailed overview of these platforms outside the scope of our study, we let the readers investigate further themselves to determine the most suitable ones for their research. Nevertheless, these tables provide the most relevant features and the capabilities of the simulator and emulator platforms. The simulators may have more capabilities than what we listed in the tables.
\section{Open Research Challenges in FACOM}
\label{sec: openresearchchallenges}

In this section, we highlight the future challenges and open problems to provide connectivity to the use cases of \gls{FACOM}.

\subsection{Artificial Intelligence}
\label{sec: openresearchchallenges_AI_Arial}
\gls{AI} recently gained attraction in wireless communications. Number of works applied \gls{AI} in various topics such as finding optimal deployment location for aerial \glspl{BS} \cite{MLPredODUAVWC, LRMUAVWCSRA}, prediction of the location of an aerial \gls{BS} \cite{TPUAVSCRNN} and aerial channel model predictionaaa \cite{AtAPLPMLUE, GNNCMmmWUAVC}.

\gls{RL} is based on trial and error method to learn from past decisions. It is also utilized in various studies to optimize routing \cite{ANUAVRTRL, AUAVNDDPGDRL} as well as ground coverage \cite{QLAQoEUAVCN, MADRLTPUAVMEC}. Nevertheless, we can utilize \gls{AI}-based techniques in \gls{FACOM} in future studies. Also, the available computational power on \glspl{AV} makes it possible to deploy \gls{AI} in the air. We can consider \gls{AI} as the facilitator to realize heterogeneous multilink networking. Novel techniques can enable intelligent routing to efficiently utilize the available channel resources to meet the \gls{QoS} demands of different aerial application. In a multilink networking environment, \gls{AI} can help predict vertical and horizontal handovers that occur in different links and optimize the link selection. Such methods can support meeting the stringent reliability and latency demands of the \gls{CNPC} traffic.

\subsection{Diversity and Unified Network Design}
\label{sec: openresearchchallenges_diversity_integratednetworkdesign}

One of the main challenges in \gls{FACOM} is the diversity of the \glspl{AV}, their use cases and the wireless communication links. \glspl{AV} differ in vehicle and flight characteristics, cruising altitude, and the communication and \gls{SWaP} requirements. Thus, the design of \gls{FACOM} network architectures requires particular considerations to provide connectivity at different altitudes, mobility speeds along with the variant \gls{QoS} metrics.

In \autoref{table: technologies_performance}, we showed the diversity of the network performances of the candidate wireless technologies per each \gls{FACOM} use case. We further emphasized the need of the heterogeneous connectivity for the \gls{RPO} of the \glspl{AV}. Although combining multiple technologies can increase the overall network robustness, its realization in real life is complex and costly. We must consider novel \gls{MAC} and \gls{IP} layer architectures to enable the coexistence of different wireless technologies. Furthermore, we need to consider the adaptation of the already-existing multi-link transport protocols or the development of novel protocols designed with the consideration of \gls{FACOM} applications. Additionally, heterogeneous connectivity implies multiple radio interfaces on the \glspl{AV}, which can challenge the \gls{SWaP} requirements.

Heterogeneous networking in \gls{FACOM} can demand smart coordination of the mixture of horizontal handovers on a same network and vertical handovers between different networks. Ensuring seamless connectivity in these scenarios can be challenging due to the limited information about handovers at \glspl{UE}. Nevertheless, having the capability to predict an upcoming handover can enable the development of intelligent routing and congestion control algorithms to minimize the disruptions on connectivity. Additionally, the \gls{3GPP} introduce conditional handovers to 5G standards in Release 16, which allow \glspl{UE} to make handover decisions at particular scenarios to improve the network robustness \cite{3GPP_21916}. New research items can emerge from this feature to utilize the knowledge about handovers at \glspl{UE}.

We should also scrutinize the emerging virtualization technologies to facilitate the integration of heterogeneous technologies in \gls{FACOM}. \gls{SDN} can help control the network resources in a unified manner and dynamically respond to the varying network conditions. Additionally, network slicing can help isolate priority traffic and can be one of the enablers to ensure high network availability for safety-critical applications. 

\subsection{Network Capacity}
\label{sec: openresearchchallenges_networkcapacity}

The capacity of today's aerial networks is low and can only support a limited number of \glspl{AV} at the same time. For instance, \gls{EAN} can provide only 75 Mbps per aircraft \cite{ean_dt_inmarsat_nokia}. The demand for high throughput applications is also on the rise, especially due to the vision-based applications of \glspl{UAV}. Thus, we need to improve the capacity of the next-generation \gls{FACOM} networks. One possibility is to upgrade the existing systems by deploying more ground stations or launching additional high throughput satellites. Moreover, we can improve the robustness of the communication links by adopting beamforming and higher order modulation schemes. We can also investigate the techniques to reduce the required capacity such as data caching on airplanes or \gls{MEC} for \glspl{RPO}. 

We can further enhance the network capacity by reducing interference. This can be the case mainly in the \gls{LoS} channels of the already-existing cellular infrastructures as well as in the \gls{A2A} links due to the large number of \glspl{AV} sharing the same spectrum. Thus, we should further explore the interference mitigation schemes. Additionally, we must ensure the coexistence of the ground and aerial \glspl{UE}. Additionally, \glspl{AV} operating at different altitudes can influence the capacity of the other vehicles, and we might observe the shadowing effects during vehicle maneuvers. In the end, the available spectrum determines the upper bound on the achievable network capacity, which we elaborate in the following section. 

\subsection{Spectrum Regulation}
\label{sec: openresearchchallenges_spectrumregulation}

The spectrum requirements in \gls{FACOM} are diverse due to the different safety metrics posed by use cases of the \glspl{AV}. In general, we can realize three types of spectrum \cite{ITU_spectrumtypes}: 
\begin{enumerate}
    \item Aeronautical safety spectrum;
    \item Licensed spectrum;
    \item Unlicensed spectrum.
\end{enumerate}

We provide the list of aviation safety spectrum in \autoref{table: aviation_spectrum}. While \gls{ICAO} requires \gls{CNPC} for \glspl{RPO} under the aeronautical safety spectrum \cite{icao_rpas_rules}, we are unclear about the authorization of the usage of this spectrum in the cellular ecosystem. Aeronautical safety spectrum is favorable since it is dedicated only for aerial services and globally harmonized. However, the available safety spectrum is scarce thus, enabling high-rate services is challenging. Aeronautical safety spectrum requires a dedicated aviation network to provide service, therefore the overall connectivity cost can be high.

\begin{savenotes}
	\begin{table}[t]
		\caption {List of the Aviation Safety Spectrum \cite{easa_airtoground_evolution}.\vspace{-0.225cm}}
		\begin{center}
			{\renewcommand{\arraystretch}{1.5} 
				\begin{tabular}{l c c}
					\hline
					\textbf{\makecell[l]{Frequency\\Band}} & \textbf{\makecell{Frequency\\Range\\(MHz)}} & \textbf{Service} \\ \hline
					HF & 2.85 - 22 & Voice, Data \\
					VHF & 117.975 - 137 & Voice, Data \\
					UHF & 235 - 267 & Voice, Data \\
					L-band & 960 - 1164 & Voice, Data \\
					C-band & 5030 - 5091 & Voice, Data \\
					C-band & 5091 - 5150 & Data \\
					\hline
				\end{tabular}}
				\label{table: aviation_spectrum}
		\end{center}
	\end{table}
\end{savenotes}

The large availability of licensed spectrum can enable high-rate services and meet the diverse \gls{QoS} demands of the \glspl{AV}. We can utilize the already-existing networks for \gls{FACOM}, and they can provide cost-efficient networking \cite{GSMA_spectrum}. National authorities regulate the licensed bands, which increases the safety of the spectrum. However, licensed bands are globally not harmonized, and this can be problematic on international routes, especially for commercial airplanes. 

As for the unlicensed spectrum, we can consider it only for the non-safety applications in \gls{FACOM}. Although the majority of the current \gls{UAV} operations take place in this spectrum for \gls{VLoS} operations, we cannot ensure guaranteed resources in a publicly shared spectrum for the safety applications, such as \gls{BVLoS} \glspl{RPO}. Moreover, the need for the regulations of airspace with \gls{UTM} for \gls{AV} operations requires a licensed spectrum. We must take these trade-offs into account for the selection of the pertinent spectrum with respect to the use case.

\subsection{Hardware and Certification}
\label{sec: openresearchchallenges_hardwareandcertification}

\gls{SWaP} specifications of the onboard parts of the communication system play an essential role on the overall efficiency of the \glspl{AV}. Size requirements are challenging, especially on \glspl{UAV} due to the limited available space. Heavier weight leads to higher fuel consumption and thus increased operational cost. Power is a key parameter due to the limited battery size on \glspl{AV}, and the power consumption of the onboard processors directly influences the achievable flight time. Therefore, certain wireless technologies may not be suitable for particular \glspl{AV}. For instance, small-scale \glspl{UAV} cannot carry heavy satellite terminals and the dish antennas are not integrable to the payload. Thus, we need to optimize the \gls{SWaP} specifications of the onboard elements. One fundamental method is to develop unified onboard systems that can operate different wireless technologies at the same time. However, the standardization and the commercialization of the multi-purpose hardware can be time-demanding. For this reason, flexible hardware architectures, such as \glspl{SDR}, can be the enabler of the multi-purpose onboard \gls{RF} systems. 

Certification is a paramount process in the safety-oriented nature of aviation. As \glspl{AV} cannot simply stop or pull off the road in the air, just like ground vehicles, the failures of safety-critical functions are intolerable. Thus, the onboard hardware must be rigid and resilient against failures.  

\subsection{Other Research Challenges}
Data dissemination becomes a relevant topic in the context of \glspl{UAV} since various works consider these vehicles as a means to disseminate data to remote \gls{IoT} and other devices. Although a number of works evaluate different aspects such as the optimization of energy consumption \cite{9174931}, supporting \gls{V2X} networks \cite{9373692} as well as the dissemination of real-time surveillance data on highways \cite{9045425}, we can apply this concept applied in various use cases in \gls{FACOM}. Future studies may consider the data dissemination in \glspl{eVTOL} and airplanes to support their \gls{MTC} use-cases.

Energy-efficiency is a paramount keyword in wireless communications due to the limited battery life of the mobile devices. Number of studied recently considered \glspl{UAV} to assist energy-efficient wireless communications. Such instances include different resource allocation frameworks to optimize the power consumption of ground \glspl{UE} \cite{9417539, 9417458}, the power consumption of \gls{BS} and \gls{UAV} \cite{9402734} as well as the power consumption of the overall system \cite{9348068}. Energy-efficient communication is a significant topic especially for \glspl{eVTOL} considering their limited energy in the air. Thus, future studies should consider the optimization of \gls{A2G} networks to minimize the power consumption of these vehicles. Lastly, several works utilized \glspl{UAV} to enable secure communications for different use-cases such as military \gls{IoT} \cite{9289282} and terrestrial cognitive radio networks \cite{9364745}. Enabling secure communications should also be considered in \gls{FACOM} since the \glspl{AV} will perform critical missions and their malfunction can cause threatful events to air passengers as well as the public on the ground.

\section{Conclusion}
\label{section: conclusion}

\comment{
\begin{figure}[t]
\begin{center}
	\centerline{\includegraphics[width=0.5\textwidth,keepaspectratio]{Images/facom_components-eps-converted-to.pdf}}
	\caption {The building elements of \gls{FACOM}. As aerospace demands cause the emergence of aerial use cases, new types of \glspl{AV} along with off-board systems come out with a demand of wireless connectivity. This survey specifies the connectivity demands emerging from the use cases and evaluates the capabilities of the communication technologies to meet these demands in \gls{FACOM}.}
	\label{fig: facom_components}
\end{center}
\end{figure}

The passion of humanity toward a more globally-connected world pushes new demands to the aerospace community to travel anywhere and anytime. These demands trigger the emergence of novel aerial use cases. \gls{FACOM} use cases pave the way for developing new \gls{AV} platforms, wireless communication in the sky and off-board systems to realize end-to-end connectivity. This are the building elements in \gls{FACOM} and how they influence each other, as illustrated in \autoref{fig: facom_components}. With the focus on connectivity, we discussed the contribution of each of these elements in our study. 
} 
The passion of humanity toward a more globally-connected world requires new means of aerial travel and transportation. Increasing widespread use of \glspl{AV} complicates civil air traffic, which demands wireless connectivity for various use cases. In this survey, we studied the wireless connectivity demands emerging from the connectivity use cases of the \glspl{AV}, and evaluated the potential ways to meet these demands utilizing different wireless communication technologies along with the identified challenges. 

The findings along with the main takeaways of this survey are summarized as follows: 

\begin{itemize}
    \item Each type of \gls{AV} hosts diverse use cases with specific \gls{QoS} metrics, which bring challenges toward providing connectivity solutions.
    \item The \gls{eVTOL} operations introduce new connectivity demands at higher altitudes (1 km) than that of commercial \glspl{UAV} (150 m). This further complicates the design of a unified architecture for \gls{UTM} networks. Moreover, the multi-pilot operations of \glspl{eVTOL} require novel networking and handover mechanism to ensure robustness. 
    \item \gls{UTM} demands connectivity to different entities to regulate the operations at low-altitude airspace since inter-region, and cross-border flights require novel handoff mechanisms between different \gls{UTM} entities.
    \item Our literature survey highlights the attention of academia on cellular networks for \gls{FACOM}. We also consider cellular systems as one of the prominent technologies to support the use cases in urban areas. The upcoming \gls{LDACS} standard is another potential technology for \glspl{RPO}, as it utilizes the aviation safety spectrum and the modified physical layer to enhance robustness. Satellite systems can also complement the ground-based architectures, particularly with the \gls{LEO} constellations. However, their performance under high mobility is uncertain. Furthermore, the main disadvantages of IEEE standards for \gls{FACOM} are the unlicensed spectrum and the lack of multi-hop architecture. 
    \item Dedicated \gls{A2G} network solutions already exist \cite{ean_dt_inmarsat_nokia}; however, we need to improve the network capacity to address the increasing demands. 
    \item Our match study on the communication technologies versus aerial use cases shows the lack of a unified networking solution to meet all the connectivity demands of the use cases. Therefore, we need to realize multi-technology heterogeneous architectures to enable \gls{FACOM}. 
    \item Although we observe the demand for heterogeneous networking, we did not come to a conclusion on which technology combinations are most suitable, particularly for \glspl{RPO}. Furthermore, it is significant to determine the required number of technologies or parallel links to ensure ultra reliable and seamless connectivity. 
    \item Although several works, such as \cite{9213109, 9184022, 8368236}, have studied heterogeneous network architectures, multi-technology architectures require further investigation, particularly with the potential technology combinations specific to each use case, as shown in \autoref{table: technologies_matchstudy}.
\end{itemize}

Overall, \gls{FACOM} is the enabler of future aerial use cases. Thus, the telecommunication industry is on the edge of a substantial opportunity to expand its wings to the sky. The open research challenges need to be addressed to enable the use cases of \gls{FACOM} in the near future. 
\section*{Acknowledgment}
\label{sec: acknowledgment}

The authors would like to thank Dr. S. Hofmann and Dr. V. Bajpai for their useful review on the paper, Mr. S. Duhovnikov for his helpful discussion about the concept of operations of eVTOLs with the traffic management entities, and Dr. M. Klügel for the discussion on the definition of the communication reliability. 

\bibliographystyle{IEEEtran}
\bibliography{IEEEabrv,Bibliography}

\begin{thebibliography}{100}
\providecommand{\url}[1]{#1}
\csname url@samestyle\endcsname
\providecommand{\newblock}{\relax}
\providecommand{\bibinfo}[2]{#2}
\providecommand{\BIBentrySTDinterwordspacing}{\spaceskip=0pt\relax}
\providecommand{\BIBentryALTinterwordstretchfactor}{4}
\providecommand{\BIBentryALTinterwordspacing}{\spaceskip=\fontdimen2\font plus
\BIBentryALTinterwordstretchfactor\fontdimen3\font minus
  \fontdimen4\font\relax}
\providecommand{\BIBforeignlanguage}[2]{{%
\expandafter\ifx\csname l@#1\endcsname\relax
\typeout{** WARNING: IEEEtran.bst: No hyphenation pattern has been}%
\typeout{** loaded for the language `#1'. Using the pattern for}%
\typeout{** the default language instead.}%
\else
\language=\csname l@#1\endcsname
\fi
#2}}
\providecommand{\BIBdecl}{\relax}
\BIBdecl

\bibitem{7463007}
S.~{Hayat}, E.~{Yanmaz}, and R.~{Muzaffar}, ``Survey on unmanned aerial vehicle
  networks for civil applications: A communications viewpoint,'' \emph{IEEE
  Communications Surveys \& Tutorials}, vol.~18, no.~4, pp. 2624--2661, 2016,
  \url{https://doi.org/10.1109/COMST.2016.2560343}.

\bibitem{plass12}
S.~{Plass}, ``Seamless networking for aeronautical communications: One major
  aspect of the sandra concept,'' \emph{IEEE Aerospace and Electronic Systems
  Magazine}, vol.~27, no.~9, pp. 21--27, 2012,
  \url{https://doi.org/10.1109/MAES.2012.6366089}.

\bibitem{DLR_Uspace}
D.~Geister and B.~Korn, ``Concept for urban airspace integration {DLR}
  {U-Space} blueprint,'' DLR, Tech. Rep., Dec. 2017, [Online]. Available
  at:\url{https://www.dlr.de/fl/Portaldata/14/Resources/dokumente/veroeffentlichungen/Concept_for_Urban_Airspace_Integration.pdf}
  [Accessed on 16/01/2021].

\bibitem{icao_annex8}
ICAO, ``Annex 8 to the convention on international civil aviation -
  airworthiness of aircraft,'' Tech. Rep., Jul. 2018, twelfth Edition,
  [Online]. Available
  at:\url{https://store.icao.int/en/annex-8-airworthiness-of-aircraft}
  [Accessed on 16/01/2021].

\bibitem{8746290}
J.~{Sae}, R.~{Wiren}, J.~{Kauppi}, H.~{Maattanen}, J.~{Torsner}, and
  M.~{Valkama}, ``Public {LTE} network measurements with drones in rural
  environment,'' in \emph{2019 IEEE 89th Vehicular Technology Conference
  (VTC2019-Spring)}, 2019, pp. 1--5,
  \url{https://doi.org/10.1109/VTCSpring.2019.8746290}.

\bibitem{sita_costeffectiveip}
E.~Mitchell, ``New generation aircraft communications - a unified solution for
  congested skies, white paper,'' SITA FOR AIRCRAFT, Tech. Rep., [Online].
  Available
  at:\url{https://www.sitaonair.aero/Resources/Documents/UnifiedAircraft/4c22a80b43de482686b56bfe0005ae9a.pdf}
  [Accessed on 16/01/2021].

\bibitem{nasa_spo_conops}
NASA, ``Concept of operations for {RCO} {SPO},'' Tech. Rep. ARC-E-DAA-TN44254,
  2017, [Online]. Available
  at:\url{https://ntrs.nasa.gov/citations/20170007262} [Accessed on
  16/01/2021].

\bibitem{3GPP_36777}
3GPP, ``Study on enhanced {LTE} support for aerial vehicles (release 15),''
  Tech. Rep. 36.777, 2017, [Online]. Available
  at:\url{https://www.3gpp.org/ftp//specs/archive/36_series/36.777/} [Accessed
  on 16/01/2021.

\bibitem{3GPP_38811}
------, ``Study on new radio (nr) to support non terrestrial networks (release
  15),'' Tech. Rep. 38.811, 2018, [Online]. Available
  at:\url{https://www.3gpp.org/ftp//specs/archive/38_series/38.811/} [Accessed
  on 16/01/2021].

\bibitem{3GPP_22125}
------, ``Unmanned aerial system ({UAS}) support in {3GPP} (release 17),''
  Tech. Rep. TS 22.125, 2019,
  \url{https://www.3gpp.org/ftp/specs/archive/22_series/22.125/}.

\bibitem{3GPP_23754}
------, ``Study on supporting unmanned aerial systems ({UAS}) connectivity,
  identification and tracking (release 17),'' Tech. Rep. TR 23.754, 2020,
  [Online]. Available
  at:\url{https://www.3gpp.org/ftp/specs/archive/23_series/23.754/} [Accessed
  on 16/01/2021].

\bibitem{3GPP_23755}
------, ``Study on application layer support for unmanned aerial systems
  ({UAS}); (release 17),'' Tech. Rep. TR 23.755, 2020, [Online]. Available
  at:\url{https://www.3gpp.org/ftp/specs/archive/23_series/23.755/} [Accessed
  on 16/01/2021].

\bibitem{3GPP_22829}
------, ``Enhancement for unmanned aerial vehicles; stage 1 (release 17),''
  Tech. Rep. TR 22.829, 2019, [Online]. Available
  at:\url{https://www.3gpp.org/ftp/specs/archive/22_series/22.829/} [Accessed
  on 16/01/2021].

\bibitem{3GPP_36785}
------, ``Vehicle to vehicle ({V2V}) services based on {LTE} sidelink; user
  equipment ({UE}) radio transmission and reception (release 14),'' Tech. Rep.
  TR 36.785, 2016, [Online]. Available
  at:\url{https://www.3gpp.org/ftp//specs/archive/36_series/36.785/} [Accessed
  on 16/01/2021].

\bibitem{3GPP_22261}
------, ``Service requirements for the {5G} system; stage 1 (release 18),''
  Tech. Rep. TS 22.261, 2021, [Online]. Available
  at:\url{https://www.3gpp.org/ftp/specs/archive/22_series/22.261/} [Accessed
  on 16/01/2021].

\bibitem{oneweb}
OneWeb, ``Oneweb - an introduction,'' 2019, [Online]. Available
  at:\url{https://www.oneweb.world/assets/news/media/OneWeb-Introduction-May2019.pdf}
  [Accessed on 16/01/2021].

\bibitem{McDowell_2020}
J.~C. McDowell, ``The low earth orbit satellite population and impacts of the
  spacex starlink constellation,'' \emph{The Astrophysical Journal}, vol. 892,
  no.~2, p. L36, Apr 2020, \url{http://dx.doi.org/10.3847/2041-8213/ab8016}.

\bibitem{wifi6}
A.~S. George and A.~S.~H. George, ``A review of {Wi-Fi} 6 : The revolution of
  6th generation {Wi-Fi} technology,'' \emph{Research Inventy: International
  Journal of Engineering and Science}, vol.~10, pp. 56--65, 2020,
  \url{http://www.researchinventy.com/papers/v10i9/I10095665.pdf}.

\bibitem{9090146}
E.~{Khorov}, I.~{Levitsky}, and I.~F. {Akyildiz}, ``Current status and
  directions of {IEEE} 802.11be, the future {Wi-Fi} 7,'' \emph{IEEE Access},
  vol.~8, pp. 88\,664--88\,688, 2020,
  \url{https://doi.org/10.1109/ACCESS.2020.2993448}.

\bibitem{wifi_alliance_halow}
Wi-Fi\;Alliance, ``{Wi-Fi} {HaLow} use case for wireless field network on
  industry automation,'' 2016, [Online]. Available
  at:\url{https://www.ieee802.org/1/files/public/docs2016/liaison-WFA-Wi-Fi-HaLow-0516.pdf}
  [Accessed on 16/01/2021].

\bibitem{5gpp_reliabilityranges}
5G-PPP, ``{Euro-5G} – supporting the european {5G} initiative {D2.6} - final
  report on programme progress and {KPI}s,'' Tech. Rep. H2020-ICT-2014-2, 2017,
  version: V0.2. [Online]. Available
  at:\url{https://5g-ppp.eu/wp-content/uploads/2017/10/Euro-5G-D2.6_Final-report-on-programme-progress-and-KPIs.pdf}
  [Accessed on 16/01/2021].

\bibitem{9149163}
Y.~{Wang}, C.~{Feng}, T.~{Zhang}, Y.~{Liu}, and A.~{Nallanathan}, ``{QoE} based
  network deployment and caching placement for cache-enabling {UAV} networks,''
  in \emph{ICC 2020 - 2020 IEEE International Conference on Communications
  (ICC)}, 2020, pp. 1--6, \url{https://doi.org/10.1109/ICC40277.2020.9149163}.

\bibitem{dt_airplaneprovideconnectivity}
A.~Burkitt-Gray, ``{LDACS}, aviation’s future terrestrial datalink, takes a
  big step forward,'' 2020, [Online]. Available
  at:\url{https://www.capacitymedia.com/articles/3826670/remote-aircraft-can-cover-area-140km-across-with-4g-and-5g-says-dt}
  [Accessed on 16/01/2021].

\bibitem{ITU_G911}
ITU, ``Series g: Transmission systems and media, digital systems and networks -
  digital transmission systems – digital sections and digital line system –
  parameters for optical fibre cable systems - parameters and calculation
  methodologies for reliability and availability of fibre optic systems,''
  Tech. Rep., 1997, [Online]. Available
  at:\url{https://www.itu.int/rec/T-REC-G.911-199704-I/en} [Accessed on
  03/02/2021].

\bibitem{easa_airtoground_evolution}
EASA, ``{EUROCAE} study: Analysis of the evolution of air/ground
  communication,'' Tech. Rep. EASA.2012.FC02 SC.006, 2016, [Online]. Available
  at:\url{https://www.easa.europa.eu/sites/default/files/dfu/SC%202012-006%20Analysis%20of%20the%20evolution%20of%20airground%20communication.pdf}
  [Accessed on 16/01/2021].

\bibitem{wirelesscomm_book}
W.~Saad, M.~Bennis, M.~Mozaffari, and X.~Lin, \emph{Wireless Communications and
  Networking for Unmanned Aerial Vehicles}.\hskip 1em plus 0.5em minus
  0.4em\relax New York NY: Cambridge University Press, 2020, iSBN:
  9781108691017 \url{https://doi.org/10.1017/9781108691017}.

\bibitem{Vinogradov_2018}
E.~Vinogradov, H.~Sallouha, S.~D. Bast, M.~M. Azari, and S.~Pollin, ``Tutorial
  on {UAVs}: A blue sky view on wireless communication,'' \emph{Journal of
  Mobile Multimedia}, vol.~14, no.~4, p. 395–468, 2018,
  \url{http://dx.doi.org/10.13052/jmm1550-4646.1443}.

\bibitem{7317490}
L.~{Gupta}, R.~{Jain}, and G.~{Vaszkun}, ``Survey of important issues in {UAV}
  communication networks,'' \emph{IEEE Communications Surveys \& Tutorials},
  vol.~18, no.~2, pp. 1123--1152, 2016,
  \url{https://doi.org/10.1109/COMST.2015.2495297}.

\bibitem{8682048}
H.~{Shakhatreh}, A.~H. {Sawalmeh}, A.~{Al-Fuqaha}, Z.~{Dou}, E.~{Almaita},
  I.~{Khalil}, N.~S. {Othman}, A.~{Khreishah}, and M.~{Guizani}, ``Unmanned
  aerial vehicles ({UAVs}): A survey on civil applications and key research
  challenges,'' \emph{IEEE Access}, vol.~7, pp. 48\,572--48\,634, 2019,
  \url{https://doi.org/10.1109/ACCESS.2019.2909530}.

\bibitem{8660516}
M.~{Mozaffari}, W.~{Saad}, M.~{Bennis}, Y.~{Nam}, and M.~{Debbah}, ``A tutorial
  on {UAVs} for wireless networks: Applications, challenges, and open
  problems,'' \emph{IEEE Communications Surveys \& Tutorials}, vol.~21, no.~3,
  pp. 2334--2360, 2019, \url{https://doi.org/10.1109/COMST.2019.2902862}.

\bibitem{8579209}
B.~{Li}, Z.~{Fei}, and Y.~{Zhang}, ``{UAV} communications for {5G} and beyond:
  Recent advances and future trends,'' \emph{IEEE Internet of Things Journal},
  vol.~6, no.~2, pp. 2241--2263, 2019,
  \url{https://doi.org/10.1109/JIOT.2018.2887086}.

\bibitem{8470897}
Y.~{Zeng}, J.~{Lyu}, and R.~{Zhang}, ``Cellular-connected {UAV}: Potential,
  challenges, and promising technologies,'' \emph{IEEE Wireless
  Communications}, vol.~26, no.~1, pp. 120--127, 2019,
  \url{https://doi.org/10.1109/MWC.2018.1800023}.

\bibitem{8675384}
A.~{Fotouhi}, H.~{Qiang}, M.~{Ding}, M.~{Hassan}, L.~G. {Giordano},
  A.~{Garcia-Rodriguez}, and J.~{Yuan}, ``Survey on {UAV} cellular
  communications: Practical aspects, standardization advancements, regulation,
  and security challenges,'' \emph{IEEE Communications Surveys \& Tutorials},
  vol.~21, no.~4, pp. 3417--3442, 2019,
  \url{https://doi.org/10.1109/COMST.2019.2906228}.

\bibitem{8359865}
X.~{Fan}, W.~{Cai}, and J.~{Lin}, ``A survey of routing protocols for highly
  dynamic mobile ad hoc networks,'' in \emph{2017 IEEE 17th International
  Conference on Communication Technology (ICCT)}, 2017, pp. 1412--1417,
  \url{https://doi.org/10.1109/ICCT.2017.8359865}.

\bibitem{8741010}
O.~S. {Oubbati}, M.~{Atiquzzaman}, P.~{Lorenz}, M.~H. {Tareque}, and M.~S.
  {Hossain}, ``Routing in flying ad hoc networks: Survey, constraints, and
  future challenge perspectives,'' \emph{IEEE Access}, vol.~7, pp.
  81\,057--81\,105, 2019, \url{https://doi.org/10.1109/ACCESS.2019.2923840}.

\bibitem{9044378}
D.~{Shumeye Lakew}, U.~{Sa’ad}, N.~{Dao}, W.~{Na}, and S.~{Cho}, ``Routing in
  flying ad hoc networks: A comprehensive survey,'' \emph{IEEE Communications
  Surveys \& Tutorials}, vol.~22, no.~2, pp. 1071--1120, 2020,
  \url{https://doi.org/10.1109/COMST.2020.2982452}.

\bibitem{8368236}
J.~{Liu}, Y.~{Shi}, Z.~M. {Fadlullah}, and N.~{Kato}, ``Space-air-ground
  integrated network: A survey,'' \emph{IEEE Communications Surveys \&
  Tutorials}, vol.~20, no.~4, pp. 2714--2741, 2018.

\bibitem{8438489}
X.~{Cao}, P.~{Yang}, M.~{Alzenad}, X.~{Xi}, D.~{Wu}, and H.~{Yanikomeroglu},
  ``Airborne communication networks: A survey,'' \emph{IEEE Journal on Selected
  Areas in Communications}, vol.~36, no.~9, pp. 1907--1926, 2018,
  \url{https://doi.org/10.1109/JSAC.2018.2864423}.

\bibitem{8935306}
M.~{Zolanvari}, R.~{Jain}, and T.~{Salman}, ``Potential data link candidates
  for civilian unmanned aircraft systems: A survey,'' \emph{IEEE Communications
  Surveys \& Tutorials}, vol.~22, no.~1, pp. 292--319, 2020,
  \url{https://doi.org/10.1109/COMST.2019.2960366}.

\bibitem{5935265}
W.~{Kampichler} and D.~{Eier}, ``Satellite based voice communication for air
  traffic management and airline operation,'' in \emph{2011 Integrated
  Communications, Navigation, and Surveillance Conference Proceedings}, 2011,
  pp. B6--1--B6--10, \url{https://doi.org/10.1109/ICNSURV.2011.5935265}.

\bibitem{helfrick07}
A.~Helfrick, \emph{Principles of Avionics}.\hskip 1em plus 0.5em minus
  0.4em\relax Avionics Communications, 2007, iSBN:188554426X.

\bibitem{sesar_euatmmasterplan}
SESAR, ``European {ATM} master plan,'' Tech. Rep., 2020, [Online]. Available
  at:\url{https://www.sesarju.eu/masterplan2020/} [Accessed on 16/01/2021].

\bibitem{schnell04}
M.~Schnell, E.~Haas, M.~Sajatovic, C.~Rijacek, and B.~Haindl, ``{B-VHF}: An
  overlay system concept for future {ATC} communications in the {VHF} band,''
  in \emph{Digital Avionics Systems Conference}, 2004, pp. 1--9,
  \url{https://doi.org/10.1109/DASC.2004.1391252}.

\bibitem{sesar_4dtrajectory}
SESAR, ``Business trajectory / ‘{4D}’ trajectory,'' 2010, [Online].
  Available
  at:\url{https://www.sesarju.eu/sites/default/files/SESAR_Factsheet_4DTrajectory__2_.pdf}
  [Accessed on 16/01/2021].

\bibitem{mayr14}
T.~Graupl, M.~Mayr, and O.~Lucke, ``Selected results for {IPv6} based {SWIM},
  {CPDLC} and voip in the {SANDRA} flight trial campaign,'' in \emph{IEEE
  Digital Avionics Systems Conference (DASC)}.\hskip 1em plus 0.5em minus
  0.4em\relax CO Pp. 3A-3A2-10: Colorado Springs, 2014, pp. 2--1,
  \url{https://doi.org/10.1109/DASC.2014.6979440}.

\bibitem{bauer11}
C.~Bauer and M.~Zitterbart, ``A survey of protocols to support {IP} mobility in
  aeronautical communications,'' \emph{IEEE Communications Surveys \&
  Tutorials}, vol.~13, no.~4, pp. 642--657, 2011,
  \url{https://doi.org/10.1109/SURV.2011.111510.00016}.

\bibitem{ICAO_9896}
ICAO, ``Aeronautical telecommunication network ({ATN}) manual for the {ATN}
  using {IPS} standards and protocols draft version 21,'' Tech. Rep. Doc 9896,
  2012, [Online]. Available
  at:\url{https://www.icao.int/safety/acp/acpwgf/acp-wg-i-15/acf2b58.doc}
  [Accessed on 16/01/2021].

\bibitem{eurocontrol_swim}
S.~Wilson, ``{EUROCONTROL} specification for {SWIM} information definition,''
  EUROCONTROL, Tech. Rep. SPEC-169, Dec 2017, [Online]. Available
  at:\url{https://www.eurocontrol.int/sites/default/files/content/documents/single-sky/specifications/EUROCONTROL-SPEC-169%20SWIM%20INFO%20Ed%201.0.pdf}
  [Accessed on 16/01/2021].

\bibitem{icao_swim}
ICAO, ``{EUROCONTROL} specification for {SWIM} information definition,'' Tech.
  Rep. Doc 10039, [Online]. Available
  at:\url{https://www.icao.int/airnavigation/IMP/Documents/SWIM%20Concept%20V2%20Draft%20with%20DISCLAIMER.pdf}
  [Accessed on 16/01/2021].

\bibitem{icao_ganp}
------, ``2016–2030 global air navigation plan,'' Tech. Rep. Doc 9750-AN/963,
  2016, [Online]. Available
  at:\url{https://www.icao.int/airnavigation/Documents/GANP-2016-interactive.pdf}
  [Accessed on 16/01/2021].

\bibitem{icao_ganp_2019}
A.~Niknejad\;-\;{ICAO}, ``Aviation system block upgrade ({ASBU}) {GANP} 6th
  edition (2019),'' 2020, [Online]. Available
  at:\url{https://www.icao.int/MID/Documents/2020/AIM%20SG6/PPT%203%20-%20ASBU%202019%20DAIM.pdf}
  [Accessed on 16/01/2021].

\bibitem{9155746}
K.~{Morioka}, J.~{Naganawa}, X.~{Lu}, N.~{Miyazaki}, A.~{Kohmura},
  N.~{Yonemoto}, and Y.~{Sumiya}, ``{QoS} validation tests for aircraft access
  to {SWIM} by ground taxiing experiments,'' in \emph{2019 IEEE 14th
  International Symposium on Autonomous Decentralized System (ISADS)}, 2019,
  pp. 1--6, \url{https://doi.org/10.1109/ISADS45777.2019.9155746}.

\bibitem{rula16}
J.~P. Rula, F.~E. Bustamante, and D.~R. Choffnes, ``When {IP}s fly: A case for
  redefining airline communication,'' \emph{HotMobile'16}, pp. 23--24, February
  2016, \url{https://doi.org/10.1145/2873587.2873605}.

\bibitem{liu11}
Y.~Liu, J.~Li, and H.~Zhang, ``{NEMO} route optimization for aeronautical
  passenger communications,'' in \emph{International Conference on
  Transportation, Mechanical, and Electrical Engineering (TMEE)}.\hskip 1em
  plus 0.5em minus 0.4em\relax Changchun, 2011, pp. 2310--2314,
  \url{https://doi.org/10.1109/TMEE.2011.6199682}.

\bibitem{acars_commsgroup}
I.~C. Group, ``Introduction to {ACARS} messaging services,'' Tech. Rep.
  ICS-200-01, Apr. 2006, [Online]. Available
  at:\url{https://www.icao.int/safety/acp/inactive%20working%20groups%20library/acp-wg-m-iridium-7/ird-swg07-wp08%20-%20acars%20app%20note.pdf}
  [Accessed on 16/01/2021].

\bibitem{icao_handbook_freqspectrum}
ICAO, ``Handbook on radio frequency spectrum requirements for civil aviation -
  volume i {ICAO} spectrum strategy, policy statements and related
  information,'' Tech. Rep. Doc 9718 AN/957, 2018, second Edition. [Online].
  Available
  at:\url{https://www.icao.int/safety/FSMP/Documents/Doc9718/Doc9718_Vol_I_2nd_ed_(2018)corr1.pdf}
  [Accessed on 16/01/2021].

\bibitem{8735356}
M.~{Niraula}, ``Security consideration for the {IPV6} based air to ground
  safety service communication,'' in \emph{2019 Integrated Communications,
  Navigation and Surveillance Conference (ICNS)}, 2019, pp. 1--11,
  \url{https://doi.org/10.1109/ICNSURV.2019.8735356}.

\bibitem{9256659}
C.~{Perner} and C.~{Schmitt}, ``Security concept for unoccupied aerial
  systems,'' in \emph{2020 AIAA/IEEE 39th Digital Avionics Systems Conference
  (DASC)}, 2020, pp. 1--8,
  \url{https://doi.org/10.1109/DASC50938.2020.9256659}.

\bibitem{8735185}
A.~A. {Perez} and F.~{Behrend}, ``A holographic checklist assistant for the
  single pilot,'' in \emph{2019 Integrated Communications, Navigation and
  Surveillance Conference (ICNS)}, 2019, pp. 1--10,
  \url{https://doi.org/10.1109/ICNSURV.2019.8735185}.

\bibitem{8569803}
S.~M. {Sprengart}, S.~M. {Neis}, and J.~{Schiefele}, ``Role of the human
  operator in future commercial reduced crew operations,'' in \emph{2018
  IEEE/AIAA 37th Digital Avionics Systems Conference (DASC)}, 2018, pp. 1--10,
  \url{https://doi.org/10.1109/DASC.2018.8569803}.

\bibitem{8039185}
Y.~{Lim}, V.~{Bassien-Capsa}, S.~{Ramasamy}, J.~{Liu}, and R.~{Sabatini},
  ``Commercial airline single-pilot operations: System design and pathways to
  certification,'' \emph{IEEE Aerospace and Electronic Systems Magazine},
  vol.~32, no.~7, pp. 4--21, 2017,
  \url{https://doi.org/10.1109/MAES.2017.160175}.

\bibitem{ITUM2197}
ITU-R, ``Technical characteristics and operational objectives for wireless
  avionics intra-communications ({WAIC}),'' Tech. Rep. M.2197, 2010, [Online].
  Available at:\url{https://www.itu.int/pub/R-REP-M.2197} [Accessed on
  16/01/2021].

\bibitem{skywise}
Airbus, ``Airbus launches skywise – aviation’s open data platform,'' 2017,
  [Online]. Available
  at:\url{https://www.airbus.com/newsroom/press-releases/en/2017/06/airbus-launches-new-open-aviation-data-platform--skywise--to-sup.html}
  [Accessed on 16/01/2021].

\bibitem{airbus_flightdata}
------, ``Big data: Airbus is mining the wealth of knowledge for aviation,''
  2018, [Online]. Available
  at:\url{https://www.airbus.com/public-affairs/berlin/en/our-topics/digitalisation.html}
  [Accessed on 16/01/2021].

\bibitem{sandraspaper_mdt}
S.~{Hofmann}, S.~{Duhovnikov}, and D.~{Schupke}, ``Massive data transfer from
  and to aircraft on ground: Feasibility and challenges,'' in \emph{IEEE
  Aerospace \& Electronic Systems Magazine}, 2021,
  \url{https://doi.org/10.1109/MAES.2021.3053119}.

\bibitem{evtol_range}
A.~Bacchini and E.~Cestino, ``Electric {VTOL} configurations comparison,''
  \emph{MDPI Aerospace}, vol.~6, p.~26, 02 2019,
  \url{https://doi.org/10.3390/aerospace6030026}.

\bibitem{luftfahrt_uam_2030}
J.~Michelmann, A.~Straubinger, A.~Becker, C.~A. Haddad, K.~Ploetner, and
  M.~Hornung, ``Urban air mobility 2030+: Pathways for {UAM} – scenario-based
  analysis,'' \emph{Deutscher Luft- Und Raumfahrtkongress 2020}, 09 2020,
  [Online]. Available
  at:\url{https://mediatum.ub.tum.de/doc/1576768/1576768.pdf} [Accessed on
  16/01/2021].

\bibitem{easa_SC-VTOL-01}
EASA, ``Special condition for small-category {VTOL} aircraft,'' Tech. Rep.
  SC-VTOL-01, 2019, [Online]. Available
  at:\url{https://www.easa.europa.eu/sites/default/files/dfu/SC-VTOL-01.pdf}
  [Accessed on 16/01/2021].

\bibitem{nasa_uam}
NASA, ``Executive briefing: Urban air mobility ({UAM}) market study,'' Tech.
  Rep., Nov 2018, [Online]. Available
  at:\url{https://www.nasa.gov/sites/default/files/atoms/files/uam-market-study-executive-summary-v2.pdf}
  [Accessed on 16/01/2021].

\bibitem{intel_evtol_dasc}
J.~Athavale, A.~Baldovin, S.~Mo, and M.~Paulitsch, ``Chip-level considerations
  to enable dependability for {EVTOL} and urban air mobility systems,'' in
  \emph{2020 AIAA/IEEE Digital Avionics Systems Conference (DASC)}, 2020,
  \url{https://doi.org/10.1109/DASC50938.2020.9256436}.

\bibitem{uberair_vehiclerequirements}
Uber, ``Uberair vehicle requirements and missions,'' [Online]. Available
  at:\url{https://s3.amazonaws.com/uber-static/elevate/Summary+Mission+and+Requirements.pdf}
  [Accessed on 16/01/2021].

\bibitem{8941429}
J.~D. {Littell}, ``Challenges in vehicle safety and occupant protection for
  autonomous electric vertical take-off and landing ({eVTOL}) vehicles,'' in
  \emph{2019 AIAA/IEEE Electric Aircraft Technologies Symposium (EATS)}, 2019,
  pp. 1--16, \url{https://doi.org/10.2514/6.2019-4504}.

\bibitem{9081685}
A.~{Bauranov} and J.~{Rakas}, ``Urban air mobility and manned {eVTOLs}: safety
  implications,'' in \emph{2019 IEEE/AIAA 38th Digital Avionics Systems
  Conference (DASC)}, 2019, pp. 1--8,
  \url{https://doi.org/10.1109/DASC43569.2019.9081685}.

\bibitem{8569645}
I.~C. {Kleinbekman}, M.~A. {Mitici}, and P.~{Wei}, ``{eVTOL} arrival sequencing
  and scheduling for on-demand urban air mobility,'' in \emph{2018 IEEE/AIAA
  37th Digital Avionics Systems Conference (DASC)}, 2018, pp. 1--7,
  \url{https://doi.org/10.1109/DASC.2018.8569645}.

\bibitem{8569225}
P.~{Pradeep} and P.~{Wei}, ``Heuristic approach for arrival sequencing and
  scheduling for {eVTOL} aircraft in on-demand urban air mobility,'' in
  \emph{2018 IEEE/AIAA 37th Digital Avionics Systems Conference (DASC)}, 2018,
  pp. 1--7, \url{https://doi.org/10.1109/DASC.2018.8569225}.

\bibitem{icao_rpas_rules}
ICAO, ``Remotely piloted aircraft system ({RPAS}) concept of operations
  (conops) for international {IFR} operations,'' Tech. Rep., [Online].
  Available
  at:\url{https://www.icao.int/safety/UA/Documents/ICAO%20RPAS%20CONOPS.pdf}
  [Accessed on 16/01/2021].

\bibitem{rtca_daa}
RTCA, ``Minimum operational performance standards ({MOPS}) for detect and avoid
  ({DAA}) systems,'' Tech. Rep. DO-365A, 045-20/PMC-1986, Mar 2020, [Online].
  Available
  at:\url{https://global.ihs.com/doc_detail.cfm?document_name=RTCA%20DO%2D365&item_s_key=00713351}
  [Accessed on 16/01/2021].

\bibitem{9081631}
L.~E. {Alvarez}, I.~{Jessen}, M.~P. {Owen}, J.~{Silbermann}, and P.~{Wood},
  ``{ACAS sXu}: Robust decentralized detect and avoid for small unmanned
  aircraft systems,'' in \emph{2019 IEEE/AIAA 38th Digital Avionics Systems
  Conference (DASC)}, 2019, pp. 1--9,
  \url{https://doi.org/10.1109/DASC43569.2019.9081631}.

\bibitem{8904324}
I.~A. {Meer}, M.~{Ozger}, M.~{Lundmark}, K.~W. {Sung}, and C.~{Cavdar},
  ``Ground based sense and avoid system for air traffic management,'' in
  \emph{2019 IEEE 30th Annual International Symposium on Personal, Indoor and
  Mobile Radio Communications (PIMRC)}, 2019, pp. 1--6,
  \url{https://doi.org/10.1109/PIMRC.2019.8904324}.

\bibitem{8539587}
M.~{Ki}, J.~{Cha}, and H.~{Lyu}, ``Detect and avoid system based on multi
  sensor fusion for {UAV},'' in \emph{2018 International Conference on
  Information and Communication Technology Convergence (ICTC)}, 2018, pp.
  1107--1109, \url{https://doi.org/10.1109/ICTC.2018.8539587}.

\bibitem{9166607}
S.~{Suherman}, R.~A. {Putra}, and M.~{Pinem}, ``Ultrasonic sensor assessment
  for obstacle avoidance in quadcopter-based drone system,'' in \emph{2020 3rd
  International Conference on Mechanical, Electronics, Computer, and Industrial
  Technology (MECnIT)}, 2020, pp. 50--53,
  \url{https://doi.org/10.1109/MECnIT48290.2020.9166607}.

\bibitem{deloitte_uam}
R.~Lineberger, A.~Hussain, M.~Metcalfe, and V.~Rutgers, ``Infrastructure
  barriers to the elevated future of mobility - are cities ready with the
  infrastructure needed for urban air transportation?'' Deloitte, Tech. Rep.,
  2019, [Online]. Available
  at:\url{https://www2.deloitte.com/content/dam/insights/us/articles/5103_Infrastructure-barriers-to-elevated-FOM/DI_Infrastructure-barriers-to-elevated-FOM.pdf}
  [Accessed on 16/01/2021].

\bibitem{MITRE_conops}
B.~Lascara, A.~Lacher, M.~DeGarmo, D.~Maroney, R.~Niles, and L.~Vempati,
  ``Urban air mobility airspace integration concepts,'' MITRE, Tech. Rep.,
  2019, [Online]. Available
  at:\url{https://www.mitre.org/sites/default/files/publications/pr-19-00667-9-urban-air-mobility-airspace-integration.pdf}
  [Accessed on 16/01/2021].

\bibitem{faa_airspaceregulations}
FAA, ``Chapter 15 - airspace,'' Tech. Rep., [Online]. Available
  at:\url{https://www.faa.gov/regulations_policies/handbooks_manuals/aviation/phak/media/17_phak_ch15.pdf}
  [Accessed on: 16/01/2021].

\bibitem{uam_icrat}
P.~Vascik, H.~Balakrishnan, and R.~J. Hansman, ``Assessment of air traffic
  control for urban air mobility and unmanned systems,'' in \emph{International
  Conference on Research in Air Transportation (ICRAT)}, 09 2018, [Online].
  Available
  at:\url{http://www.icrat.org/ICRAT/seminarContent/2018/papers/ICRAT_2018_paper_26.pdf}
  [Accessed on 16/01/2021].

\bibitem{sesar_corus}
SESAR and Corus, ``{U-Space} concept of operations,'' Tech. Rep., 2019,
  [Online]. Available
  at:\url{https://www.sesarju.eu/sites/default/files/documents/u-space/CORUS%20ConOps%20vol2.pdf}
  [Accessed on 16/01/2021].

\bibitem{nasa_utm_uasOperations}
P.~Kopardekar, ``Unmanned aerial system ({UAS}) traffic management ({UTM}):
  Enabling low-altitude airspace and {UAS} operations,'' NASA, Tech. Rep.,
  2014, [Online]. Available
  at:\url{https://ntrs.nasa.gov/api/citations/20140013436/downloads/20140013436.pdf}
  [Accessed on 16/01/2021].

\bibitem{JUTM}
H.~Ushijima, ``{UTM} project in japan,'' 2017, [Online]. Available
  at:\url{https://gutma.org/montreal-2017/wp-content/uploads/sites/2/2017/07/UTM-Project-in-Japan_METI.pdf}
  [Accessed on 16/01/2021].

\bibitem{airbusutm2}
A.~D. Evans, M.~Egorov, and S.~Munn, ``Fairness in decentralized strategic
  deconfliction in {UTM},'' in \emph{AIAA Scitech 2020 Forum}, Jan 2020,
  \url{https://doi.org/10.2514/6.2020-2203}.

\bibitem{airbusutm3}
S.~Li, M.~Egorov, and M.~Kochenderfer, ``Optimizing collision avoidance in
  dense airspace using deep reinforcement learning,'' in \emph{Thirteenth
  USA/Europe Air Traffic Management Research and Development Seminar
  (ATM2019)}, 2019, [Online]. Available
  at:\url{https://arxiv.org/pdf/1912.10146.pdf} [Accessed on 16/01/2021].

\bibitem{airbusutm4}
M.~{Egorov}, V.~{Kuroda}, and P.~{Sachs}, ``Encounter aware flight planning in
  the unmanned airspace,'' in \emph{2019 Integrated Communications, Navigation
  and Surveillance Conference (ICNS)}, 2019, pp. 1--15,
  \url{https://doi.org/10.1109/ICNSURV.2019.8735399}.

\bibitem{faa_nextgen_conops}
NASA, FAA, and NextGen, ``Unmanned aircraft system ({UAS}) traffic management
  ({UTM}) concept of operations v1.0,'' Tech. Rep., 2018, [Online]. Available
  at:\url{https://utm.arc.nasa.gov/docs/2018-UTM-ConOps-v1.0.pdf} [Accessed on
  16/01/2021].

\bibitem{nasa_utm_conopsv2}
NextGEN, ``Concept of operations v2.0 - unmanned aircraft system ({UAS})
  traffic management ({UTM}),'' FAA and NASA, Tech. Rep., 2020, [Online].
  Available
  at:\url{https://www.faa.gov/uas/research_development/traffic_management/media/UTM_ConOps_v2.pdf}
  [Accessed on: 16/01/2021].

\bibitem{nasa_uasServiceSupplierDevelopment}
J.~L. Rios, I.~S. Smith, P.~Venkatesen, D.~R. Smith, V.~Baskaran, S.~Jurcak,
  R.~Strauss, S.~Iyer, and P.~Verma, ``{UTM} {UAS} service supplier
  development,'' NASA, Tech. Rep., 2018, [Online]. Available
  at:\url{https://utm.arc.nasa.gov/docs/UTM_UAS_TCL4_Sprint1_Report.pdf}
  [Accessed on 16/01/2021].

\bibitem{3GPP_22825}
3GPP, ``Remote identification of unmanned aerial systems; stage 1 (release
  16),'' techreport TR 22.825, 2018, [Online]. Available
  at:\url{https://www.3gpp.org/ftp//Specs/archive/22_series/22.825/} [Accessed
  on 16/01/2021].

\bibitem{easa_drone_operator_pilot}
EASA, ``Drones ({UAS}),'' [Online]. Available
  at:\url{https://www.easa.europa.eu/the-agency/faqs/drones-uas#category-training-requirements-in-the-open-category-}
  [Accessed on 16/01/2021].

\bibitem{EASA_UAS_definition}
------, ``Easy access rules for unmanned aircraft systems (regulations ({EU})
  2019/947 and ({EU}) 2019/945),'' Tech. Rep., 2020, [Online]. Available
  at:\url{https://www.easa.europa.eu/document-library/easy-access-rules/easy-access-rules-unmanned-aircraft-systems-regulation-eu}
  [Accessed on 16/01/2021].

\bibitem{FAA_UAS_definition}
ICAO, ``Unmanned aircraft administration systems ({UAS}) 101,'' [Online].
  Available
  at:\url{https://www.icao.int/WACAF/Documents/RASG%20AFI/RASG-AFI-5/IP%2005%20-%20UAS%20101.pdf}
  [Accessed on 16/01/2021].

\bibitem{ICAO_UAS_RPAS_definition}
------, ``Frequently asked questions,'' [Online]. Available
  at:\url{https://www.icao.int/safety/UA/UASToolkit/Pages/FAQ.aspx#Q1}
  [Accessed on 16/01/2021.

\bibitem{3GPP_UAS_definition}
3GPP, ``{UAS} - {UAV},'' 2019, [Online]. Available
  at:\url{https://www.3gpp.org/UAS-UAV#:~:text=Tutorials%2C%20Videos%20%26%20Tools-,UAS%20%2D%20UAV,of%20Unmanned%20Aerial%20Vehicles%20(aka.}
  [Accessed on 16/01/2021].

\bibitem{DroneRegulations}
G.~L.~R. Center, ``Regulation of drones,'' The Law Library of Congress, Tech.
  Rep., 2016, [Online]. Available
  at:\url{https://www.loc.gov/law/help/regulation-of-drones/regulation-of-drones.pdf}
  [Accessed on 16/01/2021.

\bibitem{ngmn_urllc}
NGMN, ``Verticals {uRLLC} use cases and requirements by {NGMN} alliance,''
  Tech. Rep., 2020, v2.5.4. [Online]. Available
  at:\url{https://www.ngmn.org/wp-content/uploads/200210-Verticals-URLLC-Requirements-v2.5.4.pdf}
  [Accessed on 16/01/2021].

\bibitem{baiocchi15}
A.~Baiocchi, F.~Cuomo, M.~D. Felice, and G.~Fusco, ``Vehicular ad-hoc networks
  sampling protocols for traffic monitoring and incident detection in
  intelligent transportation systems,'' \emph{Transportation Research Part C:
  Emerging Technologies}, vol.~56, 2015,
  \url{https://doi.org/10.1016/j.trc.2015.03.041}.

\bibitem{8636976}
D.~{Falanga}, S.~{Kim}, and D.~{Scaramuzza}, ``How fast is too fast? the role
  of perception latency in high-speed sense and avoid,'' \emph{IEEE Robotics
  and Automation Letters}, vol.~4, no.~2, pp. 1884--1891, 2019,
  \url{https://doi.org/10.1109/LRA.2019.2898117}.

\bibitem{lilium}
Lilium, ``Certifying the lilium jet and its operations,'' October 2020,
  [Online]. Available
  at:\url{https://lilium.com/newsroom-detail/certifying-the-lilium-jet-and-its-operations}
  [Accessed on 16/01/2021].

\bibitem{airbus_uam_vehicles}
Airbus, ``Urban air mobility by airbus,'' [Online]. Available
  at:\url{https://www.airbus.com/content/dam/corporate-topics/publications/backgrounders/Urban-Air-Mobility-Backgrounder-E.pdf}
  [Accessed on 16/01/2021].

\bibitem{Airbus_BluePrint}
K.~Balakrishnan, J.~Polastre, J.~Mooberry, R.~Golding, and P.~Sachs,
  ``Blueprint for the sky,'' Airbus, Tech. Rep., 2018, [Online]. Available
  at:\url{https://storage.googleapis.com/blueprint/Airbus_UTM_Blueprint.pdf}
  [Accessed on 16/01/2021].

\bibitem{ITU_M2171}
ITU-R, ``Characteristics of unmanned aircraft systems and spectrum requirements
  to support their safe operation in non-segregated airspace,'' Tech. Rep.
  M.2171, Dec. 2009, [Online]. Available
  at:\url{https://www.itu.int/en/ITU-R/space/snl/Documents/R-REP-M.2171-2009-PDF-E.pdf}
  [Accessed on 16/01/2021].

\bibitem{baltaci_trafficgeneratorpaper}
A.~{Baltaci}, M.~{Klügel}, F.~{Geyer}, S.~{Duhovnikov}, V.~{Bajpai}, J.~{Ott},
  and D.~{Schupke}, ``Experimental {UAV} data traffic modeling and network
  performance analysis,'' in \emph{IEEE International Conference on Computer
  Communications (INFOCOM)}, 2021, (Accepted for publication.).

\bibitem{commVsAutonomy}
\BIBentryALTinterwordspacing
A.~Zolich, D.~Palma, K.~Kansanen, K.~Fj{\o}rtoft, J.~Sousa, K.~H. Johansson,
  Y.~Jiang, H.~Dong, and T.~A. Johansen, ``Survey on communication and networks
  for autonomous marine systems,'' \emph{Journal of Intelligent {\&} Robotic
  Systems}, vol.~95, no.~3, pp. 789--813, 2019. [Online]. Available:
  \url{https://doi.org/10.1007/s10846-018-0833-5}
\BIBentrySTDinterwordspacing

\bibitem{easa_moc_sc-vtol}
EASA, ``Proposed means of compliance with the special condition {VTOL},'' Tech.
  Rep. MOC SC-VTOL, 2020, [Online]. Available
  at:\url{https://www.easa.europa.eu/sites/default/files/dfu/proposed_moc_sc_vtol_issue_1.pdf}
  [Accessed on 16/01/2021].

\bibitem{cadence_do254}
Cadence, ``Do-254 explained,'' Tech. Rep., 2019, [Online]. Available
  at:\url{https://www.cadence.com/content/dam/cadence-www/global/en_US/documents/solutions/aerospace-and-defense/accelerating-do-254-approval-wp.pdf}
  [Accessed on 16/01/2021].

\bibitem{faa_failuredescriptions}
U.S.\;Department\;of\;Transportation\;-\;Federal\;Aviation\;Administration,
  ``System safety analysis and assessment for part 23 airplanes,'' Tech. Rep.
  23.1309-1E, 2011, [Online]. Available
  at:\url{https://www.faa.gov/documentLibrary/media/Advisory_Circular/AC_23_1309-1E.pdf}
  [Accessed on 16/01/2021].

\bibitem{sae_ARP4754A}
SAE, ``Guidelines for development of civil aircraft and systems,'' Tech. Rep.
  ARP4754A, 2010, [Online]. Available
  at:\url{https://en.wikipedia.org/wiki/ARP4754} [Accessed on 16/01/2021].

\bibitem{8918497}
Y.~{Zeng}, Q.~{Wu}, and R.~{Zhang}, ``Accessing from the sky: A tutorial on
  {UAV} communications for {5G} and beyond,'' \emph{Proceedings of the IEEE},
  vol. 107, no.~12, pp. 2327--2375, 2019,
  \url{https://doi.org/10.1109/JPROC.2019.2952892}.

\bibitem{iot_datarate}
G.~Castellanos, M.~Deruyck, L.~Martens, and W.~Joseph, ``System assessment of
  {WUSN} using {NB-IoT} {UAV}-aided networks in potato crops,'' \emph{IEEE
  Access}, vol.~PP, pp. 1--1, 03 2020,
  \url{https://doi.org/10.1109/ACCESS.2020.2982086}.

\bibitem{iot_datarate2}
M.~Marchese, A.~Moheddine, and F.~Patrone, ``{IoT} and {UAV} integration in
  {5G} hybrid terrestrial-satellite networks,'' \emph{Sensors}, vol.~19, p.
  3704, 08 2019, \url{https://doi.org/10.3390/s19173704}.

\bibitem{ngmn}
NGMN\;Alliance, ``{5G} white paper v1.0,'' Tech. Rep., 2015, [Online].
  Available
  at:\url{https://www.ngmn.org/wp-content/uploads/NGMN_5G_White_Paper_V1_0.pdf}
  [Accessed on 16/01/2021].

\bibitem{easa_SCLight-UAS}
EASA, ``Proposed special condition for light {UAS},'' Tech. Rep. SC Light-UAS,
  2020, [Online]. Available
  at:\url{https://www.easa.europa.eu/sites/default/files/dfu/special_condition_light_uas.pdf}
  [Accessed on 16/01/2021].

\bibitem{Liu_2020}
Y.~Liu, H.-N. Dai, Q.~Wang, M.~K. Shukla, and M.~Imran, ``Unmanned aerial
  vehicle for internet of everything: Opportunities and challenges,''
  \emph{Computer Communications}, vol. 155, p. 66–83, Apr 2020,
  \url{http://dx.doi.org/10.1016/j.comcom.2020.03.017}.

\bibitem{eurocae_rpaslatency}
EUROCAE, ``Minimum aviation systems performance standards for remote pilot
  stations conducting {IFR} operations in controlled airspace,'' Tech. Rep.,
  2019, (We used a draft version of this document.) [Online]. Available
  at:\url{https://standards.globalspec.com/std/14316531/eurocae-ed-272}
  [Accessed on 16/01/2021].

\bibitem{rtca_c2link}
RTCA, ``Minimum aviation system performance standards for {C2} link systems
  supporting operations of unmanned aircraft systems in {U.S.} airspace,''
  Tech. Rep. DO-377, Mar 2019, [Online]. Available
  at:\url{https://global.ihs.com/doc_detail.cfm?item_s_key=00783295&item_key_date=810229&rid=GS}
  [Accessed on 16/01/2021].

\bibitem{9107776}
H.~{Hu}, H.~{Zhou}, J.~{Li}, K.~{Li}, and B.~{Pan}, ``Automatic and intelligent
  line inspection using {UAV} based on beidou navigation system,'' in
  \emph{2019 6th International Conference on Information Science and Control
  Engineering (ICISCE)}, 2019, pp. 1004--1008,
  \url{https://doi.org/10.1109/ICISCE48695.2019.00202}.

\bibitem{9142230}
M.~{Car}, L.~{Markovic}, A.~{Ivanovic}, M.~{Orsag}, and S.~{Bogdan},
  ``Autonomous wind-turbine blade inspection using {LiDAR}-equipped unmanned
  aerial vehicle,'' \emph{IEEE Access}, vol.~8, pp. 131\,380--131\,387, 2020,
  \url{https://doi.org/10.1109/ACCESS.2020.3009738}.

\bibitem{8853298}
D.~{Sadykova}, D.~{Pernebayeva}, M.~{Bagheri}, and A.~{James}, ``{IN-YOLO}:
  Real-time detection of outdoor high voltage insulators using {UAV} imaging,''
  \emph{IEEE Transactions on Power Delivery}, vol.~35, no.~3, pp. 1599--1601,
  2020, \url{https://doi.org/10.1109/TPWRD.2019.2944741}.

\bibitem{8756125}
V.~T. {Hoang}, M.~D. {Phung}, T.~H. {Dinh}, and Q.~P. {Ha}, ``System
  architecture for real-time surface inspection using multiple {UAVs},''
  \emph{IEEE Systems Journal}, vol.~14, no.~2, pp. 2925--2936, 2020,
  \url{https://doi.org/10.1109/JSYST.2019.2922290}.

\bibitem{10.5555/3237383.3237462}
D.~Albani, T.~Manoni, D.~Nardi, and V.~Trianni, ``Dynamic {UAV} swarm
  deployment for non-uniform coverage,'' in \emph{Proceedings of the 17th
  International Conference on Autonomous Agents and MultiAgent Systems}, ser.
  AAMAS ’18.\hskip 1em plus 0.5em minus 0.4em\relax Richland, SC:
  International Foundation for Autonomous Agents and Multiagent Systems, 2018,
  p. 523–531, \url{https://dl.acm.org/doi/abs/10.5555/3237383.3237462}.

\bibitem{9059403}
C.~{Kinzel}, J.~{Marchio}, S.~{Biaz}, and R.~{Chapman}, ``Multiplatform
  stereoscopic {3D} terrain mapping for {UAV} localization,'' in \emph{2019
  IEEE 16th International Conference on Mobile Ad Hoc and Sensor Systems
  Workshops (MASSW)}, 2019, pp. 67--71,
  \url{https://doi.org/10.1109/MASSW.2019.00020}.

\bibitem{9072891}
L.~{Ambata}, A.~F. {Chavez}, N.~{Devnani}, J.~{Ingeniero}, and A.~K.
  {Paggabao}, ``Three-dimensional mapping system for environment and object
  detection on an unmanned aerial vehicle,'' in \emph{2019 IEEE 11th
  International Conference on Humanoid, Nanotechnology, Information Technology,
  Communication and Control, Environment, and Management ( HNICEM )}, 2019, pp.
  1--6, \url{https://doi.org/10.1109/HNICEM48295.2019.9072891}.

\bibitem{8949363}
S.~W. {Chen}, G.~V. {Nardari}, E.~S. {Lee}, C.~{Qu}, X.~{Liu}, R.~A.~F.
  {Romero}, and V.~{Kumar}, ``{SLOAM}: Semantic {LiDAR} odometry and mapping
  for forest inventory,'' \emph{IEEE Robotics and Automation Letters}, vol.~5,
  no.~2, pp. 612--619, 2020, \url{https://doi.org/10.1109/LRA.2019.2963823}.

\bibitem{motlagh16A}
N.~{Hossein Motlagh}, T.~{Taleb}, and O.~{Arouk}, ``Low-altitude unmanned
  aerial vehicles-based internet of things services: Comprehensive survey and
  future perspectives,'' \emph{IEEE Internet of Things Journal}, vol.~3, no.~6,
  pp. 899--922, 2016, \url{https://doi.org/10.1109/JIOT.2016.2612119}.

\bibitem{10.1145/3347713}
I.~Mademlis, N.~Nikolaidis, A.~Tefas, I.~Pitas, T.~Wagner, and A.~Messina,
  ``Autonomous {UAV} cinematography: A tutorial and a formalized shot-type
  taxonomy,'' \emph{ACM Comput. Surv.}, vol.~52, no.~5, Sep. 2019,
  \url{https://doi.org/10.1145/3347713}.

\bibitem{9096075}
J.~N.~C. {Hayton}, T.~{Barros}, C.~{Premebida}, M.~J. {Coombes}, and U.~J.
  {Nunes}, ``{CNN}-based human detection using a {3D} {LiDAR} onboard a
  {UAV},'' in \emph{2020 IEEE International Conference on Autonomous Robot
  Systems and Competitions (ICARSC)}, 2020, pp. 312--318,
  \url{https://doi.org/10.1109/ICARSC49921.2020.9096075}.

\bibitem{wang17}
X.~{Wang}, A.~{Chowdhery}, and M.~{Chiang}, ``Networked drone cameras for
  sports streaming,'' in \emph{2017 IEEE 37th International Conference on
  Distributed Computing Systems (ICDCS)}, 2017, pp. 308--318,
  \url{https://doi.org/10.1109/ICDCS.2017.200}.

\bibitem{lim15}
J.~Lim, ``Flirtey demos world's first drone delivery tech in {U.S.}'' Jul 2015,
  [Online] Available
  at:\url{https://www.forbes.com/sites/jlim/2015/07/23/flirtey-demos-worlds-first-drone-delivery-tech-in-u-s/#5b4951d04368}
  [Accessed on 16/01/2021].

\bibitem{layne15}
N.~Layne, ``Exclusive: Wal-mart seeks to test drones for home delivery,
  pickup,'' \emph{Reuters}, Oct 2015, [Online]. Available
  at:\url{http://www.reuters.com/article/us-wal-mart-stores-drones-exclusive/exclusive-wal-mart-seeks-to-test-drones-for-home-delivery-pickup-idUSKCN0SK2IQ20151027}
  [Accessed on 16/01/2021].

\bibitem{DHL_UAV}
G.~L.~R. Center, ``Unmanned aerial vehicles in logistics - a {DHL} perspective
  on implications and use cases for the logistics industry,'' DHL Trend
  Research, Tech. Rep., 2014, [Online]. Available
  at:\url{https://www.dhl.com/content/dam/downloads/g0/about_us/logistics_insights/DHL_TrendReport_UAV.pdf}
  [Accessed on 16/01/2021].

\bibitem{7557769}
N.~K. {Yang}, K.~T. {San}, and Y.~S. {Chang}, ``A novel approach for real time
  monitoring system to manage {UAV} delivery,'' in \emph{2016 5th IIAI
  International Congress on Advanced Applied Informatics (IIAI-AAI)}, 2016, pp.
  1054--1057, \url{https://doi.org/10.1109/IIAI-AAI.2016.195}.

\bibitem{8972296}
H.~{Huang}, A.~V. {Savkin}, and C.~{Huang}, ``Control of a novel parcel
  delivery system consisting of a {UAV} and a public train,'' in \emph{2019
  IEEE 17th International Conference on Industrial Informatics (INDIN)},
  vol.~1, 2019, pp. 1047--1050,
  \url{https://doi.org/10.1109/INDIN41052.2019.8972296}.

\bibitem{8972170}
------, ``When drones take public transport: Towards low cost and large range
  parcel delivery*,'' in \emph{2019 IEEE 17th International Conference on
  Industrial Informatics (INDIN)}, vol.~1, 2019, pp. 1657--1660,
  \url{https://doi.org/10.1109/INDIN41052.2019.8972170}.

\bibitem{9126258}
H.~{Chung}, S.~{Maharjan}, Y.~{Zhang}, F.~{Eliassen}, and K.~{Strunz},
  ``Placement and routing optimization for automated inspection with {UAVs}: A
  study in offshore wind farm,'' \emph{IEEE Transactions on Industrial
  Informatics}, pp. 1--1, 2020, \url{https://doi.org/10.1109/TII.2020.3004816}.

\bibitem{9014596}
X.~{Zhen}, Z.~{Enze}, and C.~{Qingwei}, ``Rotary unmanned aerial vehicles path
  planning in rough terrain based on multi-objective particle swarm
  optimization,'' \emph{Journal of Systems Engineering and Electronics},
  vol.~31, no.~1, pp. 130--141, 2020,
  \url{https://doi.org/10.21629/JSEE.2020.01.14}.

\bibitem{app10124362}
J.~Kim, H.~Moon, and H.~Jung, ``Drone-based parcel delivery using the rooftops
  of city buildings: Model and solution,'' \emph{Applied Sciences}, vol.~10,
  no.~12, 2020, \url{https://doi.org/10.3390/app10124362}.

\bibitem{motlagh17}
N.~H. {Motlagh}, M.~{Bagaa}, and T.~{Taleb}, ``{UAV}-based {IoT} platform: A
  crowd surveillance use case,'' \emph{IEEE Communications Magazine}, vol.~55,
  no.~2, pp. 128--134, 2017, \url{https://doi.org/10.1109/MCOM.2017.1600587CM}.

\bibitem{9114970}
J.~{Cui}, Z.~{Ding}, Y.~{Deng}, A.~{Nallanathan}, and L.~{Hanzo}, ``Adaptive
  {UAV}-trajectory optimization under quality of service constraints: A
  model-free solution,'' \emph{IEEE Access}, vol.~8, pp. 112\,253--112\,265,
  2020, \url{https://doi.org/10.1109/ACCESS.2020.3001752}.

\bibitem{9054054}
D.~B. {Licea}, E.~{Nurellari}, and M.~{Ghogho}, ``Energy-efficient {3D} {UAV}
  trajectory design for data collection in wireless sensor networks,'' in
  \emph{ICASSP 2020 - 2020 IEEE International Conference on Acoustics, Speech
  and Signal Processing (ICASSP)}, 2020, pp. 8329--8333,
  \url{https://doi.org/10.1109/ICASSP40776.2020.9054054}.

\bibitem{8842600}
M.~{Samir}, S.~{Sharafeddine}, C.~M. {Assi}, T.~M. {Nguyen}, and A.~{Ghrayeb},
  ``{UAV} trajectory planning for data collection from time-constrained {IoT}
  devices,'' \emph{IEEE Transactions on Wireless Communications}, vol.~19,
  no.~1, pp. 34--46, 2020, \url{https://doi.org/10.1109/TWC.2019.2940447}.

\bibitem{9098850}
Y.~{Wang}, Z.~{Hu}, X.~{Wen}, Z.~{Lu}, and J.~{Miao}, ``Minimizing data
  collection time with collaborative {UAVs} in wireless sensor networks,''
  \emph{IEEE Access}, vol.~8, pp. 98\,659--98\,669, 2020,
  \url{https://doi.org/10.1109/ACCESS.2020.2996665}.

\bibitem{8432487}
J.~{Gong}, T.~{Chang}, C.~{Shen}, and X.~{Chen}, ``Flight time minimization of
  {UAV} for data collection over wireless sensor networks,'' \emph{IEEE Journal
  on Selected Areas in Communications}, vol.~36, no.~9, pp. 1942--1954, 2018,
  \url{https://doi.org/10.1109/JSAC.2018.2864420}.

\bibitem{9121255}
Y.~{Liu}, K.~{Liu}, J.~{Han}, L.~{Zhu}, Z.~{Xiao}, and X.~{Xia}, ``Resource
  allocation and {3D} placement for {UAV}-enabled energy-efficient {IoT}
  communications,'' \emph{IEEE Internet of Things Journal}, pp. 1--1, 2020,
  \url{https://doi.org/10.1109/JIOT.2020.3003717}.

\bibitem{8894454}
Z.~{Wang}, R.~{Liu}, Q.~{Liu}, J.~S. {Thompson}, and M.~{Kadoch},
  ``Energy-efficient data collection and device positioning in {UAV}-assisted
  {IoT},'' \emph{IEEE Internet of Things Journal}, vol.~7, no.~2, pp.
  1122--1139, 2020, \url{https://doi.org/10.1109/JIOT.2019.2952364}.

\bibitem{7036813}
Y.~{Pang}, Y.~{Zhang}, Y.~{Gu}, M.~{Pan}, Z.~{Han}, and P.~{Li}, ``Efficient
  data collection for wireless rechargeable sensor clusters in harsh terrains
  using {UAVs},'' in \emph{2014 IEEE Global Communications Conference}, 2014,
  pp. 234--239, \url{https://doi.org/10.1109/GLOCOM.2014.7036813}.

\bibitem{WPT_IoT}
J.~{Baek}, S.~I. {Han}, and Y.~{Han}, ``Optimal {UAV} route in wireless
  charging sensor networks,'' \emph{IEEE Internet of Things Journal}, vol.~7,
  no.~2, pp. 1327--1335, 2020, \url{https://doi.org/10.1109/JIOT.2019.2954530}.

\bibitem{10.1145/3382756}
G.~Faraci, C.~Grasso, and G.~Schembra, ``Fog in the clouds: {UAVs} to provide
  edge computing to {IoT} devices,'' \emph{ACM Trans. Internet Technol.},
  vol.~20, no.~3, Aug. 2020, \url{https://doi.org/10.1145/3382756}.

\bibitem{9031732}
Q.~{Wang}, H.~{Dai}, Q.~{Wang}, M.~K. {Shukla}, W.~{Zhang}, and C.~G. {Soares},
  ``On connectivity of {UAV}-assisted data acquisition for underwater internet
  of things,'' \emph{IEEE Internet of Things Journal}, vol.~7, no.~6, pp.
  5371--5385, 2020, \url{https://doi.org/10.1109/JIOT.2020.2979691}.

\bibitem{nixon17}
A.~Nixon, ``Best drones for agriculture 2017: the ultimate buyer's guide,''
  [Online]. Available
  at:\url{http://bestdroneforthejob.com/drone-buying-guides/agriculture-drone-buyers-guide/}
  [Accessed on 16/01/2021].

\bibitem{8077607}
C.~M. {Gevaert}, J.~{Tang}, F.~J. {García-Haro}, J.~{Suomalainen}, and
  L.~{Kooistra}, ``Combining hyperspectral {UAV} and multispectral formosat-2
  imagery for precision agriculture applications,'' in \emph{2014 6th Workshop
  on Hyperspectral Image and Signal Processing: Evolution in Remote Sensing
  (WHISPERS)}, 2014, pp. 1--4,
  \url{https://doi.org/10.1109/WHISPERS.2014.8077607}.

\bibitem{8959613}
E.~{Fathelrahman}, E.~{Neumann}, M.~{Hussein}, A.~{Jalil}, F.~{Hassan},
  A.~{Dirir}, and S.~{Muhammad}, ``Unmanned arial vehicle ({UAV}) imagery and
  manual sampling for parasitic weed recognition and measurements,'' in
  \emph{2019 International Conference on Electrical and Computing Technologies
  and Applications (ICECTA)}, 2019, pp. 1--4,
  \url{https://doi.org/10.1109/ICECTA48151.2019.8959613}.

\bibitem{9128695}
V.~{Czymmek}, R.~{Schramm}, and S.~{Hussmann}, ``Vision based crop row
  detection for low cost {UAV} imagery in organic agriculture,'' in \emph{2020
  IEEE International Instrumentation and Measurement Technology Conference
  (I2MTC)}, 2020, pp. 1--6,
  \url{https://doi.org/10.1109/I2MTC43012.2020.9128695}.

\bibitem{9140588}
N.~{Al-Thani}, A.~{Albuainain}, F.~{Alnaimi}, and N.~{Zorba}, ``Drones for
  sheep livestock monitoring,'' in \emph{2020 IEEE 20th Mediterranean
  Electrotechnical Conference ( MELECON)}, 2020, pp. 672--676,
  \url{https://doi.org/10.1109/MELECON48756.2020.9140588}.

\bibitem{8995212}
G.~D.~S. {Vieira}, B.~M. {Rocha}, F.~{Soares}, J.~C. {Lima}, H.~{Pedrini},
  R.~{Costa}, and J.~{Ferreira}, ``Extending the aerial image analysis from the
  detection of tree crowns,'' in \emph{2019 IEEE 31st International Conference
  on Tools with Artificial Intelligence (ICTAI)}, 2019, pp. 1681--1685,
  \url{https://doi.org/10.1109/ICTAI.2019.00247}.

\bibitem{8805152}
E.~C. {Tetila}, B.~B. {Machado}, G.~K. {Menezes}, A.~{Da Silva Oliveira},
  M.~{Alvarez}, W.~P. {Amorim}, N.~A. {De Souza Belete}, G.~G. {Da Silva}, and
  H.~{Pistori}, ``Automatic recognition of soybean leaf diseases using {UAV}
  images and deep convolutional neural networks,'' \emph{IEEE Geoscience and
  Remote Sensing Letters}, vol.~17, no.~5, pp. 903--907, 2020,
  \url{https://doi.org/10.1109/LGRS.2019.2932385}.

\bibitem{9086204}
H.~{Wu}, H.~{Li}, A.~{Shamsoshoara}, A.~{Razi}, and F.~{Afghah}, ``Transfer
  learning for wildfire identification in {UAV} imagery,'' in \emph{2020 54th
  Annual Conference on Information Sciences and Systems (CISS)}, 2020, pp.
  1--6, \url{https://doi.org/10.1109/CISS48834.2020.1570617429}.

\bibitem{8921148}
C.~D. {López} and L.~F. {Giraldo}, ``Optimization of energy and water
  consumption on crop irrigation using {UAVs} via path design,'' in \emph{2019
  IEEE 4th Colombian Conference on Automatic Control (CCAC)}, 2019, pp. 1--5,
  \url{https://doi.org/10.1109/CCAC.2019.8921148}.

\bibitem{9107681}
S.~{Fengjie}, W.~{Xianchang}, and Z.~{Rui}, ``A new optimization method
  application to agricultural plant protection {UAV} scheduling,'' in
  \emph{2019 6th International Conference on Information Science and Control
  Engineering (ICISCE)}, 2019, pp. 80--84,
  \url{https://doi.org/10.1109/ICISCE48695.2019.00026}.

\bibitem{8934989}
A.~{Mokrane}, A.~C. {Braham}, and B.~{Cherki}, ``{UAV} coverage path planning
  for supporting autonomous fruit counting systems,'' in \emph{2019
  International Conference on Applied Automation and Industrial Diagnostics
  (ICAAID)}, vol.~1, 2019, pp. 1--5,
  \url{https://doi.org/10.1109/ICAAID.2019.8934989}.

\bibitem{8988024}
F.~{Fuentes-Peñailillo}, S.~{Ortega-Farías}, D.~D.~L. {Fuente-Sáiz}, and
  M.~{Rivera}, ``Digital count of sunflower plants at emergence from very low
  altitude using {UAV} images,'' in \emph{2019 IEEE CHILEAN Conference on
  Electrical, Electronics Engineering, Information and Communication
  Technologies (CHILECON)}, 2019, pp. 1--5,
  \url{https://doi.org/10.1109/CHILECON47746.2019.8988024}.

\bibitem{8943842}
S.~A. {Shvorov}, N.~A. {Pasichnyk}, S.~D. {Kuznichenko}, I.~V. {Tolok}, S.~V.
  {Lienkov}, and L.~A. {Komarova}, ``Using {UAV} during planned harvesting by
  unmanned combines,'' in \emph{2019 IEEE 5th International Conference Actual
  Problems of Unmanned Aerial Vehicles Developments (APUAVD)}, 2019, pp.
  252--257, \url{https://doi.org/10.1109/APUAVD47061.2019.8943842}.

\bibitem{RysavyResearch}
R.~Research, ``{LTE} to {5G}: Cellular and broadband innovation,'' Tech. Rep.,
  Aug. 2017, [Online]. Available
  at:\url{https://www.5gamericas.org/wp-content/uploads/2019/07/2017_5G_Americas_Rysavy_LTE_5G_Innovation__Final_for_Upload_v2.pdf}
  [Accessed on 16/01/2021].

\bibitem{chen20205g}
Y.~Chen, X.~Lin, T.~Khan, M.~Afshang, and M.~Mozaffari, ``{5G} air-to-ground
  network design and optimization: A deep learning approach,'' 2020, [Online].
  Available at:\url{https://arxiv.org/abs/2011.08379} [Accessed on 16/01/2021].

\bibitem{8911712}
A.~E. {Garcia}, M.~{Ozger}, A.~{Baltaci}, S.~{Hofmann}, D.~{Gera}, M.~{Nilson},
  C.~{Cavdar}, and D.~{Schupke}, ``Direct air to ground communications for
  flying vehicles: Measurement and scaling study for {5G},'' in \emph{2019 IEEE
  2nd 5G World Forum (5GWF)}, 2019, pp. 310--315,
  \url{https://doi.org/10.1109/5GWF.2019.8911712}.

\bibitem{8030548}
M.~{Vondra}, E.~{Dinc}, M.~{Prytz}, M.~{Frodigh}, D.~{Schupke}, M.~{Nilson},
  S.~{Hofmann}, and C.~{Cavdar}, ``Performance study on seamless {DA2GC} for
  aircraft passengers toward {5G},'' \emph{IEEE Communications Magazine},
  vol.~55, no.~11, pp. 194--201, 2017,
  \url{https://doi.org/10.1109/MCOM.2017.1700188}.

\bibitem{7107747}
D.~{Rosigkeit}, S.~V. {Baumgartner}, and A.~{Nottensteiner}, ``Usability of
  long term evolution ({LTE}) in {DLR}'s research aircraft {DO} 228-212,'' in
  \emph{2015 German Microwave Conference}, 2015, pp. 40--43,
  \url{https://doi.org/10.1109/GEMIC.2015.7107747}.

\bibitem{8384882}
W.~{Yun-sheng} and L.~{Yanxiao}, ``A flexible airborne datalink system
  architecture for civil helicopters,'' in \emph{2018 Integrated
  Communications, Navigation, Surveillance Conference (ICNS)}, 2018, pp.
  4A4--1--4A4--7, \url{https://doi.org/10.1109/ICNSURV.2018.8384882}.

\bibitem{10.1145/3325421.3329765}
S.~Hayat, C.~Bettstetter, A.~Fakhreddine, R.~Muzaffar, and D.~Emini, ``An
  experimental evaluation of {LTE-A} throughput for drones,'' in
  \emph{Proceedings of the 5th Workshop on Micro Aerial Vehicle Networks,
  Systems, and Applications}, ser. DroNet'19.\hskip 1em plus 0.5em minus
  0.4em\relax New York, NY, USA: Association for Computing Machinery, 2019, p.
  3–8, \url{https://doi.org/10.1145/3325421.3329765}.

\bibitem{8858453}
H.~{Marques}, P.~{Marques}, J.~{Ribeiro}, T.~{Alves}, and L.~{Pereira},
  ``Experimental evaluation of cellular networks for {UAV} operation and
  services,'' in \emph{2019 IEEE 24th International Workshop on Computer Aided
  Modeling and Design of Communication Links and Networks (CAMAD)}, 2019, pp.
  1--6, \url{https://doi.org/10.1109/CAMAD.2019.8858453}.

\bibitem{8539117}
L.~{Chen}, Z.~{Huang}, Z.~{Liu}, D.~{Liu}, and X.~{Huang}, ``{4G} network for
  air-ground data transmission: A drone based experiment,'' in \emph{2018 IEEE
  International Conference on Industrial Internet (ICII)}, 2018, pp. 167--168,
  \url{https://doi.org/10.1109/ICII.2018.00028}.

\bibitem{8568838}
M.~C. {Batistatos}, G.~E. {Athanasiadou}, D.~A. {Zarbouti}, G.~V. {Tsoulos},
  and N.~C. {Sagias}, ``{LTE} ground-to-air measurements for {UAV}-assisted
  cellular networks,'' in \emph{12th European Conference on Antennas and
  Propagation (EuCAP 2018)}, 2018, pp. 1--5,
  \url{https://doi.org/10.1049/cp.2018.1160}.

\bibitem{8377373}
R.~{Amorim}, H.~{Nguyen}, J.~{Wigard}, I.~Z. {Kovács}, T.~B. {Sorensen}, and
  P.~{Mogensen}, ``{LTE} radio measurements above urban rooftops for aerial
  communications,'' in \emph{2018 IEEE Wireless Communications and Networking
  Conference (WCNC)}, 2018, pp. 1--6,
  \url{https://doi.org/10.1109/WCNC.2018.8377373}.

\bibitem{8891526}
M.~{Lopez}, T.~B. {Sorensen}, P.~{Mogensen}, J.~{Wigard}, and I.~Z. {Kovacs},
  ``Shadow fading spatial correlation analysis for aerial vehicles: Ray tracing
  vs. measurements,'' in \emph{2019 IEEE 90th Vehicular Technology Conference
  (VTC2019-Fall)}, 2019, pp. 1--5,
  \url{https://doi.org/10.1109/VTCFall.2019.8891526}.

\bibitem{8891422}
M.~{Bucur}, T.~{Sorensen}, R.~{Amorim}, M.~{Lopez}, I.~Z. {Kovacs}, and
  P.~{Mogensen}, ``Validation of large-scale propagation characteristics for
  {UAVs} within urban environment,'' in \emph{2019 IEEE 90th Vehicular
  Technology Conference (VTC2019-Fall)}, 2019, pp. 1--6,
  \url{https://doi.org/10.1109/VTCFall.2019.8891422}.

\bibitem{9001975}
M.~{Naveed}, S.~{Qazi}, B.~A. {Khawaja}, and M.~{Mustaqim}, ``Evaluation of
  video streaming capacity of {UAVs} with respect to channel variation in
  {4G-LTE} surveillance architecture,'' in \emph{2019 8th International
  Conference on Information and Communication Technologies (ICICT)}, 2019, pp.
  149--154, \url{https://doi.org/10.1109/ICICT47744.2019.9001975}.

\bibitem{7396417}
S.~{Qazi}, A.~S. {Siddiqui}, and A.~I. {Wagan}, ``{UAV} based real time video
  surveillance over {4G LTE},'' in \emph{2015 International Conference on Open
  Source Systems Technologies (ICOSST)}, 2015, pp. 141--145,
  \url{https://doi.org/10.1109/ICOSST.2015.7396417}.

\bibitem{8102112}
L.~M. {Schalk} and M.~{Herrmann}, ``Suitability of {LTE} for
  drone-to-infrastructure communications in very low level airspace,'' in
  \emph{2017 IEEE/AIAA 36th Digital Avionics Systems Conference (DASC)}, 2017,
  pp. 1--7, \url{https://doi.org/10.1109/DASC.2017.8102112}.

\bibitem{8756738}
A.~E. {Garcia}, S.~{Hofmann}, C.~{Sous}, L.~{Garcia}, A.~{Baltaci}, C.~{Bach},
  R.~{Wellens}, D.~{Gera}, D.~{Schupke}, and H.~E. {Gonzalez}, ``Performance
  evaluation of network slicing for aerial vehicle communications,'' in
  \emph{2019 IEEE International Conference on Communications Workshops (ICC
  Workshops)}, 2019, pp. 1--6, \url{https://doi.org/10.1109/ICCW.2019.8756738}.

\bibitem{8746579}
R.~{Amorim}, I.~Z. {Kovacs}, J.~{Wigard}, G.~{Pocovi}, T.~B. {Sorensen}, and
  P.~{Mogensen}, ``Improving drone's command and control link reliability
  through dual-network connectivity,'' in \emph{2019 IEEE 89th Vehicular
  Technology Conference (VTC2019-Spring)}, 2019, pp. 1--6,
  \url{https://doi.org/10.1109/VTCSpring.2019.8746579}.

\bibitem{9145477}
J.~{Säe}, R.~{Wirén}, J.~{Kauppi}, J.~{Torsner}, S.~{Andreev}, and
  M.~{Valkama}, ``Reliability of {UAV} connectivity in dual-{MNO} networks: A
  performance measurement campaign,'' in \emph{2020 IEEE International
  Conference on Communications Workshops (ICC Workshops)}, 2020, pp. 1--5,
  \url{https://doi.org/10.1109/ICCWorkshops49005.2020.9145477}.

\bibitem{8998329}
R.~{Amer}, W.~{Saad}, and N.~{Marchetti}, ``Mobility in the sky: Performance
  and mobility analysis for cellular-connected {UAVs},'' \emph{IEEE
  Transactions on Communications}, vol.~68, no.~5, pp. 3229--3246, 2020,
  \url{https://doi.org/10.1109/TCOMM.2020.2973629}.

\bibitem{9151343}
Y.~{Li}, N.~I. {Miridakis}, T.~A. {Tsiftsis}, G.~{Yang}, and M.~{Xia},
  ``Air-to-air communications beyond {5G}: A novel {3D} {CoMP} transmission
  scheme,'' \emph{IEEE Transactions on Wireless Communications}, pp. 1--1,
  2020, \url{https://doi.org/10.1109/TWC.2020.3010569}.

\bibitem{8406959}
X.~{Zhang}, H.~{Wang}, and H.~{Zhao}, ``An {SDN} framework for {UAV} backbone
  network towards knowledge centric networking,'' in \emph{IEEE INFOCOM 2018 -
  IEEE Conference on Computer Communications Workshops (INFOCOM WKSHPS)}, 2018,
  pp. 456--461, \url{https://doi.org/10.1109/INFCOMW.2018.8406959}.

\bibitem{7983162}
K.~J.~S. {White}, E.~{Denney}, M.~D. {Knudson}, A.~K. {Mamerides}, and D.~P.
  {Pezaros}, ``A programmable {SDN}+{NFV}-based architecture for {UAV}
  telemetry monitoring,'' in \emph{2017 14th IEEE Annual Consumer
  Communications Networking Conference (CCNC)}, 2017, pp. 522--527,
  \url{https://doi.org/10.1109/CCNC.2017.7983162}.

\bibitem{8385471}
K.~{Heimann}, J.~{Tiemann}, S.~{Boecker}, and C.~{Wietfeld}, ``On the potential
  of {5G} mmwave pencil beam antennas for {UAV} communications: An experimental
  evaluation,'' in \emph{WSA 2018; 22nd International ITG Workshop on Smart
  Antennas}, 2018, pp. 1--6.

\bibitem{9149027}
H.~{Wu}, R.~{Hou}, and B.~{Sun}, ``Location information assisted mmwave hybrid
  beamforming scheme for {5G}-enabled {UAVs},'' in \emph{ICC 2020 - 2020 IEEE
  International Conference on Communications (ICC)}, 2020, pp. 1--6,
  \url{https://doi.org/10.1109/ICC40277.2020.9149027}.

\bibitem{9162928}
P.~{Foroughi}, H.~{Beyranvand}, M.~{Gagnaire}, and S.~A. {Zahr}, ``User
  association in hybrid {UAV}-cellular networks for massive real-time {IoT}
  applications,'' in \emph{IEEE INFOCOM 2020 - IEEE Conference on Computer
  Communications Workshops (INFOCOM WKSHPS)}, 2020, pp. 243--248,
  \url{https://doi.org/10.1109/INFOCOMWKSHPS50562.2020.9162928}.

\bibitem{9093794}
C.~{Chen}, T.~{Zhang}, Y.~{Liu}, G.~Y. {Li}, and Z.~{Zeng}, ``Joint user
  association and caching placement for cache-enabled {UAVs} in cellular
  networks,'' in \emph{IEEE INFOCOM 2019 - IEEE Conference on Computer
  Communications Workshops (INFOCOM WKSHPS)}, 2019, pp. 1--6.

\bibitem{9162853}
Z.~{Lv}, J.~{Hao}, and Y.~{Guo}, ``Energy minimization for {MEC}-enabled
  cellular-connected {UAV}: Trajectory optimization and resource scheduling,''
  in \emph{IEEE INFOCOM 2020 - IEEE Conference on Computer Communications
  Workshops (INFOCOM WKSHPS)}, 2020, pp. 478--483,
  \url{https://doi.org/10.1109/INFOCOMWKSHPS50562.2020.9162853}.

\bibitem{9249665}
S.~{Singh}, H.~{Narasimhan}, I.~{Güvenç}, A.~{Bhuyan}, H.~{Dai}, and M.~L.
  {Sichitiu}, ``Coverage analysis for ground and aerial users in mmwave
  cellular networks in urban settings,'' in \emph{2020 SoutheastCon}, 2020, pp.
  1--8, \url{https://doi.org/10.1109/SoutheastCon44009.2020.9249665}.

\bibitem{9149403}
Y.~{Zhang}, T.~{Arakawa}, J.~V. {Krogmeier}, C.~R. {Anderson}, D.~J. {Love},
  and D.~R. {Buckmaster}, ``Large-scale cellular coverage analyses for {UAV}
  data relay via channel modeling,'' in \emph{ICC 2020 - 2020 IEEE
  International Conference on Communications (ICC)}, 2020, pp. 1--6,
  \url{https://doi.org/10.1109/ICC40277.2020.9149403}.

\bibitem{9148841}
R.~{Amer}, W.~{Saad}, B.~{Galkin}, and N.~{Marchetti}, ``Performance analysis
  of mobile cellular-connected drones under practical antenna configurations,''
  in \emph{ICC 2020 - 2020 IEEE International Conference on Communications
  (ICC)}, 2020, pp. 1--7, \url{https://doi.org/10.1109/ICC40277.2020.9148841}.

\bibitem{8969181}
X.~{Wang} and M.~C. {Gursoy}, ``Coverage analysis for cellular-connected {UAVs}
  with {3D} antenna patterns,'' in \emph{2019 IEEE Global Conference on Signal
  and Information Processing (GlobalSIP)}, 2019, pp. 1--5,
  \url{https://doi.org/10.1109/GlobalSIP45357.2019.8969181}.

\bibitem{8692749}
M.~M. {Azari}, F.~{Rosas}, and S.~{Pollin}, ``Cellular connectivity for {UAVs}:
  Network modeling, performance analysis, and design guidelines,'' \emph{IEEE
  Transactions on Wireless Communications}, vol.~18, no.~7, pp. 3366--3381,
  2019, \url{https://doi.org/10.1109/TWC.2019.2910112}.

\bibitem{6391922}
T.~{Chi}, Y.~{Ming}, {Tseng}, S.~{Kuo}, and C.~{Liao}, ``Civil {UAV} path
  planning algorithm for considering connection with cellular data network,''
  in \emph{2012 IEEE 12th International Conference on Computer and Information
  Technology}, 2012, pp. 327--331, \url{https://doi.org/10.1109/CIT.2012.83}.

\bibitem{9149190}
O.~{Esrafilian}, R.~{Gangula}, and D.~{Gesbert}, ``{3D}-map assisted {UAV}
  trajectory design under cellular connectivity constraints,'' in \emph{ICC
  2020 - 2020 IEEE International Conference on Communications (ICC)}, 2020, pp.
  1--6, \url{https://doi.org/10.1109/ICC40277.2020.9149190}.

\bibitem{9005434}
S.~{Zhang} and R.~{Zhang}, ``Trajectory optimization for cellular-connected
  {UAV} under outage duration constraint,'' \emph{Journal of Communications and
  Information Networks}, vol.~4, no.~4, pp. 55--71, 2019,
  \url{https://doi.org/10.23919/JCIN.2019.9005434}.

\bibitem{9013177}
------, ``Radio map based path planning for cellular-connected {UAV},'' in
  \emph{2019 IEEE Global Communications Conference (GLOBECOM)}, 2019, pp. 1--6,
  \url{https://doi.org/10.1109/GLOBECOM38437.2019.9013177}.

\bibitem{9145194}
P.~{Susarla}, Y.~{Deng}, G.~{Destino}, J.~{Saloranta}, T.~{Mahmoodi},
  M.~{Juntti}, and O.~{Sílven}, ``Learning-based trajectory optimization for
  {5G} mmwave uplink {UAVs},'' in \emph{2020 IEEE International Conference on
  Communications Workshops (ICC Workshops)}, 2020, pp. 1--7,
  \url{https://doi.org/10.1109/ICCWorkshops49005.2020.9145194}.

\bibitem{9014041}
Y.~{Zeng} and X.~{Xu}, ``Path design for cellular-connected {UAV} with
  reinforcement learning,'' in \emph{2019 IEEE Global Communications Conference
  (GLOBECOM)}, 2019, pp. 1--6,
  \url{https://doi.org/10.1109/GLOBECOM38437.2019.9014041}.

\bibitem{9253515}
A.~{Exposito}, D.~{Schupke}, and H.~{Esteban}, ``Route optimisation for maximum
  air to ground channel quality,'' \emph{IEEE Access}, vol.~8, pp.
  203\,619--203\,630, 2020, \url{https://doi.org/10.1109/ACCESS.2020.3037075}.

\bibitem{8685130}
F.~{Cui}, Y.~{Cai}, Z.~{Qin}, M.~{Zhao}, and G.~Y. {Li}, ``Multiple access for
  mobile-{UAV} enabled networks: Joint trajectory design and resource
  allocation,'' \emph{IEEE Transactions on Communications}, vol.~67, no.~7, pp.
  4980--4994, 2019, \url{https://doi.org/10.1109/TCOMM.2019.2910263}.

\bibitem{8647671}
S.~{Zhang}, H.~{Zhang}, B.~{Di}, and L.~{Song}, ``Resource allocation and
  trajectory design for cellular {UAV-to-X} communication networks in {5G},''
  in \emph{2018 IEEE Global Communications Conference (GLOBECOM)}, 2018, pp.
  1--6, \url{https://doi.org/10.1109/GLOCOM.2018.8647671}.

\bibitem{10.1145/3376897.3377860}
L.~Bertizzolo, T.~X. Tran, B.~Amento, B.~Balasubramanian, R.~Jana, H.~Purdy,
  Y.~Zhou, and T.~Melodia, ``Live and let live: Flying {UAVs} without affecting
  terrestrial {UE}s,'' in \emph{Proceedings of the 21st International Workshop
  on Mobile Computing Systems and Applications}, ser. HotMobile '20.\hskip 1em
  plus 0.5em minus 0.4em\relax New York, NY, USA: Association for Computing
  Machinery, 2020, p. 21–26, \url{https://doi.org/10.1145/3376897.3377860}.

\bibitem{9037325}
M.~M.~U. {Chowdhury}, S.~J. {Maeng}, E.~{Bulut}, and I.~{Güvenc}, ``{3D}
  trajectory optimization in {UAV}-assisted cellular networks considering
  antenna radiation pattern and backhaul constraint,'' \emph{IEEE Transactions
  on Aerospace and Electronic Systems}, pp. 1--1, 2020,
  \url{https://doi.org/10.1109/TAES.2020.2981233}.

\bibitem{9149835}
C.~{Zhan} and Y.~{Zeng}, ``Energy-efficient data uploading for
  cellular-connected {UAV} systems,'' \emph{IEEE Transactions on Wireless
  Communications}, pp. 1--1, 2020,
  \url{https://doi.org/10.1109/TWC.2020.3010320}.

\bibitem{8647820}
W.~{Mei}, Q.~{Wu}, and R.~{Zhang}, ``Cellular-connected {UAV}: Uplink
  association, power control and interference coordination,'' in \emph{2018
  IEEE Global Communications Conference (GLOBECOM)}, 2018, pp. 206--212,
  \url{https://doi.org/10.1109/GLOCOM.2018.8647820}.

\bibitem{8913615}
W.~{Mei} and R.~{Zhang}, ``Cooperative downlink interference transmission and
  cancellation for cellular-connected {UAV}: A divide-and-conquer approach,''
  \emph{IEEE Transactions on Communications}, vol.~68, no.~2, pp. 1297--1311,
  2020, \url{https://doi.org/10.1109/TCOMM.2019.2955953}.

\bibitem{8964329}
Z.~{Ullah}, F.~{Al-Turjman}, and L.~{Mostarda}, ``Cognition in {UAV}-aided {5G}
  and beyond communications: A survey,'' \emph{IEEE Transactions on Cognitive
  Communications and Networking}, pp. 1--1, 2020,
  \url{https://doi.org/10.1109/TCCN.2020.2968311}.

\bibitem{7470934}
B.~{Van Der Bergh}, A.~{Chiumento}, and S.~{Pollin}, ``{LTE} in the sky:
  Trading off propagation benefits with interference costs for aerial nodes,''
  \emph{IEEE Communications Magazine}, vol.~54, no.~5, pp. 44--50, 2016,
  \url{https://doi.org/10.1109/MCOM.2016.7470934}.

\bibitem{Qualcomm_ltetrials}
Qualcomm\;Technologies\;Inc., ``{LTE} unmanned aircraft systems, trial report,
  v1.0.1,'' Qualcomm Technologies, Inc., Tech. Rep., Dec. 2017, [Online].
  Available
  at:\url{https://www.qualcomm.com/media/documents/files/lte-unmanned-aircraft-systems-trial-report.pdf}
  [Accessed on 16/01/2021].

\bibitem{9128682}
T.~Z.~H. {Ernest}, A.~S. {Madhukumar}, R.~P. {Sirigina}, and A.~K. {Krishna},
  ``Impact of cellular interference on uplink {UAV} communications,'' in
  \emph{2020 IEEE 91st Vehicular Technology Conference (VTC2020-Spring)}, 2020,
  pp. 1--5, \url{https://doi.org/10.1109/VTC2020-Spring48590.2020.9128682}.

\bibitem{9099899}
W.~{Mei} and R.~{Zhang}, ``Uplink cooperative interference cancellation for
  cellular-connected {UAV}: A quantize-and-forward approach,'' \emph{IEEE
  Wireless Communications Letters}, pp. 1--1, 2020,
  \url{https://doi.org/10.1109/LWC.2020.2997837}.

\bibitem{9069873}
R.~M. {de Amorim}, J.~{Wigard}, I.~Z. {Kovacs}, T.~B. {Sorensen}, and P.~E.
  {Mogensen}, ``Enabling cellular communication for aerial vehicles: Providing
  reliability for future applications,'' \emph{IEEE Vehicular Technology
  Magazine}, vol.~15, no.~2, pp. 129--135, 2020,
  \url{https://doi.org/10.1109/MVT.2020.2980438}.

\bibitem{8976147}
W.~{Mei} and R.~{Zhang}, ``{UAV}-sensing-assisted cellular interference
  coordination: A cognitive radio approach,'' \emph{IEEE Wireless
  Communications Letters}, vol.~9, no.~6, pp. 799--803, 2020,
  \url{https://doi.org/10.1109/LWC.2020.2970416}.

\bibitem{8641388}
------, ``Uplink cooperative {NOMA} for cellular-connected {UAV},'' \emph{IEEE
  Journal of Selected Topics in Signal Processing}, vol.~13, no.~3, pp.
  644--656, 2019, \url{https://doi.org/10.1109/JSTSP.2019.2899208}.

\bibitem{8906143}
X.~{Pang}, G.~{Gui}, N.~{Zhao}, W.~{Zhang}, Y.~{Chen}, Z.~{Ding}, and
  F.~{Adachi}, ``Uplink precoding optimization for {NOMA} cellular-connected
  {UAV} networks,'' \emph{IEEE Transactions on Communications}, vol.~68, no.~2,
  pp. 1271--1283, 2020, \url{https://doi.org/10.1109/TCOMM.2019.2954136}.

\bibitem{8301389}
H.~C. {Nguyen}, R.~{Amorim}, J.~{Wigard}, I.~Z. {KováCs}, T.~B. {Sørensen},
  and P.~E. {Mogensen}, ``How to ensure reliable connectivity for aerial
  vehicles over cellular networks,'' \emph{IEEE Access}, vol.~6, pp.
  12\,304--12\,317, 2018, \url{https://doi.org/10.1109/ACCESS.2018.2808998}.

\bibitem{8369158}
R.~{Amorim}, H.~{Nguyen}, J.~{Wigard}, I.~Z. {Kovács}, T.~B. {Sørensen},
  D.~Z. {Biro}, M.~{Sørensen}, and P.~{Mogensen}, ``Measured uplink
  interference caused by aerial vehicles in {LTE} cellular networks,''
  \emph{IEEE Wireless Communications Letters}, vol.~7, no.~6, pp. 958--961,
  2018, \url{https://doi.org/10.1109/LWC.2018.2841386}.

\bibitem{9238986}
W.~{Dong}, X.~{Mao}, R.~{Hou}, X.~{Lv}, and H.~{Li}, ``An enhanced handover
  scheme for cellular-connected {UAVs},'' in \emph{2020 IEEE/CIC International
  Conference on Communications in China (ICCC)}, 2020, pp. 418--423,
  \url{https://doi.org/10.1109/ICCC49849.2020.9238986}.

\bibitem{9221328}
B.~{Galkin}, R.~{Amer}, E.~{Fonseca}, and L.~A. {DaSilva}, ``Intelligent base
  station association for {UAV} cellular users: A supervised learning
  approach,'' in \emph{2020 IEEE 3rd 5G World Forum (5GWF)}, 2020, pp.
  383--388, \url{https://doi.org/10.1109/5GWF49715.2020.9221328}.

\bibitem{9295084}
G.~{Geraci}, A.~{Garcia‐Rodriguez}, L.~{Galati Giordano}, and
  D.~{López‐Pérez}, \emph{Enhanced Cellular Support for {UAVs} with Massive
  {MIMO}}, 2021, pp. 181--201, \url{https://doi.org/10.1002/9781119575795.ch7}.

\bibitem{9261835}
C.~{Fan}, S.~{Bao}, B.~{Li}, and C.~{Zhao}, ``Joint resource allocation for
  dynamic cellular-enabled {UAVs} communication,'' \emph{IET Communications},
  vol.~14, no.~18, pp. 3161--3168, 2020,
  \url{https://doi.org/10.1049/iet-com.2019.1121}.

\bibitem{9129453}
A.~{Azari}, F.~{Ghavimi}, M.~{Ozger}, R.~{Jantti}, and C.~{Cavdar}, ``Machine
  learning assisted handover and resource management for cellular connected
  drones,'' in \emph{2020 IEEE 91st Vehicular Technology Conference
  (VTC2020-Spring)}, 2020, pp. 1--7,
  \url{https://doi.org/10.1109/VTC2020-Spring48590.2020.9129453}.

\bibitem{8761670}
L.~{Liu}, S.~{Zhang}, and R.~{Zhang}, ``Exploiting {NOMA} for multi-beam {UAV}
  communication in cellular uplink,'' in \emph{ICC 2019 - 2019 IEEE
  International Conference on Communications (ICC)}, 2019, pp. 1--6,
  \url{https://doi.org/10.1109/ICC.2019.8761670}.

\bibitem{8636798}
A.~N. {Sakib} and R.~{Mostafa}, ``Improvement in reverse link capacity and
  handoff for a {UAV} in a {CDMA} cellular network using directional
  beampattern,'' in \emph{2018 10th International Conference on Electrical and
  Computer Engineering (ICECE)}, 2018, pp. 133--136,
  \url{https://doi.org/10.1109/ICECE.2018.8636798}.

\bibitem{9145089}
M.~M.~U. {Chowdhury}, W.~{Saad}, and I.~{Güvenç}, ``Mobility management for
  cellular-connected {UAVs}: A learning-based approach,'' in \emph{2020 IEEE
  International Conference on Communications Workshops (ICC Workshops)}, 2020,
  pp. 1--6, \url{https://doi.org/10.1109/ICCWorkshops49005.2020.9145089}.

\bibitem{9296324}
C.~{Liu}, W.~{Yuan}, Z.~{Wei}, X.~{Liu}, and D.~W.~K. {Ng}, ``Location-aware
  predictive beamforming for {UAV} communications: A deep learning approach,''
  \emph{IEEE Wireless Communications Letters}, pp. 1--1, 2020,
  \url{https://doi.org/10.1109/LWC.2020.3045150}.

\bibitem{9214955}
W.~{Miao}, C.~{Luo}, G.~{Min}, Y.~{Mi}, and Z.~{Yu}, ``Location-based robust
  beamforming design for cellular-enabled {UAV} communications,'' \emph{IEEE
  Internet of Things Journal}, pp. 1--1, 2020,
  \url{https://doi.org/10.1109/JIOT.2020.3028853}.

\bibitem{9093884}
W.~{Miao}, C.~{Luo}, G.~{Min}, and Z.~{Zhao}, ``Lightweight {3-D} beamforming
  design in {5G} {UAV} broadcasting communications,'' \emph{IEEE Transactions
  on Broadcasting}, vol.~66, no.~2, pp. 515--524, 2020,
  \url{https://doi.org/10.1109/TBC.2020.2990564}.

\bibitem{8964589}
Y.~{Huang}, Q.~{Wu}, T.~{Wang}, G.~{Zhou}, and R.~{Zhang}, ``{3D} beam tracking
  for cellular-connected {UAV},'' \emph{IEEE Wireless Communications Letters},
  vol.~9, no.~5, pp. 736--740, 2020,
  \url{https://doi.org/10.1109/LWC.2020.2968312}.

\bibitem{9356531}
S.~{Alfattani}, W.~{Jaafar}, Y.~{Hmamouche}, H.~{Yanikomeroglu},
  A.~{Yongaçoglu}, N.~D. {Dao}, and P.~{Zhu}, ``Aerial platforms with
  reconfigurable smart surfaces for 5g and beyond,'' \emph{IEEE Communications
  Magazine}, vol.~59, no.~1, pp. 96--102, 2021.

\bibitem{9120632}
D.~{Ma}, M.~{Ding}, and M.~{Hassan}, ``Enhancing cellular communications for
  {UAVs} via intelligent reflective surface,'' in \emph{2020 IEEE Wireless
  Communications and Networking Conference (WCNC)}, 2020, pp. 1--6,
  \url{https://doi.org/10.1109/WCNC45663.2020.9120632}.

\bibitem{9145273}
Z.~{Mohamed} and S.~{Aïssa}, ``Leveraging {UAVs} with intelligent reflecting
  surfaces for energy-efficient communications with cell-edge users,'' in
  \emph{2020 IEEE International Conference on Communications Workshops (ICC
  Workshops)}, 2020, pp. 1--6,
  \url{https://doi.org/10.1109/ICCWorkshops49005.2020.9145273}.

\bibitem{8902713}
R.~{Amorim}, I.~Z. {Kovács}, J.~{Wigard}, T.~B. {Sorensen}, and P.~{Mogensen},
  ``Forecasting spectrum demand for {UAVs} served by dedicated allocation in
  cellular networks,'' in \emph{2019 IEEE Wireless Communications and
  Networking Conference Workshop (WCNCW)}, 2019, pp. 1--6,
  \url{https://doi.org/10.1109/WCNCW.2019.8902713}.

\bibitem{8493128}
H.~{Hellaoui}, O.~{Bekkouche}, M.~{Bagaa}, and T.~{Taleb}, ``Aerial control
  system for spectrum efficiency in {UAV}-to-cellular communications,''
  \emph{IEEE Communications Magazine}, vol.~56, no.~10, pp. 108--113, 2018,
  \url{https://doi.org/10.1109/MCOM.2018.1800078}.

\bibitem{9045290}
S.~{Singh}, S.~L. {Sunkara}, I.~{Güvenç}, A.~{Bhuyan}, H.~{Dai}, and M.~L.
  {Sichitiu}, ``Spectrum reuse among aerial and ground users in mmwave cellular
  networks in urban settings,'' in \emph{2020 IEEE 17th Annual Consumer
  Communications Networking Conference (CCNC)}, 2020, pp. 1--6,
  \url{https://doi.org/10.1109/CCNC46108.2020.9045290}.

\bibitem{dinc17}
E.~{Dinc}, M.~{Vondra}, and C.~{Cavdar}, ``Multi-user beamforming and ground
  station deployment for {5G} direct air-to-ground communication,'' in
  \emph{GLOBECOM 2017 - 2017 IEEE Global Communications Conference}, 2017, pp.
  1--7, \url{https://doi.org/10.1109/GLOCOM.2017.8254571}.

\bibitem{dinc2020total}
E.~Dinc, M.~Vondra, and C.~Cavdar, ``Total cost of ownership optimization for
  direct air-to-ground system design,'' \emph{arXiv Preprint ArXiv:2008.08728},
  2020, [Online]. Available at:\url{https://arxiv.org/abs/2008.08728} [Accessed
  on 16/01/2021].

\bibitem{atg4}
Gogo, ``Gogo {ATG},'' [Online]. Available
  at:\url{http://learn.business.gogoair.com/rs/097-RTO-429/images/Gogo_Brochure_ATG.pdf}
  [Accessed on 16/01/2021].

\bibitem{smartskywhite}
Smartsky, ``Smartsky’s connectivity ecosystem for best inflight user
  experiences,'' 2014, [Online]. Available
  at:\url{http://smartskynetworks.com/wp-content/uploads/2015/09/160104-Connectivity-Ecosystem-1.pdf}
  [Accessed on 16/01/2021].

\bibitem{ecc}
ECC, ``Broadband direct-air-to-ground communications ({DA2GC}),'' Tech. Rep.
  Report 214, June 2014, [Online]. Available
  at:\url{https://docdb.cept.org/download/27d4b5f0-025c/ECCREP214.PDF}
  [Accessed on 16/01/2021].

\bibitem{nokia}
Alcatel\;Lucent, ``Using air-to-ground {LTE} for in-flight ultra-broadband,
  strategic white paper,'' Tech. Rep., 2015, [Online]. Available
  at:\url{http://www.tmcnet.com/tmc/whitepapers/documents/whitepapers/2015/11529-using-air-to-ground-lte-in-flight-ultra.pdf}
  [Accessed on 16/01/2021].

\bibitem{telecomperspective_lte5G}
G.~Yang, X.~Lin, Y.~Li, H.~Cui, M.~Xu, D.~Wu, H.~Rydén, and S.~Redhwan, ``A
  telecom perspective on the internet of drones: From {LTE}-advanced to {5G},''
  03 2018, \url{https://arxiv.org/abs/1803.11048}.

\bibitem{GSMA_5GGuide}
GSMA, ``The {5G} guide - a reference for operators,'' Tech. Rep., Apr. 2019,
  [Online] Available
  at:\url{https://www.gsma.com/wp-content/uploads/2019/04/The-5G-Guide_GSMA_2019_04_29_compressed.pdf}
  [Accessed on 16/01/2021].

\bibitem{qualcomm_urllc}
G.~Brown, ``Ultra-reliable low-latency {5G} for industrial-automation,'' Heavy
  Reading and Qualcomm, Tech. Rep., 2016, [Online]. Available
  at:\url{https://www.qualcomm.com/media/documents/files/read-the-white-paper-by-heavy-reading.pdf}
  [Accessed on 16/01/2021].

\bibitem{8647652}
M.~{Ozger}, M.~{Vondra}, and C.~{Cavdar}, ``Towards beyond visual line of sight
  piloting of {UAVs} with ultra reliable low latency communication,'' in
  \emph{2018 IEEE Global Communications Conference (GLOBECOM)}, 2018, pp. 1--6,
  \url{https://doi.org/10.1109/GLOCOM.2018.8647652}.

\bibitem{9295186}
R.~M. {de Amorim}, J.~{Wigard}, I.~Z. {Kovács}, and T.~B. {Sørensen},
  \emph{Performance Enhancements for {LTE}‐Connected {UAVs}: Experiments and
  Simulations}, 2021, pp. 139--161,
  \url{https://doi.org/10.1002/9781119575795.ch5}.

\bibitem{9183797}
R.~{Gupta}, A.~{Kumari}, S.~{Tanwar}, and N.~{Kumar}, ``Blockchain-envisioned
  softwarized multi-swarming {UAVs} to tackle {COVID}-i9 situations,''
  \emph{IEEE Network}, pp. 1--8, 2020,
  \url{https://doi.org/10.1109/MNET.011.2000439}.

\bibitem{8741314}
F.~{Zhou}, W.~{Li}, L.~{Meng}, and M.~{Kadoch}, ``Capacity enhancement for
  hotspot area in {5G} cellular networks using mmwave aerial base station,''
  \emph{IEEE Wireless Communications Letters}, vol.~8, no.~3, pp. 677--680,
  2019, \url{https://doi.org/10.1109/LWC.2018.2882445}.

\bibitem{9133208}
J.~{Ji}, K.~{Zhu}, D.~{Niyato}, and R.~{Wang}, ``Probabilistic cache placement
  in {UAV}-assisted networks with {D2D} connections: Performance analysis and
  trajectory optimization,'' \emph{IEEE Transactions on Communications}, pp.
  1--1, 2020, \url{https://doi.org/10.1109/TCOMM.2020.3006908}.

\bibitem{8368062}
E.~{Turgut} and M.~C. {Gursoy}, ``Downlink analysis in unmanned aerial vehicle
  ({UAV}) assisted cellular networks with clustered users,'' \emph{IEEE
  Access}, vol.~6, pp. 36\,313--36\,324, 2018,
  \url{https://doi.org/10.1109/ACCESS.2018.2841655}.

\bibitem{9153823}
Y.~{Qin}, M.~A. {Kishk}, and M.~{Alouini}, ``Performance evaluation of
  {UAV}-enabled cellular networks with battery-limited drones,'' \emph{IEEE
  Communications Letters}, pp. 1--1, 2020,
  \url{https://doi.org/10.1109/LCOMM.2020.3013286}.

\bibitem{ecc_report309}
ECC, ``Analysis of the usage of aerial {UE} for communication in current {MFCN}
  harmonised bands,'' Tech. Rep. Report 309, 2020, [Online]. Available
  at:\url{https://docdb.cept.org/download/9f4305fb-aa77/ECC%20Report%20309.pdf}
  [Accessed on 16/01/2021].

\bibitem{8871183}
S.~{Zhu}, L.~{Gui}, N.~{Cheng}, F.~{Sun}, and Q.~{Zhang}, ``Joint design of
  access point selection and path planning for {UAV}-assisted cellular
  networks,'' \emph{IEEE Internet of Things Journal}, vol.~7, no.~1, pp.
  220--233, 2020, \url{https://doi.org/10.1109/JIOT.2019.2947718}.

\bibitem{8654727}
U.~{Challita}, W.~{Saad}, and C.~{Bettstetter}, ``Interference management for
  cellular-connected {UAVs}: A deep reinforcement learning approach,''
  \emph{IEEE Transactions on Wireless Communications}, vol.~18, no.~4, pp.
  2125--2140, 2019, \url{https://doi.org/10.1109/TWC.2019.2900035}.

\bibitem{8417707}
T.~{Cogalan}, S.~{Videv}, and H.~{Haas}, ``Inflight connectivity: Deploying
  different communication networks inside an aircraft,'' in \emph{2018 IEEE
  87th Vehicular Technology Conference (VTC Spring)}, 2018, pp. 1--6,
  \url{https://doi.org/10.1109/VTCSpring.2018.8417707}.

\bibitem{7503863}
M.~{Gürsu}, M.~{Vilgelm}, S.~{Zoppi}, and W.~{Kellerer}, ``Reliable
  co-existence of 802.15.4e {TSCH}-based {WSN} and {Wi-Fi} in an aircraft
  cabin,'' in \emph{2016 IEEE International Conference on Communications
  Workshops (ICC)}, 2016, pp. 663--668,
  \url{https://doi.org/10.1109/ICCW.2016.7503863}.

\bibitem{7986413}
F.~{Fabra}, C.~T. {Calafate}, J.~{Cano}, and P.~{Manzoni}, ``On the impact of
  inter-{UAV} communications interference in the 2.4 {GHz} band,'' in
  \emph{2017 13th International Wireless Communications and Mobile Computing
  Conference (IWCMC)}, 2017, pp. 945--950,
  \url{https://doi.org/10.1109/IWCMC.2017.7986413}.

\bibitem{multicastvideostreaming_uav}
R.~Muzaffar, E.~Yanmaz, C.~Raffelsberger, C.~Bettstetter, and A.~Cavallaro,
  ``Live multicast video streaming from drones: an experimental study,'' in
  \emph{Autonomous Robots}, vol.~44, 2020, p. 75–91,
  \url{https://doi.org/10.1007/s10514-019-09851-6}.

\bibitem{rahman14}
M.~A. {Rahman}, ``Enabling drone communications with {WiMAX} technology,'' in
  \emph{IISA 2014, The 5th International Conference on Information,
  Intelligence, Systems and Applications}, 2014, pp. 323--328,
  \url{https://doi.org/10.1109/IISA.2014.6878796}.

\bibitem{6928291}
J.~{Zhang}, X.~{Li}, and J.~{Qian}, ``{ZigBee} network-based detection of
  anomaly detection runway on airport,'' in \emph{2014 IEEE Far East Forum on
  Nondestructive Evaluation/Testing}, 2014, pp. 335--339,
  \url{https://doi.org/10.1109/FENDT.2014.6928291}.

\bibitem{7830305}
H.~{Liang}, ``Design of aircraft hydraulic leakage detection system based on
  {GPRS} {ZigBee},'' in \emph{2016 9th International Symposium on Computational
  Intelligence and Design (ISCID)}, vol.~1, 2016, pp. 101--104,
  \url{https://doi.org/10.1109/ISCID.2016.1031}.

\bibitem{7625728}
S.~{Min} and H.~{Nam}, ``A formation flight control of {UAVs} using {ZigBee},''
  in \emph{2016 13th International Conference on Ubiquitous Robots and Ambient
  Intelligence (URAI)}, 2016, pp. 163--165,
  \url{https://doi.org/10.1109/URAI.2016.7625728}.

\bibitem{Ankit_WAIC}
A.~{Dwivedi}, S.~{Zoppi}, W.~{Kellerer}, F.~{Neubauer}, and D.~{Schupke},
  ``Wireless avionics intra-communication ({WAIC}) {QoS} measurements of an
  ultra wideband ({UWB}) device for low-data rate transmissions,'' in
  \emph{2020 AIAA/IEEE 39th Digital Avionics Systems Conference (DASC)}, 2020,
  pp. 1--10, \url{https://doi.org/10.1109/DASC50938.2020.9256542}.

\bibitem{9119755}
W.~{Khawaja}, O.~{Ozdemir}, F.~{Erden}, I.~{Güvenç}, and D.~W. {Matolak},
  ``Ultra-wideband air-to-ground propagation channel characterization in an
  open area,'' \emph{IEEE Transactions on Aerospace and Electronic Systems},
  pp. 1--1, 2020, \url{https://doi.org/10.1109/TAES.2020.3003104}.

\bibitem{8741964}
W.~{Khawaja}, O.~{Ozdemir}, F.~{Erden}, I.~{Guvenç}, and D.~W. {Matolak},
  ``{UWB} air-to-ground propagation channel measurements and modeling using
  {UAVs},'' in \emph{2019 IEEE Aerospace Conference}, 2019, pp. 1--10,
  \url{https://doi.org/10.1109/AERO.2019.8741964}.

\bibitem{9153116}
Z.~{Cui}, C.~{Briso-Rodríguez}, K.~{Guan}, I.~{Güvenç}, and Z.~{Zhong},
  ``Wideband air-to-ground channel characterization for multiple propagation
  environments,'' \emph{IEEE Antennas and Wireless Propagation Letters},
  vol.~19, no.~9, pp. 1634--1638, 2020,
  \url{https://doi.org/10.1109/LAWP.2020.3012889}.

\bibitem{9148768}
Z.~{Cui}, C.~{Briso-Rodríguez}, K.~{Guan}, and Z.~{Zhong}, ``Ultra-wideband
  air-to-ground channel measurements and modeling in hilly environment,'' in
  \emph{ICC 2020 - 2020 IEEE International Conference on Communications (ICC)},
  2020, pp. 1--6, \url{https://doi.org/10.1109/ICC40277.2020.9148768}.

\bibitem{8986049}
Y.~{Wang}, L.~{Ma}, and Z.~{Meng}, ``Effects of {UWB} electromagnetic pulse on
  {UAV} data link system,'' in \emph{2019 IEEE 6th International Symposium on
  Electromagnetic Compatibility (ISEMC)}, 2019, pp. 1--4,
  \url{https://doi.org/10.1109/ISEMC48616.2019.8986049}.

\bibitem{8824115}
L.~I. {Balderas}, A.~{Reyna}, M.~A. {Panduro}, C.~{Del Rio}, and A.~R.
  {Gutiérrez}, ``Low-profile conformal {UWB} antenna for {UAV} applications,''
  \emph{IEEE Access}, vol.~7, pp. 127\,486--127\,494, 2019,
  \url{https://doi.org/10.1109/ACCESS.2019.2939511}.

\bibitem{8661597}
S.~{Gao}, F.~{Shang}, and C.~{Du}, ``Design of multichannel and multihop
  low-power wide-area network for aircraft vibration monitoring,'' \emph{IEEE
  Transactions on Instrumentation and Measurement}, vol.~68, no.~12, pp.
  4887--4895, 2019, \url{https://doi.org/10.1109/TIM.2019.2897037}.

\bibitem{8471484}
J.~A. {Godoy}, F.~{Cabrera}, V.~{Araña}, D.~{Sánchez}, I.~{Alonso}, and
  N.~{Molina}, ``A new approach of {V2X} communications for long range
  applications in {UAVs},'' in \emph{2018 2nd URSI Atlantic Radio Science
  Meeting (AT-RASC)}, 2018, pp. 1--4,
  \url{https://doi.org/10.23919/URSI-AT-RASC.2018.8471484}.

\bibitem{7593502}
K.~{Kainrath}, M.~{Gruber}, H.~{Flühr}, and E.~{Leitgeb}, ``Communication
  techniques for remotely piloted aircraft with integrated modular avionics,''
  in \emph{2016 International Conference on Broadband Communications for Next
  Generation Networks and Multimedia Applications (CoBCom)}, 2016, pp. 1--6,
  \url{https://doi.org/10.1109/COBCOM.2016.7593502}.

\bibitem{8570043}
Z.~{Yuan}, J.~{Jin}, L.~{Sun}, K.~{Chin}, and G.~{Muntean}, ``Ultra-reliable
  {IoT} communications with {UAVs}: A swarm use case,'' \emph{IEEE
  Communications Magazine}, vol.~56, no.~12, pp. 90--96, 2018,
  \url{https://doi.org/10.1109/MCOM.2018.1800161}.

\bibitem{8094705}
V.~{Tikhvinskiy}, P.~{Korchagin}, G.~{Bochechka}, A.~{Gryazev}, and
  A.~{Aytmagambetov}, ``Spectrum sharing in 800 mhz band: Experimental,
  estimation of {LoRa} networks and air traffic control radars co-existence,''
  in \emph{2017 International Symposium on Electromagnetic Compatibility - EMC
  EUROPE}, 2017, pp. 1--6,
  \url{https://doi.org/10.1109/EMCEurope.2017.8094705}.

\bibitem{8453394}
M.~{Zolanvari}, M.~A. {Teixeira}, and R.~{Jain}, ``Analysis of {AeroMACS} data
  link for unmanned aircraft vehicles,'' in \emph{2018 International Conference
  on Unmanned Aircraft Systems (ICUAS)}, 2018, pp. 752--759,
  \url{https://doi.org/10.1109/ICUAS.2018.8453394}.

\bibitem{8288087}
E.~{Dinc}, M.~{Vondra}, and C.~{Cavdar}, ``Seamless gate-to-gate connectivity
  concept: Onboard {LTE}, {Wi-Fi} and {LAA},'' in \emph{2017 IEEE 86th
  Vehicular Technology Conference (VTC-Fall)}, 2017, pp. 1--7,
  \url{https://doi.org/10.1109/VTCFall.2017.8288087}.

\bibitem{wifihalow}
G.~{Gurtner} and C.~{Bongiorno}, ``{IEEE} 802.11ah - sub 1{GHz} {WLAN} for
  {IoT},'' [Online]. Available
  at:\url{https://upcommons.upc.edu/bitstream/handle/2117/182119/802.11ah_wi-fi_iot-5709.pdf?sequence=1&isAllowed=y}
  [Accessed on 16/01/2021].

\bibitem{garciarodriguez2020ieee}
A.~Garcia-Rodriguez, D.~Lopez-Perez, L.~Galati-Giordano, and G.~Geraci,
  ``{IEEE} 802.11be: {Wi-Fi} 7 strikes back,'' 2020,
  \url{https://arxiv.org/abs/2008.02815}.

\bibitem{Song_2014}
S.~Song and B.~Issac, ``Analysis of {WiFi} and {WiMAX} and wireless network
  coexistence,'' \emph{International Journal of Computer Networks \&
  Communications}, vol.~6, no.~6, p. 63–77, Nov 2014,
  \url{http://dx.doi.org/10.5121/ijcnc.2014.6605}.

\bibitem{WiMAX_uldatarate}
WiMAX\;Forum, ``Mobile {WiMAX} – part i: A technical overview and performance
  evaluation,'' Tech. Rep., 2006, [Online]. Available
  at:\url{https://wimaxforum.org/news/downloads/mobile_wimax_part1_overview_and_performance.pdf}
  [Accessed on 16/01/2021].

\bibitem{element14_zigbee}
A.~Tomar, ``Introduction to {Zibgbee} technology,'' 2016, [Online]. Available
  at:\url{https://www.cs.odu.edu/~cs752/papers/zigbee-001.pdf} [Accessed on
  16/01/2021].

\bibitem{8242220}
M.~{Hirose} and T.~{Kobayashi}, ``Measurements of ultra-wideband propagation
  within a small aircraft for replacing wire harnesses,'' in \emph{2017 IEEE
  Radio and Antenna Days of the Indian Ocean (RADIO)}, 2017, pp. 1--2,
  \url{https://doi.org/10.23919/RADIO.2017.8242220}.

\bibitem{8733537}
P.~{Sedlacek}, M.~{Slanina}, and P.~{Masek}, ``An overview of the {IEEE}
  802.15.4z standard its comparison and to the existing {UWB} standards,'' in
  \emph{2019 29th International Conference Radioelektronika
  (RADIOELEKTRONIKA)}, 2019, pp. 1--6,
  \url{https://doi.org/10.1109/RADIOELEK.2019.8733537}.

\bibitem{AeroMACS_specifications}
SESAR, ``{AeroMACS} system requirements document,'' Tech. Rep., 2010, [Online].
  Available
  at:\url{https://www.icao.int/safety/acp/ACPWGF/acp%20s%20test/AeroMACS_System_Requirements_Document_V1.0.pdf}
  [Accessed on 16/01/2021].

\bibitem{aeromacswimax}
WiMAX\;Forum, ``{AeroMACS}: A common platform for air traffic management
  applications, white paper,'' Senza Fili Consulting, Tech. Rep., 2015,
  [Online]. Available
  at:\url{http://files.wimaxforum.org/Document/Download/SenzaFili_AeroMACS_White_Paper_2015}
  [Accessed on 16/01/2021].

\bibitem{wilson10}
J.~D. {Wilson} and R.~J. {Kerczewski}, ``Interference analysis for an
  aeronautical mobile airport communications system,'' in \emph{2011 Aerospace
  Conference}, 2011, pp. 1--8, \url{https://doi.org/10.1109/AERO.2011.5747373}.

\bibitem{ietf_raw_ldacs}
N.~Maeurer, T.~Graeupl, and C.~Schmitt, ``L-band digital aeronautical
  communications system ({LDACS}),'' IETF, techreport
  draft-maeurer-raw-ldacs-00, Nov. 2019, [Online]. Available
  at:\url{https://tools.ietf.org/html/draft-maeurer-raw-ldacs-00#page-4}
  [Accessed on 16/01/2021].

\bibitem{ldacs_flighttest}
EUROCONTROL, ``{LDACS}, aviation’s future terrestrial datalink, takes a big
  step forward,'' 2019, [Online]. Available
  at:\url{https://www.eurocontrol.int/news/ldacs-aviations-future-terrestrial-datalink-takes-big-step-forward}
  [Accessed on 16/01/2021].

\bibitem{8496839}
A.~{Mancini}, R.~M. {Lebrón}, and J.~L. {Salazar}, ``The impact of a wet $s$
  -band radome on dual-polarized phased-array radar system performance,''
  \emph{IEEE Transactions on Antennas and Propagation}, vol.~67, no.~1, pp.
  207--220, 2019, \url{https://doi.org/10.1109/TAP.2018.2876733}.

\bibitem{8286975}
J.~{Zhao}, F.~{Gao}, Q.~{Wu}, S.~{Jin}, Y.~{Wu}, and W.~{Jia}, ``Beam tracking
  for {UAV} mounted satcom on-the-move with massive antenna array,'' \emph{IEEE
  Journal on Selected Areas in Communications}, vol.~36, no.~2, pp. 363--375,
  2018, \url{https://doi.org/10.1109/JSAC.2018.2804239}.

\bibitem{7482031}
J.~{Lee}, J.~{Lim}, and E.~{Kim}, ``Comparison between multimode-monopulse and
  step-tracking techniques for a {UAV} satellite terminal,'' in \emph{2016
  Wireless Telecommunications Symposium (WTS)}, 2016, pp. 1--5,
  \url{https://doi.org/10.1109/WTS.2016.7482031}.

\bibitem{7026641}
H.~{Tsuji}, T.~{Orikasa}, A.~{Miura}, M.~{Toyoshima}, and R.~{Miura},
  ``On-board ka-band satellite tracking antenna for unmanned aircraft system,''
  in \emph{2014 International Symposium on Antennas and Propagation Conference
  Proceedings}, 2014, pp. 283--284,
  \url{https://doi.org/10.1109/ISANP.2014.7026641}.

\bibitem{9130836}
J.~{Yu}, X.~{Liu}, Y.~{Gao}, and X.~{Shen}, ``{3D} channel tracking for
  {UAV}-satellite communications in space-air-ground integrated networks,''
  \emph{IEEE Journal on Selected Areas in Communications}, pp. 1--1, 2020,
  \url{https://doi.org/10.1109/JSAC.2020.3005490}.

\bibitem{8713061}
H.~{Tsuji}, T.~{Okura}, T.~{Kan}, T.~{Takahashi}, and M.~{Toyoshima}, ``Ka-band
  airborne array antenna development for satellite communications,'' in
  \emph{2018 21st International Symposium on Wireless Personal Multimedia
  Communications (WPMC)}, 2018, pp. 443--446,
  \url{https://doi.org/10.1109/WPMC.2018.8713061}.

\bibitem{Catalani_2020}
A.~Catalani, G.~Toso, P.~Angeletti, M.~Albertini, and P.~Russo, ``Development
  of enabling technologies for ku-band airborne satcom phased-arrays,''
  \emph{Electronics}, vol.~9, no.~3, p. 488, Mar 2020,
  \url{http://dx.doi.org/10.3390/electronics9030488}.

\bibitem{8761108}
S.~{Hofmann}, A.~E. {Garcia}, D.~{Schupke}, H.~E. {Gonzalez}, and F.~H.~P.
  {Fitzek}, ``Connectivity in the air: Throughput analysis of air-to-ground
  systems,'' in \emph{ICC 2019 - 2019 IEEE International Conference on
  Communications (ICC)}, 2019, pp. 1--6,
  \url{https://doi.org/10.1109/ICC.2019.8761108}.

\bibitem{5286353}
E.~H. {Fazli} and M.~{Werner}, ``View angle statistics of aircraft airborne
  antenna to {GEO} satellites,'' in \emph{2009 International Workshop on
  Satellite and Space Communications}, 2009, pp. 341--345,
  \url{https://doi.org/10.1109/IWSSC.2009.5286353}.

\bibitem{8690876}
D.~{Stolpmann}, C.~{Petersen}, V.~{Eichhorn}, and A.~{Timm-Giel}, ``Extending
  on-the-fly network coding by interleaving for avionic satellite links,'' in
  \emph{2018 IEEE 88th Vehicular Technology Conference (VTC-Fall)}, 2018, pp.
  1--5, \url{https://doi.org/10.1109/VTCFall.2018.8690876}.

\bibitem{8886850}
T.~{Jost}, W.~{Wang}, M.~{Schwinzerl}, F.~{Pérez-Fontán}, M.~{Schönhuber},
  N.~{Floury}, and S.~{Dimitrov}, ``Measurements and model for the
  satellite-to-aircraft channel in l-band,'' \emph{IET Microwaves, Antennas
  Propagation}, vol.~13, no.~13, pp. 2207--2215, 2019,
  \url{https://doi.org/10.1049/iet-map.2018.6166}.

\bibitem{8587496}
T.~{Kojima} and G.~{Muto}, ``A low inter-code interference walsh-hadamard code
  division multiplexing for helicopter satellite communications,'' in
  \emph{2018 International Conference on Advanced Technologies for
  Communications ({ATC})}, 2018, pp. 1--4,
  \url{https://doi.org/10.1109/ATC.2018.8587496}.

\bibitem{8587471}
D.~{Sato} and T.~{Kojima}, ``An accurate time diversity combining with a novel
  channel estimation for helicopter satellite communications,'' in \emph{2018
  International Conference on Advanced Technologies for Communications
  ({ATC})}, 2018, pp. 89--93, \url{https://doi.org/10.1109/ATC.2018.8587471}.

\bibitem{8924539}
G.~{Muto} and T.~{Kojima}, ``Inter-code interference reduction in
  walsh-hadamard code division multiplexing for helicopter satellite
  communications,'' in \emph{2019 International Conference on Advanced
  Technologies for Communications ({ATC})}, 2019, pp. 19--22,
  \url{https://doi.org/10.1109/ATC.2019.8924539}.

\bibitem{7601547}
A.~{Tato}, C.~{Mosquera}, and I.~{Gomez}, ``Link adaptation in mobile satellite
  links: Field trials results,'' in \emph{2016 8th Advanced Satellite
  Multimedia Systems Conference and the 14th Signal Processing for Space
  Communications Workshop (ASMS/SPSC)}, 2016, pp. 1--8,
  \url{https://doi.org/10.1109/ASMS-SPSC.2016.7601547}.

\bibitem{9391325}
H.~Suzuki, ``Haps to ground propagation model considering general terrestrial
  path,'' in \emph{2020 International Symposium on Antennas and Propagation
  (ISAP)}, 2021, pp. 683--684,
  \url{https://doi.org/10.23919/ISAP47053.2021.9391325}.

\bibitem{9391216}
A.~SATO, S.~KIMURA, H.~LIN, and H.~OMOTE, ``Propagation loss model of human
  body shielding in haps communications,'' in \emph{2020 International
  Symposium on Antennas and Propagation (ISAP)}, 2021, pp. 65--66,
  \url{https://doi.org/10.23919/ISAP47053.2021.9391216}.

\bibitem{9391285}
H.~Omote, S.~Kimura, H.-Y. Lin, and A.~Sato, ``Haps propagation loss model for
  urban and suburban environments,'' in \emph{2020 International Symposium on
  Antennas and Propagation (ISAP)}, 2021, pp. 681--682,
  \url{https://doi.org/10.23919/ISAP47053.2021.9391285}.

\bibitem{9378524}
Y.~Albagory, ``An adaptive bidirectional multibeam high-altitude platforms
  aeronautical telecommunication network using dual concentric conical
  arrays,'' \emph{IEEE Access}, vol.~9, pp. 43\,324--43\,338, 2021,
  \url{https://doi.org/10.1109/ACCESS.2021.3066091}.

\bibitem{6581252}
S.~{Temel} and I.~{Bekmezci}, ``On the performance of flying ad hoc networks
  ({FANET}s) utilizing near space high altitude platforms ({HAP}s),'' in
  \emph{2013 6th International Conference on Recent Advances in Space
  Technologies (RAST)}, 2013, pp. 461--465,
  \url{https://doi.org/10.1109/RAST.2013.6581252}.

\bibitem{faa_adsb}
FAA, ``{ADS-B} in strategy document - closing the loop on the flight deck,''
  Tech. Rep., Dec. 2019, version 1.1, [Online]. Available
  at:\url{https://www.faa.gov/nextgen/programs/adsb/media/ADS-B_In_Strategy.pdf}
  [Accessed on 16/01/2021].

\bibitem{aopa_adsb}
M.~Collins\;-\;AOPA, ``{ADS-B}: The nextgen big picture - are you ready to
  swim?'' 2018, [Online]. Available
  at:\url{https://www.aopa.org/news-and-media/all-news/2018/september/pilot/ads-b-nextgen-big-picture}
  [Accessed on 16/01/2021].

\bibitem{kim16}
Y.~Kim, J.~Y. Jo, and S.~Lee, ``A secure location verification method for
  {ADS-B},'' \emph{IEEE/AIAA 35th Digital Avionics Systems Conference (DASC)},
  pp. 1--10, 2016, \url{https://doi.org/10.1109/DASC.2016.7778003}.

\bibitem{icao_adsb}
ICAO, ``Report of {ADS-B} seminar and the fourteenth meeting of the automatic
  dependent surveillance – broadcast ({ADS-B}) study and implementation task
  force ({ADS-B} sitf/14),'' Tech. Rep., 2015, [Online]. Available
  at:\url{https://www.icao.int/APAC/Meetings/2015%20ADSBSITF14/Final%20Report%20of%20ADS-B%20SITF14.pdf}
  [Accessed on 16/01/2021].

\bibitem{sesar_adsb}
S.~D. Manager, ``{ADS-B} and other means of surveillance implementation
  status,'' Tech. Rep., 2018, [Online]. Available
  at:\url{https://ec.europa.eu/transport/sites/transport/files/20180515-sesar-ads-b-report.pdf}
  [Accessed on 16/01/2021].

\bibitem{uavionix_adsb}
uAvionix, ``Urban air mobility shaping the future of avionics,'' 2020,
  [Online]. Available at:\url{https://uavionix.com/uam/} [Accessed on
  16/01/2021].

\bibitem{mitre_adsb}
M.~{Guterres}, S.~{Jones}, G.~{Orrell}, and R.~{Strain}, ``{ADS-B} surveillance
  system performance with small {UAS} at low altitudes,'' in \emph{AIAA
  Information Systems-AIAA Infotech @ Aerospace}, jan 2017, pp. 1--15,
  \url{https://doi.org/10.2514/6.2017-1154}.

\bibitem{ADSB2020}
D.~of~Transportation Federal Aviation~Administration, ``Automatic dependent
  surveillance broadcast ({ADS-B}) out performance requirements to support
  {ATC} service,'' FAA, Tech. Rep. 14 CFR Part 91, 2015, [Online]. Available
  at:\url{https://www.federalregister.gov/documents/2010/05/28/2010-12645/automatic-dependent-surveillance-broadcast-ads-b-out-performance-requirements-to-support-air-traffic}
  [Accessed on 16/01/2021].

\bibitem{8714514}
J.~{Budroweit}, M.~P. {Jaksch}, and T.~{Delovski}, ``Design of a multi-channel
  {ADS-B} receiver for small satellite-based aircraft surveillance,'' in
  \emph{2019 IEEE Radio and Wireless Symposium (RWS)}, 2019, pp. 1--4,
  \url{https://doi.org/10.1109/RWS.2019.8714514}.

\bibitem{8712093}
R.~N. {Pahlevy}, A.~{Dwi Prasetyo}, and {Edwar}, ``Nanosatellite {ADS-B}
  receiver prototype for commercial aircraft detection,'' in \emph{2018
  International Conference on Control, Electronics, Renewable Energy and
  Communications (ICCEREC)}, 2018, pp. 6--12,
  \url{https://doi.org/10.1109/ICCEREC.2018.8712093}.

\bibitem{inmarsat_satfrequencies}
P.~Revillion\;-\;Euroconsult, ``Fundamentals and dynamics of the satellite
  communications business - {Euroconsult} for {Inmarsat} capital markets day,''
  2016, [Online]. Available
  at:\url{https://pdfslide.net/documents/fundamentals-and-dynamics-of-the-satellite-communications-capital-markets-2016.html}
  [Accessed on 16/01/2021].

\bibitem{inmarsat_sband_satellite}
Inmarsat, ``{Inmarsat} s-band services,'' 03 2015, [Online]. Available
  at:\url{https://web.archive.org/web/20141222012307/https://www.inmarsat.com/wp-content/uploads/2014/06/Inmarsat-S-band-services-June-2014.pdf}
  [Accessed on 16/01/2021].

\bibitem{9124728}
N.~{Cassiau}, G.~{Noh}, S.~{Jaeckel}, L.~{Raschkowski}, J.~{Houssin},
  L.~{Combelles}, M.~{Thary}, J.~{Kim}, J.~{Dore}, and M.~{Laugeois},
  ``Satellite and terrestrial multi-connectivity for {5G}: Making spectrum
  sharing possible,'' in \emph{2020 IEEE Wireless Communications and Networking
  Conference Workshops (WCNCW)}, 2020, pp. 1--6,
  \url{https://doi.org/10.1109/WCNCW48565.2020.9124728}.

\bibitem{10.1117/12.2536091}
M.~A. Piqueras, J.~Martí, S.~Delgado, R.~Singh, D.~Beeder, R.~Parish,
  G.~Turgeon, A.~Castells, and L.~Roux, ``A flight demonstration photonic
  payload for up to q/v-band implemented in a satellite ka-band hosted payload
  aimed at broadband high throughput satellites,'' in \emph{International
  Conference on Space Optics — ICSO 2018}, Z.~Sodnik, N.~Karafolas, and
  B.~Cugny, Eds., vol. 11180, International Society for Optics and
  Photonics.\hskip 1em plus 0.5em minus 0.4em\relax SPIE, 2019, pp. 1796 --
  1804, \url{https://doi.org/10.1117/12.2536091}.

\bibitem{hughes_qvband}
Y.~Antia, S.~Morrar, and D.~Roos, ``Jupiter high throughput satellite system
  — 500 gbps from space,'' in \emph{Microwave Journal}, 2019, [Online].
  Available
  at:\url{https://www.hughes.com/sites/hughes.com/files/2019-10/Microwave%20Journal%20HTS%20article_191021.pdf}
  [Accessed on 16/01/2021].

\bibitem{inmarsat_powerconsumption}
Inmarsat, ``M2m terminals - a quick reference guide,'' 2015, [Online].
  Available
  at:\url{https://web.archive.org/web/20170629015134/https://www.inmarsat.com/wp-content/uploads/2013/10/Inmarsat_M2M_terminals_September_2015_EN_LowRes.pdf.pdf}
  [Accessed on 16/01/2021].

\bibitem{4062836}
P.~K. {Chowdhury}, M.~{Atiquzzaman}, and W.~{Ivancic}, ``Handover schemes in
  satellite networks: State-of-the-art and future research directions,''
  \emph{IEEE Communications Surveys \& Tutorials}, vol.~8, no.~4, pp. 2--14,
  2006, \url{https://doi.org/10.1109/COMST.2006.283818}.

\bibitem{satellite_bands}
L.~A. {Belov}, S.~M. {Smolskiy}, and V.~N. {Kochemasov}, \emph{Handbook of
  {RF}, Microwave, and Milli-Meter Wave Components}.\hskip 1em plus 0.5em minus
  0.4em\relax Boston/London: Artech House, 2012, iSBN: 978-1-60807-209-5.

\bibitem{hps_qbandantennasize}
High\;Performance\;Space\;Structure\;Systems\;GmbH, ``Ka-band \& q/v-band
  antennas for satcom and science applications,'' 2017, [Online]. Available
  at:\url{http://www.hps-gmbh.com/wp-content/uploads/2017/05/HPS_Ka-Band-QV-Band-Antennas-Flyer.pdf}
  [Accessed on 16/01/2021].

\bibitem{ASatTELEBBS}
V.~Point, ``Analysis of satellite-based telecommunications and broadband
  services,'' November 2013, [Online]. Available
  at:\url{https://ecfsapi.fcc.gov/file/7520956711.pdf} [Accessed on
  16/01/2021].

\bibitem{inmarsat_servicescomparison}
Inmarsat, ``Enterprise services comparison - a quick reference guide,''
  [Online]. Available
  at:\url{http://www.pacificrim.com.au/media/custom/upload/File-1413073790.pdf}
  [Accessed on 16/01/2021].

\bibitem{Curzi_2020}
G.~Curzi, D.~Modenini, and P.~Tortora, ``Large constellations of small
  satellites: A survey of near future challenges and missions,''
  \emph{Aerospace}, vol.~7, no.~9, p. 133, Sep 2020,
  \url{http://dx.doi.org/10.3390/aerospace7090133}.

\bibitem{o3bnetworks}
O3b, ``{O3b} networks - our technology,'' [Online]. Available
  at:\url{https://web.archive.org/web/20150317220715/http://www.o3bnetworks.com/o3b-advantage/our-technology}
  [Accessed on 16/01/2021].

\bibitem{o3bnetworks2}
------, ``{O3b} maritime - first-class connectivity at sea,'' [Online].
  Available
  at:\url{https://web.archive.org/web/20121024025342/http://o3bnetworks.com/media/60982/o3b_maritime_brochure.pdf}
  [Accessed on 16/01/2021].

\bibitem{DELPORTILLO2019123}
N.~{del Portillo}, B.~G. Cameron, and E.~F. Crawley, ``A technical comparison
  of three low earth orbit satellite constellation systems to provide global
  broadband,'' \emph{Acta Astronautica}, vol. 159, pp. 123 -- 135, 2019,
  \url{https://doi.org/10.1016/j.actaastro.2019.03.040}.

\bibitem{viasat_geocapacity}
ViaSat, ``Viasat {VS}-2,'' 2017, [Online]. Available
  at:\url{https://www.viasat.com/space-innovation/satellite-fleet/viasat-2/}
  [Accessed on 16/01/2021].

\bibitem{ses_meo}
J.~Parkinson, ``Our connectivity - {MEO} evolution,'' 2019, [Online]. Available
  at:\url{https://www.ioag.org/Public%20Documents/07_SES.pdf} [Accessed on
  16/01/2021].

\bibitem{geo_availability}
Inmarsat, ``Global reach global impact - inmarsat plc annual report and
  accounts 2015,'' 2016, [Online]. Available
  at:\url{https://www.annualreports.com/HostedData/AnnualReportArchive/i/LSE_ISAT_2015.pdf}
  [Accessed on 16/01/2021].

\bibitem{meo_datarate}
S.~Malloy\;-\;O3b\;Networks, ``Broadband opportunities council comment,'' 2015,
  [Online]. Available
  at:\url{https://www.ntia.doc.gov/files/ntia/o3b_networks_boc.pdf} [Accessed
  on 16/01/2021].

\bibitem{8933127}
S.~{Xia}, Q.~{Jiang}, C.~{Zou}, and G.~{Li}, ``Beam coverage comparison of
  {LEO} satellite systems based on user diversification,'' \emph{IEEE Access},
  vol.~7, pp. 181\,656--181\,667, 2019,
  \url{https://doi.org/10.1109/ACCESS.2019.2959824}.

\bibitem{8700141}
Y.~{Su}, Y.~{Liu}, Y.~{Zhou}, J.~{Yuan}, H.~{Cao}, and J.~{Shi}, ``Broadband
  {LEO} satellite communications: Architectures and key technologies,''
  \emph{IEEE Wireless Communications}, vol.~26, no.~2, pp. 55--61, 2019,
  \url{https://doi.org/10.1109/MWC.2019.1800299}.

\bibitem{iridiumdataplan}
Bluecosmo, ``Iridium go! unlimited data plan,'' [Online]. Available
  at:\url{https://www.bluecosmo.com/iridium-go-unlimited-data-plan.html#:~:text=the%20Iridium%20Unlimited%20Monthly%20Service,or%20worry%20about%20expiry%20dates.}
  [Accessed on 16/01/2021].

\bibitem{ISNOV}
Iridium, ``Iridium satellite network overview,'' 2014, [Online]. Available
  at:\url{https://www.icao.int/Meetings/GTM/Documents/Iridium.pdf} [Accessed on
  16/01/2021].

\bibitem{LEOSATOV}
C.~Redding, ``Overview of {LEO} satellite systems,'' in \emph{1999
  International Symposium on Advanced Radio Technologies}, 1999,
  \url{https://www.its.bldrdoc.gov/media/30335/red_s.pdf}.

\bibitem{spacexnongeo}
M.~A. FCC, ``Spacex non-geostationary satellite system,'' 2017, [Online].
  Available
  at:\url{https://licensing.fcc.gov/myibfs/download.do?attachment_key=1252848}
  [Accessed on 16/01/2021].

\bibitem{onewebnongeo}
R.~J. Barnett, ``Oneweb non-geostationary satellite system ({LEO}),'' January
  2018, [Online]. Available
  at:\url{http://licensing.fcc.gov/myibfs/download.do?attachment_key=1357110}
  [Accessed on 16/01/2021].

\bibitem{leo_lifeexpectancy}
M.~A. Hallex and T.~S. Cottom, ``Proliferated commercial satellite
  constellations implications for national security,'' National Defense
  University Press, Tech. Rep., 2020, [Online]. Available
  at:\url{https://ndupress.ndu.edu/Portals/68/Documents/jfq/jfq-97/jfq-97_20-29_Hallex-Cottom.pdf?ver=2020-03-31-130614-940}
  [Accessed on 16/01/2021].

\bibitem{CHOLEOSATMNET}
H.~Uzunalioğlu, W.~Yen, and I.~F. Akyildiz, ``A connection handover protocol
  for {LEO} satellite {ATM} networks,'' in \emph{3rd Annual ACM/IEEE MobiCom
  '97}, September 1997, pp. 204--214,
  \url{https://doi.org/10.1145/262116.262148}.

\bibitem{2kuGOGO}
Gogo, ``Gogo 2ku global connectivity solution,'' 2019, [Online]. Available
  at:\url{https://www.gogoair.com/learning-center/gogo-2ku-brochure/?download=true}
  [Accessed on 16/01/2021].

\bibitem{ElbertABCA2KU}
B.~E. GoGo, ``Aeronautical broadband for commercial aviation: Evaluating the
  2ku solution,'' 2014, [Online]. Available
  at:\url{https://concourse.gogoair.com/whitepaper-aeronautical-broadband-commercial-aviation-evaluating-2ku-solution/}
  [Accessed on 16/01/2021].

\bibitem{HAPslides}
A.~Hummelholm, ``High altitude platforms for wireless communications
  ({HAP}s),'' 2006, postgraduate Seminar on Wideband Radio Communications.
  [Online]. Available
  at:\url{http://www.comlab.hut.fi/opetus/4210/presentations/17_haps.pdf}
  [Accessed on 16/01/2021].

\bibitem{8439347}
J.~{Liu} and D.~W. {Matolak}, ``Worst month tropospheric attenuation
  variability analysis: Itu model vs. rain gauge data for air-satellite
  links,'' in \emph{2018 11th Global Symposium on Millimeter Waves (GSMM)},
  2018, pp. 1--5, \url{https://doi.org/10.1109/GSMM.2018.8439347}.

\bibitem{1423332}
S.~{Karapantazis} and F.~{Pavlidou}, ``Broadband communications via
  high-altitude platforms: A survey,'' \emph{IEEE Communications Surveys
  Tutorials}, vol.~7, no.~1, pp. 2--31, 2005,
  \url{https://doi.org/10.1109/COMST.2005.1423332}.

\bibitem{ITU_F1500}
ITU, ``Preferred characteristics of systems in the fixed service using high
  altitude platforms operating in the bands 47.2-47.5 {GHz} and 47.9-48.2
  {GHz},'' Tech. Rep., 2000, [Online]. Available
  at:\url{https://www.itu.int/dms_pubrec/itu-r/rec/f/R-REC-F.1500-0-200005-I%21%21PDF-E.pdf}
  [Accessed on 16/01/2021].

\bibitem{kurt2020vision}
G.~K. Kurt, M.~G. Khoshkholgh, S.~Alfattani, A.~Ibrahim, T.~S.~J. Darwish,
  M.~S. Alam, H.~Yanikomeroglu, and A.~Yongacoglu, ``A vision and framework for
  the high altitude platform station (haps) networks of the future,''
  \emph{IEEE Communications Surveys Tutorials}, pp. 1--1, 2021.

\bibitem{ITU_WRC19}
ITUPublications, ``World radiocommunication conference 2019 (wrc-19) final
  acts,'' ITU, Tech. Rep., 2019, [Online]. Available
  at:\url{https://www.itu.int/dms_pub/itu-r/opb/act/R-ACT-WRC.14-2019-PDF-E.pdf}
  [Accessed on 16/01/2021].

\bibitem{HAPPSTT}
F.~A. d'Oliveira, F.~C.~L. De~Melo, and T.~C. Devezas,
  ``\BIBforeignlanguage{en}{High-altitude platforms - present situation and
  technology trends},'' \emph{\BIBforeignlanguage{en}{Journal of Aerospace
  Technology and Management}}, vol.~8, pp. 249 -- 262, 09 2016, [Online].
  Available
  at:\url{http://www.scielo.br/scielo.php?script=sci_arttext&pid=S2175-91462016000300249&nrm=iso}
  [Accessed on 16/01/2021].

\bibitem{GRENA2013215}
R.~Grena, ``Solar balloons as mixed solar–wind power systems,'' \emph{Solar
  Energy}, vol.~88, pp. 215 -- 226, 2013,
  \url{https://doi.org/10.1016/j.solener.2012.11.021}.

\bibitem{googleloon}
S.~Kaur and S.~Randhawa, ``Google loon: Balloon-powered internet for
  everyone,'' in \emph{AIP Conference Proceedings}, vol. 2034, 10 2018, p.
  020006, \url{https://doi.org/10.1063/1.5067349}.

\bibitem{ARUM2020232}
S.~C. Arum, D.~Grace, and P.~D. Mitchell, ``A review of wireless communication
  using high-altitude platforms for extended coverage and capacity,''
  \emph{Computer Communications}, vol. 157, pp. 232 -- 256, 2020,
  \url{https://doi.org/10.1016/j.comcom.2020.04.020}.

\bibitem{airbus_zephyr}
Airbus, ``Zephyr - focus of an aircraft. endurance of a satellite,'' [Online].
  Available
  at:\url{https://www.airbus.com/content/dam/corporate-topics/publications/brochures/0612_17_zephyr_datasheet_e_horizontal_a4_lowres.pdf}
  [Accessed on 16/01/2021].

\bibitem{EHELINETBCAPP03}
D.~Grace, J.~Thornton, G.~White, C.~Spillard, and D.~A.~J. Pearce, ``The
  european helinet broadband communications application - an update on
  progress,'' in \emph{4th Stratospheric Platform Systems Workshop}, Tokyo,
  Japan, February 2003, [Online]. Available
  at:\url{https://www.semanticscholar.org/paper/The-European-HeliNet-Broadband-Communications-%E2%80%93-An-Grace-Thornton/38e3784f67462efa5bd9d20e0cf20cb97cf5e22b?p2df}
  [Accessed on 16/01/2021].

\bibitem{ITURIULTxFSHAPs}
ITU, ``Impact of uplink transmission in the fixed service using high altitude
  platform stations on the earth exploration-satellite service (passive) in the
  31.3-31.8 {GHz} band,'' Tech. Rep. F.1570, April 2010, [Online]. Available
  at:\url{https://www.itu.int/rec/R-REC-F.1570/en} [Accessed on 16/01/2021].

\bibitem{HAP_book}
D.~Grace and M.~Mohorcic, \emph{Broadband Communications via High Altitude
  Platforms}.\hskip 1em plus 0.5em minus 0.4em\relax Wiley Publishing, 2010,
  iSBN: 0470697164, 9780470697160.

\bibitem{WCSSPRDPTELEBSHAP01}
R.~{Miura} and M.~{Oodo}, ``Wireless communications system using stratospheric
  platforms: R and d program on telecom and broadcasting system using high
  altitude platform stations,'' \emph{Journal of the Communication Research
  Laboratory}, vol.~48, no.~4, pp. 33--48, Dec. 2001,
  \url{https://ui.adsabs.harvard.edu/abs/2001JCRL...48...33M}.

\bibitem{HAPSWCOMM01}
T.~C. Tozer and D.~Grace, ``High-altitude platforms for wireless
  communications,'' \emph{IEEE Electronics \& Communications Engineering
  Journal}, vol.~13, no.~3, p. 127–137, June 2001,
  \url{https://doi.org/10.1049/ecej:20010303}.

\bibitem{CFHAPEUROP00}
D.~Grace, T.~C. Tozer, and N.~E. Daly, ``Communications from high altitude
  platforms a european perspective,'' in \emph{2nd Stratospheric Platforms
  System Workshop}, Tokyo, Japan, September 2000.

\bibitem{CCAP01}
B.~El-Jabu and S.~Steele, ``Cellular communications using aerial platforms,''
  \emph{IEEE Transactions on Vehicular Technology}, vol.~50, no.~3, p.
  686–700, May 2001, \url{https://doi.org/10.1109/25.933305}.

\bibitem{9178753}
Y.~{Shibata}, N.~{Kanazawa}, M.~{Konishi}, K.~{Hoshino}, Y.~{Ohta}, and
  A.~{Nagate}, ``System design of gigabit {HAPS} mobile communications,''
  \emph{IEEE Access}, pp. 1--1, 2020,
  \url{https://doi.org/10.1109/ACCESS.2020.3019820}.

\bibitem{EWCSHAPCWTHC97}
G.~M. {Djuknic}, J.~{Freidenfelds}, and Y.~{Okunev}, ``Establishing wireless
  communications services via high-altitude aeronautical platforms: A concept
  whose time has come?'' \emph{IEEE Communications Magazine}, vol.~35, no.~9,
  pp. 128--135, 1997, \url{https://doi.org/10.1109/35.620534}.

\bibitem{IHAPTUMTSNIACD03}
E.~Falletti, M.~Mondin, F.~Dovis, and D.~Grace, ``Integration of a {HAP} within
  a terrestrial {UMTS} network: Interference analysis and cell dimensioning,''
  \emph{Wireless Personal Communications}, vol.~24, no.~2, p. 291–325,
  February 2003, \url{https://doi.org/10.1023/A:1022502924226}.

\bibitem{HALONet00}
M.~J. Colella, J.~N. Martin, and F.~Akyildiz, ``The {HALO} network,''
  \emph{IEEE Communications Magazine}, vol.~38, no.~6, p. 142–148, June 2000,
  \url{https://doi.org/10.1109/35.846086}.

\bibitem{CPHAPNOmmWB00}
T.~Tozer and A.~Burr, ``Communications performance of high altitude platform
  networks operating in the mm-wave bands,'' in \emph{3rd Int’l. Airship
  Convention and Exhibition}, Friedrichshafen, Germany, July 2000, [Online].
  Available
  at:\url{https://www.researchgate.net/publication/215650825_Communications_Performance_of_High_Altitude_Platforms_Networks_operating_in_the_mm-Wave_Bands}
  [Accessed on 16/01/2021.].

\bibitem{PMCSHAP01}
D.~Grace, N.~E. Daly, T.~C. Tozer, A.~G. Burr, and D.~A.~J. Pearce, ``Providing
  multimedia communications services from high altitude platforms,''
  \emph{International Journal of Satellite Communications}, vol.~19, no.~6, pp.
  559--580, 2001, \url{https://doi.org/10.1002/sat.698}.

\bibitem{8891546}
K.~{Hoshino}, S.~{Sudo}, and Y.~{Ohta}, ``A study on antenna beamforming method
  considering movement of solar plane in {HAPS} system,'' in \emph{2019 IEEE
  90th Vehicular Technology Conference (VTC2019-Fall)}, 2019, pp. 1--5,
  \url{https://doi.org/10.1109/VTCFall.2019.8891546}.

\bibitem{PFTPTBUHAPS01}
R.~Miura and M.~Suzuki, ``Preliminary flight test program on telecom and
  broadcasting using high altitude platform stations,'' \emph{Wireless Personal
  Communications}, vol.~24, no.~2, pp. 341--361, 2003,
  \url{https://doi.org/10.1023/A:1022507025134}.

\bibitem{WNBHAPPINCS02}
D.~{Avagnina}, F.~{Dovis}, A.~{Ghiglione}, and P.~{Mulassano}, ``Wireless
  networks based on high-altitude platforms for the provision of integrated
  navigation/communication services,'' \emph{IEEE Communications Magazine},
  vol.~40, no.~2, pp. 119--125, 2002, \url{https://doi.org/10.1109/35.983918}.

\bibitem{8746381}
M.~Q. {Vu}, N.~T. {Dang}, and A.~T. {Pham}, ``{HAP}-aided relaying satellite
  {FSO} {QKD} systems for secure vehicular networks,'' in \emph{2019 IEEE 89th
  Vehicular Technology Conference (VTC2019-Spring)}, 2019, pp. 1--6,
  \url{https://doi.org/10.1109/VTCSpring.2019.8746381}.

\bibitem{TDWCSSPK02}
J.~{Park}, B.~{Ku}, Y.~{Kim}, and D.~{Ahn}, ``Technology development for
  wireless communications system using stratospheric platform in {Korea},'' in
  \emph{13th IEEE International Symposium on Personal, Indoor and Mobile Radio
  Communications}, vol.~4, 2002, pp. 1577--1581 Vol.4,
  \url{https://doi.org/10.1109/PIMRC.2002.1045444}.

\bibitem{9356529}
M.~S. {Alam}, G.~K. {Kurt}, H.~{Yanikomeroglu}, P.~{Zhu}, and N.~D. {Dao},
  ``High altitude platform station based super macro base station
  constellations,'' \emph{IEEE Communications Magazine}, vol.~59, no.~1, pp.
  103--109, 2021.

\bibitem{O1}
J.~Yan, C.~Hua, C.~Chen, and X.~Guan, ``The capacity of aeronautical ad-hoc
  networks,'' \emph{Wireless Networks}, vol.~20, no.~7, p. 2123–2130, Oct.
  2014, \url{https://doi.org/10.1007/s11276-014-0737-7}.

\bibitem{FANETS}
I.~Bekmezci, O.~K. Sahingoz, and S.~Temel, ``Flying ad-hoc networks ({FANET}s):
  A survey,'' \emph{Ad Hoc Networks}, vol.~11, no.~3, pp. 1254--1270, 2013,
  \url{https://doi.org/10.1016/j.adhoc.2012.12.004}.

\bibitem{O4-14}
E.~Yanmaz, C.~Costanzo, C.~Bettstetter, and W.~Elmenreich, ``A discrete
  stochastic process for coverage analysis of autonomous {UAV} networks,'' in
  \emph{IEEE Globecom Workshops}, 2010,
  \url{https://doi.org/10.1109/GLOCOMW.2010.5700247}.

\bibitem{O4-13}
E.~Yanmaz, R.~Kuschnig, and C.~Bettstetter, ``Channel measurements over
  802.11a-based {UAV}-to-ground links,'' in \emph{GLOBECOM Wi-UAV Workshop},
  2011, p. 1280–1284, \url{https://doi.org/10.1109/GLOCOMW.2011.6162389}.

\bibitem{O2}
E.~Sakhaee and A.~Jamalipour, ``The global in-flight internet,'' \emph{IEEE
  Journal on Selected Areas in Communications}, vol.~24, no.~9, pp. 1748--1757,
  September 2006, \url{https://doi.org/10.1109/JSAC.2006.875122}.

\bibitem{9115898}
M.~M. {Azari}, G.~{Geraci}, A.~{Garcia-Rodriguez}, and S.~{Pollin},
  ``{UAV-to-UAV} communications in cellular networks,'' \emph{IEEE Transactions
  on Wireless Communications}, pp. 1--1, 2020,
  \url{https://doi.org/10.1109/TWC.2020.3000303}.

\bibitem{8904448}
------, ``Cellular {UAV-to-UAV} communications,'' in \emph{2019 IEEE 30th
  Annual International Symposium on Personal, Indoor and Mobile Radio
  Communications (PIMRC)}, 2019, pp. 1--7,
  \url{https://doi.org/10.1109/PIMRC.2019.8904448}.

\bibitem{9055054}
S.~{Zhang}, H.~{Zhang}, and L.~{Song}, ``Beyond {D2D}: Full dimension
  {UAV}-to-everything communications in {6G},'' \emph{IEEE Transactions on
  Vehicular Technology}, vol.~69, no.~6, pp. 6592--6602, 2020,
  \url{https://doi.org/10.1109/TVT.2020.2984624}.

\bibitem{8735159}
J.~{Lieb} and G.~{Peklar}, ``Evaluation of an unique communication interface
  system {D2X} for {UAVs} intercommunicating with air and ground {UTM} users,''
  in \emph{2019 Integrated Communications, Navigation and Surveillance
  Conference (ICNS)}, 2019, pp. 1--9,
  \url{https://doi.org/10.1109/ICNSURV.2019.8735159}.

\bibitem{9198731}
H.~A. {Win}, R.~{Dantu}, and P.~{Shrestha}, ``On-road performance evaluation of
  {IEEE} 802.11p/{WAVE} in {BSM} signalling and video streaming using {WSMP},''
  in \emph{2020 Wireless Telecommunications Symposium (WTS)}, 2020, pp. 1--8,
  \url{https://doi.org/10.1109/WTS48268.2020.9198731}.

\bibitem{8766809}
T.~{Liu}, Z.~{Zhang}, H.~{Jiang}, Y.~{Qian}, K.~{Liu}, J.~{Dang}, and L.~{Wu},
  ``Measurement-based characterization and modeling for low-altitude {UAV}
  air-to-air channels,'' \emph{IEEE Access}, vol.~7, pp. 98\,832--98\,840,
  2019, \url{https://doi.org/10.1109/ACCESS.2019.2929955}.

\bibitem{7414180}
N.~{Goddemeier} and C.~{Wietfeld}, ``Investigation of air-to-air channel
  characteristics and a {UAV} specific extension to the rice model,'' in
  \emph{2015 IEEE Globecom Workshops (GC Wkshps)}, 2015, pp. 1--5,
  \url{https://doi.org/10.1109/GLOCOMW.2015.7414180}.

\bibitem{10.1145/3412060.3418431}
M.~Polese, L.~Bertizzolo, L.~Bonati, A.~Gosain, and T.~Melodia, ``An
  experimental mmwave channel model for {UAV-to-UAV} communications,'' in
  \emph{Proceedings of the 4th ACM Workshop on Millimeter-Wave Networks and
  Sensing Systems}, ser. mmNets'20.\hskip 1em plus 0.5em minus 0.4em\relax New
  York, NY, USA: Association for Computing Machinery, 2020,
  \url{https://doi.org/10.1145/3412060.3418431}.

\bibitem{9133075}
R.~{Molina-Masegosa}, J.~{Gozalvez}, and M.~{Sepulcre}, ``Comparison of {IEEE}
  802.11p and {LTE-V2X}: An evaluation with periodic and aperiodic messages of
  constant and variable size,'' \emph{IEEE Access}, vol.~8, pp.
  121\,526--121\,548, 2020, \url{https://doi.org/10.1109/ACCESS.2020.3007115}.

\bibitem{8891313}
W.~{Anwar}, N.~{Franchi}, and G.~{Fettweis}, ``Physical layer evaluation of
  {V2X} communications technologies: {5G} {NR-V2X}, {LTE-V2X}, {IEEE} 802.11bd,
  and {IEEE} 802.11p,'' in \emph{2019 IEEE 90th Vehicular Technology Conference
  (VTC2019-Fall)}, 2019, pp. 1--7,
  \url{https://doi.org/10.1109/VTCFall.2019.8891313}.

\bibitem{mbb_coverage}
Mobilenetworkguide, ``Mobile base stations,'' 2020, [Online]. Available
  at:\url{https://mobilenetworkguide.com.au/mobile_base_stations.html}
  [Accessed on 16/01/2021].

\bibitem{nbiot_ratelatency}
Ministry\;of\;Digital\;Economy\;and\;Society, Depa, Huawei, and Quectel,
  ``Thailand {IoT} industry white paper - {IoT} technologies, ecosystem and
  application development guide,'' Tech. Rep., 2019, [Online]. Available
  at:\url{https://www.huawei.com/minisite/iot/pdf/thailand_iot_white_paper.pdf}
  [Accessed on 16/01/2021].

\bibitem{gsma_ltem}
GSMA, ``{LTE-M} deployment guide to basic feature set requirements,'' Tech.
  Rep., 2019, [Online]. Available
  at:\url{https://www.gsma.com/IoT/wp-content/uploads/2019/08/201906-GSMA-LTE-M-Deployment-Guide-v3.pdf}
  [Accessed on 16/01/2021].

\bibitem{huawei_wifi6}
Huawei, ``Huawei {Wi-Fi} 6 (802.11ax) technology white paper,'' Tech. Rep.,
  2019, [Online]. Available
  at:\url{https://www.messe.de/apollo/hannover_messe_2020/obs/Binary/A1032221/Huawei%20Wi-Fi%206%20%28802.11ax%29%20Technology%20White%20Paper.pdf}
  [Accessed on 16/01/2021].

\bibitem{oughton2020revisiting}
E.~J. Oughton, W.~Lehr, K.~Katsaros, I.~Selinis, D.~Bubley, and J.~Kusuma,
  ``Revisiting wireless internet connectivity: {5G} vs {Wi-Fi} 6,'' 2020,
  \url{https://arxiv.org/abs/2010.11601}.

\bibitem{wifi_halow_datarate}
IMEC, ``{IEEE}802.11ah {Wi-Fi} {HaLow} radio in {TSMC} 40nm {CMOS},'' 2016,
  [Online]. Available
  at:\url{https://www.ieee802.org/1/files/public/docs2016/liaison-WFA-Wi-Fi-HaLow-0516.pdf}
  [Accessed on 16/01/2021].

\bibitem{5654506}
J.~M. {Westall} and J.~J. {Martin}, ``Performance characteristics of an
  operational {WiMAX} network,'' \emph{IEEE Transactions on Mobile Computing},
  vol.~10, no.~7, pp. 941--953, 2011,
  \url{https://doi.org/10.1109/TMC.2010.226}.

\bibitem{itu_lora}
S.~Tabbane\;-\;ITU, ``{IoT} systems overview,'' 2019, [Online]. Available
  at:\url{https://www.itu.int/en/ITU-D/Regional-Presence/AsiaPacific/SiteAssets/Pages/Events/2019/ITU-ASP-CoE-Training-on-/IoT%20systems%20overview.pdf}
  [Accessed on 16/01/2021].

\bibitem{orangelabs_lora}
N.~Ducrot, D.~Ray, A.~Saadani, O.~Hersent, G.~Pop, G.~R. Orange, and Actility,
  ``{LoRa} device developer guide - orange connected objects \& partnerships,''
  2016, [Online]. Available
  at:\url{https://developer.orange.com/od-uploads/LoRa-Device-Developer-Guide-Orange.pdf}
  [Accessed on 16/01/2021].

\bibitem{ldacs_latency}
N.~Mäurer, M.~Schnell, T.~Gräupl, and C.~Schmitt, ``L-band digital
  aeronautical communications system ({LDACS}) future communications in
  aviation,'' 2019, [Online]. Available
  at:\url{https://www.ietf.org/proceedings/104/slides/slides-104-paw-08-ldacs-00}
  [Accessed on 16/01/2021].

\bibitem{LDACS_specifications}
SESAR, ``Updated {LDACS}1 system specification,'' Tech. Rep., 2011, [Online].
  Available
  at:\url{http://www.ldacs.com/wp-content/uploads/2014/02/LDACS1-Updated-Specification-Proposal-D2-Deliverable.pdf}
  [Accessed on 16/01/2021].

\bibitem{ETSI_122185}
ETSI, ``{LTE}; service requirements for {V2X} services ({3GPP TS} 22.185
  version 14.3.0 release 14),'' Tech. Rep. TS 122.185, 2017, [Online].
  Available
  at:\url{https://www.etsi.org/deliver/etsi_ts/122100_122199/122185/14.03.00_60/ts_122185v140300p.pdf}
  [Accessed on 16/01/2021].

\bibitem{v2x_80211p_reliability}
5GAA, ``An assessment of {LTE-V2X} ({PC5}) and 802.11p direct communications
  technologies for improved road safety in the {EU},'' 2017, [Online].
  Available
  at:\url{https://5gaa.org/wp-content/uploads/2017/12/5GAA-Road-safety-FINAL2017-12-05.pdf}
  [Accessed on 16/01/2021].

\bibitem{v2x_coverage}
J.~Springer\;-\;5GAA, ``Connectivity standards in the automotive industry,''
  2019, [Online]. Available
  at:\url{https://www.itu.int/en/ITU-T/extcoop/cits/Documents/Meeting-20191028-e-meeting/18_5GAA_ITS-status-update.pdf}
  [Accessed on 16/01/2021].

\bibitem{80211p_reliability}
L.~Miao, K.~Djouani, B.~Van~Wyk, and Y.~Hamam, ``Evaluation and enhancement of
  {IEEE} 802.11p standard: A survey,'' \emph{Mobile Computing}, vol.~1, pp.
  15--30, 11 2012, \url{https://archive.org/details/MC10022}.

\bibitem{7222906}
U.~{Hernandez-Jayo} and I.~{De-la-Iglesia}, ``Reliability analysis of {IEEE}
  802.11p wireless communication and vehicle safety applications,'' in
  \emph{2013 International Conference on Wireless Information Networks and
  Systems (WINSYS)}, 2013, pp. 1--8.

\bibitem{9045408}
J.~{Lin}, W.~{Cai}, S.~{Zhang}, X.~{Fan}, S.~{Guo}, and J.~{Dai}, ``A survey of
  flying ad-hoc networks: Characteristics and challenges,'' in \emph{2018
  Eighth International Conference on Instrumentation Measurement, Computer,
  Communication and Control (IMCCC)}, 2018, pp. 766--771,
  \url{https://doi.org/10.1109/IMCCC.2018.00165}.

\bibitem{O2-6}
T.~Avakian, ``Here's how many planes are in the air at any moment,'' [Online].
  Available
  at:\url{https://www.travelandleisure.com/airlines-airports/number-of-planes-in-air}
  [Accessed on 16/01/2021].

\bibitem{ICAO_airplaneseparation}
ICAO, ``Procedures for air navigation services - air traffic management,
  sixteenth edition,'' Tech. Rep. Doc 4444, Nov. 2016, [Online]. Available
  at:\url{https://ops.group/blog/wp-content/uploads/2017/03/ICAO-Doc4444-Pans-Atm-16thEdition-2016-OPSGROUP.pdf}
  [Accessed on 16/01/2021].

\bibitem{8473483}
M.~{Vondra}, M.~{Ozger}, D.~{Schupke}, and C.~{Cavdar}, ``Integration of
  satellite and aerial communications for heterogeneous flying vehicles,''
  \emph{IEEE Network}, vol.~32, no.~5, pp. 62--69, 2018,
  \url{https://doi.org/10.1109/MNET.2018.1800055}.

\bibitem{O7}
D.~Medina, F.~Hoffmann, F.~Rossetto, and C.~H. Rokitansky, ``A geographic
  routing strategy for north atlantic in-flight internet access via airborne
  mesh networking,'' \emph{IEEE/ACM Transactions on Networking}, vol.~20,
  no.~4, pp. 1231--1244, August 2012,
  \url{https://doi.org/10.1109/TNET.2011.2175487}.

\bibitem{8761882}
S.~{Hofmann}, V.~{Megas}, M.~{Ozger}, D.~{Schupke}, F.~H.~P. {Fitzek}, and
  C.~{Cavdar}, ``Combined optimal topology formation and rate allocation for
  aircraft to aircraft communications,'' in \emph{ICC 2019 - 2019 IEEE
  International Conference on Communications (ICC)}, 2019, pp. 1--6,
  \url{https://doi.org/10.1109/ICC.2019.8761882}.

\bibitem{O6-18}
{Nguyen Thi Xuan My}, {Yoshikazu Miyanaga}, and {Chaiyachet Saivichit},
  ``Connectivity analytical modelling for a single flight path ad hoc
  aeronautical network,'' in \emph{ECTI International Confernce on Electrical
  Engineering/Electronics, Computer, Telecommunications and Information
  Technology (ECTI-CON)}, 2010, pp. 51--55.

\bibitem{O6}
H.~Zhang, X.~Chen, B.~Zheng, and Y.~Wang, ``Analysis of connectivity
  requirement for aeronautical ad hoc networks,'' in \emph{International
  Conference on Electronic \& Mechanical Engineering and Information
  Technology}, Harbin, Heilongjiang, 2011, pp. 3943--3946,
  \url{https://doi.org/10.1109/EMEIT.2011.6023090}.

\bibitem{9081678}
M.~A. {Bellido-Manganell} and M.~{Schnell}, ``Towards modern air-to-air
  communications: the {LDACS} {A2A} mode,'' in \emph{2019 IEEE/AIAA 38th
  Digital Avionics Systems Conference (DASC)}, 2019, pp. 1--10,
  \url{https://doi.org/10.1109/DASC43569.2019.9081678}.

\bibitem{8932658}
B.~{Du}, R.~{Xue}, L.~{Zhao}, and V.~C.~M. {Leung}, ``Coalitional graph game
  for air-to-air and air-to-ground cognitive spectrum sharing,'' \emph{IEEE
  Transactions on Aerospace and Electronic Systems}, vol.~56, no.~4, pp.
  2959--2977, 2020, \url{https://doi.org/10.1109/TAES.2019.2958162}.

\bibitem{O2-9}
G.~Aggelou, \emph{Mobile Ad Hoc Networks: From Wireless {LAN}s to {4G}
  Networks}.\hskip 1em plus 0.5em minus 0.4em\relax New York: McGraw-Hill
  Professional, 2004, iSBN: 978-0071413053.

\bibitem{O3}
Q.~Vey, ``Access and routing in aeronautical ad-hoc networks,'' \emph{16e
  Congrès Des Doctorants EDSYS (École Doctorale Systèmes), École Doctorale
  Systèmes}, May 2015,
  \url{https://hal-enac.archives-ouvertes.fr/hal-01355743}.

\bibitem{O3-14}
D.~Niculescu and B.~Nath, ``Trajectory based forwarding and its applications,''
  in \emph{Annual International Conference on Mobile Computing and Networking,
  (MobiCom)}, New York, NY, USA, 2003, p. 260–272,
  \url{https://doi.org/10.1145/938985.939012}.

\bibitem{O5}
Q.~Vey, A.~Pirovano, J.~Radzik, and F.~Garcia, ``Aeronautical ad hoc network
  for civil aviation,'' in \emph{Communication Technologies for
  Vehicles}.\hskip 1em plus 0.5em minus 0.4em\relax Springer International
  Publishing, 05 2014, \url{https://doi.org/10.1007/978-3-319-06644-8_8}.

\bibitem{MPMBSACAAS}
R.~J. Kumpfbeck, J.~F. Pedersen, and J.~T. Merenda, ``Minimum protrusion
  mechanically beam steered aircraft array antenna systems,'' Tech. Rep., July
  2001, uS Patent, 6,259,415. [Online]. Available
  at:\url{https://portal.unifiedpatents.com/patents/patent/US-6259415-B1}
  [Accessed on 16/01/2021].

\bibitem{8739859}
A.~{Reyna}, M.~A. {Panduro}, A.~{Mendez}, L.~{Balderas}, and C.~{Del-Río},
  ``Distributed antenna array for {FANET}’s wireless links using time
  modulation,'' in \emph{2019 13th European Conference on Antennas and
  Propagation (EuCAP)}, 2019, pp. 1--3.

\bibitem{8737798}
M.~A. {Khan}, A.~{Khalid}, and F.~{Khanzada}, ``Dual-radio dual-band
  configuration for flexible communication in flying ad-hoc network
  ({FANET}),'' in \emph{2019 International Conference on Communication
  Technologies (ComTech)}, 2019, pp. 108--113,
  \url{https://doi.org/10.1109/COMTECH.2019.8737798}.

\bibitem{8676034}
B.~{Zeng}, T.~{Song}, and J.~{An}, ``A dual-antenna collaborative communication
  strategy for flying ad hoc networks,'' \emph{IEEE Communications Letters},
  vol.~23, no.~5, pp. 913--917, 2019,
  \url{https://doi.org/10.1109/LCOMM.2019.2908153}.

\bibitem{6006004}
C.~{Lin}, H.~T. {Kung}, T.~{Lin}, S.~J. {Tarsa}, and D.~{Vlah}, ``Achieving
  high throughput ground-to-uav transport via parallel links,'' in \emph{2011
  Proceedings of 20th International Conference on Computer Communications and
  Networks (ICCCN)}, 2011, pp. 1--7,
  \url{https://doi.org/10.1109/ICCCN.2011.6006004}.

\bibitem{8923123}
J.~{Gueldenring}, L.~{Koring}, P.~{Gorczak}, and C.~{Wietfeld}, ``Heterogeneous
  multilink aggregation for reliable {UAV} communication in maritime search and
  rescue missions,'' in \emph{2019 International Conference on Wireless and
  Mobile Computing, Networking and Communications (WiMob)}, 2019, pp. 215--220,
  \url{https://doi.org/10.1109/WiMOB.2019.8923123}.

\bibitem{ean_dt_inmarsat_nokia}
EAN, ``European aviation network - online press conference,'' 2018, [Online].
  Available
  at:\url{https://www.telekom.com/resource/blob/514350/a230e29da337c92fe8efa2261b4ce41b/dl-press-conference-ean-5-feb-18-data.pdf}
  [Accessed on 16/01/2021].

\bibitem{8700598}
S.~R. {Pokhrel}, J.~{Jin}, and H.~L. {Vu}, ``Mobility-aware multipath
  communication for unmanned aerial surveillance systems,'' \emph{IEEE
  Transactions on Vehicular Technology}, vol.~68, no.~6, pp. 6088--6098, 2019,
  \url{https://doi.org/10.1109/TVT.2019.2912851}.

\bibitem{8735265}
A.~{Volkert}, H.~{Hackbarth}, T.~J. {Lieb}, and S.~{Kern}, ``Flight tests of
  ranges and latencies of a threefold redundant {C2} multi-link solution for
  small drones in {VLL} airspace,'' in \emph{2019 Integrated Communications,
  Navigation and Surveillance Conference (ICNS)}, 2019, pp. 1--14,
  \url{https://doi.org/10.1109/ICNSURV.2019.8735265}.

\bibitem{9162943}
P.~{Saxena}, T.~{Dreibholz}, H.~{Skinnemoen}, O.~{Alay}, M.~A.
  {Vazquez-Castro}, S.~{Ferlin}, and G.~{Acar}, ``Resilient hybrid satcom and
  terrestrial networking for unmanned aerial vehicles,'' in \emph{IEEE INFOCOM
  2020 - IEEE Conference on Computer Communications Workshops (INFOCOM
  WKSHPS)}, 2020, pp. 418--423,
  \url{https://doi.org/10.1109/INFOCOMWKSHPS50562.2020.9162943}.

\bibitem{8010762}
E.~{Dinc}, M.~{Vondra}, S.~{Hofmann}, D.~{Schupke}, M.~{Prytz}, S.~{Bovelli},
  M.~{Frodigh}, J.~{Zander}, and C.~{Cavdar}, ``In-flight broadband
  connectivity: Architectures and business models for high capacity
  air-to-ground communications,'' \emph{IEEE Communications Magazine}, vol.~55,
  no.~9, pp. 142--149, 2017, \url{https://doi.org/10.1109/MCOM.2017.1601181}.

\bibitem{7073483}
S.~J. {Nawaz}, N.~M. {Khan}, M.~I. {Tiwana}, N.~{Hassan}, and S.~I. {Shah},
  ``Airborne internet access through submarine optical fiber cables,''
  \emph{IEEE Transactions on Aerospace and Electronic Systems}, vol.~51, no.~1,
  pp. 167--177, 2015, \url{https://doi.org/10.1109/TAES.2014.130416}.

\bibitem{8552136}
A.~{Numani}, S.~J. {Nawaz}, and M.~A. {Javed}, ``Architecture and routing
  protocols for airborne internet access,'' in \emph{2018 IEEE International
  Conference on Consumer Electronics - Asia (ICCE-Asia)}, 2018, pp. 206--212,
  \url{https://doi.org/10.1109/ICCE-ASIA.2018.8552136}.

\bibitem{5172851}
S.~{Ayaz}, C.~{Bauer}, C.~{Kissling}, F.~{Schreckenbach}, F.~{Arnal},
  C.~{Baudoin}, K.~{Leconte}, M.~{Ehammer}, and T.~{Graeupl}, ``Architecture of
  an {IP}-based aeronautical network,'' in \emph{2009 Integrated
  Communications, Navigation and Surveillance Conference}, 2009, pp. 1--9,
  \url{https://doi.org/10.1109/ICNSURV.2009.5172851}.

\bibitem{8569862}
Y.~{Hu}, K.~{Abdo}, F.~{BenSlama}, M.~{Ali}, Q.~{Cormbe}, F.~{Benamrane},
  D.~{Luong}, and R.~{Barossi}, ``A {SDN}-based aeronautical communications
  network architecture,'' in \emph{2018 IEEE/AIAA 37th Digital Avionics Systems
  Conference (DASC)}, 2018, pp. 1--10,
  \url{https://doi.org/10.1109/DASC.2018.8569862}.

\bibitem{8569540}
C.~{Cavdar}, D.~{Gera}, S.~{Hofmann}, D.~{Schupke}, A.~{Ghosh}, and
  A.~{Nordlöw}, ``Demonstration of an integrated {5G} network in an aircraft
  cabin environment,'' in \emph{2018 IEEE/AIAA 37th Digital Avionics Systems
  Conference (DASC)}, 2018, pp. 1--10,
  \url{https://doi.org/10.1109/DASC.2018.8569540}.

\bibitem{9142706}
L.~{Tomaszewski}, R.~{Kołakowski}, and S.~{Kukliński}, ``Integration of
  {U-Space} and {5GS} for {UAV} services,'' in \emph{2020 IFIP Networking
  Conference (Networking)}, 2020, pp. 767--772.

\bibitem{9076122}
O.~{Bekkouche}, K.~{Samdanis}, M.~{Bagaa}, and T.~{Taleb}, ``A service-based
  architecture for enabling {UAV} enhanced network services,'' \emph{IEEE
  Network}, vol.~34, no.~4, pp. 328--335, 2020,
  \url{https://doi.org/10.1109/MNET.001.1900556}.

\bibitem{7763289}
T.~C. {Hong}, K.~{Kang}, K.~{Lim}, and J.~Y. {Ahn}, ``Network architecture for
  control and non-payload communication of {UAV},'' in \emph{2016 International
  Conference on Information and Communication Technology Convergence (ICTC)},
  2016, pp. 762--764, \url{https://doi.org/10.1109/ICTC.2016.7763289}.

\bibitem{8550873}
N.~{Vanitha} and G.~{Padmavathi}, ``A comparative study on communication
  architecture of unmanned aerial vehicles and security analysis of false data
  dissemination attacks,'' in \emph{2018 International Conference on Current
  Trends Towards Converging Technologies (ICCTCT)}, 2018, pp. 1--8,
  \url{https://doi.org/10.1109/ICCTCT.2018.8550873}.

\bibitem{6825193}
J.~{Li}, Y.~{Zhou}, and L.~{Lamont}, ``Communication architectures and
  protocols for networking unmanned aerial vehicles,'' in \emph{2013 IEEE
  Globecom Workshops (GC Wkshps)}, 2013, pp. 1415--1420,
  \url{https://doi.org/10.1109/GLOCOMW.2013.6825193}.

\bibitem{8277614}
M.~A. {Khan}, A.~{Safi}, I.~M. {Qureshi}, and I.~U. {Khan}, ``Flying ad-hoc
  networks ({FANET}s): A review of communication architectures, and routing
  protocols,'' in \emph{2017 First International Conference on Latest Trends in
  Electrical Engineering and Computing Technologies (INTELLECT)}, 2017, pp.
  1--9, \url{https://doi.org/10.1109/INTELLECT.2017.8277614}.

\bibitem{9199627}
S.~{Al-Emadi} and A.~{Al-Mohannadi}, ``Towards enhancement of network
  communication architectures and routing protocols for {FANET}s: A survey,''
  in \emph{2020 3rd International Conference on Advanced Communication
  Technologies and Networking (CommNet)}, 2020, pp. 1--10,
  \url{https://doi.org/10.1109/CommNet49926.2020.9199627}.

\bibitem{8599015}
Q.~{Zhang}, M.~{Jiang}, Z.~{Feng}, W.~{Li}, W.~{Zhang}, and M.~{Pan}, ``{IoT}
  enabled {UAV}: Network architecture and routing algorithm,'' \emph{IEEE
  Internet of Things Journal}, vol.~6, no.~2, pp. 3727--3742, 2019,
  \url{https://doi.org/10.1109/JIOT.2018.2890428}.

\bibitem{9184022}
S.~{Aggarwal}, N.~{Kumar}, and S.~{Tanwar}, ``Blockchain envisioned {UAV}
  communication using {6G} networks: Open issues, use cases, and future
  directions,'' \emph{IEEE Internet of Things Journal}, pp. 1--1, 2020,
  \url{https://doi.org/10.1109/JIOT.2020.3020819}.

\bibitem{9213109}
H.~{Qu}, X.~{Xu}, J.~{Zhao}, and P.~{Yue}, ``An {SDN}-based space-air-ground
  integrated network architecture and controller deployment strategy,'' in
  \emph{2020 IEEE 3rd International Conference on Computer and Communication
  Engineering Technology (CCET)}, 2020, pp. 138--142,
  \url{https://doi.org/10.1109/CCET50901.2020.9213109}.

\bibitem{3GPP_23501}
3GPP, ``System architecture for the {5G} system ({5GS}); stage 2 (release 16),
  v16.6.0,'' 650 Route Des Lucioles - Sophia Antipolis Valbonne - FRANCE, Tech.
  Rep. TS 23.501, Sep. 2020, [Online]. Available
  at:\url{https://www.3gpp.org/ftp//specs/archive/23_series/23.501/} [Accessed
  on 16/01/2021].

\bibitem{avenssim}
E.~Marconato, M.~Rodrigues, R.~Pires, D.~Pigatto, L.~Q. Filho, A.~Pinto, and
  K.~C. Branco, ``{AVENS} – a novel flying ad hoc network simulator with
  automatic code generation for unmanned aircraft system,'' in
  \emph{Proceedings of the 50th Hawaii International Conference on System
  Sciences}, 01 2017, \url{https://doi.org/10.24251/HICSS.2017.760}.

\bibitem{6825196}
A.~Y. {Javaid}, W.~{Sun}, and M.~{Alam}, ``{UAVSim}: A simulation testbed for
  unmanned aerial vehicle network cyber security analysis,'' in \emph{2013 IEEE
  Globecom Workshops (GC Wkshps)}, 2013, pp. 1432--1436,
  \url{https://doi.org/10.1109/GLOCOMW.2013.6825196}.

\bibitem{baidya2018flynetsim}
S.~Baidya, Z.~Shaikh, and M.~Levorato, ``{FlyNetSim}: An open source
  synchronized {UAV} network simulator based on ns-3 and ardupilot,'' in
  \emph{MSWIM '18}.\hskip 1em plus 0.5em minus 0.4em\relax New York, NY, USA:
  Association for Computing Machinery, 2018, p. 37–45,
  \url{https://doi.org/10.1145/3242102.3242118}.

\bibitem{airsim2017fsr}
S.~Shah, D.~Dey, C.~Lovett, and A.~Kapoor, ``Airsim: High-fidelity visual and
  physical simulation for autonomous vehicles,'' in \emph{Field and Service
  Robotics}, 2017, \url{https://arxiv.org/abs/1705.05065}.

\bibitem{Guerra_2019}
W.~Guerra, E.~Tal, V.~Murali, G.~Ryou, and S.~Karaman, ``Flightgoggles:
  Photorealistic sensor simulation for perception-driven robotics using
  photogrammetry and virtual reality,'' \emph{2019 IEEE/RSJ International
  Conference on Intelligent Robots and Systems (IROS)}, Nov 2019,
  \url{http://dx.doi.org/10.1109/IROS40897.2019.8968116}.

\bibitem{8880644}
V.~{Sanchez-Aguero}, F.~{Valera}, B.~{Nogales}, L.~F. {Gonzalez}, and
  I.~{Vidal}, ``{VENUE}: Virtualized environment for multi-{UAV} network
  emulation,'' \emph{IEEE Access}, vol.~7, pp. 154\,659--154\,671, 2019,
  \url{https://doi.org/10.1109/ACCESS.2019.2949119}.

\bibitem{10.1177/1756829319837668}
J.~Modares, N.~Mastronarde, and K.~Dantu, ``Simulating unmanned aerial vehicle
  swarms with the {UB-ANC} emulator,'' \emph{International Journal of Micro Air
  Vehicles}, vol.~11, p. 1756829319837668, 2019,
  \url{https://doi.org/10.1177/1756829319837668}.

\bibitem{interusssim}
B.~{Pelletier}, ``{InterUSS} platform,'' 2016, [Online]. Available
  at:\url{https://github.com/interuss} [Accessed on 26/02/2021].

\bibitem{blueskysim}
J.~Hoekstra and J.~Ellerbroek, ``Bluesky {ATC} simulator project: An open data
  and open source approach,'' in \emph{7th International Conference for
  Research on Air Transportation}, 06 2016,
  \url{https://repository.tudelft.nl//islandora/object/uuid:d1131a90-f0ea-4489-a217-ad29987689a1}.

\bibitem{elsasim}
G.~{Gurtner} and C.~{Bongiorno}, ``{ELSA} air traffic simulator,'' [Online].
  Available at:\url{https://github.com/ELSA-Project/ELSA-ABM} [Accessed on
  16/01/2021].

\bibitem{euroscopesim}
E.~Bocaneanu, P.~Selmeci, I.~Rado, A.~Abraham, A.~Orban, D.~Vertesy, I.~Nagy,
  T.~Atanasov, R.~Carlson, B.~Supnik, R.~Stefan, B.~Candela, S.~Boerner,
  S.~Ylismaeki, J.~Holopainen, T.~Reimann, C.~Phillips, A.~Bocaneanu,
  J.~Bencsik, A.~Grama, G.~Lauenstein, and Z.~Daniel, ``Euroscope,'' [Online].
  Available at:\url{https://www.euroscope.hu/wp/} [Accessed on 16/01/2021].

\bibitem{openscopesim}
openScope, ``openscope air traffic control simulator,'' [Online]. Available
  at:\url{https://github.com/openscope/openscope} [Accessed on 16/01/2021].

\bibitem{airsim_link}
AirSim, ``Welcome to airsim,'' 2018, [Online]. Available
  at:\url{https://microsoft.github.io/AirSim/} [Accessed on 03/02/2021].

\bibitem{cdssim}
A.~Mohini, ``{CDSSim} - multi {UAV} communication and control simulation
  framework,'' Master's thesis, University of Cincinnati, 3 2019,
  \url{https://etd.ohiolink.edu/apexprod/rws_etd/send_file/send?accession=ucin1554373574457271&disposition=inline}.

\bibitem{7983185}
N.~R. {Zema}, A.~{Trotta}, G.~{Sanahuja}, E.~{Natalizio}, M.~{Di Felice}, and
  L.~{Bononi}, ``{CUSCUS}: Communications-control distributed simulator,'' in
  \emph{2017 14th IEEE Annual Consumer Communications Networking Conference
  (CCNC)}, 2017, pp. 601--602, \url{https://doi.org/10.1109/CCNC.2017.7983185}.

\bibitem{cuscus_link}
N.~R. {Zema}, ``Communication-control distributed simulator (cuscus),'' 2018,
  [Online]. Available
  at:\url{https://gitlab.utc.fr/zemanico/CUSCUS/-/wikis/home} [Accessed on
  03/02/2021].

\bibitem{8942250}
V.~T. {Nguyen}, K.~{Jung}, and T.~{Dang}, ``{DroneVR}: A web virtual reality
  simulator for drone operator,'' in \emph{2019 IEEE International Conference
  on Artificial Intelligence and Virtual Reality (AIVR)}, 2019, pp. 257--2575,
  \url{https://doi.org/10.1109/AIVR46125.2019.00060}.

\bibitem{flairsim}
Laboratoire\;Heudiasyc, UTC, Institut\;Des\;Sciences, and
  De\;L'information\;Et\;De\;Leurs\;Interactions, ``Software flair,'' [Online].
  Available at:\url{https://UAV.hds.utc.fr/software-flair/} [Accessed on
  16/01/2021].

\bibitem{flairsim_userguide}
R.~{Chennouf} and S.~{Lamrous}, ``Fl-air framework user's guide,'' 2017,
  [Online]. Available
  at:\url{https://devel.hds.utc.fr/software/flair/raw-attachment/wiki/documentation/flair_tx17.pdf}
  [Accessed on 03/02/2021].

\bibitem{hectorquadratorsim}
J.~{Meyer}, A.~{Sendobry}, S.~{Kohlbrecher}, U.~{Klingauf}, and O.~{Von Stryk},
  ``Comprehensive simulation of quadrotor {UAVs} using {ROS} and gazebo,'' vol.
  7628, 11 2012, pp. 400--411,
  \url{https://doi.org/10.1007/978-3-642-34327-8_36}.

\bibitem{8880832}
G.~{Grieco}, R.~{Artuso}, P.~{Boccadoro}, G.~{Piro}, and L.~A. {Grieco}, ``An
  open source and system-level simulator for the internet of drones,'' in
  \emph{2019 IEEE 30th International Symposium on Personal, Indoor and Mobile
  Radio Communications (PIMRC Workshops)}, 2019, pp. 1--6,
  \url{https://doi.org/10.1109/PIMRCW.2019.8880832}.

\bibitem{jmavsim_link}
A.~{Babushkin} and J.~{Oes}, ``jmavsim - simple multirotor simulator with
  mavlink protocol support,'' 2015, [Online]. Available
  at:\url{https://github.com/DrTon/jMAVSim} [Accessed on 16/01/2021].

\bibitem{jmavsim_multivehiclesim}
JMAVSim, ``Multi-vehicle simulation with jmavsim,'' 2020, [Online]. Available
  at:\url{https://docs.px4.io/master/en/simulation/multi_vehicle_jmavsim.html}
  [Accessed on 03/02/2021].

\bibitem{jmavsim_hilsim}
------, ``Hardware in the loop simulation (hitl),'' 2020, [Online]. Available
  at:\url{https://docs.px4.io/master/en/simulation/hitl.html} [Accessed on
  03/02/2021].

\bibitem{Sliwa_2019}
B.~Sliwa, M.~Patchou, and C.~Wietfeld, ``Lightweight simulation of hybrid
  aerial- and ground-based vehicular communication networks,'' \emph{2019 IEEE
  90th Vehicular Technology Conference (VTC2019-Fall)}, Sep 2019,
  \url{http://dx.doi.org/10.1109/VTCFall.2019.8891340}.

\bibitem{limosim_link}
B.~{Zoltan}, B.~{Sliwa}, and A.~{Varga}, ``Limosim: Lightweight ict-centric
  mobility simulation,'' 2017, [Online]. Available
  at:\url{https://github.com/BenSliwa/LIMoSim_ns3} [Accessed on 16/01/2021].

\bibitem{8574594}
B.~{Boroujerdian}, H.~{Genc}, S.~{Krishnan}, W.~{Cui}, A.~{Faust}, and
  V.~{Reddi}, ``{MAVBench}: Micro aerial vehicle benchmarking,'' in \emph{2018
  51st Annual IEEE/ACM International Symposium on Microarchitecture (MICRO)},
  2018, pp. 894--907, \url{https://doi.org/10.1109/MICRO.2018.00077}.

\bibitem{multiuavsimulation_link}
T.~{Dietrich} and M.~{Sommer}, ``multiuav simulation in {OMNeT++},'' 2016,
  [Online]. Available
  at:\url{https://github.com/ThomDietrich/multiUAV-simulation} [Accessed on
  16/01/2021].

\bibitem{7918926}
S.~{Habib}, M.~{Malik}, S.~{Ur Rahman}, and M.~A. {Raja}, ``Nuav - a testbed
  for developing autonomous unmanned aerial vehicles,'' in \emph{2017
  International Conference on Communication, Computing and Digital Systems
  (C-CODE)}, 2017, pp. 185--192,
  \url{https://doi.org/10.1109/C-CODE.2017.7918926}.

\bibitem{nuav_link}
A.~{Raja}, ``Nuav,'' 2017, [Online]. Available
  at:\url{https://gitlab.com/adilraja/NUAV} [Accessed on 16/01/2021].

\bibitem{bhagat2020uav}
S.~{Bhagat} and P.~B. {Sujit}, ``{UAV} target tracking in urban environments
  using deep reinforcement learning,'' in \emph{2020 International Conference
  on Unmanned Aircraft Systems (ICUAS)}, 2020, pp. 694--701,
  \url{https://doi.org/10.1109/ICUAS48674.2020.9213856}.

\bibitem{obstacle_avoidance_for_uav_link}
S.~{Bhagat} and S.~{Uppal}, ``Obstacle avoidance simulator for unmanned aerial
  vehicles ({UAVs}),'' 2018, [Online]. Available
  at:\url{https://github.com/sarthak268/Obstacle_Avoidance_for_UAV} [Accessed
  on 16/01/2021].

\bibitem{openamasesim}
M.~Duquette, D.~Kingston, J.~Richardson, and B.~Hocking, ``About amase,''
  [Online]. Available
  at:\url{https://github.com/afrl-rq/OpenAMASE/wiki/About-AMASE} [Accessed on
  16/01/2021].

\bibitem{8443728}
M.~{Schmittle}, A.~{Lukina}, L.~{Vacek}, J.~{Das}, C.~P. {Buskirk}, S.~{Rees},
  J.~{Sztipanovits}, R.~{Grosu}, and V.~{Kumar}, ``{OpenUAV}: A {UAV} testbed
  for the {CPS} and robotics community,'' in \emph{2018 ACM/IEEE 9th
  International Conference on Cyber-Physical Systems (ICCPS)}, 2018, pp.
  130--139, \url{https://doi.org/10.1109/ICCPS.2018.00021}.

\bibitem{openuav_link}
A.~J. {Poruthukaran}, S.~{Dcunha}, H.~{Anand}, and M.~{Schmittle}, ``Openuav
  project,'' [Online]. Available at:\url{https://github.com/Open-UAV} [Accessed
  on 16/01/2021].

\bibitem{boccadoro2019plane}
P.~Boccadoro and A.~Cardellicchio, ``{PLANE}: An extensible open source
  framework for modeling the internet of drones,'' 2019,
  \url{https://arxiv.org/abs/1906.08085}.

\bibitem{rosquadratorsim}
W.~Selby, ``Ros integration,'' [Online]. Available
  at:\url{https://www.wilselby.com/research/ros-integration/} [Accessed on
  16/01/2021].

\bibitem{rotorssim}
F.~Furrer, M.~Burri, M.~Achtelik, and R.~Siegwart, ``{RotorS} – a modular
  gazebo mav simulator framework,'' \emph{Studies in Computational
  Intelligence}, vol. 625, pp. 595--625, 01 2016,
  \url{https://doi.org/10.1007/978-3-319-26054-9_23}.

\bibitem{6920935}
J.~{Waterman}, B.~{Kate}, K.~{Dantu}, and M.~{Welsh}, ``Demo abstract:
  {Simbeeotic}: A simulation-emulation platform for large scale micro-aerial
  swarms,'' in \emph{2012 ACM/IEEE 11th International Conference on Information
  Processing in Sensor Networks (IPSN)}, 2012, pp. 139--140,
  \url{https://doi.org/10.1109/IPSN.2012.6920935}.

\bibitem{soria2020swarmlab}
E.~Soria, F.~Schiano, and D.~Floreano, ``Swarmlab: a matlab drone swarm
  simulator,'' in \emph{2017 Annual IEEE International Systems Conference
  (SysCon)}, 2020, [Online]. Available
  at:\url{https://arxiv.org/pdf/2005.02769.pdf} [Accessed on 04/02/2021].

\bibitem{10.1007/978-3-319-46448-0_27}
M.~Mueller, N.~Smith, and B.~Ghanem, ``A benchmark and simulator for {UAV}
  tracking,'' in \emph{Computer Vision - ECCV 2016}, B.~Leibe, J.~Matas,
  N.~Sebe, and M.~Welling, Eds.\hskip 1em plus 0.5em minus 0.4em\relax Springer
  International Publishing, 2016, pp. 445--461, iSBN: 978-3-319-46448-0.

\bibitem{8607252}
M.~{Theile}, O.~D. {Dantsker}, R.~{Nai}, and M.~{Caccamo}, ``{uavEE}: A
  modular, power-aware emulation environment for rapid prototyping and testing
  of {UAVs},'' in \emph{2018 IEEE 24th International Conference on Embedded and
  Real-Time Computing Systems and Applications (RTCSA)}, 2018, pp. 217--224,
  \url{https://doi.org/10.1109/RTCSA.2018.00034}.

\bibitem{crrcsim}
CRRCSim, ``Crrcsim flight simulator,'' 2008, [Online]. Available
  at:\url{http://wiki.flightgear.org/CRRCSim} [Accessed on 16/01/2021].

\bibitem{flightgearsim}
J.~{Berndt}, F.~{Ledbury}, J.~{Turner}, and J.~{Pitman}, ``Flightgear flight
  simulator,'' 2010, [Online]. Available
  at:\url{https://home.flightgear.org/about/} [Accessed on 16/01/2021].

\bibitem{geofssim}
GeoFS, ``The accessible flight simulator,'' [Online]. Available
  at:\url{https://www.geo-fs.com/} [Accessed on 16/01/2021].

\bibitem{MLPredODUAVWC}
Q.~{Zhang}, M.~{Mozaffari}, W.~{Saad}, M.~{Bennis}, and M.~{Debbah}, ``Machine
  learning for predictive on-demand deployment of uavs for wireless
  communications,'' in \emph{2018 IEEE Global Communications Conference
  (GLOBECOM)}, 2018, pp. 1--6.

\bibitem{LRMUAVWCSRA}
J.~{Chen}, U.~{Yatnalli}, and D.~{Gesbert}, ``Learning radio maps for uav-aided
  wireless networks: A segmented regression approach,'' in \emph{2017 IEEE
  International Conference on Communications (ICC)}, 2017, pp. 1--6.

\bibitem{TPUAVSCRNN}
K.~{Xiao}, J.~{Zhao}, Y.~{He}, and S.~{Yu}, ``Trajectory prediction of uav in
  smart city using recurrent neural networks,'' in \emph{ICC 2019 - 2019 IEEE
  International Conference on Communications (ICC)}, 2019, pp. 1--6.

\bibitem{AtAPLPMLUE}
\BIBentryALTinterwordspacing
Y.~Zhang, J.~Wen, G.~Yang, Z.~He, X.~Luo, and L.~Liu, ``Air-to-air path loss
  prediction based on machine learning methods in urban environments,''
  \emph{Wirel. Commun. Mob. Comput.}, vol. 2018, Jan. 2018. [Online].
  Available: \url{https://doi.org/10.1155/2018/8489326}
\BIBentrySTDinterwordspacing

\bibitem{GNNCMmmWUAVC}
W.~Xia, S.~Rangan, M.~Mezzavillla, A.~Lozano, G.~Geraci, V.~Semkin, and
  G.~Loianno, ``Generative neural network channel modeling for millimeter-wave
  uav communication,'' 2020.

\bibitem{ANUAVRTRL}
N.~{Imanberdiyev}, C.~{Fu}, E.~{Kayacan}, and I.~{Chen}, ``Autonomous
  navigation of uav by using real-time model-based reinforcement learning,'' in
  \emph{2016 14th International Conference on Control, Automation, Robotics and
  Vision (ICARCV)}, 2016, pp. 1--6.

\bibitem{AUAVNDDPGDRL}
O.~Bouhamed, H.~Ghazzai, H.~Besbes, and Y.~Massoud, ``Autonomous uav
  navigation: A ddpg-based deep reinforcement learning approach,'' 2020.

\bibitem{QLAQoEUAVCN}
\BIBentryALTinterwordspacing
S.~Colonnese, F.~Cuomo, G.~Pagliari, and L.~Chiaraviglio, ``Q-square: A
  q-learning approach to provide a qoe aware uav flight path in cellular
  networks,'' \emph{Ad Hoc Networks}, vol.~91, p. 101872, 2019. [Online].
  Available:
  \url{https://www.sciencedirect.com/science/article/pii/S1570870518309508}
\BIBentrySTDinterwordspacing

\bibitem{MADRLTPUAVMEC}
L.~{Wang}, K.~{Wang}, C.~{Pan}, W.~{Xu}, N.~{Aslam}, and L.~{Hanzo},
  ``Multi-agent deep reinforcement learning-based trajectory planning for
  multi-uav assisted mobile edge computing,'' \emph{IEEE Transactions on
  Cognitive Communications and Networking}, vol.~7, no.~1, pp. 73--84, 2021.

\bibitem{3GPP_21916}
3GPP, ``Summary of rel-16 work items (release 16),'' techreport TR 21.916,
  2021, [Online]. Available
  at:\url{https://www.3gpp.org/ftp/Specs/archive/21_series/21.916/} [Accessed
  on 29/07/2021].

\bibitem{ITU_spectrumtypes}
N.~Vassiliev\;-\;ITU, ``Potential spectrum and telecom technologies for small
  {UAS},'' 2017, [Online]. Available
  at:\url{https://www.icao.int/Meetings/UAS2017/Documents/Nikolai%20Vassiliev_Background_Day%201.pdf}
  [Accessed on 16/01/2021].

\bibitem{GSMA_spectrum}
GSMA, ``Mobile spectrum for unmanned aerial vehicles {GSMA} public policy
  position,'' Tech. Rep., Oct. 2017, [Online]. Available
  at:\url{https://www.gsma.com/spectrum/wp-content/uploads/2018/12/Mobile-spectrum-for-Unmanned-Aerial-Vehicles.pdf}
  [Accessed on 16/01/2021].

\bibitem{9174931}
A.~A. Al-Habob, O.~A. Dobre, S.~Muhaidat, and H.~Vincent~Poor,
  ``Energy-efficient data dissemination using a uav: An ant colony approach,''
  \emph{IEEE Wireless Communications Letters}, vol.~10, no.~1, pp. 16--20,
  2021, \url{https://doi.org/10.1109/LWC.2020.3019001}.

\bibitem{9373692}
R.~Zhang, R.~Lu, X.~Cheng, N.~Wang, and L.~Yang, ``A uav-enabled data
  dissemination protocol with proactive caching and file sharing in v2x
  networks,'' \emph{IEEE Transactions on Communications}, pp. 1--1, 2021,
  \url{https://doi.org/10.1109/TCOMM.2021.3064569}.

\bibitem{9045425}
N.~Bashir and S.~Boudjit, ``An energy-efficient collaborative scheme for uavs
  and vanets for dissemination of real-time surveillance data on highways,'' in
  \emph{2020 IEEE 17th Annual Consumer Communications and Networking Conference
  (CCNC)}, 2020, pp. 1--6,
  \url{https://doi.org/10.1109/CCNC46108.2020.9045425}.

\bibitem{9417539}
Z.~Mohamed and S.~Aïssa, ``Resource allocation for energy-efficient cellular
  communications via aerial irs,'' in \emph{2021 IEEE Wireless Communications
  and Networking Conference (WCNC)}, 2021, pp. 1--6,
  \url{https://doi.org/10.1109/WCNC49053.2021.9417539}.

\bibitem{9417458}
X.~Xi, X.~Cao, P.~Yang, J.~Chen, and D.~wu, ``Energy-efficient resource
  allocation in a multi-uav-aided noma network,'' in \emph{2021 IEEE Wireless
  Communications and Networking Conference (WCNC)}, 2021, pp. 1--7,
  \url{https://doi.org/10.1109/WCNC49053.2021.9417458}.

\bibitem{9402734}
S.~Ahmed, M.~Z. Chowdhury, S.~R. Sabuj, M.~I. Alam, and Y.~M. Jang,
  ``Energy-efficient uav relaying robust resource allocation in uncertain
  adversarial networks,'' \emph{IEEE Access}, vol.~9, pp. 59\,920--59\,934,
  2021, \url{https://doi.org/10.1109/ACCESS.2021.3073015}.

\bibitem{9348068}
S.~Najmeddin, S.~Aïssa, and S.~Tahar, ``Energy-efficient resource allocation
  for uav-enabled information and power transfer with noma,'' in \emph{GLOBECOM
  2020 - 2020 IEEE Global Communications Conference}, 2020, pp. 1--6,
  \url{https://doi.org/10.1109/GLOBECOM42002.2020.9348068}.

\bibitem{9289282}
M.~Golam, J.-M. Lee, and D.-S. Kim, ``A uav-assisted blockchain based secure
  device-to-device communication in internet of military things,'' in
  \emph{2020 International Conference on Information and Communication
  Technology Convergence (ICTC)}, 2020, pp. 1896--1898,
  \url{https://doi.org/10.1109/ICTC49870.2020.9289282}.

\bibitem{9364745}
P.~X. Nguyen, V.-D. Nguyen, H.~V. Nguyen, and O.-S. Shin, ``Uav-assisted secure
  communications in terrestrial cognitive radio networks: Joint power control
  and 3d trajectory optimization,'' \emph{IEEE Transactions on Vehicular
  Technology}, vol.~70, no.~4, pp. 3298--3313, 2021,
  \url{https://doi.org/10.1109/TVT.2021.3062283}.

\end{thebibliography}

\end{document}